\documentclass[citeauthoryear,runningheads]{svmult}

\usepackage{makeidx}   
\usepackage{graphicx}  
\usepackage{subeqnar}  
\usepackage{multicol}  
\usepackage{physmubb}  
\makeindex             

\newcommand{\bea}{\begin{eqnarray}}
\newcommand{\eea}{\end{eqnarray}}
\newcommand{\be}{\begin{equation}}
\newcommand{\ee}{\end{equation}}
\newcommand{\ben}{\begin{enumerate}}
\newcommand{\een}{\end{enumerate}}
\newcommand{\bi}{\begin{itemize}}
\newcommand{\ei}{\end{itemize}}
\newcommand{\bc}{\begin{center}}
\newcommand{\ec}{\end{center}}

\def\llabel#1{\label{sc:#1}}
\def\elabel#1{\label{eq:#1}}
\def\flabel#1{\label{fig:#1}}
\def\NOTE#1{{ }}

\def\eck#1{\left\lbrack #1 \right\rbrack}

\def\rund#1{\left( #1 \right)}
\def\abs#1{\left\vert #1 \right\vert}

\def\ave#1{\left\langle #1 \right\rangle}
\def\Re{{\cal R}\hbox{e}}
\def\Im{{\cal I}\hbox{m}}
\def\A{{\cal A}}

\def\D{{\cal D}}
\def\I{{\cal I}}

\def\E{{\cal E}}

\def\U{{\cal U}}
\def\d{{\rm d}}

\def\eps{{\epsilon}}
\def\vp{\varphi}
\def\vt{{\vartheta}}

\def\kB{k_{\rm B}}
\def\Real{{\rm I\mathchoice{\kern-0.70mm}{\kern-0.70mm}{\kern-0.65mm}%
  {\kern-0.50mm}R}}
\def\C{\rm C\kern-.42em\vrule width.03em height.58em depth-.02em
       \kern.4em}
\font \bolditalics = cmmib10
\def\bx#1{\leavevmode\thinspace\hbox{\vrule\vtop{\vbox{\hrule\kern1pt
        \hbox{\vphantom{\tt/}\thinspace{\bf#1}\thinspace}}
      \kern1pt\hrule}\vrule}\thinspace}
\def\Rm#1{{\rm #1}}
\def \vc #1{{\textfont1=\bolditalics \hbox{$\bf#1$}}}
{\catcode`\@=11
\gdef\SchlangeUnter#1#2{\lower2pt\vbox{\baselineskip 0pt \lineskip0pt
  \ialign{$\m@th#1\hfil##\hfil$\crcr#2\crcr\sim\crcr}}}
}
\def\gtrsim{\mathrel{\mathpalette\SchlangeUnter>}}
\def\lesssim{\mathrel{\mathpalette\SchlangeUnter<}}
\def\arcsecf {\hbox{$.\!\!^{\prime\prime}$}} 
\def\arcminf {\hbox{$.\!\!^{\prime}$}}                                     
\newcommand{\bmi}[1]{\begin{minipage}{#1 cm}}
\newcommand{\emi}{\end{minipage}}

\begin{document}
\title*{Weak Gravitational Lensing}
\toctitle{Weak Gravitational Lensing}
\titlerunning{Weak Gravitational Lensing}
%
\author{Peter Schneider
\\
Institut f. Astrophysik, Universit\"at Bonn, D-53121 Bonn, Germany}

\authorrunning{P. Schneider}
%
%

\maketitle              

\large

\noindent
This review was written in the fall of 2004.

\noindent
{\bf To appear as:}

\noindent
Schneider, P.\ 2005, in: {\it Kochanek, C.S., Schneider, P.,
Wambsganss, J.: Gravitational Lensing: Strong, Weak \& Micro}.
G. Meylan, P. Jetzer \& P. North (eds.),  Springer-Verlag: Berlin, 
p.273

\normalsize


\section{\llabel{WL-1}Introduction}
Multiple images, microlensing (with appreciable magnifications) and arcs in
clusters are phenomena of {\em strong lensing}. In {\em weak gravitational
  lensing}, the Jacobi matrix $\A$ is very close to the unit matrix, which
implies weak distortions and small magnifications.  Those cannot be identified
in individual sources, but only in a statistical sense. Because of that, the
accuracy of any weak lensing study will depend on the number of sources which
can be used for the weak lensing analysis.  This number can be made large
either by having a large number density of sources, or to observe a large
solid angle on the sky, or both. Which of these two aspects is more relevant
depends on the specific application. Nearly without exception, the sources
employed in weak lensing studies up to now are distant galaxies observed in
the optical or near-IR passband, since they form the densest population of
distant objects in the sky (which is a statement both about the source
population in the Universe and the sensitivity of detectors employed in
astronomical observations). To observe large number densities of sources, one
needs deep observations to probe the faint (and thus more numerous) population
of galaxies. Faint galaxies, however, are small, and therefore their observed
shape is strongly affected by the Point Spread Function, caused by atmospheric
seeing (for ground-based observations) and telescope effects. These effects
need to be well understood and corrected for, which is the largest challenge
of observational weak lensing studies. On the other hand, observing large
regions of the sky quickly leads to large data sets, and the problems
associated with handling them. We shall discuss some of the most important
aspects of weak lensing observations in Sect.\ \ref{sc:WL-3}.

The effects just mentioned have prevented the detection of weak lensing effects
in early studies with photographic plates (e.g., Tyson et al.\ 1984); they are
not linear detectors (so correcting for PSF effects is not reliable), nor are
they sensitive enough for obtaining sufficiently deep images. Weak lensing
research came through a number of observational and technical advances. Soon
after the first giant arcs in clusters were discovered (see Sect.\ 1.2 of
Schneider, this volume; hereafter referred to as IN)
by Soucail et al.\ (1987) and Lynds \& Petrosian (1989), Fort et al.\ (1988)
observed objects in the lensing cluster Abell 370 which were less extremely
stretched than the giant arc, but still showed a large axis ratio and was
aligned in the direction tangent to its separation vector to the cluster
center; they termed these images `arclets'. Indeed, with the spectroscopic
verification (Mellier et al.\ 1991) of the arclet A5 in A\ 370 being located
at much larger distance from us than the lensing cluster, the gravitational
lens origin of these arclets was proven. When the images of a few background
galaxies are deformed so strongly that they can be identified as distorted by
lensing, there should be many more galaxy images where the distortion is much
smaller, and where it can only be detected by averaging over many such
images. Tyson et al.\ (1990) reported this statistical distortion
effect in two clusters, thereby initiating the weak lensing studies of the
mass distribution of clusters of galaxies. This very fruitful field of
research was put on a rigorous theoretical basis by Kaiser \& Squires (1993)
who showed that from the measurement of the (distorted) shapes of galaxies one
can obtain a parameter-free map of the projected mass distribution in
clusters. 

The flourishing of weak lensing in the past ten years was mainly due
to three different developments. First, the potential of weak lensing
was realized, and theoretical methods were worked out for using weak
lensing measurements in a large number of applications, many of which
will be described in later sections. This realization, reaching out of
the lensing community, also slowly changed the attitude of time
allocation committees, and telescope time for such studies was
granted. Second, returning to the initial remark, one requires large
fields-of-views for many weak lensing application, and the development
of increasingly large wide-field cameras installed at the best
astronomical sites has allowed large observational progress to be
made. Third, quantitative methods for the correction of observations
effects, like the blurring of images by the atmosphere and telescope
optics, have been developed, of which the most frequently used one
came from Kaiser et al.\ (1995). We shall describe this technique, its
extensions, tests and alternative methods in Sect.\ \ref{sc:WL-3.5}.

We shall start by describing the basics of weak lensing in Sect.\
\ref{sc:WL-2}, namely how the shear, or the projected tidal gravitational
field of the intervening matter distribution can be determined from measuring
the shapes of images of distant galaxies. Practical aspects of observations
and the measurements of image shapes are discussed in Sect.\
\ref{sc:WL-3}. The next two sections are devoted to clusters of galaxies; in
Sect.\ \ref{sc:WL-4}, some general properties of clusters are described, and
their strong lensing properties are considered, whereas in Sect.\
\ref{sc:WL-5} weak lensing by clusters is treated. As already mentioned, this
allows us to obtain a parameter-free map of the projected (2-D) mass
distribution of clusters.

We then turn to lensing by the inhomogeneously distributed matter
distribution in the Universe, the large-scale structure. Starting with
Gunn (1967), the observation of the distortion of light bundles by the
inhomogeneously distributed matter in the Universe was
realized as a unique probe to study the properties of the cosmological
(dark) matter distribution. The theory of this cosmic shear effect,
and its applications, was worked out in the early 1990's (e.g.,
Blandford et al.\ 1991).  In contrast to the lensing situations
studied in the rest of this book, here the deflecting mass is
manifestly three-dimensional; we therefore need to generalize the
theory of geometrically-thin mass distributions and consider the
propagation of light in an inhomogeneous Universe. As will be shown,
to leading order this situation can again be described in terms of an
`equivalent' surface mass density. The theoretical aspects of this
large-scale structure lensing, or cosmic shear, are contained in
Sect.\ \ref{sc:WL-6}. Although the theory of cosmic shear was well in
place for quite some time, it took until the year 2000 before it was
observationally discovered, independently and simultaneously by four
groups. These early results, as well as the much more extensive
studies carried out in the past few years, are presented and discussed
in Sect.\ \ref{sc:WL-7}. In Sect.\ \ref{sc:WL-8}, we consider the weak
lensing effects of galaxies, which can be used to investigate the mass
profile of galaxies. As we shall see, this galaxy-galaxy lensing,
first detected by Brainerd et al.\ (1996), is directly related to the
connection between the galaxy distribution in the Universe and the
underlying (dark) matter distribution; this lensing effect is
therefore ideally suited to study the biasing of galaxies; we shall
also describe alternative lensing effects for investigating the
relation between luminous and dark matter. In the final Sect.\
\ref{sc:WL-9} we discuss higher-order cosmic shear statistics and how
lensing by the large-scale structure affects the lens properties of
localized mass concentrations. Some final remarks are given in
Sect.\ts\ref{sc:WL-10}.

Until very recently, weak lensing has been considered by a
considerable fraction of the community as `black magic' (or to quote
one member of a PhD examination committee: ``You have a mass
distribution about which you don't know anything, and then you
observe sources which you don't know either, and then you claim to
learn something about the mass distribution?''). Most likely the
reason for this is that weak lensing is indeed weak. One cannot `see'
the effect, nor can it be graphically displayed easily. Only by investigating
many faint galaxy images can a signal be extracted from the data, and
the human eye is not sufficient to perform this analysis. This is
different even from the analysis of CMB anisotropies which, similarly,
need to be analyzed by statistical means, but at least one can display a
temperature map of the sky. However, in recent years weak lensing has
gained a lot of credibility, not only because it has contributed
substantially to our knowledge about the mass distribution in the
Universe, but also because different teams, with different data set
and different data analysis tools, agree on their results. 

Weak lensing has been reviewed before; we shall mention only five
extensive reviews. Mellier (1999) provides a detailed compilation of
the weak lensing results before 1999, whereas Bartelmann \& Schneider
(2001; hereafter BS01) present a detailed account of the theory and
technical aspects of weak lensing.\footnote{We follow here the
notation of BS01, except that we denote the angular diameter distance
explicitly by $D^{\rm ang}$, whereas $D$ is the comoving angular
diameter distance, which we also write as $f_K$, depending on the
context; see Sect.\ts 4.3 of IN for more details. In most cases, the
distance ratio $D_{\rm ds}/D_{\rm s}$ is used, which is the same for
both distance definitions.}  More recent summaries of results can also
be found in Wittman (2002) and Refregier (\cite{Refre03a}), as well as
the cosmic shear review by van Waerbeke \& Mellier
(\cite{vW-M-Review03}).

The coverage of topics in this review has been a subject of choice; no claim
is made about completeness of subjects or references. In particular, due to
the lack of time during the lectures, the topic of weak lensing of the CMB
temperature fluctuations has not been covered at all, and is also not included
in this written version. Apart from this increasingly important subject, I
hope that most of the currently actively debated aspects of weak lensing are
mentioned, and the interested reader can find her way to more details through
the references provided. 
 
\section{\llabel{WL-2}The principles of weak gravitational lensing}
\subsection{\llabel{WL-2.1}Distortion of faint galaxy images}
Images of distant sources are distorted in shape and size, owing to
the tidal gravitational field through which light bundles from these
sources travel to us.  Provided the angular size of a lensed image of
a source is much smaller than the characteristic angular scale on
which the tidal field varies, the distortion can be described by the
linearized lens mapping, i.e., the Jacobi matrix $\A$. The invariance
of the surface brightness by gravitational light deflection,
$I(\vc\theta) = I^{(\rm s)}[\vc\beta(\vc\theta)]$, together with the
locally linearized lens equation,
\be
\vc \beta-\vc\beta_0=\mathcal{A}(\vc\theta_0)
    \cdot(\vc\theta-\vc\theta_0)\;,
\ee
where $\vc\beta_0=\vc\beta(\vc\theta_0)$,
then describes the distortion of small lensed images as
\be
I(\vc\theta) = I^{(\rm s)}[\vc\beta_0+\mathcal{A}(\vc\theta_0)
    \cdot(\vc\theta-\vc\theta_0)]\;.
\ee
We recall (see IN) that the Jacobi matrix can be written as
\be
  \mathcal{A}(\vc\theta) = (1-\kappa)
 \left(
    \begin{array}{cc}
      1-g_1 & -g_2 \\ 
      -g_2 & 1+g_1 \\
    \end{array}
  \right)\;, \; {\rm where} \;\; g(\vc\theta)={\gamma(\vc\theta)\over 
[1-\kappa(\vc\theta)]}
\elabel{redshearM}
\end{equation}
is the reduced shear, and the $g_\alpha$, $\alpha=1,2$, are its
Cartesian components. The reduced shear describes the shape distortion
of images through gravitational light deflection.  The (reduced) shear
is a 2-component quantity, most conveniently written as a complex
number,
\be
\gamma=\gamma_1 +{\rm i}\gamma_2=|\gamma|\,{\rm e}^{2{\rm i}\vp}\;;
\quad g=g_1+{\rm i} g_2=|g|\,{\rm e}^{2{\rm i}\vp}\;;
\ee
its amplitude describes the degree of distortion, whereas its phase
$\vp$ yields the direction of distortion. The reason for the factor
`2' in the phase is the fact that an ellipse transforms into itself
after a rotation by $180^\circ$.  Consider a circular source with
radius $R$ (see Fig.\ts\ref{fig:WL-scetch1}); mapped by the local 
Jacobi matrix, its image is an ellipse, with semi-axes
\[
{R\over 1-\kappa-|\gamma|}={R\over (1-\kappa)(1-|g|)}
\quad ;\quad{R\over 1-\kappa+|\gamma|}={R\over (1-\kappa)(1+|g|)} \;
\]
and the major axis encloses an angle $\vp$ with the positive
$\theta_1$-axis.  Hence, if sources with circular isophotes could be
identified, the measured image ellipticities would immediately yield
the value of the reduced shear, through the axis ratio
\[
|g|={1-b/a \over 1+b/a}\quad \Leftrightarrow \quad
{b\over a}={1-|g|\over 1+|g|}
\]
and the orientation of the major axis $\vp$. In these relations it was
assumed that $b\le a$, and $|g|<1$. We shall discuss the case $|g|>1$
later. 

\begin{figure}
\bc
\includegraphics[width=11.5cm]{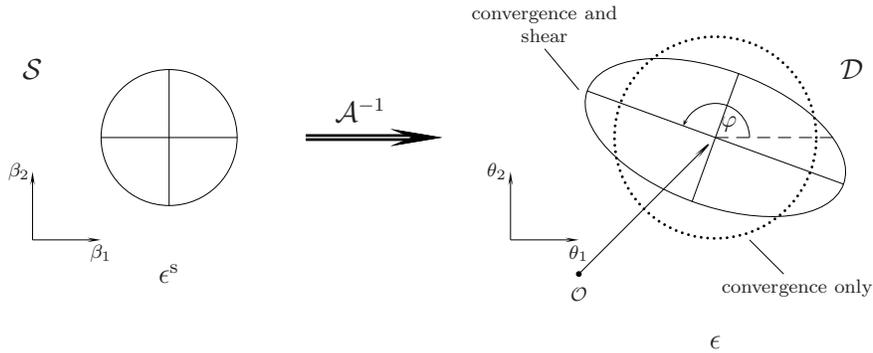}
\ec
\caption{A circular source, shown at the left, is mapped by the
inverse Jacobian $\A^{-1}$ onto an ellipse. In the absence of shear,
the resulting image is a circle with modified radius, depending on
$\kappa$. Shear causes an axis ratio different from unity, and the
orientation of the resulting ellipse depends on the phase of the shear
(source: M. Bradac)
}
\flabel{WL-scetch1}
\end{figure}

However, faint galaxies
are not intrinsically round, so that the observed image ellipticity is
a combination of intrinsic ellipticity and shear.  The strategy to
nevertheless obtain an estimate of the (reduced) shear consists in
locally averaging over many galaxy images, assuming that the intrinsic
ellipticities are {\em randomly oriented}. In order to follow this
strategy, one needs to clarify first how to define `ellipticity' for a
source with arbitrary isophotes (faint galaxies are not simply
elliptical); in addition, seeing by the atmospheric turbulence will
blur -- and thus circularize -- observed images, together with other
effects related to the observation procedure. We will consider
these issues in turn.

\subsection{\llabel{WL-2.2}Measurements of shapes and shear}
\subsubsection{Definition of image ellipticities.}
Let $I(\vc\theta)$ be the brightness distribution of an image, assumed to
be isolated on the sky; the
center of the image can be defined as 
\begin{equation}
  \bar{\vc\theta} \equiv
  \frac{\int\!\d^2\theta\;I(\vc\theta)\,q_I[I(\vc\theta)]\,\vc\theta}
       {\int\!\d^2\theta\;I(\vc\theta)\,q_I[I(\vc\theta)]}\;,
\elabel{4.1}
\end{equation}
where $q_I(I)$ is a suitably chosen weight function; e.g., if $q_I(I)={\rm
H}(I-I_{\rm th})$, where ${\rm H}(x)$ is the Heaviside step function,
$\bar{\vc\theta}$ would be the center of light within a limiting isophote of
the image.  We next define the tensor of second brightness moments,
\begin{equation}
  Q_{ij} = \frac{
    \int\!\d^2\theta\;I(\vc\theta)\,q_I[I(\vc\theta)]\,
    (\theta_i-\bar\theta_i)\,(\theta_j-\bar\theta_j)
  }{
    \int\!\d^2\theta\;I(\vc\theta)\,q_I[I(\vc\theta)]
  }\;,\quad i,j\in \{1,2\}\;.
\elabel{4.2}
\end{equation}
Note that 
for an image with circular isophotes, $Q_{11}=Q_{22}$, and $Q_{12}=0$.
The trace of $Q$ describes the size of the image, whereas the
traceless part of $Q_{ij}$ contains the ellipticity information.
From $Q_{ij}$, one defines two complex ellipticities,
\begin{equation}
  \chi \equiv \frac{Q_{11}-Q_{22}+2{\rm i}Q_{12}}
  {Q_{11}+Q_{22}}\; \;\;
{\rm and} \;\;
  \epsilon \equiv \frac{Q_{11}-Q_{22}+2{\rm i}Q_{12}}
  {Q_{11}+Q_{22}+2(Q_{11}Q_{22}-Q_{12}^2)^{1/2}}\;.
\elabel{4.10}
\end{equation}
Both of them have the same phase (because of the same numerator), but a
different absolute value.
Fig.\ \ref{fig:imageshapes} illustrates the shape of images as a function of
their complex ellipticity $\chi$.
For an image with elliptical isophotes of axis ratio $r\le 1$, one
obtains 
\be
|\chi|={1-r^2\over 1+r^2}\quad ;\quad
|\eps|={1-r\over 1+r}\;.
\ee
\begin{figure}
\bmi{6.3}
\includegraphics[width=6cm]{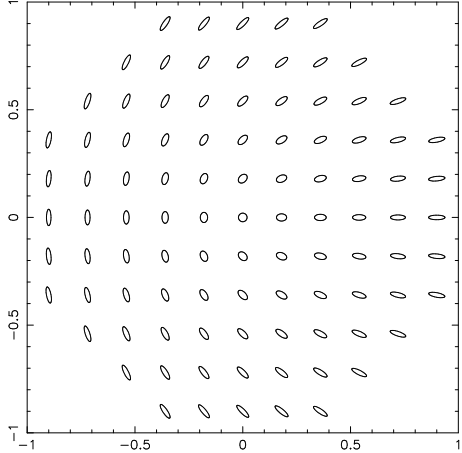}
\emi
\bmi{5.4}
\caption{The shape of image ellipses for a circular source, in dependence on
  their two ellipticity components $\chi_1$ and $\chi_2$; a
  corresponding plot in term of the ellipticity components $\eps_i$
  would look quite similar. Note that the ellipticities are rotated by
  $90^\circ$ when $\chi \to -\chi$ (source: D. Clowe)}
\flabel{imageshapes}
\emi
\end{figure}
Which of these two definitions is more convenient depends on the
context; one can easily transform one into the other,
\be
  \epsilon = \frac{\chi}{1+(1-|\chi|^2)^{1/2}}\;,\quad
  \chi = \frac{2\epsilon}{1+|\epsilon|^2}\;.
\elabel{4.11}
\end{equation}
In fact, other (but equivalent) ellipticity definitions have been used
in the literature (e.g., Kochanek 1990; Miralda--Escud\'e 1991; Bonnet
\& Mellier 1995), but the two given above appear to be most
convenient.

\subsubsection{From source to image ellipticities.}
In total analogy, one defines the second-moment brightness tensor $Q^{(\rm
s)}_{ij}$, and the complex ellipticities $\chi^{(\rm s)}$ and
$\epsilon^{(\rm s)}$ for the unlensed source.
From 
\be  
Q^{(\rm s)}_{ij} = \frac{
    \int\!\d^2\beta\;I^{(\rm s)}(\vc\theta)\,q_I[I^{(\rm s)}(\vc\beta)]\,
    (\beta_i-\bar\beta_i)\,(\beta_j-\bar\beta_j)
  }{
    \int\!\d^2\beta\;I^{(\rm
      s)}(\vc\theta)\,q_I[I^{(\rm s)}(\vc\beta)] 
  }\;,\quad i,j\in \{1,2\}\;,
\elabel{4.2a}
\end{equation}
one finds with $\d^2\beta=\det\A\,\d^2\theta$,
$\vc\beta-\bar{\vc\beta} =\A\rund{\vc\theta-\bar{\vc\theta}}$, 
that 
\begin{equation}
  Q^{(\rm s)} = \A \,Q\,\A^T = 
  \A \,Q\,\A\;,
\elabel{4.5}
\end{equation}
where $\A\equiv\A(\bar{\vc\theta})$.
Using the definitions of the complex ellipticities, one finds the
transformations (e.g., Schneider \& Seitz 1995; Seitz \& Schneider
1997)
\begin{equation}
  \chi^{(\rm s)} = \frac{\chi-2g+g^2\chi^*}{1+|g|^2-2\Re(g\chi^*)}
\; ;\quad
  \epsilon^{(\rm s)} = \left\{\begin{array}{ll}
    \displaystyle\frac{\epsilon-g}{1-g^*\epsilon} &
    \quad\hbox{if}\; |g|\le1\;; \\
    \\
    \displaystyle\frac{1-g\epsilon^*}{\epsilon^*-g^*} & 
    \quad\hbox{if}\; |g|>1 \;.\\
  \end{array}\right.
\elabel{4.6}
\end{equation}
The inverse transformations are
obtained by interchanging source and image
ellipticities, and $g\to -g$ in the foregoing equations.

\subsubsection{Estimating the (reduced) shear.}
In the following we make the assumption that the intrinsic orientation
of galaxies is random,
\begin{equation}
  \mathrm{E}\rund{\chi^{(\rm s)}} = 0 = \mathrm{E}\rund{\epsilon^{(\rm s)}}\;,
\elabel{4.13}
\end{equation}
which is expected to be valid since there should be no direction
singled out in the Universe. This then implies that the expectation
value of $\eps$ is [as obtained by averaging the transformation law
(\ref{eq:4.6}) over the intrinsic source orientation]
\be
{\rm E}(\epsilon) = \left\{\begin{array}{ll}
    \displaystyle g &
    \quad\hbox{if}\; |g|\le1 \\
    \\
    \displaystyle 1/g^* & 
    \quad\hbox{if}\; |g|>1 \;.\\
  \end{array}\right.
\elabel{4.6a}
\end{equation}
This is a remarkable result (Schramm \& Kaiser 1995; Seitz \& Schneider
1997), since it shows that each image ellipticity provides an unbiased
estimate of the local shear, though a very noisy one. The noise is
determined by the intrinsic ellipticity dispersion
\[
\sigma_\eps=\sqrt{\ave{\eps^{(\rm s)}\eps^{(\rm s)*}}}\;,
\]
in the sense that, when averaging over $N$ galaxy images all subject
to the same reduced shear, the 1-$\sigma$ deviation of their mean
ellipticity from the true shear is $\sigma_\eps/\sqrt{N}$. A more
accurate estimate of this error is 
\be
\sigma = \sigma_\eps\eck{1-\min\rund{|g|^2,|g|^{-2}}}/\sqrt{N} 
\elabel{noisesti}
\ee
(Schneider et al.\ 2000).  Hence, the
noise can be beaten down by averaging over many galaxy images;
however, the region over which the shear can be considered roughly
constant is limited, so that averaging over galaxy images is always
related to a smoothing of the shear.  Fortunately, we live in a
Universe where the sky is `full of faint galaxies', as was
impressively demonstrated by the Hubble Deep Field images (Williams et al.\
1996) and previously from ultra-deep ground-based observations (Tyson
1987). Therefore, 
the accuracy of a shear estimate depends on the local number
density of galaxies for which a shape can be measured. In order to
obtain a high density, one requires deep imaging observations. As a
rough guide, on a 3 hour exposure with a 4-meter class telescope,
about 30 galaxies per arcmin$^2$ can be used for a shape
measurement.

In fact, considering (\ref{eq:4.6a}) we conclude that the expectation
value of the observed ellipticity is the same for a reduced shear $g$
and for $g'=1/g^*$. Schneider \& Seitz (1995) have shown that one
cannot distinguish between these two values of the reduced shear from
a purely local measurement, and term this fact the `local degeneracy';
this also explains the symmetry between $|g|$ and $|g|^{-1}$ in
(\ref{eq:noisesti}).  Hence, from a local weak lensing observation one
cannot tell the case $|g|<1$ (equivalent to $\det \A>0$) from the one
of $|g|>1$ or $\det \A<0$. This local degeneracy is, however, broken
in large-field observations, as the region of negative parity of any
lens is small (the Einstein radius inside of which $|g|>1$ of massive
lensing clusters is typically $\lesssim 30''$, compared to data fields
of several arcminutes used for weak lensing studies of clusters), and
the reduced shear must be a smooth function of position on the sky.

Whereas the transformation between source and image ellipticity appears
simpler in the case of $\chi$ than $\eps$ -- see (\ref{eq:4.6}), the
expectation value of $\chi$ cannot be easily calculated and depends explicitly
on the intrinsic ellipticity distribution of the sources. In particular, the
expectation value of $\chi$ is not simply related to the reduced shear
(Schneider \& Seitz 1995). However, in the weak lensing regime, $\kappa\ll 1$,
$|\gamma|\ll 1$, one finds
\begin{equation}
  \gamma \approx g \approx 
  \langle\epsilon\rangle \approx
  \frac{\langle\chi\rangle}{2}\;.
\elabel{4.18}
\end{equation}

\subsection{\llabel{WL-2.3}Tangential and cross component of shear}
\subsubsection{Components of the shear.}
The shear components $\gamma_1$ and $\gamma_2$ are defined relative to
a reference Cartesian coordinate frame. Note that the shear is {\em
not} a vector (though it is often wrongly called that way in the literature),
owing to its transformation properties under rotations: Whereas the components
of a vector are multiplied by $\cos\vp$ and $\sin\vp$ when the coordinate
frame is rotated by an angle $\vp$, the shear components are multiplied by
$\cos(2\vp)$ and $\sin(2\vp)$, or simply, the complex shear gets multiplied by
${\rm e}^{-2{\rm i}\vp}$. The reason for this transformation behavior of the
shear traces back to its original definition as the traceless part of the
Jacobi matrix $\A$. This transformation behavior is the
same as that of the linear polarization; the shear is therefore a {\em
polar}. In analogy with vectors, it is often useful to consider the
shear components in a rotated reference frame, that is, to measure
them w.r.t.\ a different direction; for example, the arcs in clusters
are tangentially aligned, and so their ellipticity is oriented tangent
to the radius vector in the cluster.

\begin{figure}
\bmi{7}
\includegraphics[width=6.8cm]{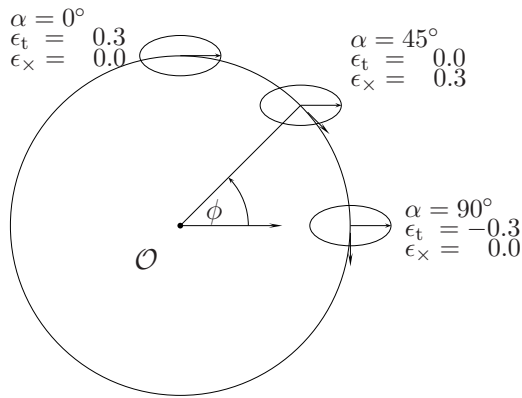}
\emi
\bmi{4.7}
\caption{Illustration of the tangential and cross-components of the
shear, for an image with $\eps_1=0.3$, $\eps_2=0$, and three different
directions $\phi$ with respect to a reference point (source: M.\
Bradac)}
\flabel{WL-scetch2}
\emi
\end{figure} 

If $\phi$ specifies a direction, one defines the {\em tangential} and
{\em cross components} of the shear {\em relative to this direction} as
\be
\gamma_{\rm t}=-\Re\eck{\gamma\,{\rm e}^{-2{\rm i}\phi}} \quad,
\quad
\gamma_\times=-\Im\eck{\gamma\,{\rm e}^{-2{\rm i}\phi}} \;;
\elabel{sheart+c}
\ee
For example, in case of a circularly-symmetric matter distribution,
the shear at any point will be oriented tangent to the direction
towards the center of symmetry.  Thus in this case choose $\phi$ to be
the polar angle of a point; then, $\gamma_\times=0$. In full analogy
to the shear, one defines the tangential and cross components of an
image ellipticity, $\eps_{\rm t}$ and $\eps_\times$. An illustration
of these definitions is provided in Fig.\ts\ref{fig:WL-scetch2}.

The sign in (\ref{eq:sheart+c}) is easily explained (and memorized) as
follows: consider a circular mass distribution and a point on the
$\theta_1$-axis outside the Einstein radius. The image of a circular source
there will be stretched in the direction of the $\theta_2$-axis. In this case,
$\phi=0$ in (\ref{eq:sheart+c}), the shear is real and negative, and in order
to have the tangential shear positive, and thus to define tangential shear in
accordance with the intuitive understanding of the word, a minus sign is
introduced. Negative tangential ellipticity implies that the image is oriented
in the radial direction. We warn the reader that sign conventions and
notations have undergone several changes in the literature, and the current
author had his share in this. 

\subsubsection{Minimum lens strength for its weak lensing detection.}
As a first application of this decomposition, we consider how massive
a lens needs to be in order that it produces a detectable weak lensing
signal. For this purpose, 
consider a lens modeled as an SIS with one-dimensional velocity
dispersion $\sigma_v$.
In the annulus $\theta_{\rm in}\le\theta\le\theta_{\rm out}$,
centered on the lens, let there be $N$ galaxy images with positions
$\vc\theta_i=\theta_i(\cos\phi_i,\sin\phi_i)$ and 
(complex) ellipticities $\eps_i$.
For each one of them, consider the tangential ellipticity
\be
\eps_{{\rm t}i}=-\Re\rund{\eps_i\,{\rm e}^{-2 {\rm i}\phi_i}}\;.
\ee
The weak lensing signal-to-noise for the
detection of the lens obtained by considering a weighted average over
the tangential ellipticity is (see BS01, Sect.\ 4.5) 
\begin{eqnarray}
 \frac{{\rm S}}{{\rm N}} &=&
 \frac{\theta_{\rm E}}{\sigma_\epsilon}\,\sqrt{\pi n}\,
 \sqrt{\ln(\theta_{\rm out}/\theta_{\rm in})} 
\nonumber
 \\ &=&
 8.4\,
 \left(\frac{n}{30\,{\rm arcmin}^{-2}}\right)^{1/2}
 \left(\frac{\sigma_\epsilon}{0.3}\right)^{-1}
 \left(\frac{\sigma_v}{600\,{\rm km\,s}^{-1}}\right)^2 
\elabel{4.55} 
 \\ &\times& \phantom{8.4\,}
 \left(
   \frac{\ln(\theta_{\rm out}/\theta_{\rm in})}{\ln10}
 \right)^{1/2} 
 \left\langle\frac{D_{\rm ds}}{D_{\rm s}}\right\rangle\;,
 \nonumber 
\end{eqnarray}
where $\theta_{\rm E}=4\pi (\sigma_v/c)^2 (D_{\rm ds}/D_{\rm s})$ is
the Einstein radius of an SIS, $n$ the mean number density of
galaxies, and the average of the distance ratio is taken 
over the source population from which the shear measurements are obtained.
Hence, the S/N is proportional to the
lens strength (as measured by $\theta_{\rm E}$), the square root of
the number density, and inversely proportional to $\sigma_\eps$, as
expected.  From this consideration we conclude that clusters of
galaxies with $\sigma_v\gtrsim 600\,{\rm km/s}$ can be detected with
sufficiently large S/N by weak lensing, but individual galaxies
($\sigma_v\lesssim 200\,{\rm km/s}$) are too weak as lenses to be
detected individually. Furthermore, the final factor in
(\ref{eq:4.55}) implies that, for a given source population, 
the cluster detection will be more
difficult for increasing lens redshift.

\subsubsection{Mean tangential shear on circles.}
In the case of axi-symmetric mass distributions, the tangential shear
is related to the surface mass density $\kappa(\theta)$ and the mean
surface mass density $\bar\kappa(\theta)$ inside the radius $\theta$
by $\gamma_{\rm t}=\bar\kappa-\kappa$, as can be easily shown by the
relation in Sect.\ts 3.1 of IN. It is remarkable that a very similar
expression holds for general matter distributions. To see this, we
start from Gauss' theorem, which states that
\[
\int_0^\theta \d^2\vt \;
\nabla\cdot\nabla\psi=\theta \oint \d\vp\; \nabla\psi\cdot \vc n\;,
\]
where the integral on the left-hand side extends over the area of a
circle of radius $\theta$ (with its center chosen as the origin of the
coordinate system), $\psi$ is an arbitrary scalar function, the
integral on the right extends over the circle with radius $\theta$,
and $\vc n$ is the outward directed normal on this circle. 
Taking $\psi$ to be the deflection potential and
noting that $\nabla^2\psi=2\kappa$, one obtains
\be
m(\theta)\equiv{1\over \pi}\int_0^\theta \d^2\vt \;\kappa(\vc\vt)
=
{\theta\over 2\pi}\oint \d\vp\;{\partial\psi\over\partial \theta}\;,
\ee
where we used that $\nabla\psi\cdot \vc n=\psi_{,\theta}$. 
Differentiating this equation with respect to $\theta$ yields
\be
{\d m\over \d\theta}={m\over\theta}+{\theta\over 2\pi}
\oint\d\vp\;{\partial^2\psi\over \partial \theta^2}\;.
\elabel{mass-rela1}
\ee
Consider a point on the $\theta_1$-axis; there,
$\psi_{,\theta\theta}=\psi_{11}= \kappa+\gamma_1=\kappa-\gamma_{\rm
t}$. This last expression is independent on the choice of coordinates
and must therefore hold for all $\vp$. Denoting by
$\ave{\kappa(\theta)}$ and $\ave{\gamma_{\rm t}(\theta)}$ the mean
surface mass density and mean tangential shear on the circle of radius
$\theta$, (\ref{eq:mass-rela1}) becomes
\be
{\d m\over
\d\theta}={m\over\theta}+\theta\eck{\ave{\kappa(\theta)}
-\ave{\gamma_{\rm t}(\theta)} }\;.
\elabel{mass-rela2}
\ee
The dimensionless
mass $m(\theta)$ in the circle is related to the mean surface mass
density inside the circle $\bar \kappa(\theta)$ by
\be
m(\theta)= \theta^2\,\bar\kappa(\theta)
=2\int_0^\theta\d\vt\;\vt\, \ave{\kappa(\vt)}\;.
\elabel{mass-rela3}
\ee
Together with $\d m /\d\theta=2\theta\ave{\kappa(\theta)}$,
(\ref{eq:mass-rela2}) becomes, after dividing through $\theta$, 
\be
\ave{\gamma_{\rm t}}=\bar\kappa-\ave{\kappa}\;,
\elabel{mass-rela4}
\ee
a relation which very closely matches the result mentioned above for
axi-symmetric mass distributions (Bartelmann 1995). One important
immediate implication of this result is that from a measurement of
the tangential shear, averaged over concentric circles, one can determine the
azimuthally-averaged mass profile of lenses, even if the density is
not axi-symmetric.

\subsection{\llabel{WL-2.4}Magnification effects}
Recall from IN that a magnification $\mu$ changes source
counts according to 
\begin{equation}
  n(>S,\vc\theta,z) = \frac{1}{\mu(\vc\theta,z)}\,
  n_0\left(>\frac{S}{\mu(\vc\theta,z)},z\right)\;,
\elabel{4.38}
\end{equation}
where $n(>S,z)$ and $n_0(>S,z)$ are the lensed and unlensed cumulative
number densities of sources, respectively. The first argument of $n_0$
accounts for the change of the flux (which implies that a
magnification $\mu>1$ allows the detection of intrinsically fainter
sources), whereas the prefactor in (\ref{eq:4.38}) stems from the
change of apparent solid angle. In the case that $n_0(S)\propto
S^{-\alpha}$, this yields
\be  
\frac{n(>S)}{n_0(>S)} = \mu^{\alpha-1}\;,
\elabel{4.43}
\end{equation}
and therefore, if $\alpha>1$ ($<1$), source counts are enhanced
(depleted); the steeper the counts, the stronger the effect. In the
case of weak lensing, where $|\mu-1|\ll 1$, one probes the source
counts only over a small range in flux, so that they can always be
approximated (locally) by a power law.
Provided that $\kappa\ll 1$, $|\gamma|\ll 1$, a further
approximation applies,
\be
 \mu \approx 1+2\kappa \; ;\quad{\rm and} \quad \frac{n(>S)}{n_0(>S)}
 \approx 1+2(\alpha-1)\kappa \;.
\elabel{mu-simple}
\ee
Thus, from a measurement of the local number density $n(>S)$ of galaxies,
$\kappa$ can in principle be inferred directly.  It should be noted that
$\alpha\sim 1$ for galaxies in the B-band, but in redder bands, $\alpha<1$
(e.g., Ellis 1997); therefore, one expects a depletion of their counts in
regions of magnification $\mu>1$. Broadhurst et al.\ (1995) have
discussed in detail the effects of magnification in weak lensing. Not only are
the number counts affected, but since this is a redshift-dependent effect
(since both $\kappa$ and $\gamma$ depend, for a given physical surface mass
density, on the source redshift), the redshift distribution of galaxies is
locally changed by magnification. 

Since magnification is merely a stretching of solid angle, Bartelmann \&
Narayan (1995) pointed out that magnified images at fixed surface brightness
have a larger solid angle than unlensed ones; in addition, the surface
brightness of a galaxy is expected to be a strong function of redshift
[$I\propto (1+z)^{-4}$], owing to the Tolman effect.  Hence, if this effect
could be harnessed, a (redshift-dependent) magnification could be measured
statistically.  Unfortunately, this method is hampered by observational
difficulties; it seems that estimating a reliable estimate for the surface
brightness from seeing-convolved images (see Sect.\ \ref{sc:WL-3.5}) is even
more difficult than determining image shapes.

\section{\llabel{WL-3}Observational issues and challenges}

\begin{figure}
\bmi{7}
\includegraphics[width=7.0cm]{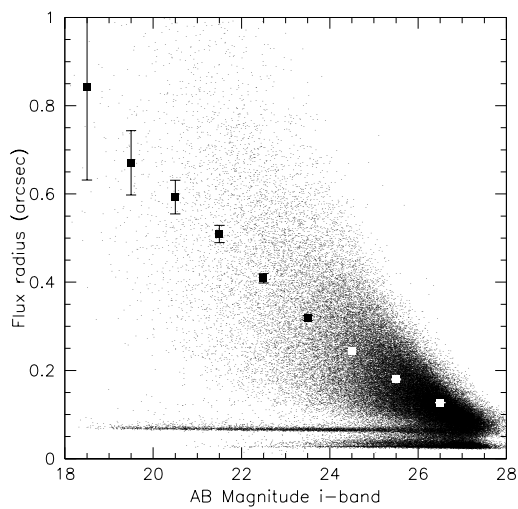}
\emi
\bmi{4.7}
\caption{The size of galaxies observed with the ACS camera on-board
HST. Small dots denote the half-light radius of individual galaxies,
bigger points with error bars show the mean size in a magnitude
bin. The horizontal line of point at $r_{\rm h}\approx 0\arcsecf 08$
correspond to stellar images in the ACS fields, as they have all the
same size but vary in magnitude, and points at even smaller size are
noise artefacts which are not used for any lensing analysis (source:
T. Schrabback)}
\flabel{STIS-sizes}
\emi
\end{figure}

Weak lensing, employing the shear method, relies on the shape
measurements of faint galaxy images.  Since the noise due to intrinsic
ellipticity dispersion is $\propto \sigma_\eps/\sqrt{n}$, one needs a
high number density $n$ to beat this noise component down. However,
the only way to increase the number density of galaxies is to observe
to fainter magnitudes. As it turns out, galaxies at faint magnitudes
are small, in fact typically smaller than the size of the point-spread
function (PSF), or the seeing disk (see Fig.\
\ref{fig:STIS-sizes}). Hence, for them one needs usually large
correction factors between the true ellipticity and that of the
seeing-convolved image. On the other hand, fainter galaxies tend to
probe higher-redshift galaxies, which increases the lensing signal due
to $D_{\rm ds}/D_{\rm s}$-dependence of the `lensing efficiency'.

\subsection{\llabel{WL-3.1}Strategy}
In the present observational situation, only the optical sky is
densely populated with sources; therefore, weak lensing observations
are performed with optical (or near-IR) CCD-cameras (photometric
plates are not linear enough to measure these subtle effects).  In
order to substantiate this comment, note that the Hubble Deep Field
North contains about 3000 galaxies, but only seven radio sources are
detected in a very deep integration with the VLA (Richards et al.\
1998).\footnote{The source density on the radio sky will become at least
comparable to that currently on the optical sky with the future Square
Kilometer Array (SKA).} In order to obtain a high number density of
sources, long exposures are needed: as an illustrative example, to get
a number density of useful galaxies (i.e., those for which a shape can
be measured reliably) of $n\sim 20\,{\rm arcmin}^{-2}$, one needs
$\sim 2\,{\rm hours}$ integration on a 4-m class telescope in good
seeing $\sigma\lesssim 1''$.

Furthermore, large solid angles are desired, either to get large areas around
clusters for their mass reconstruction, or to get good statistics of lenses on
blank field surveys, such as they are needed for galaxy-galaxy lensing and
cosmic shear studies. It is now possible to cover large area in reasonable
amounts of observing time, since large format CCD cameras have recently become
available; for example, the Wide-Field Imager (WFI) at the ESO/MPG 2.2-m
telescope at La Silla has (8K)$^2$ pixels and covers an area of $\sim
(0.5\,{\rm deg})^2$. Until recently, the CFH12K camera with $8{\rm K}\times
12{\rm K}$ pixels and field $\sim 30'\times 45'$ was mounted at the
Canada-French-Hawaii Telescope (CFHT) on Mauna Kea and was arguably the most
efficient wide-field imaging instrument hitherto.  In 2003, MegaCam has been
put into operation on the CFTH which has (18K)$^2$ pixels and covers $\sim
1\,{\rm deg}^2$. Several additional cameras of comparable size will become
operational in the near future, including the 1\ deg$^2$ instrument OmegaCAM
on the newly built VLT Survey Telescope on Paranal. The largest field camera
on a 10-m class telescope is SuprimeCAM, a $34'\times 27'$ multi-chip camera
on the Subaru 8.2-meter telescope.  Unfortunately, many optical astronomers
(and decision making panels of large facilities) consider the prime use of
large telescopes to be spectroscopy; for example, although the four ESO VLT
unit telescopes are equipped with a total of ten instruments, the largest
imagers on the VLT are the two FORS instruments, with a $\sim 6\arcminf 7$
field-of-view.\footnote{Nominally, the VIMOS instrument has a four times
  larger f.o.v., but our analysis of early VIMOS imaging data indicates that
  it is totally useless for weak lensing observations, owing to its highly
  anisotropic PSF, which even seems to show discontinuities on chips, and its
  large variation of the seeing size across chips. It may be hoped that some
  of these image defects are improved after a complete overhaul of the
  instrument which occurred recently.}

\begin{figure}
\includegraphics[width=11.7cm]{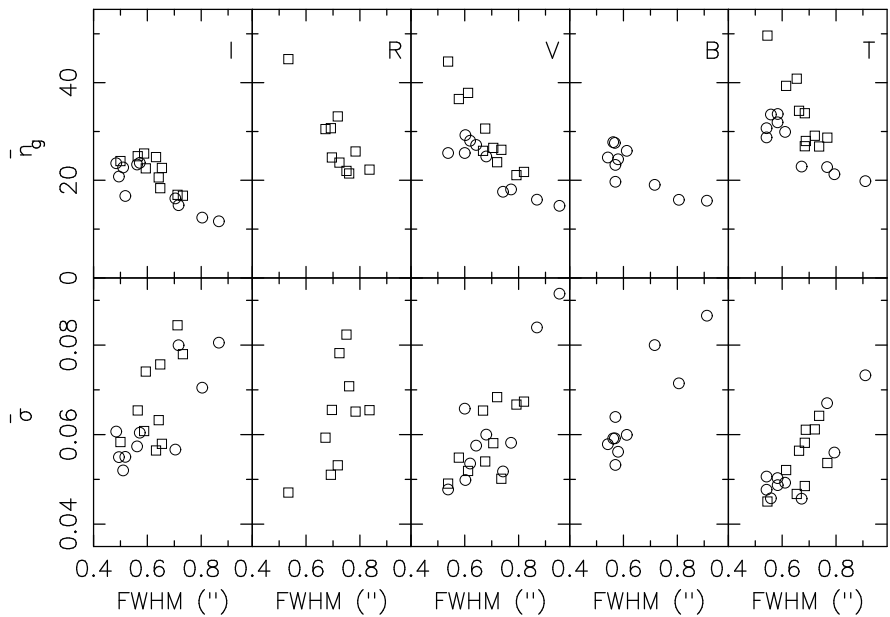}
\caption{Mean number density of galaxy images for which a shape can be
measured (upper row) and the r.m.s. noise of a shear measurement in an
area of $1\, {\rm arcmin}^2$ as a function of the full width at half
maximum (FWHM) of the point-spread function (PSF) -- i.e., the
seeing. The data were taken on 20 different fields with the FORS2
instrument at the VLT, with different filters (I, R, V and R). Squares
show data taken with about 2 hours integration time, circles those
with $\sim 45\,{\rm min}$ exposure. The right-most panels show the
coadded data of I,R,V for the long exposures, and I,V,B for the
$45\,{\rm min}$ fields. The useful number of galaxy images is seen to
be a strong function of the seeing, except for the I-band (which is
related to the higher sky brightness and the way objects are
detected). But even more dramatically, the noise due to intrinsic
source ellipticity decreases strongly for better seeing conditions,
which is due to (1) higher number density of galaxies for which a
shape can be measured, and (2) smaller corrections for PSF blurring,
reducing the associated noise of this correction. In
fact, this figure shows that seeing is a more important quantity than
the total exposure time (from Clowe et al.\ \cite{CloweEdics})}
\flabel{FWHM}
\end{figure}

The typical pixel size of these cameras is $\sim 0\arcsecf 2$, which
is needed to sample the seeing disk in times of good seeing. From
Fig.\ \ref{fig:STIS-sizes} one concludes immediately that the seeing
conditions are absolutely critical for weak lensing: an image with
$0\arcsecf6$ is substantially more useful than one with taken under
the more typical condition of $0\arcsecf8$ (see
Fig.\ts\ref{fig:FWHM}).  There are two separate reasons why the seeing
is such an important factor. First, seeing blurs the images and make
them rounder; accordingly, to correct for the seeing effect, a larger
correction factor is needed in the worse seeing conditions. In
addition, since the galaxy images from which the shear is to be
determined are faint, a larger seeing smears the light from these
galaxies over a larger area on the sky, reducing its contrast relative
to the sky noise, and therefore leads to noisier estimates of the
ellipticities even before the correction.

Deep observations of a field require multiple exposures. As a
characteristic number, the exposure time for an R-band image on a 4-m
class telescope is not longer than $\sim 10\,{\rm min}$ to avoid the
non-linear part of the CCD sensitivity curve (exposures in shorter
wavelength bands can be longer, since the night sky is fainter in
these filters).  Therefore, these large-format cameras imply a high
data rate; e.g., one night of observing with the WFI yields $\sim
30\,{\rm GB}$ of science and calibration data. This number will
increase by a factor $\sim 6$ for MegaCam. Correspondingly, handling
this data requires large disk space for efficient data reduction.

\subsection{\llabel{WL-3.2}Data reduction: Individual frames}
We shall now consider a number of issues concerning the reduction of
imaging data, starting here with the steps needed to treat individual
chips on individual frames, and later consider aspects of combining
them into a coadded image.

\subsubsection{Flatfielding.} The pixels of a CCD have different
sensitivity, i.e., they yield different counts for a given amount of
light falling onto them. In order to calibrate the pixel sensitivity,
one needs flatfielding.
Three standard methods for this are in use: 
\ben
\item
Dome-flats: a uniformly illuminated screen in the telescope dome is
exposed; the counts in the pixels are then proportional to their
sensitivity. The problem here is that the screen is not really of
uniform brightness.
\item
Twilight-flats: in the period of twilight after sunset, or before
sunrise, the cloudless sky is nearly uniformly bright. Short exposures
of regions of the sky without bright stars are then used to calibrate
the pixel sensitivity.
\item Superflats: if many exposures with different pointings are taken with a
  camera during a night, then any given pixel is not covered by a source for
  most of the exposures (because the fraction of the sky at high galactic
  latitudes which is covered by objects is fairly small, as demonstrated by
  the deep fields taken by the HST). Hence, the (exposure-time normalized)
  counts of any pixel will show, in addition to a little tail due to those
  exposures when a source has covered it, a distribution around its
  sensitivity to the uniform night-sky brightness; from that distribution, the
  flat-field can be constructed, by taking its mode or its median.
\een

\begin{figure}
\includegraphics[width=7.8cm,angle=90]{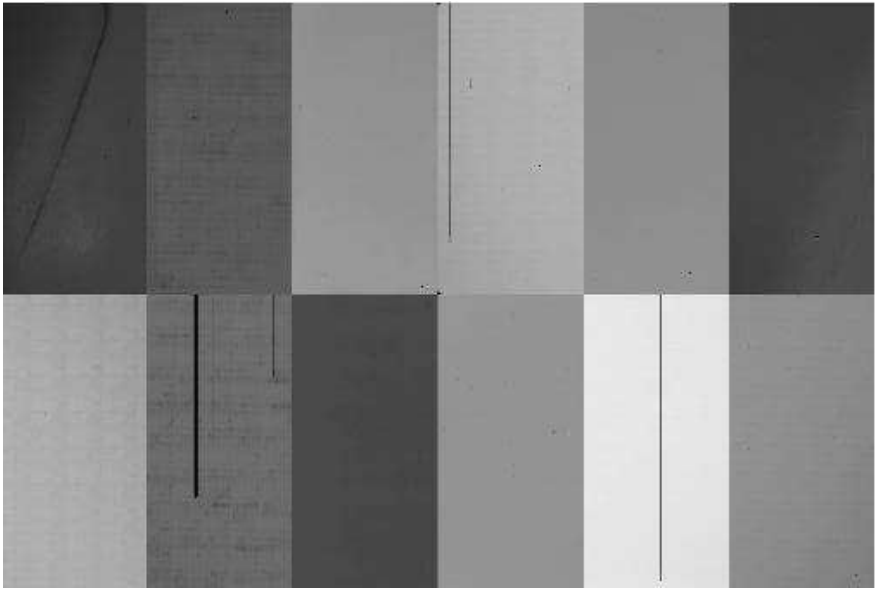}
\caption{A flat field for the CFH12K camera, showing the sensitivity
variations between pixels and in particular between chips. Also, bad
columns are clearly seen}
\flabel{CFHT-flat}
\end{figure}

\subsubsection{Bad pixels.} Each CCD has defects, in that some pixels
are dead or show a signal unrelated to their illumination. This can
occur as individual pixels, or whole pixel columns. No information of
the sky image is available at these pixel positions. One therefore
employs dithering: several exposures of the same field, but with
slightly different pointings (dither positions) are taken. Then, any
position of the field falls on bad pixels only in a small fraction of
exposures, so that the full two-dimensional brightness distribution
can be recovered. 

\subsubsection{Cosmic rays.} Those mimic groups of bad pixels; they
can be removed owing to the fact that a given point of the image will
most likely be hit by a cosmic only once, so that by comparison
between the different exposures, cosmic rays can be removed (or more
precisely, masked). Another signature of a cosmic ray is that the
width of its track is typically much smaller than the seeing disk, the
minimum size of any real source. 

\subsubsection{Bright stars.} Those cause large diffraction spikes,
and depending on the optics and the design of the camera, reflection
rings, ghost images and other unwanted features.  It is therefore best
to choose fields where no or very few bright stars are present. The
diffraction spikes of stars need to be masked, as well as the other
features just mentioned.

\begin{figure}
\includegraphics[width=7.8cm,angle=90]{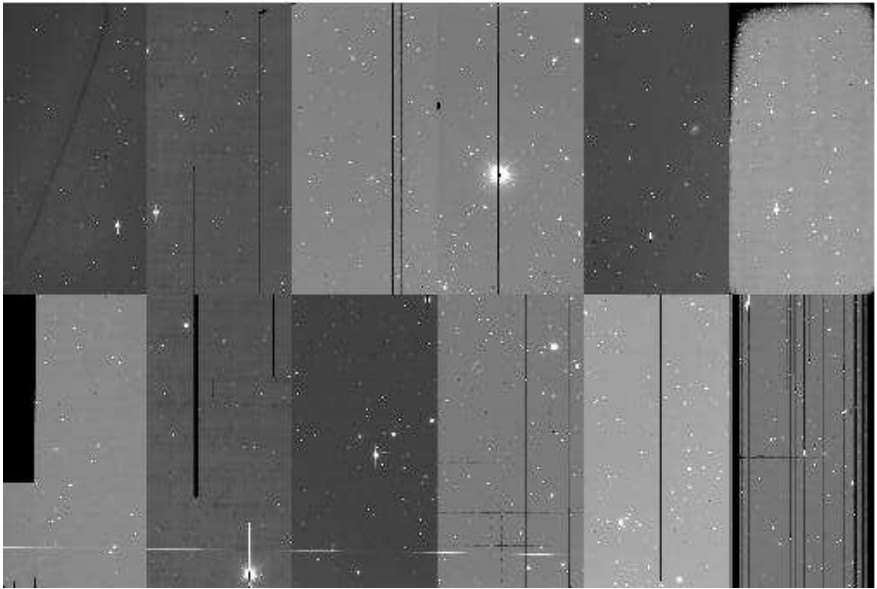}
\caption{A raw frame from the CFH12K camera, showing quite a number of
effects mentioned in the text: bad column, saturation of bright stars,
bleeding, and sensitivity variations across the field and in particular
between chips}
\flabel{CFHT-raw}
\end{figure}

\subsubsection{Fringes.}
Owing to light reflection within the CCD, patterns of illumination
across the field can be generated (see Fig.\ts\ref{fig:fringes}); this
is particularly 
true for thin chips when rather long wavelength filters are used. In
clear nights, the fringe pattern is stable, i.e., essentially the same
for all images taken during the night; in that case, it can be deduced
from the images and subtracted off the individual exposures. However,
if the nights are not clear, this procedure no longer works well; it
is then safer to observe at shorter wavelength. For example, for the
WFI, fringing is a problem for I-band images, but for the R-band
filter, the amplitude of fringing is small. For the FORS instruments
at the VLT, essentially no fringing occurs even in the I band (Maoli
et al.\ 2001). 

\begin{figure}
\bmi{6}
\includegraphics[width=5.7cm]{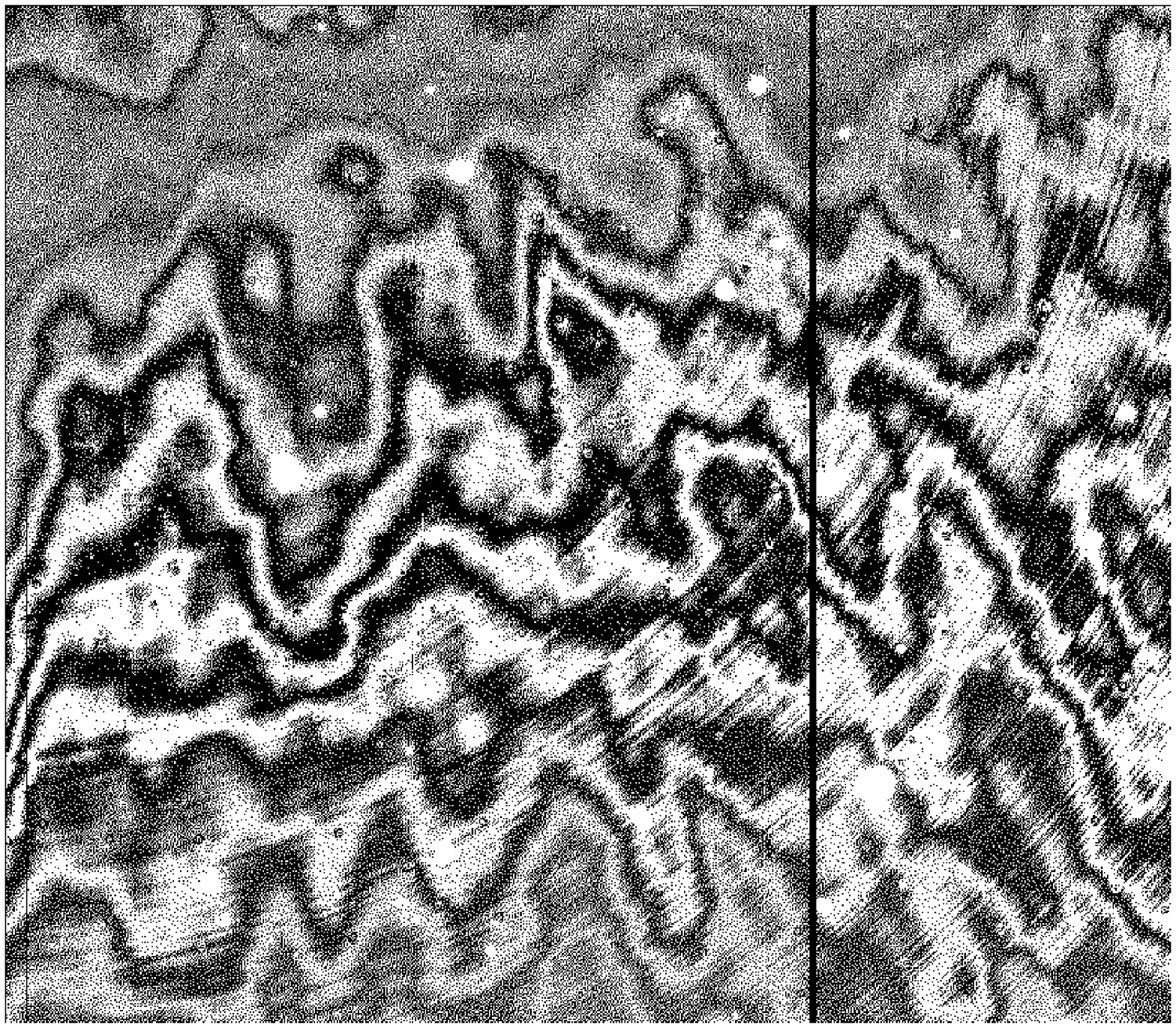}

\vspace{0.2cm}
\includegraphics[width=5.7cm]{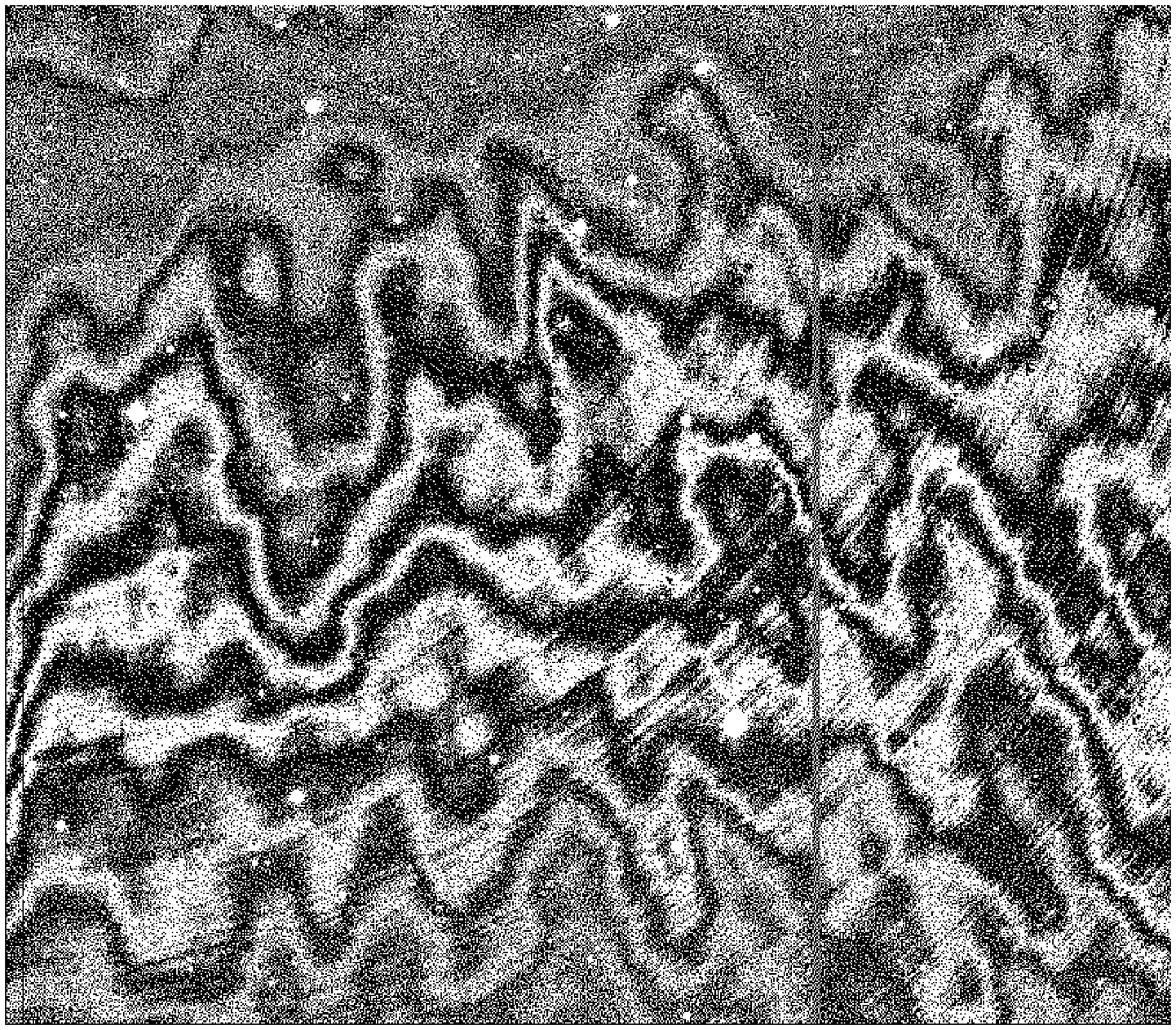}
\emi
\bmi{5.7}
\includegraphics[width=5.7cm]{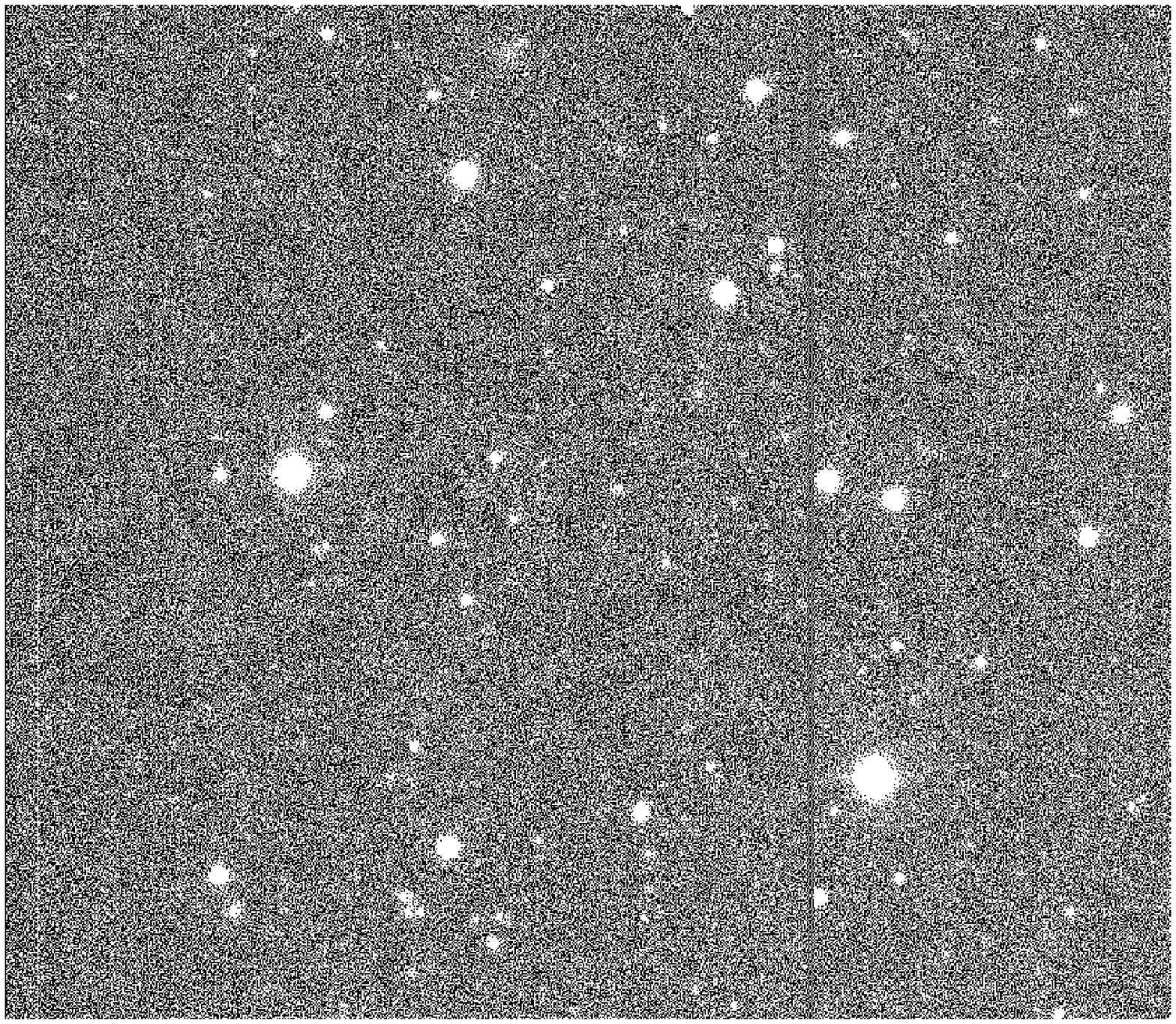}

\vspace{0.2cm}
\includegraphics[width=5.7cm]{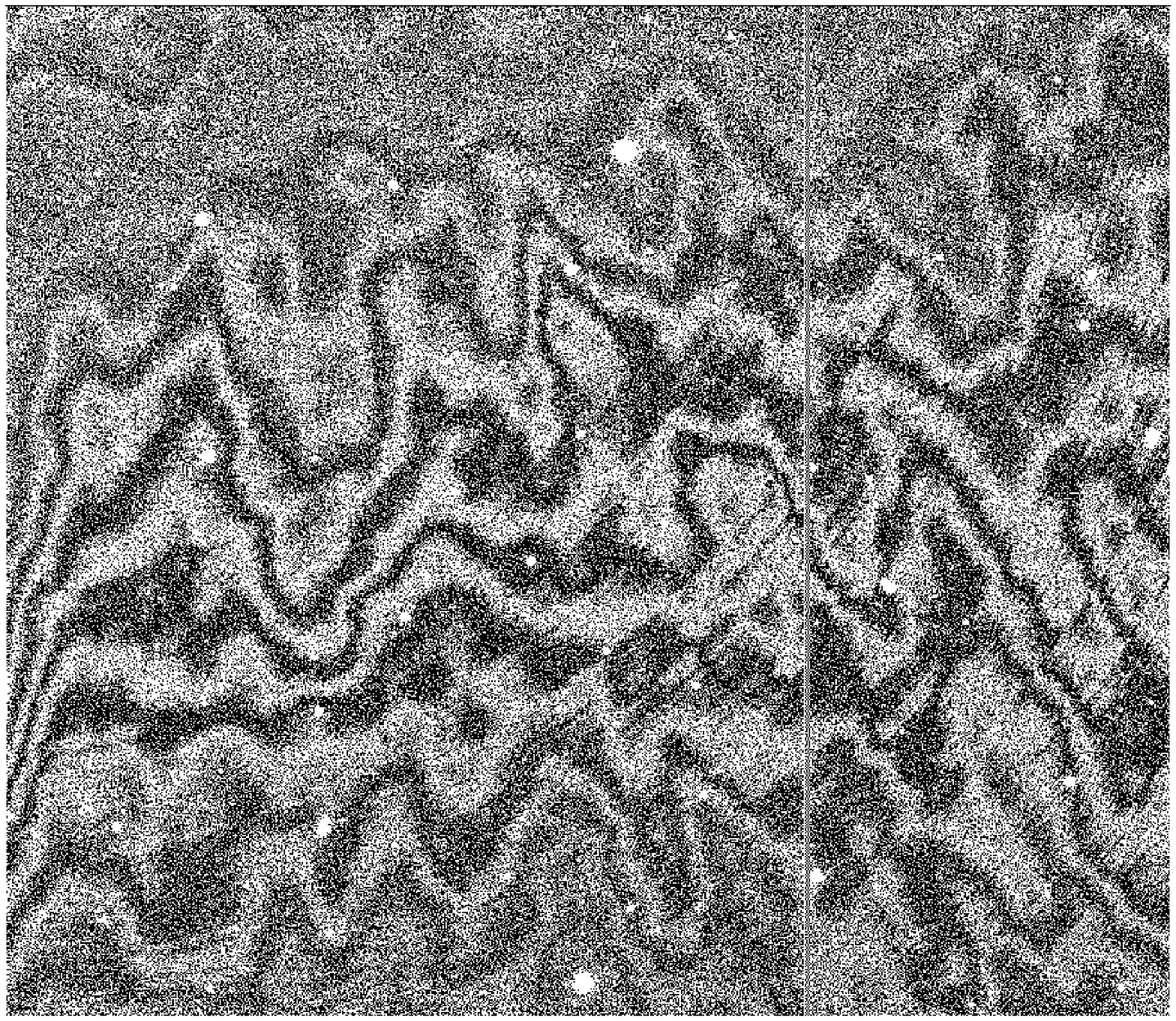}
\emi

\caption{The two left panels show the fringe patterns of images taken
with the WFI in the I-band; the upper one was taken during photometric
conditions, the lower one under non-photometric conditions. Since the
fringe pattern is spatially stable, it can be corrected for (left
panels), but the result is satisfactory only in the former case
(source: M. Schirmer \& T. Erben)}
\flabel{fringes}
\end{figure}

\subsubsection{Gaps.} 
The individual CCDs in multi-chip cameras cannot be brought together
arbitrarily close; hence, there are gaps between the CCDs (see
Fig.\ts\ref{fig:WFI-layout} for an example).  In order to cover the
gaps, the dither pattern can be chosen such as to cover the gaps, so
that they fall on different parts of the sky in different
exposures. As we shall see, such relatively large dither patterns also
provide additional advantages.

\begin{figure}
\bmi{7}
\includegraphics[width=6.9cm]{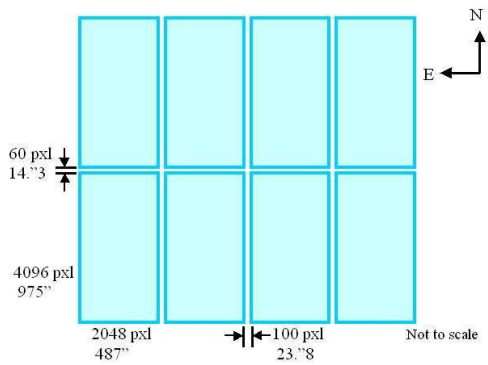}
\emi
\bmi{4.7}
\caption{Layout of the Wide Field Imager (WFI) at the ESO/MPG 2.2m
telescope at La Silla. The eight chips each have $\sim 2048\times
4096$ pixels and cover $\sim 7\arcminf 5\times 15'$}
\flabel{WFI-layout}
\emi
\end{figure}

\subsubsection{Satellite trails, asteroid trails.} Those have to be
identified, either by visual inspection (currently the default) or by
image recognition software which can detect these linear features
which occur either only once, or at different positions on different
exposures. These are then masked, in the same way as some of the other
features mentioned above.

\subsection{\llabel{WL-3.3}Data reduction: coaddition}
After taking several exposures with slightly different pointing
positions (for the reasons given above), frames shall be coadded to a
sum-frame; some of the major steps in this coaddition procedure are:

\subsubsection{Astrometric solution.} 
One needs to coadd data from the same true (or sky) position, not the same
pixel position. Therefore, one needs a very precise mapping from sky
coordinates to pixel coordinates. Field distortions, which occur in every
camera (and especially so in wide-field cameras), make this mapping
non-linear (see Fig.\ts\ref{fig:WFI-distort}). Whereas the distortion
map of the telescope/camera system is to a 
large degree constant and therefore one of the known features, it is not
stable to the sub-pixel accuracy needed for weak lensing work, owing to its
dependence on the zenith angle (geometrical distortions of the telescope due
to gravity), temperature etc. Therefore, the pixel-to-sky mapping has to be
obtained from the data itself. Two methods are used to achieve this: one of
them makes use of an external reference catalog, such as the US Naval
Observatory catalogue for point sources; it contains about 2 point sources per
arcmin$^2$ (at high Galactic latitudes) with $\sim 0.3$\ts arcsec positional
accuracy.  Matching point sources on the exposures with those in the USNO
catalog therefore yields the mapping with sub-arcsecond accuracy.  Far higher
accuracy of the relative astrometry is achieved (and needed) from internal
astrometry, which is obtained by matching objects which appear at different
pixel coordinates, and in particular, on different CCDs for the various
dithering positions. Whereas the sky coordinates are constant, the pixel
coordinates change between dithering positions. Since the distortion map can
be described by a low-order polynomial, the comparison of many objects
appearing at (substantially) different pixel positions yield many more
constraints than the free parameters in the distortion map and thus yields the
distortion map with much higher relative accuracy than external data.  The
corresponding astrometric solution can routinely achieve an accuracy of 0.1
pixel, or typically $0\arcsecf02$ -- compared with a typical field size of
$\sim 30'$.

\begin{figure}
\bc
\includegraphics[width=10cm]{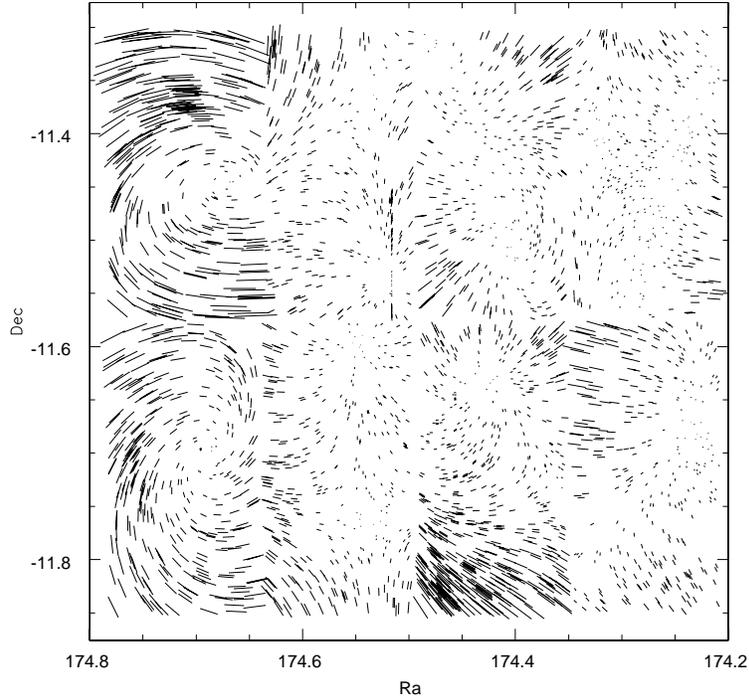}
\ec
\caption{This figure shows the geometric distortion of the
WFI. Plotted is the difference of the positions of stars as obtained
from a simple translation, and a third-order astrometric correction
obtained in the process of image reduction. The patterns in the two
left chips is due to their rotation relative to the other six
chips. Whereas this effect looks dramatic at first sight, the maximum
length of the sticks corresponds to about 6 pixels, or $1\arcsecf2$.
Given that the WFI covers a field of $\sim 33'$, the
geometrical distortions are remarkably small -- however, they are
sufficiently large that they have to be taken into account in the
coaddition process (source: T. Erben \& M. Schirmer)} 
\flabel{WFI-distort}
\end{figure}

\subsubsection{Photometric solution.} 
Flatfielding corrects for the different sensitivities of the pixels and
therefore yields accurate relative photometry across individual
exposures. The different exposures are tied together by matching the
brightness of joint objects, in particular across chip boundaries. To
achieve an absolute photometric calibration, one needs external data
(e.g., standard star observations).

\subsubsection{The coaddition process.}
Coaddition has to happen with sub-pixel accuracy; hence, one cannot just shift
pixels from different exposures on top of each other, although this procedure
is still used by some groups. The by-now standard method is drizzling
(Fruchter \& Hook 2002), in which a new pixel frame is defined which usually
has smaller pixel size than the original image pixels (typically by a factor
of two) and which is linearly related to the sky coordinates. The
astrometrically and photometrically calibrated individual frames are now
remapped onto this new pixel grid, and the pixel values are summed up into the
sub-pixel grid, according to the overlap area between exposure pixel and
drizzle pixel (see Fig.\ts\ref{fig:drizzle}).
By that, drizzling automatically is flux conserving. In the
coaddition process, weights are assigned, accounting for the noise properties
of the individual exposures (including the masks, of course).

\begin{figure}
\bc
\includegraphics[width=10cm]{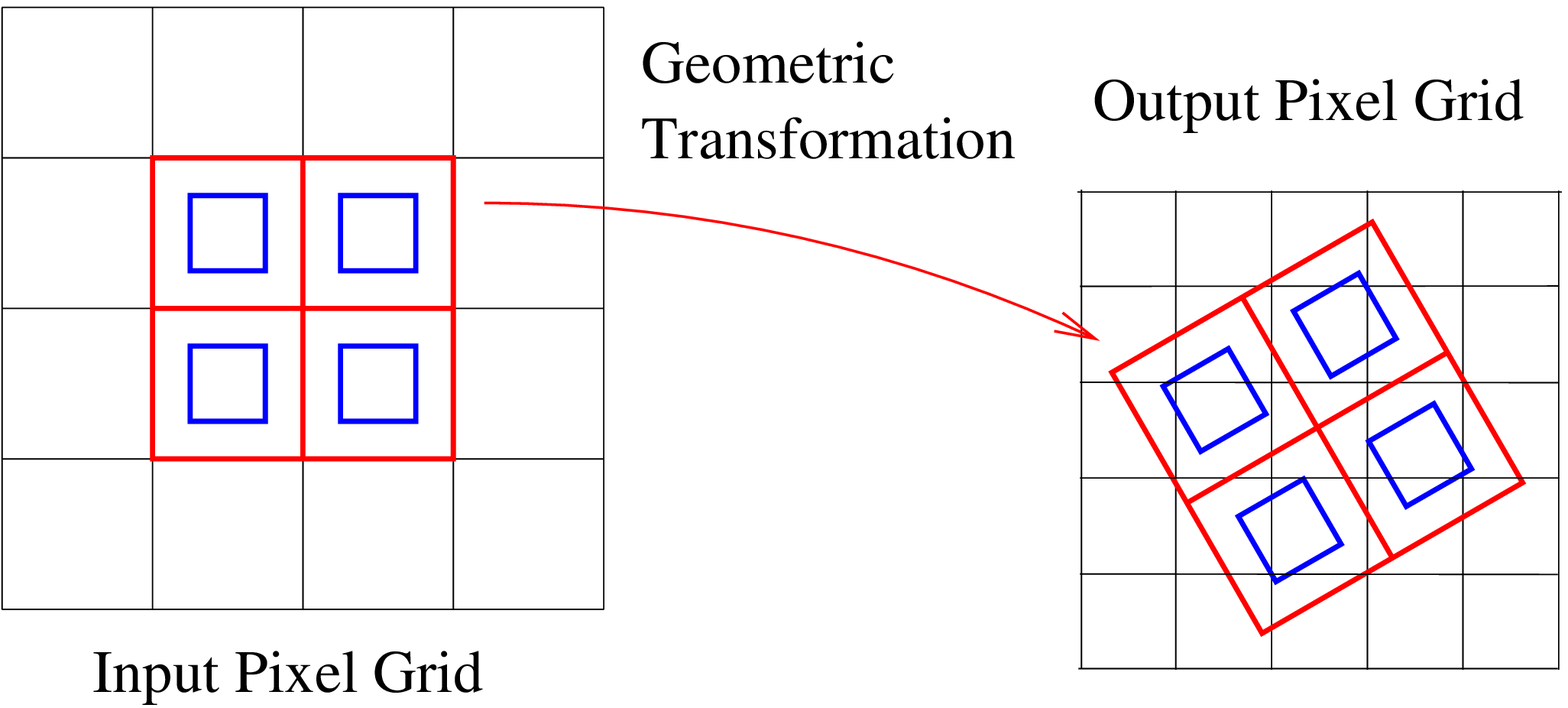}
\ec
\caption{The principle of drizzling in the process of coaddition is
shown. The pixel grid of each individual exposure is mapped onto an
output grid, where the shifts and geometric distortions obtained
during the astrometric solutions are applied. The counts of the input
pixel, multiplied by the relative weight of this pixel, are then
dropped onto the output pixels, according to the relative overlap
area, where the output pixels can be chosen smaller than the input
pixels. The same procedure is applied to the weight maps of the
individual exposures.  If many exposures are coadded, the input pixel
can also be shrunk before dropping onto the output pixel. After
processing all individual exposures in this way, a coadded image and a
coadded weight map is obtained (source: T. Schrabback)}
\flabel{drizzle}
\end{figure}

The result of the coaddition procedure is then a science frame, plus
a weight map which contains information about the pixel noise, which
is of course spatially varying, owing to the masks, CCD gaps, removed
cosmic rays and bad pixels. Fig.\ \ref{fig:coaddframe} shows a typical
example of a coadded image and its corresponding weight map.

\begin{figure}
\bc
\includegraphics[width=11.7cm,angle=-90]{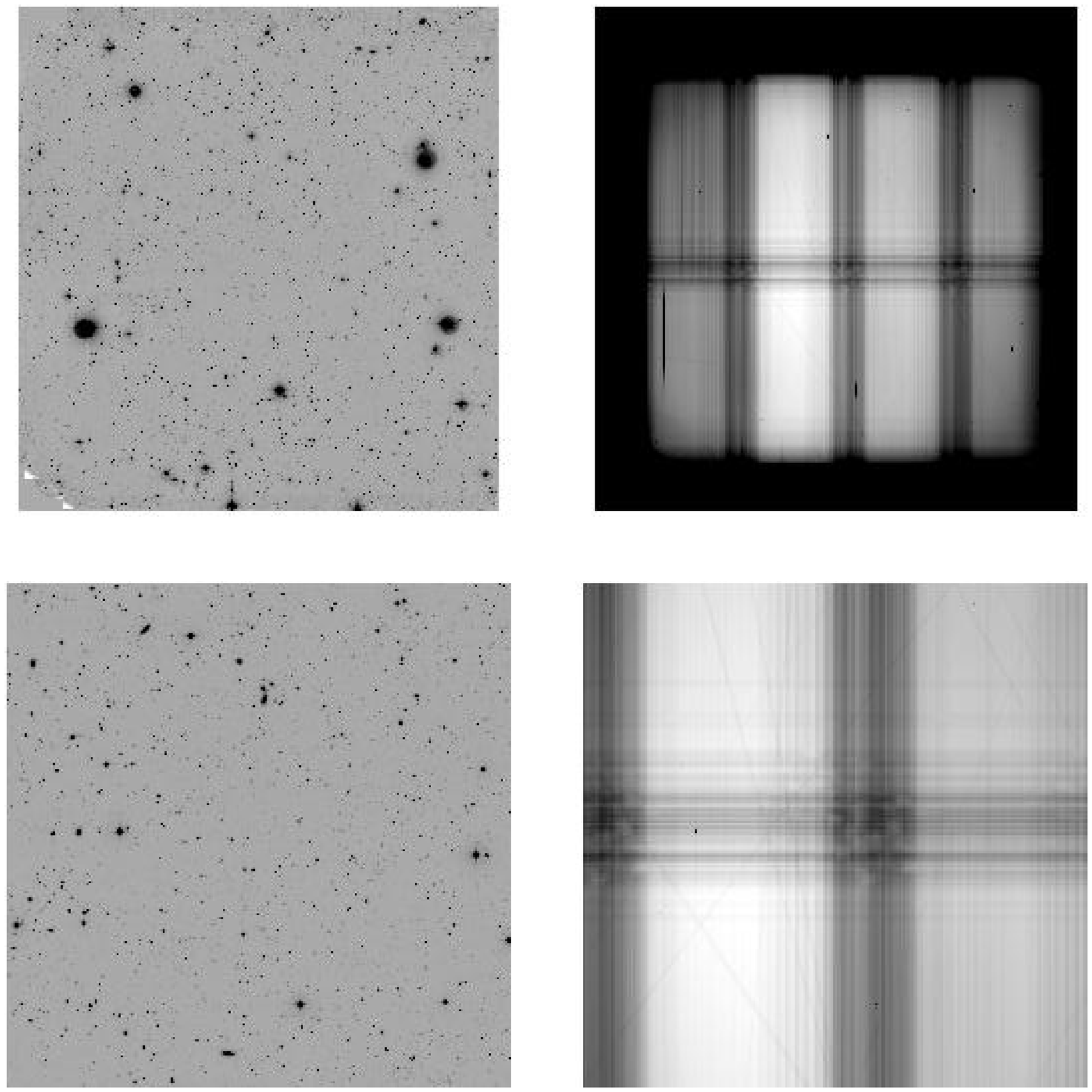}
\ec
\caption{A final coadded frame from a large number of individual
exposures with the WFI is shown in the upper left panel, with the corresponding
weight map at the upper right. The latter clearly shows the large-scale
inhomogeneity of the chip sensitivity and the illumination, together
with the different number of exposures contributing to various regions
in the output image due to dithering and the gaps between CCDs. The
two lower panels show a blow-up of the central part. Despite the
highly inhomogeneous weight, the coadded image apparently shows no
tracer of the gaps, which indicates that a highly accurate relative
photometric solution was obtained (source: T. Erben \& M. Schirmer)}
\flabel{coaddframe}
\end{figure}

The quality of the coadded image can be checked in a number of
ways. Coaddition should not erase information contained in the original
exposures (except, of course, the variability of sources). This means that the
PSF of the coadded image should not be larger than the weighted mean of the
PSFs of the individual frames. Insufficient relative astrometry would lead to
a blurring of images in the coaddition. Furthermore, the anisotropy of the PSF
should be similar to the weighted mean of the PSF anisotropies of the
individual frames; again, insufficient astrometry could induce an artificial
anisotropy of the PSF in the coaddition (which can be easily visualized, by
adding two round images with a slight center offset, where a finite
ellipticity would be induced). 

Probably, there does not exist the `best' coadded image from a given set of
individual exposures. This can be seen by considering a set of exposures with
fairly different individual seeing. If one is mainly interested in photometric
properties of rather large galaxies, one would prefer a coaddition which puts
all the individual exposures together, in order to maximize the total exposure
time and therefore to minimize the photometric noise of the coadded
sources. For weak lensing purposes, such a coaddition is certainly not
optimal, as adding exposures with bad seeing together with those of good
seeing creates a coadded image with a seeing intermediate between the good and
the bad. Since seeing is a much more important quantity than depth for the
shape determination of faint and small galaxy images, it would be better to
coadd only the images with the good seeing. In this respect, the fact that
large imaging instruments are operated predominantly in service observing more
employing queue scheduling is a very valuable asset: data for weak lensing
studies are then taken only if the seeing is better than a specified limit; in
this way one has a good chance to get images of homogeneously good seeing
conditions. 

As a specific example, we show in Fig.\ \ref{fig:CDFS-deep} the
`deepest wide-field image in the Southern sky', targeted towards the
Chandra Deep Field South, one of regions in the sky in which all major
observatories have agreed to obtain, and make publically available,
very deep images for a detailed multi-band study. For example, the
Hubble Ultra Deep Field (Beckwith et al.\ 2003) is located in the
CDFS, the deepest Chandra X-ray exposures are taken in this field, as
well as two ACS@HST mosaic images, one called the GOODS field (Great
Observatories Origins Deep Survey; cf.\ Giavalisco \& Mobasher 2004),
the other the GEMS survey (Rix et al.\ 2004).

\begin{figure}
\includegraphics[width=11.8cm]{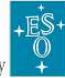}
\caption{A multi-color WFI image of the CDFS; the
field is slightly larger than one-half degree on the side. To obtain
this image, about 450 different WFI exposures were combined, resulting
in a total exposure time of 15.8\ hours in B, 15.6\ hours in V, and
17.8\ hours in R. The data were obtained in the frame of three
different projects -- the GOODS project, the public ESO
Imaging Survey, and the COMBO-17 survey. These data were reduced and
coadded by Mischa Schirmer \& Thomas Erben; more than 2 TB of disk
space were needed for the reduction.}  
\flabel{CDFS-deep}
\end{figure}

\subsection{\llabel{WL-3.4}Image analysis}
The final outcome of the data reduction steps described above is an
image of the sky, together with a weight map providing the noise
properties of the image. The next step is the scientific exploitation
of this image, which in the case of weak lensing includes the
identification of sources, and to measure their magnitude, size and
shape. 

As a first step, individual sources on the image need to be
identified, to obtain a catalog of sources for which the
ellipticities, sizes and magnitudes are to be determined later.  This
can done with by-now standard software, like SExtractor 
(Bertin \& Arnouts 1996), or may be part of specialized software packages
developed specifically for weak lensing, such as IMCAT, developed by
Nick Kaiser (see
below). Although this first step seems straightforward at first
glance, it is not: images of sources can be overlapping, the
brightness distribution of many galaxies (in particular those with
active star formation) tends to be highly structured, with a
collection of bright spots, and therefore the software must be taught
whether or not these are to be split into different sources, or be
taken as one (composite) source. This is not only a software problem;
in many cases, even visual inspection cannot decide whether a given
light distribution corresponds to one or several sources.
The shape and size of the images are affected by the point-spread
function (PSF), which results from the telescope optics, but for
ground-based images, is dominated by the blurring caused by the
atmospheric turbulence; furthermore, the PSF may be affected by telescope
guiding and the coaddition process described earlier.

\subsubsection{The point-spread function.}
Atmospheric turbulence and the other effects mentioned above smear the
image of the sky, according to
\be
I^{\rm obs}(\vc\theta)=\int\d^2\vt\;I(\vc\vt)\,P(\vc\theta-\vc\vt)\;,
\ee
where $I(\vc\vt)$ is the brightness profile outside the atmosphere, $I^{\rm
  obs}(\vc\vt)$ the observed brightness profile, and
$P$ is the PSF; it describes how point sources would appear on the
image. To first approximation, the PSF is a bell-shaped function; its
full width at half maximum (FWHM) is called the `seeing' of the
image. At excellent sites, and excellent telescopes, the seeing has a
median of $\sim 0\arcsecf 7$--$\sim 0\arcsecf 8$; exceptionally,
images with a seeing of $\sim 0\arcsecf 5$ can be obtained. 
Recall that typical faint galaxies are considerably smaller than this
seeing size, hence their appearance is dominated by the PSF.

The main effect of seeing on image shapes is that it makes an elliptical
source rounder: a small source with a large ellipticity will nevertheless
appear as a fairly round image if its size is considerably smaller than the
PSF. If not properly corrected for, this smearing effect would lead to a
serious underestimate of ellipticities, and thus of the shear estimates.
Furthermore, the PSF is not fully isotropic; small anisotropies can be
introduced by guiding errors, the coaddition, the telescope optics, bad
focusing etc. An anisotropic PSF makes round sources elliptical, and
therefore mimics a shear. Also here, the effect of the PSF anisotropy depends
on the image size and is strongest for the smallest sources.  PSF anisotropies
of several percent are typical; hence, if not corrected for, its effect can be
larger than the shear to be measured.

The PSF can be measured at the position of stars (point sources) on
the field; if it is a smooth function of position, it can be fitted by
a low-order polynomial, which then yields a model for the PSF at all
points, in particular at every image position, and one can correct for
the effects of the PSF. A potential problem occurs if the PSF jumps
between chips boundaries in multi-chip cameras, since then the
coaddition produces PSF jumps on the coadded frame; this happens in
cameras where the chips are not sufficiently planar, and thus not in
focus simultaneously. For the WFI@ESO/MPG 2.2-m, this however is not a
problem, but for some other cameras this problem exists and is
severe. There is an obvious way to deal with that problem, namely to
coadd data only from the same CCD chip. In this case, the gaps between
chips cannot be closed in the coadded image, but for most weak lensing
purposes this is not a very serious issue. In order not to lose too
much area in this coaddition, the dither pattern, i.e., the pointing
differences in the individual exposures, should be kept small;
however, it should not be smaller than, say, $20''$, since otherwise
some pixels may always fall onto a few larger galaxies in the field,
which then causes problems in constructing a superflat. Furthermore,
small shifts between exposures means that the number of objects
falling onto different chips in different exposures is small, thus
reducing the accuracy of the astrometric solution. In any case, the
dither strategy shall be constructed for each camera individually,
taken into account its detailed properties.

\subsection{\llabel{WL-3.5}Shape measurements}
Specific software has been developed to deal with the issues mentioned
above; the one that is most in use currently has been developed by 
Kaiser et al.\ (1995; hereafter KSB), with substantial
additions by Luppino \& Kaiser (1997), and later modifications by
Hoekstra et al.\ (1998). The numerical implementation of this method
is called IMCAT and is publically available. The basic features of this
method shall be outlined next.

First one notes that the definition (\ref{eq:4.2}) of the second-order moments
of the image brightness is not very practical for applying it to real data. As
the effective range of integration depends on the surface brightness of the
image (through the weight function $q_I$) the presence of noise enters the
definition the $Q_{ij}$ in a non-linear fashion. Furthermore, neighboring
images can lead to very irregularly shaped integration ranges. In addition,
this definition is hampered by the discreteness of pixels. For these reasons,
the definition is modified by introducing a weight function
$q_\theta(\vc\theta)$ which depends explicitly on the image coordinates,
\be
  Q_{ij} = \frac{
    \int\!\d^2\theta\,q_\theta(\vc\theta)\,I(\vc\theta)\,
    (\theta_i-\bar\theta_i)\,(\theta_j-\bar\theta_j)
  }{
    \int\!\d^2\theta\,q_\theta(\vc\theta)\,I(\vc\theta)
  }\;,\quad i,j\in \{1,2\}\;,
\elabel{Qmodidef}
\ee
where the size of the weight function $q_\theta$ is adapted to the size of the
galaxy image (for optimal S/N measurement). One typically chooses $q_\theta$
to be circular Gaussian. The image center $\bar\vc\theta$ is defined as
before, but also with the new weight function $q_\theta(\vc\theta)$, instead of
$q_I(I)$. However, with this definition, the transformation between image and
source brightness moments is no longer simple; in particular, the relation
(\ref{eq:4.5}) between the second-order brightness moments of source and image
no longer holds. The explicit spatial dependence of the weight, introduced for
very good practical reasons, destroys the convenient relations that we derived
earlier -- welcome to reality.

In KSB, the anisotropy of the PSF is characterized by its (complex)
ellipticity $q$, measured at the positions of the stars, and fitted by a
low-order polynomial.  Assume that the (reduced) shear $g$ and the PSF
anisotropy $q$ are small; then they both will have a small effect on the
measured ellipticity. Linearizing these two effects, one can write (employing
the Einstein summation convention)
\be
  \hat{\chi}^\mathrm{obs}_\alpha = \chi^0_\alpha
  +P^\mathrm{sm}_{\alpha\beta} q_\beta+P^g_{\alpha\beta}g_\beta  \;.
\elabel{4.85}
\end{equation}
The interpretation of the various terms is found as follows: First
consider an image in the absence of shear and the case of an isotropic
PSF; then $\hat{\chi}^\mathrm{obs} = \chi^0$; thus, $\chi^0$ is the
image ellipticity one would obtain for $q=0$ and $g=0$; it is the
source smeared by an isotropic PSF.  It is important to note that
${\rm E}({\chi}^0)=0$, due to the random orientation of sources.
The tensor $P^{\rm sm}$ describes how the image ellipticity responds
to the presence of a PSF anisotropy; similarly, the tensor $P^g$
describes the response of the image ellipticity to shear in the
presence of smearing by the seeing disk.  Both, $P^\mathrm{sm}$ and
$P^g$ have to be calculated for each image individually; they depend
on higher-order moments of the brightness distribution and the size of
the PSF. A full derivation of the explicit equations can be found in
Sect.\ 4.6.2 of BS01.

Given that $\ave{\chi^0}=0$, an estimate of the (reduced) shear is
provided by 
\be
\eps=(P^g)^{-1}\rund{\hat{\chi}^\mathrm{obs}-P^\mathrm{sm} q} \;.
\elabel{KSB-shearest} 
\ee 
If the source size is much smaller than the PSF,
the magnitude of $P^g$ can be very small, i.e., the correction factor
in (\ref{eq:KSB-shearest}) can be very large. Given that the measured
ellipticity $\hat \chi^{\rm obs}$ is affected by noise, this noise
then also gets multiplied by a large factor. Therefore, depending on
the magnitude of $P^g$, the error of the shear estimates differ
between images; this can be accounted for by specifically weighting
these estimates when using them for statistical purposes (e.g., in the
estimate of the mean shear in a given region). Different authors use
different weighting schemes when applying KSB. Also, the tensors
$P^{\rm sm}$ and $P^g$ are expected to depend mainly on the size of
the image and their signal-to-noise; therefore, it is advantageous to
average these tensors over images having the same size and S/N,
instead of using the individual tensor values which are of course also
affected by noise. Erben et al.\ (2001) and Bacon et al.\ (2001) have
tested the KSB scheme on simulated data and in particular investigated
various schemes for weighting shear estimates and for determining the
tensors in (\ref{eq:4.85}); they concluded that simulated shear values
can be recovered with a systematic uncertainty of about 10\%.

Maybe by now you are confused -- what is `real ellipticity' of an image,
independent of weights etc.? Well, this question has no answer, since only
images with conformal elliptical isophotes have a `real ellipticity'. By the
way, not necessarily the one that is the outcome of the KSB procedure. The KSB
process does not aim toward measuring `the' ellipticity of any individual
galaxy image; it tries to measure `a' ellipticity which, when averaged over a
random intrinsic orientation of the source, yields an unbiased estimate of the
reduced shear.

Given that the shape measurements of faint galaxies and their
correction for PSF effects is central for weak lensing, several
different schemes for measuring shear have been developed (e.g.,
Valdes et al.\ 1983; Bonnet \& Mellier 1995; Kuijken 1999; Kaiser
2000; Refregier \cite{Refre03b}; Bernstein \& Jarvis 2002). In the
shapelet method of Refregier (2003b; see also Refregier \& Bacon
2003), the brightness distribution of galaxy images is expanded in a
set of basis functions (`shapelets') whose mathematical properties are
particularly convenient. With a corresponding decomposition of the PSF
(the shape of stars) into these shapelets and their low-order
polynomial fit across the image, a partial deconvolution of the
measured images becomes possible, using linear algebraic relations
between the shapelet coefficients. The effect of a shear on the
shapelet coefficients can be calculated, yielding then an estimate of
the reduced shear. In contrast to the KSB scheme, higher-order
brightness moments, and not just the quadrupoles, of the images are
used for the shear estimate.

These alternative methods for measuring image ellipticities (in the sense 
mentioned above, namely to provide an unbiased estimate of the local reduced
shear) have not been tested yet to the same extent as is true for the KSB
method. Before they become a standard in the field of weak lensing, several
groups need to independently apply these techniques to real and synthetic data
sets to evaluate their strengths and weaknesses. In this regard, 
one needs to note that weak lensing has, until recently, been
regarded by many researchers as a field where the observational results are
difficult to `believe' (and sure, not all colleagues have given up this view,
yet). The difficulty to display the directly measured quantities graphically
so that they can be directly `seen' makes it difficult to convince others
about the reliability of the measurements. The fact that the way from the
coadded imaging data to the final 
result is, except for the researchers who actually do the analysis, close to a
black box with hardly any opportunity to display intermediate results (which
would provide others with a quality check) implies that the methods employed
should be standardized and well checked. 

Surprisingly enough, there are very few (published) attempts where the same
data set is analyzed by several groups independently, and intermediate and
final results being compared. Kleinheinrich (2003) in her dissertation has
taken several subsets of the data that led to the deep image shown in
Fig.\ts\ref{fig:CDFS-deep} and compared the individual image ellipticities
between the various subsets. If the subsets had comparable seeing, the
measured ellipticities could be fairly well reproduced, with an rms difference
of about $0.15$, which is small compared to the dispersion of the image
ellipticities $\sigma_\eps\sim 0.35$. Hence, these differences, which
presumably are due to the different noise realizations on the different
images, are small compared to the `shape noise' coming from the finite
intrinsic ellipticities of galaxies. If the subsets had fairly different
seeing, the smearing correction turns out to lead to a systematic bias in the
measured ellipticities. From the size of this bias, the conclusions obtained
from the simulations are confirmed -- measuring a shear with better that $\sim
10\%$ accuracy will be difficult with the KSB method, where the main problem
lies in the smearing correction.

\subsubsection{Shear observations from space.}
We conclude this section with a few comments on weak lensing
observations from space. Since the PSF is the largest problem in shear
measurements, one might be tempted to use observations from space
which are not affected by the atmosphere. At present, the Hubble Space
Telescope (HST) is the only spacecraft that can be considered for this
purpose. Weak lensing observations have been carried out using two of
its instruments, WFPC2 and STIS. The former has a field-of-view of
about 5\ arcmin$^2$, whereas STIS has a field of $51''$. These small
fields imply that the number of stars that can be found on any given
exposure at high galactic latitude is very small, in fact typically
zero for STIS. Therefore, the PSF cannot be measured from these
exposures themselves. Given that an instrument in space is expected to
be much more stable than one on the ground, one might expect that the
PSF is stable in time; then, it can be investigated by analyzing
exposures which contain many stars (e.g., from a star cluster). In
fact, Hoekstra et al.\ (1998) and H\"ammerle et al.\ (2002) have shown
that the PSFs of WFPC2 and STIS are approximately constant in
time. The situation is improved with the new camera ACS onboard HST,
where the field size of $\sim 3\arcminf4$ is large enough to contain about a
dozen stars even for high galactic latitude, and where some control
over the PSF behavior on individual images is obtained. We shall
discuss the PSF stability of the ACS in Sect.\ts \ref{sc:WL-7.3}
below. 

The PSF of a diffraction-limited telescope is much more complex than
that of the seeing-dominated one for ground-based observations. The
assumption underlying the KSB method, namely that the PSF can be
described by a axi-symmetric function convolved with a small
anisotropic kernel, is strongly violated for the HST PSF; it is
therefore less obvious how well the shear measurements with the KSB
method work in space. In addition, the HST PSF in not well sampled
with the current imaging instruments, even though STIS and ACS have a
pixel scale of $0\arcsecf 05$. The number density of cosmic rays is
much larger in space, so their removal can be more cumbersome than for
ground-based observations. The intense particle bombardment also leads
to aging of the CCD, which lose their sensitivity and attain
charge-transfer efficiency problems.  Despite these potential
problems, a number of highly interesting weak lensing results obtained
with the HST have been reported, in particular on clusters, and we
shall discuss some of them in later sections. The new Advanced Camera
for Surveys (ACS) on-board HST has a considerably larger field-of-view
than previous instruments and will most likely become a highly
valuable tool for weak lensing studies.

\section{\llabel{WL-4}Clusters of galaxies: Introduction, 
and strong lensing}
\subsection{\llabel{WL-4.1}Introduction}
Galaxies are not distributed randomly, but they cluster together,
forming groups 
and clusters of galaxies. Those can be identified as overdensities of galaxies
projected onto the sky, and this has of course been the original method for
the detection of clusters, e.g., leading to the famous and still heavily used
Abell (1958) catalog and its later Southern extension (Abell et al.\ 1989;
ACO).  Only later -- with the exception of Zwicky's early insight 9n
1933 that
the Coma cluster must contain a lot of missing mass --
it was realized that the visible galaxies are but a minor
contribution to the clusters since they are dominated by dark matter. From
X-ray observations we know that clusters contain a very hot intracluster gas
which emits via free-free and atomic line radiation. Many galaxies are
members of a cluster or 
a group; indeed, the Milky Way is one of them, being one of two luminous
galaxies of the Local Group (the other one is M31, the Andromeda galaxy), of
which $\sim 35$ member galaxies are known, most of them dwarfs.

In the first part of this section we shall describe general properties
of galaxy clusters, in particular methods to determine their masses,
before turning to their strong lensing properties, such as show up in
the spectacular giant luminous arcs. Very useful reviews on clusters
of galaxies are from Sarazin (1986) and in a recent proceedings volume
(Mulchaey et al.\ 2004).

\subsection{\llabel{WL-4.2}General properties of clusters}
Clusters of galaxies contain tens to hundreds of bright galaxies;
their galaxy population is dominated by early-type galaxies (E's and
S0's), i.e.\ galaxies without active star formation.  Often a very
massive cD galaxy is located at their center; these galaxies differ
from normal ellipticals in that they have a much more extended
brightness profile -- they are the largest galaxies. The morphology of
clusters as seen in their distribution of galaxies can vary a lot,
from regular, compact clusters (often dominated by a central cD galaxy)
to a bimodal distribution, or highly irregular morphologies with
strong substructure. Since clusters are at the top of the mass scale
of virialized objects, the hierarchical merging scenario of structure
growth predicts that many of them have formed only recently through
the merging of two or more lower-mass sub-clusters, and so the
irregular morphology just indicates that this happened.

X-ray observations reveal the presence of a hot (several keV)
intracluster medium (ICM) which is highly enriched in heavy elements;
hence, this gas has been processed through star-formation cycles in
galaxies. The mass of the ICM surpasses that of the baryons in
the cluster galaxies; the mass balance in clusters is approximately as
follows: stars in cluster galaxies contribute $\sim 3\%$ of the total
mass, the ICM another $\sim 15\%$, and the rest ($\gtrsim 80\%$) is
dark matter.  Hence, clusters are dominated by dark matter; as
discussed below (Sect.\ \ref{sc:WL-4.3}), the mass of clusters can be
determined with three vastly different methods which overall yield
consistent results, leadding to the aforementioned mass ratio.

We shall now quote a few characteristic values which apply to rich,
massive clusters. Their virial radius, i.e., the radius inside of
which the mass distribution is in approximate virial equilibrium (or
the radius inside of which the mean mass density of clusters is $\sim
200$ times the critical density of the Universe -- cf.\ Sect.\ 4.5 of
IN) is $r_{\rm vir}\sim 1.5\,h^{-1}\,{\rm Mpc}$. A typical value for
the one-dimensional velocity dispersion of the member galaxies is
$\sigma_v\sim 1000\,{\rm km/s}$. In equilibrium, this equals the
thermal velocity of the ICM, corresponding to a temperature of $T\sim
10^{7.5}\, {\rm K}\sim 3\,{\rm keV}$.  The mass of massive clusters
within the virial radius (i.e., the {\it virial mass}) is $\sim
10^{15}M_\odot$. The mass-to-light ratio of clusters (as measured from
the B-band luminosity) is typically of order $(M/L)\sim 300
h^{-1}\,(M_\odot/L_\odot)$.  Of course, the much more numerous typical
clusters have smaller masses (and temperatures).

\subsubsection{Cosmological interest for clusters.}
Clusters are the most massive bound and virialized structures in the
Universe; this, 
together with the (related) fact that their dynamical time scale
(e.g., the crossing time $\sim r_{\rm vir}/\sigma_v$) is not much
smaller than the Hubble time $H_0^{-1}$ -- so that they retain a
`memory' of their formation -- render them of particular interest for
cosmologists.  The evolution of their abundance, i.e., their comoving
number density as a function of mass and redshift, is an important
probe for cosmological models and traces the growth of structure;
massive clusters are expected to be much rarer at high redshift than
today.  Their present-day abundance provides one of the measures for
the normalization of the power spectrum of cosmological density
fluctuations.  Furthermore, they form (highly biased) signposts of the
dark matter distribution in the Universe, so their spatial
distribution traces the large-scale mass distribution in the Universe.
Clusters act as laboratories for studying the evolution of galaxies
and baryons in the Universe. Since the galaxy number density is
highest in clusters, mergers of their member galaxies and, more
importantly, other
interactions between them occur frequently. Therefore, the evolution
of galaxies with redshift is most easily studied in clusters. For
example, the Butcher--Oemler effect (the fact that the fraction of
blue galaxies in clusters is larger at higher redshifts than today) is
a clear sign of galaxy evolution which indicates that star formation
in galaxies is suppressed once they have become cluster members. More
generally, there exists a density-morphology relation for galaxies,
with an increasing fraction of early-types with increasing spatial
number density, with clusters being on the extreme for the latter.
Finally, clusters were (arguably) the first objects for which the
presence of dark matter has been concluded (by Zwicky in 1933). Since
they are so large, and present the gravitational collapse of a region
in space with initial comoving radius of $\sim 8 h^{-1}\,{\rm Mpc}$,
one expects that their mixture of baryonic and dark matter is
characteristic for the mean mass fraction in the Universe (White et
al.\ 1993). With the baryon fraction of $\sim 15\%$ mentioned above,
and the density parameter in baryons determined from big-bang
nucleosynthesis in connection to the determination of the deuterium
abundance in Ly$\alpha$ QSO absorption systems, $\Omega_{\rm b}\approx
0.02 h^{-2}$, one obtains a density parameter for matter of
$\Omega_{\rm m}\sim 0.3$, in agreement with results from other
methods, most noticibly from the recent WMAP CMB measurements (e.g.,
Spergel et al.\ 2003).

\subsection{\llabel{WL-4.3}The mass of galaxy clusters}
Cosmologists can predict the abundance of clusters as a function of
their mass (e.g., using numerical simulations); however, the mass of a
cluster is not directly observable, but only its luminosity, or the
temperature of the X-ray emitting intra-cluster medium. Therefore, in
order to compare observed clusters with the cosmological predictions,
one needs a way to determine their masses.  Three principal methods
for determining the mass of galaxy clusters are in use: 
\bi
\item
Assuming virial equilibrium, the observed velocity distribution
of galaxies in clusters can be converted into a mass estimate,
employing the virial theorem; this
method typically requires assumptions about the statistical
distribution of the anisotropy of the galaxy orbits.
\item
The hot intra-cluster gas, as visible through its Bremsstrahlung in
X-rays, traces the gravitational potential of the cluster. Under
certain assumptions (see below), the mass profile can be constructed
from the X-ray emission.
\item
Weak and strong gravitational lensing probes the projected mass
profile of clusters, with strong lensing confined to the central regions of
clusters, whereas weak lensing can yield mass measurements for larger radii. 
\ei
All three methods are complementary; lensing yields the line-of-sight
projected density of clusters, in contrast to the other two methods
which probe the mass inside spheres.
On the other hand, those rely on equilibrium (and symmetry)
conditions; e.g., the virial method assumes virial equilibrium
(that the cluster is dynamically relaxed) and the degree of anisotropy of the
galaxy orbit distribution.

\subsubsection{Dynamical mass estimates.}
Estimating the mass of clusters based on the virial theorem, 
\be 
2 E_{\rm
  kin}+E_{\rm pot}=0\;, 
\elabel{viritheo} 
\ee 
has been the traditional
method, employed by Zwicky in 1933 to find strong hints for the presence of
dark matter in the Coma cluster.  The specific kinetic energy of a galaxy is
$v^2/2$, whereas the potential energy is determined by the cluster mass
profile, which can thus be determined using (\ref{eq:viritheo}). One should
note that only the line-of-sight component of the galaxy velocities can be
measured; hence, in order to derive the specific kinetic energy of galaxies,
one needs to make an assumption on the distribution of orbit anisotropies in
the cluster potential. Assuming an isotropic distribution of orbits, the
l.o.s. velocity distribution can then be related to the 3-D velocity
dispersion, which in turn can be transformed into a mass estimate if spherical
symmetry is assumed.  This method requires many redshifts for an accurate mass
estimate, which are available only for a few clusters. However, a revival of
this method is expected and already seen by now, owing to the new
high-multiplex optical spectrographs.

\subsubsection{X-ray mass determination of clusters.}
The intracluster gas emits via Bremsstrahlung; the emissivity depends on the
gas density and temperature, and, at lower $T$, also on its chemical
composition, since at $T\lesssim 1\,{\rm keV}$ the line radiation from highly
ionized atomic species starts to dominate the total emissivity of a hot gas.
Investigating the properties of the ICM with X-ray observations have revealed
a wealth of information on the properties of clusters (see Sarazin 1986).
Assuming that the gas is in hydrostatic equilibrium in the potential well of
the cluster, the gas pressure $P$ must balance gravity, or
\[
\nabla P=-\rho_{\rm g}\,\nabla \Phi \;,
\]
where $\rho_{\rm g}$ is the gas density.
In the case of spherical symmetry, this becomes
\[
{1\over \rho_{\rm g}}\,{\d P\over \d r}=-{\d\Phi\over \d r}
=-{G\,M(r)\over r^2}\;.
\]
From the X-ray brightness profile and temperature measurement, $M(r)$,
the mass inside $r$, both dark and luminous, can then be determined,
\be 
M(r)=-{\kB T r^2\over G\mu m_{\rm p}}
\rund{{\d\ln\rho_{\rm g}\over \d r} 
+ {\d\ln T\over \d r}}  \;,
\elabel{6.37}
\ee
where $\mu m_{\rm p}$ is the mean particle mass in the gas.
Only for relatively few clusters are detailed X-ray brightness and
temperature profile measurements available. In the absence of a
temperature profile measurement, one often assumes that $T$ does not
vary with distance form the cluster center. In this case,
assuming that the dark matter particles also have an isothermal
distribution (with velocity traced by the galaxy velocities),
one can show that
\be
\rho_{\rm g}(r)\propto [\rho_{\rm tot}(r)]^\beta\; ;\quad
{\rm with}\quad \beta={\mu m_{\rm p}\sigma_v^2\over k_{\rm B}T_{\rm
g}}\;.
\elabel{betadef}
\ee
Hence, $\beta$ is the ratio between kinetic and thermal energy. The
mass profile corresponding to the isothermality assumption follows
from the Lame--Emden equation which, however,
has no closed-form solution.  In the King approximation, the density
and X-ray brightness profile (which is obtained by a line-of-sight
integral at projected distance $R$ from the
cluster center over the emissivity, which in turn is proportional to the
square of the electron density, or $\propto \rho_{\rm g}^2$, for an
isothermal gas) become
\[
\rho_{\rm g}(r)=\rho_{\rm g0}\eck{1+\rund{r\over r_{\rm
c}}^2}^{-3\beta/2}\;; \quad
I(R)\propto \eck{1+\rund{R\over r_{\rm
c}}^2}^{-3\beta/2+1/2}
\]
where $r_{\rm c}$ is the core radius. The observed brightness profile
can now be fitted with these $\beta$-models, yielding estimates of
$\beta$ and $r_{\rm c}$ from which the cluster mass follows.
Typical values for $r_{\rm c}$ range from $0.1$ to $0.3 h^{-1}\,{\rm
Mpc}$; and $\beta=\beta_{\rm fit}\sim 0.65$. On the other hand, one
can determine $\beta$ from the temperature $T$ and the galaxy velocity
dispersion using (\ref{eq:betadef}), which yields $\beta_{\rm
spec}\approx 1$. The discrepancy between these two estimates of
$\beta$ is not well understood and probably indicates that one of
assumptions underlying this `$\beta$-models' fails in many clusters, which is
not too surprising (see below). 

The hot ICM loses energy through its thermal radiation; the cooling
time $t_{\rm cool}$ of the gas, i.e., the ratio between the thermal
energy density and the X-ray emissivity, is larger than the Hubble
time $\sim H_0^{-1}$ for all but the innermost regions. In the center
of clusters, the gas density can be high enough to have $t_{\rm
cool}<H_0^{-1}$, so that there the gas can no longer be in hydrostatic
equilibrium. One expects that the gas flows towards the cluster
center, thereby being compressed and therefore maintain approximate
pressure balance. Such `cooling flows' (see, e.g., Fabian 1994) are
observed indirectly, through highly peaked X-ray emission in cluster
centers which indicates a strong increase of the gas density;
furthermore, these cooling-flow clusters show a decrease of $T$
towards the center. The mass-flow rate in these clusters can be as
high as $100 M_\odot\,{\rm yr}^{-1}$ or even more, so that the total
cooled mass can be larger than the baryonic mass of a massive
galaxy. However, the fate of the cooled gas is unknown. 

\subsubsection{New results from Chandra \& XMM.}
The two X-ray satellites Chandra and XMM, launched in 1999, have
greatly increased our view of the X-ray Universe, and have led to a
number of surprising results about clusters. X-ray spectroscopy
verified the presence of cool gas near the center of cooling-flow
clusters, but no indication for gas with temperature below $\sim
1\,{\rm keV}$ has been seen, whereas the cooling is expected to
rapidly proceed to very low temperatures, as the cooling function
increases for lower $T$ where atomic transitions become increasingly
important. Furthermore, the new observations have revealed that at
least the inner regions of clusters often show a considerably more
complicated structure than implied by hydrostatic equilibrium. In some
cases, the intracluster medium is obviously affected by a central AGN,
which produces additional energy and entropy input, which might
explain why no sub-keV gas has been detected. As the AGN activity of a galaxy
may be switched on and off, depending on the fueling of the central black
hole, even in clusters without a currently active AGN such heating might have
occurred in the recent past, as indicated in some cases by radio relics. 
Cold fronts with very
sharp edges (discontinuities in density and temperature, but such that
$P\propto \rho T$ is approximately constant across the front), and shocks
have been discovered, most likely showing ongoing or recent merger
events. In many clusters, the temperature and metalicity appears to
be strongly varying functions of position which invalidates the
assumption of isothermality underlying the $\beta$-model.  Therefore,
mass estimates of central parts of clusters from X-ray observations
require special care, and one  needs to revise the simplified
models used in the pre-Chandra era. In fact, has there ever been the believe
that the $\beta$-model provides an adequate description of the gas in a
cluster, the results from Chandra and XMM show that this is unjustified. 
The physics of the intracluster gas appears to be considerably more
complicated than that.

\subsection{\llabel{WL-4.4}Luminous arcs \& multiple images}
Strong lensing effects in cluster show up in the form of giant luminous arcs,
strongly distorted arclets, and multiple images of background galaxies. Since
strong lensing only occurs in the central part of clusters, it can be used
only to probe their inner mass structure. However, strong lensing yields by
far the most accurate central mass determinations in those cases where several
strong lensing features can be identified. For a detailed account of strong
lensing in clusters, the reader is referred to the review by Fort \& Mellier
(1994).

Furthermore, clusters thus act as a `natural telescope'; many of the
most distant galaxies have been found by searching behind clusters,
employing the lensing magnification. For example, the recently
discovered very high redshift galaxies at $z\approx 7$ (Kneib et al.\
2004) and $z=10$ (Pell\'o et al.\ 2004) were found through a search in
the direction of the high-magnification region in the clusters A2218
and A1835, respectively. In the first of these two cases, the multiple
imaging of the background galaxy provides not only the magnification,
but also an estimate of the redshift of the source (which is not
determined by any spectral line), whereas in the latter case, only the
implied high magnification makes the source visible on deep HST images
and allows its spectroscopy, yielding a spectral line which most
likely is due to Ly$\alpha$. The magnification is indeed a very
important asset, as can be seen from a simple example: a value of
$\mu=5$ reduces the observing time for obtaining a spectrum by a
factor 25 (in the case where the noise is sky background dominated) --
which is the difference of being doable or not. Recognizing the power
of natural telescopes, the deepest SCUBA surveys for faint
sub-millimeter sources have been conducted (e.g., Blain et al.\ 1999)
around clusters with well-constrained (from lensing) mass distribution
to reach further down the (unlensed) flux scale.

\begin{figure}
\bc
\includegraphics[width=11.7cm]{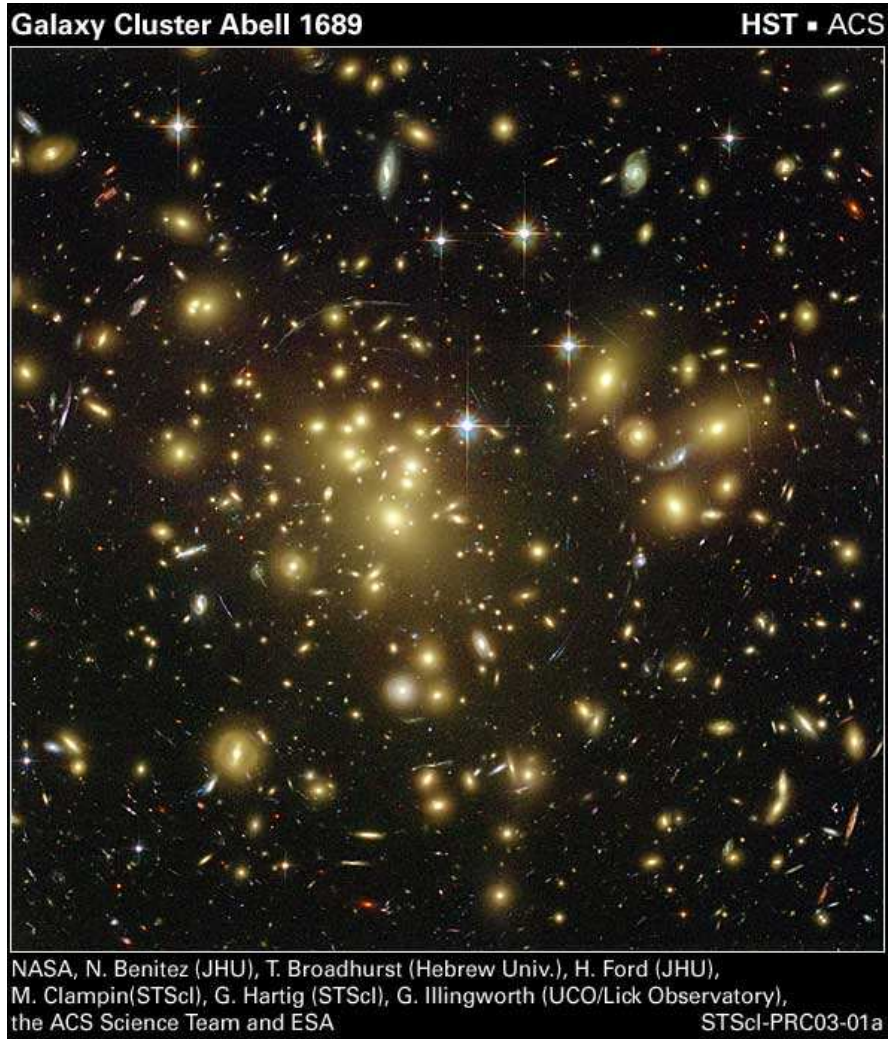}
\ec
\caption{The galaxy cluster Abell 1689 is the most impressive lensing
cluster yet found. This image has been taken with the new Advanced
Camera for Surveys (ACS) onboard HST. Numerous arcs are seen. A simple
estimate for the mass of the center of the cluster, obtained by
identifying the arcs radius with the Einstein radius, yields an
extremely large equivalent velocity dispersion. The distribution of
the arcs shown here indicates that such a simple assumption is
misleading, and more detailed modeling required}
\flabel{A1689ACS}
\end{figure}

\subsubsection{First go: $M(\le \theta_{\rm E})$.}
Giant arcs occur where the distortion (and magnification) is very
large, that is near critical curves.  To a first approximation,
assuming a spherical mass distribution, the location of the arc from
the cluster center (which usually is assumed to coincide with the
brightest cluster galaxy) yields the Einstein radius of the cluster,
so that the mass estimate (see IN, Eq.\ 43) can be applied.  
\be
M(\theta_{\rm arc})\approx \pi\,(D^{\rm ang}_{\rm d}\,\theta_{\rm
arc})^2\,\Sigma_{\rm cr}\;.
\ee
Therefore, this simple estimate yields the mass inside the arc radius.
However, this estimate not very accurate, perhaps good to within $\sim 30\%$
(Bartelmann \& Steinmetz 1996). Its reliability depends on the level of asymmetry and
substructure in the cluster mass distribution. Furthermore, it is likely to
overestimate the mass in the mean, since arcs preferentially occur along the
major axis of clusters. Of course, the method is very difficult to apply if
the center of the cluster is not readily identified or if the cluster is
obviously bimodal.  For these reasons, this simple method for mass estimates
is not regarded as particularly accurate.

\subsubsection{Detailed modeling.}
The mass determination in cluster centers becomes much more accurate
if several arcs and/or multiple images are present, since in this
case, detailed modeling can be done. This typically proceeds in an
interactive way: First, multiple images have to be identified (based
on their colors and/or detailed morphology, as available with HST
imaging).  Simple (plausible) mass models are then assumed, with
parameters fixed by matching the multiple images, and requiring the
distortion at the arc location(s) to be strong and to have the correct
orientation.  This model then predicts the presence of possible
further multiple images; they can be checked for through morphology,
surface brightness (in particular if HST images of the cluster are
available) and color. If confirmed, a new, refined model is
constructed including these new additional strong lensing constraints,
which yields further strong lensing predictions etc.  As is the case
for galaxy lensing (see SL), the components of the mass models are not
arbitrary, but chosen to be physically motivated. Typically, as major
component a ellipsoidal isothermal or NWF distribution is used to
describe the overall mass distribution of the cluster. Refinements of
the mass distribution are introduced as mass components centered on
bright cluster member galaxies or on subgroups of such galaxies,
describing massive subhalos which survived a previous merger.  Such
models have predictive power and can be trusted in quite some detail;
the accuracy of mass estimates in some favorable cases can be as high
as a few percent.

\begin{figure}
\bmi{6.5}
\includegraphics[width=6.5cm]{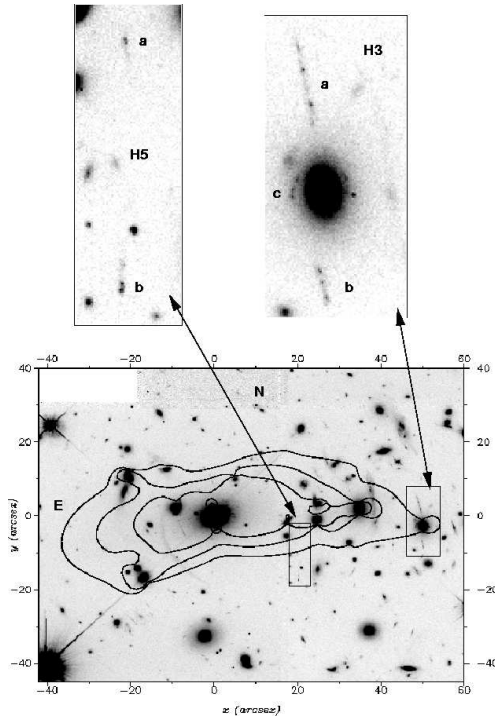}
\emi
\bmi{5.2}
\caption{The lower panel shows the critical curves of the cluster
A2390 (cluster redshift $z_{\rm d}=0.231$), for three different source
redshifts of $z_{\rm s}=1$, 2.5 and 4 (from inner to outer). The lens
model is based on the detailed HST image shown here.  Identified are
two sets of multiple images, shown in the upper two panels, which
obviously need to be at very high redshift. Indeed, spectroscopy shows
that they have $z_{\rm s}=4.04$ and $z_{\rm s}=4.05$ (from Pell\'o et
al.\ 1999)} 
\flabel{A2390-pello} 
\emi
\end{figure}

In fact, these models can be used to predict the redshift of arcs and
arclets. As an example, we mention the strong lensing analysis of the
cluster Abell 2390 based on HST imaging (Pell\'o et al.\ 1999).
Two pairs of multiple images were identified (see
Fig.\ts\ref{fig:A2390-pello}) which then implies that the critical
curve has to pass between the individual components. The location of
the critical curves depends, however, on the source redshift. As shown
in the figure, the sources have to be at a high redshift in order for
the corresponding critical curves to have the correct location. In
fact, spectroscopy placed the two sources at $z_{\rm s}=4.04$ and
$z_{\rm s}=4.05$, as predicted by the lens model.

Since the distortion of a lens also depends on the source redshift,
once a detailed mass model is available from arcs with known
redshifts for at least some of them, one can estimate the value of 
the lens strength $\propto D_{\rm ds}/D_{\rm s}$ and thus infer the
redshift of arclets. This method has been successfully applied to HST
observations of clusters (Ebbels et al.\ 1998). Of course, having
spectroscopic redshifts of the arcs available increases the
calibration of the mass models; they are therefore very useful.

\subsubsection{Lens properties from Fourier transforms.}  
Before discussing results from these detailed models, a brief
technical section shall be placed here, related to calculating lens
properties of general mass distributions.  A general method to obtain
the lensing quantities of a mass distribution is through Fourier
transformation.  We assume that we have a mass distribution of finite
mass; this is not a serious restriction even for models with formally
infinite total mass, because we can truncate them on large scales,
thus making the total mass finite, without affected any lensing
properties at smaller scales. We define the Fourier transform
$\hat\kappa(\vc \ell)$ of the surface mass density as\footnote{We
denote the Fourier variable of three-dimensional space as $\vc k$,
that of angular position by $\vc\ell$.}
\be
\hat\kappa(\vc \ell)=\int_{\Real^2}\d^2 \theta\;\kappa(\vc \theta)\,
\exp\rund{\Rm i \vc \ell \cdot \vc \theta}\; ,
\elabel{FTkappa}
\ee
and its inverse by
\be
\kappa(\vc \theta)={1\over (2\pi)^2}\int_{\Real^2}\d^2\ell\;
\hat\kappa(\vc \ell) \exp\rund{-\Rm i \vc \ell \cdot \vc \theta}\; .
\elabel{FTinvkappa}
\ee
Similarly, we define the Fourier transforms of the deflection
potential, $\hat\psi(\vc \ell)$, of the deflection angle,
$\hat{\vc\alpha}(\vc \ell)$, and of the complex shear, $\hat\gamma(\vc
\ell)$. Differentiation by $\theta_i$ in real space is replaced by
multiplication by $-\Rm i \ell_i$ in Fourier space. Therefore, the Fourier
transform of $\partial\psi/\partial \theta_j$ is $-\Rm i \ell_j
\hat\psi(\vc \ell)$. Hence, the Poisson equation as given in Sect.\ts
2.2 of IN becomes in Fourier space
\be
-|\vc\ell|^2\hat\psi(\rm\ell)=2\hat\kappa(\vc\ell) \;.
\elabel{FTPoisson}
\ee
Thus, for $\vc \ell\ne \vc 0$, the Fourier transform of the potential
which satisfies the  Poisson equation can be readily determined. The
$\vc \ell=\vc 0$ mode remains undetermined; however, since this mode
corresponds to a constant in $\psi$, it is unimportant and can be set
to zero. Once $\hat\psi$ is determined, the Fourier transform of the
deflection angle and the shear follows from their definitions in terms
of the deflection potential, given in Sect.\ts 2.2 of IN,  
\be
\hat{\vc\alpha}(\vc \ell)= -\Rm i \vc \ell \hat\psi(\vc \ell)\;,
\elabel{FTalpha}
\ee
\be
\hat\gamma(\vc \ell)=-\rund{ {\ell_1^2-\ell_2^2\over 2}+{\rm
i}\ell_1\ell_2}  \hat\psi(\vc \ell)\;.
\elabel{FTgamma}
\ee
Thus, in principle, one determines the relevant quantities by
Fourier transforming $\kappa$, then calculating the Fourier transforms
of the potential, deflection, and shear, whose real-space counterparts
are then obtained from an inverse Fourier transform, like in
(\ref{eq:FTinvkappa}). 

Up to now we have not gained anything; the Fourier transforms as
defined above are two-dimensional integrals, as are the real-space
relations between deflection angle and shear, and the
surface-mass density. However, provided $\kappa$ becomes `small
enough' for large values of $\abs{\vc \theta}$, the integral in
(\ref{eq:FTkappa}) may be approximated by one over a finite region in
$\vc \theta$-space. This finite integral is further approximated as a
sum over gridpoints, with a regular grid covering the lens
plane. Consider a square in the lens plane of side $L$, and let $N$ be
the number of gridpoints per dimension, so that $\Delta \theta=L/N$ is
the size of a gridcell. The inverse grid, i.e., the $\vc \ell$-grid,
has a gridcell of size $\Delta \ell=2\pi/L$. The discrete Fourier
transform then uses the values of $\kappa$ on the $\vc \theta$-grid to
calculate $\hat\kappa$ on the $\vc \ell$-grid. The latter, in fact, is
then the Fourier transform of the periodic continuation of the mass
distribution in $\vc \theta$-space. Because of this periodic
continuation, the deflection angle as calculated from the discrete
Fourier transform, which is performed by the Fast Fourier Transform
(FFT) method, is the sum of the input
mass distribution, plus all of its periodic continuation. Here,
finally, is why we have considered the Fourier method: the FFT is a very
efficient and quick procedure (see, e.g., Press et al.\ 1992), and
arguably the best one in cases of mass distributions for which no
analytical progress can be made. The lensing properties are calculated
on a grid; if needed, they can be obtained for other points by interpolation.

Because of the periodic continuation, the mass distribution has to
decreases sufficiently quickly for large $\abs{\vc \theta}$, or 
be truncated at large radii. In any case, $L$ should be taken
sufficiently large to minimize these periodicity effects.

Another point to mention is that a periodic mass distribution, each
element of which has positive total mass, has an infinite mass, so
that the deflection potential has to diverge; on the other hand, the
deflection potential is enforced to be periodic. This apparent
contradiction can be resolved by noting that the $\vc \ell=\vc 0$ mode of
$\hat\kappa$ is not used in the calculation of $\hat{\vc\alpha}$ and
$\hat\gamma$. Indeed, if $\hat\psi$ and $\psi$ are calculated from the
above equations, then the resulting $\psi$ {\em does not satisfy the
Poisson equation}; the $\psi$ resulting from this procedure is the one
corresponding to $\kappa-\bar\kappa$, where $\bar\kappa$ is the
average of $\kappa$ on the $\vc \theta$-grid. A similar remark is true for
the deflection angle. Thus, at the end, one has to add a term
$\bar\kappa\abs{\vc \theta}^2/2$ to $\psi$, and a term $\bar\kappa
\vc \theta$ to $\vc\alpha$. 

Since the FFT is very fast, one can choose $N$ and $L$ large, and
then consider only the central part of the $\vc \theta$-grid needed
for the actual lens modeling.

\subsection{Results from strong lensing in clusters}
The main results of the strong lensing investigations of clusters can
be summarized as follows:
\bi
\item
The mass in cluster centers is much more concentrated than predicted by
(simple) models based on X-ray observations. The latter usually predict a
relatively large core of the mass distribution. These large cores
would render clusters sub-critical to lensing, i.e., they would be
unable to produce giant arcs or multiple images. In fact, when arcs
were first discovered they came as a big surprise because of these
expectations. By now we know that the intracluster medium is much more
complicated than assumed in these `$\beta$-model' fits for the X-ray
emission. 
\item
The mass distribution in the inner region of clusters often shows strong
substructure, or multiple mass peaks. These are also seen in the
galaxy distribution of clusters, but with the arcs can be verified to
also correspond to mass peaks (examples of this include the cluster
Abell\ 2218 where arcs also curve around a secondary concentration of
bright galaxies, clearly indicating the presence of a mass
concentration, or the obviously bimodal cluster A\ 370). These are
easily understood in the frame of hierarchical mergers in a CDM model;
the merged clusters retain their multiple peaks for a dynamical time
or even longer, and are therefore not in virial equilibrium.
\item
The orientation of the (dark) matter appears to follow closely
the orientation of the light in the cD galaxy; this supports the idea that
the growth of the cD galaxy is related to the cluster as a whole,
through repeated accretion of lower-mass member galaxies. In that
case, the cD galaxy `knows' the orientation of the cluster.
\item
There is in general good agreement between lensing and X-ray mass
estimates (e.g., Ettori \& Lombardi 2003; Donahue et al.\ 2003)
for those clusters where a `cooling flow' indicates that they
are in dynamical equilibrium, provided the X-ray analysis takes the
presence of the cooling flow into account (Allen 1998).
\ei
Probably our `favourate' clusters in which strong lensing effects are
investigated in detail are biased in favor of having strong
substructure, as this increases the lensing cross section for the
occurrence of giant arcs (see below). Hence, it may be
that the most detailed results obtained from strong lensing in
clusters apply to a class of clusters which are especially selected
because of their ability to produce spectacular arcs, and thus of
their asymmetric mass distribution. Therefore, one must be careful in
generalizing conclusions drawn from the `arc clusters' to the cluster
population as a whole. 

\subsubsection{Discrepancies.}
There are a few clusters where the lensing results and those obtained from
analyzing the X-ray observations or cluster dynamics are in stong apparent
conflict. Two of the most prominent ones shall be mentioned here. The cluster
A1689 (see Fig.\ts\ref{fig:A1689ACS}) has arcs more than $\sim 40''$ away from
the cluster center, which would imply a huge mass in this cluster center. 
This high mass is apparently confirmed by the high velocity dispersion of its
member galaxies, although their distribution in redshift makes it likely that
the cluster consists of several subcomponents (see Clowe \& Schneider 2001 for
a summary of these results). Several weak lensing results of this cluster have
been published, and they are not all in agreement: whereas Tyson \& Fischer
(1995) from weak shear, and Taylor et al.\ (1998) and Dye et al.\ (2001) from
the magnification method (that will be discussed in the next Section) find
also a very high mass for this cluster, the weak lensing analysis of Clowe \&
Schneider (2001; see also King et al.\ \cite{KCS-1689}), based on deep
wide-field imaging data of this cluster, finds a more moderate mass (or
equivalent velocity dispersion) for this cluster. A new XMM-Newton X-ray
observation of this cluster (Andersson \& Madejski 2004) lends support for the
smaller mass; in fact, their estimate of the virial mass of the cluster agrees
with that obtained by Clowe \& Schneider (2001). However, the disrepancy with
the strong lensing mass in the cluster center remains at present; a
quantitative analysis of the ACS data shown in Fig.\ts\ref{fig:A1689ACS} will
hopefully shed light on this issue.

A second clear example for discrepant results in the cluster Cl\ts
0024+17. It has a prominent arc system, indicating an Einstein radius of $\sim
30''$, and thus a high mass. The X-ray properties of this cluster, however,
indicate a much smaller mass (Soucail et al.\ 2000), roughly by a factor of
three. This discrepancy has been reaffirmed by recent Chandra observations,
which confirmed this factor-of-three problem (Ota et al.\ 2004). The
resolution of this discrepancy has probably been found by Czoske et al.\
(2001, 2002), who performed an extensive spectroscopic survey of cluster
galaxies. Their result is best interpreted such that Cl\ts 0024+17 presents a
merger of two clusters along our line-of-sight, which implies that the
measured velocity dispersion cannot be easily turned into a mass, as this
system is not in virial equailibrium, and that the X-ray data cannot be
converted to a mass either, due to the likely strong deviation from spherical
symmetry and equilibrium. A wide field sparsely sampled HST observation of
this cluster (Kneib et al.\ \cite{Kneib03-804}) also indicates the presence of
a second mass concentration about $3'$ away from the main peak.
As will be mentioned below, clusters undergoing
mergers have particularly high cross sections for producing arcs (Torri et
al.\ 2004); hence, our `favourites' are most likely selected for these
non-equilibrium clusters.

\subsubsection{Arc statistics.}
The abundance of arcs is expected to be a strong function of the
cosmological parameters: they not only determine the abundance of
massive clusters (through the mass function discussed in Sect.\ts 4.5
of IN), but also the degree of relaxation of clusters, which in turn
affects their strong lensing cross section (Bartelmann et al.\ 1998).
It is therefore interesting to consider the expected abundance of arcs
as a function of cosmological parameters and compare this to the
observed abundance. In a series of papers, M.\ Bartelmann and his
colleagues have studied the expected giant arc abundance, using
analytical as well as numerical techniques (e.g., Bartelmann \& Weiss 1994; 
Bartelmann et al.\ 1995, 1998, 2002; Meneghetti et al.\ 2004; see also 
Dalal et al.\ 2003; Oguri et al.\ 2003; Wambsganss et al.\ 2004). Some of the
findings of these studies can be summarized as follows:
\bi
\item 
The formation of arcs depends very sensitively on the deviation
from spherical symmetry and the detailed substructure of the mass
distribution in the cluster. For this reason, analytical models which
cannot describe this substructure with sufficient realism (see
Bergmann \& Petrosian 1993) do not provide realiable predictions for
the arc statistics (in particular, axisymmetric mass models are
essentially useless for estimating arc statistics),
and one needs to refer to numerical simulations of
structure formation. Since the substructure and triaxiality plays such
an important role, these simulations have to be of high spatial and
mass resolution.
\item 
The frequency of arcs depends of course on the abundance of clusters,
  which in turn depends on the cosmological model and the fluctuation
  spectrum of the matter, in particular its normalization
  $\sigma_8$. Furthermore, clusters at a given redshift have different
  mean ages in different cosmological models, as the history of
  structure growth, and thus the merging history, depends on
  $\Omega_{\rm m}$ and $\Omega_\Lambda$. Since the age of a cluster is
  one of the determining parameters for its level of substructure --
  younger clusters do not have had enough time to fully relax -- this
  affects the lensing cross section of the clusters for arc formation.
  In fact, during epochs of mergers, the arc cross-section can have
  temporary excursions by large factors. Even the same cluster at the
  same epoch can have arc forming cross sections that vary by more
  than an order-of-magnitude between different projection directions
  of the cluster.  For fixed cluster abundance today, low-density
  models form clusters earlier than high-density models.
\item 
Since the largest contribution of the total cross section for arc
  formation comes from clusters at intermediate redshift ($z\sim
  0.4$), also the equation-of-state of the dark energy matters; as
  shown in Meneghetti et al.\ (2004), what matters is the dark energy
  density at the epoch of cluster formation. In addition, the earlier
  clusters form, the higher their characteristic density, which then
  makes them more efficient lenses for arc formation.
\ei
Taking these effects together, a low-density open model produces a
larger number of arcs than a flat low-density model, which in turn has
more arcs than a high-density model, for a given cluster abundance
today. Whereas the differences between these models obtained by 
Meneghetti et al.\ (2004) are smaller than claimed in Bartelmann et
al.\ (1998), they in principle allow constraining the cosmological
parameters, provided they can be compared with the observed number of
arcs. 

Unfortunately, there are only a few systematic
studies of clusters with regards to their strong
lensing contents. Luppino et al.\ (1999) report on 8 giant arcs in their
sample of the 38 most massive clusters found in the Einstein Medium
Sensitivity Survey. Zaritsky \& Gonzalez (2003) surveyed clusters in the
redshift range $0.5\lesssim z\lesssim 0.7$ over $69\, {\rm deg}^2$ and found
two giant arcs with $R<21.5$ and a length $\theta_1>10''$. Gladders at al.\
(2003) found 5 arc candidates in their Red Cluster Sequence survey of
$90\,{\rm deg}^2$, all of them being associated with high-redshift
clusters. In contrast to the claim by Bartelmann et al.\ (1998), these
observered arc frequencies can be accounted for in a standard $\Lambda$CDM
Universe, as shown by Dalal et al.\ (2003). There are several differences
between these two studies, which are based on different assumptions about the
number density of clusters and the source redshift distribution, which Dalal
et al.\ (2003) took from the Hubble Deep Field, whereas Bartelmann et al.\
(1998) assumed all sources having $z_{\rm s}=1$.

The strong dependence on the source redshift distribution has been pointed out
by Wambsganss et al.\ (2004). In contrast to the other studies, they
investigated the arc statistics using ray tracing through a three-dimensional
mass distribution obtained from cosmological simulations, whereas the other
studies mentioned considered the lensing effect of individual clusters found
in these simulations. Although the former approach is more realistic, the
assumption of Wambsganss et al.\ (2004) that the magnification of a light ray
is a good measure for the length-to-width ratio of a corresponding arc is
certainly not justified in detail, as shown in Dalal et al.\ (2003). The
agreement of the lensing probability between Wambsganss et al.\ (2004) and
Bartelmann et al.\ (1998) for all $z_{\rm s}=1$ is therefore most likely a
coincidence.

There are further difficulties in obtaining realistic predictions for the
occurrence of giant arcs that can be compared with observations. First, the
question of whether an image counts as an arc depends on a combination of
source size, lens magnification, and seeing. Seeing makes arcs rounder and
therefore reduces their length-to-width ratio. An impressive demonstration of
this effect is provided by the magnificent system of arcs in the cluster A1689
observed with the ACS onboard the HST, as shown in Fig.\ts\ref{fig:A1689ACS},
compared to earlier ground-based images of this cluster. Second, 
several of the above-mentioned papers assume the source
size to be $\theta=1''$, whereas many arcs observed with HST are essentially
unresolved in width, implying much smaller source sizes (and accordingly, a
much higher sensitivity to seeing effects). Third, magnification bias is
usually not taken into account in these theoretical studies. In fact,
accounting properly for the magnification bias is quite difficult, as the
surveys reporting on arc statistics are not really flux-limited. One might
argue that they are surface brightness-limited, but even if this were true,
the surface brightness of an arc coming from a small source depends very much
on the seeing.

Therefore at present, the abundance of arcs seem to be not in conflict with a
$\Lambda$CDM model, but more realistic simulations which take the
aforementioned effects into account are certainly needed for a definite
conclusion on this issue. On the observational side, increasing the number of
clusters for which high-quality imaging is performed is of great importance,
and the survey of luminous X-ray clusters imaged either with the ACS@HST or
with ground-based telescopes during periods of excellent seeing would improve
the observational situation dramatically. Blank-field surveys, such as they
are conducted for cosmic shear research (see Sect.\ts\ref{sc:WL-7}), could be
used for blind searches of arcs (that is, not restricted to regions around
known clusters). It may turn out, however, that the number of `false
positives' is unacceptably high, e.g., by misidentification of edge-on
spirals, or blends of sources that yield apparent images with a high
length-to-width ratio.  

\subsubsection{Constraints on collisional dark matter.}
Spergel \& Steinhardt (2000) suggested the possibility that dark
matter particles are not only weakly interacting, but may have a
larger elastic scattering cross-section. If this cross-section of such
self-interacting dark matter is sufficiently large, it may help to
explain two of the remaining apparent discrepancies between the
predictions of the Cold Dark Matter model and observations: The slowly
rising rotation curves of dwarf galaxies (e.g., de Blok et al.\ 2001)
and the substructure of galaxy-scale dark matter halos (see Sect.\ts 8
of SL). Self-interacting may soften the strength of the central
density concentration as compared to the NFW profile, and could
destroy most of the subclumps. However, there are other consequence of
such an interaction, in that the shapes of the inner parts of dark
matter halos tend to be more spherical.
Meneghetti et al.\ (2001) have investigated the influence of self
interaction of dark matter particles on clusters of galaxies, in
particular their ability to form giant arcs. From their numerical
simulations of clusters with varying cross-sections of particles, they
showed that even a relatively small cross-section is sufficient to
reduce the ability of clusters to produce giant arcs by an order of
magnitude. This is mainly due to two effects, the reduced asymmetry of
the resulting mass distribution and the shallower central density
profile. Furthermore, self-interactions destroy the ability of clusters
to form radial arcs. Therefore, the `desired' effect of
self-interaction -- to smooth the mass distribution of galaxies -- has
the same consequence for clusters, and can therefore probably be ruled
out as a possible mechanism to cure the aforementioned apparent
problems of the CDM model. From combining X-ray and lensing data of the cluster
0657$-$56, Markevitch et al.\ (2004) obtained upper limits on the
self-interaction cross section of dark matter.

\subsubsection{Do clusters follow the universal NFW profile?}
The CDM paradigm of structure formation predict a universal density
profile of dark matter halos. One might therefore investigate whether
the strong lensing properties of clusters are compatible with this
mass profile. Of particular value for such an investigation are
clusters which contain several strong lensing features, and in
particular a radial arc, as it probes the inner critical curve of the
cluster. Sand et al.\ (2004; see also Sand et al.\ 2002) claim from a
sample of three clusters with radial arcs, that the slope of the inner
mass profile must be considerably flatter than predicted by the NFW
model. However, this conclusion is derived under the assumption of an
axially-symmetric lens model. As is true for strong lensing by
galaxies (see SL), axisymmetric mass model are not generic, and
therefore conclusions derived from them are prone to the systematic of
the symmetry assumption. That was demonstrated by Bartelmann \&
Meneghetti (2004) who showed that, as expected, the conclusion about
the inner slope changes radically once a finite ellipticity of the
mass distribution is allowed for, removing the apparent discrepancy
with the predictions from CDM models.

\subsubsection{Cosmological parameters from strong lensing systems.}
The lens strength, at given physical surface mass density $\Sigma$, depends on
the redshifts of lens and source, as well as on the geometry of the Universe
which enters the distance-redshift relation. Therefore, it has been suggested
that a cluster which contains a large number of strong lensing features can be
used to constrain cosmological parameters, provided the sources of the arcs
and multiple image systems cover a large range of redshifts (Link \& Pierce
1998). Simulations of this effect, using realistic cluster models, confirmed
that such purely geometrical constraints can in principle be derived (Golse et
al.\ 2002). One of the best studied strong-lensing cluster up to now is A2218,
for which four multiple-image systems with measured (spectroscopic) redshift
have been identified which allows very tight constraints on the mass
distribution in this cluster.  Soucail et al.\ (2004) applied the
aforementioned method to this cluster and obtained first constraints on the
density parameter $\Omega_{\rm m}$, assuming a flat cosmological model. This
work can be viewed as a proof of concept; the new ACS camera onboard HST will
allow the identification of even richer strong lensing systems in clusters, of
which the one in A1689 (see Fig.\ts\ref{fig:A1689ACS}) is a particularly
impressive example.  

\section{\llabel{WL-5}Mass reconstructions from weak lensing}
Whereas strong lensing probes the mass distribution in the inner part
of clusters, weak lensing can be used to study the mass distribution
at much larger angular separations from the cluster center. In fact,
as we shall see, weak lensing can provide a parameter-free
reconstruction of the projected two-dimensional mass distribution in
clusters -- and hence offers the prospect of mapping the dark matter
distribution of clusters directly.
This discovery (Kaiser \& Squires 1993) can be viewed to
mark the beginning of quantitative weak lensing research. But even
before this discovery, weak lensing by clusters has been observed in a
number of cases. Fort et al.\ (1988) found that in addition to the
giant arc in A\ 370, there are a number of images stretched in the
direction tangent to the center of the cluster, but with much less
spectacular axis ratios than the giant arc in this cluster; they
termed these new features `arclets'. Tyson et al.\ (1990) found a
statistically significant tangential alignment of faint galaxy images
relative to the center of the clusters A\ 1689 and Cl\ 1409+52, and
obtained a mass profile from these lens distortion maps. Comparison
with numerical simulations yielded an estimate of the cluster
velocity dispersion, assuming an isothermal sphere profile. 

In this section we consider the parameter-free mass reconstruction
technique, first the original Kaiser \& Squires method, and then a
number of improvements of this method.  We then turn to the magnification
effects; the change of the number density of background sources, as
predicted from (\ref{eq:4.43}), can be turned into a local estimate of
the surface mass density, and this method has been employed in a
number of clusters. Next we shall consider inverse methods for the
reconstruction of the mass distribution, which on the one hand are
more difficult to apply than the `direct' methods, but on the other
hand are expected to yield more satisfactory results. Whereas the
two-dimensional maps yield a good visual impression on the mass
distribution in clusters, it is hard to extract quantitative
information from them. In order to get quantities that describe the
mass and that can be compared between clusters, often parameterized
mass models are more useful, which are considered next. Finally, we
consider aperture mass measures, which have been introduced originally
to obtain a mass quantity that is unaffected by the mass-sheet
degeneracy, but as will be shown, has a number of other useful
features. In particular, employing the aperture mass, one can device a
method to systematically search for mass concentrations on
cluster-mass scales, using their shear properties only, i.e. without
referring to their luminous properties.

\subsection{\llabel{WL-5.1}The Kaiser--Squires inversion}
Weak lensing yields an estimate of the local (reduced) shear, as
discussed in Sect.\ \ref{sc:WL-2.2}. Here we shall discuss how to derive
the surface mass density from a measurement of the (reduced) shear.
Recalling eq.\ (IN-26), 
the relation between shear and surface mass density is
\begin{eqnarray}
  \gamma(\vc\theta) &=& \frac{1}{\pi}\int_{\Real^2}\d^2\theta'\,
  {\mathcal D}(\vc\theta-\vc\theta')\,
  \kappa(\vc\theta')\;,\quad \hbox{with}\nonumber\\
  {\mathcal D}(\vc\theta) &\equiv&
  -\frac{\theta_1^2-\theta_2^2+2{\rm i}\theta_1\theta_2}
  {|\vc\theta|^4}
  = \frac{-1}{(\theta_1-{\rm i}\theta_2)^2}\;.
\elabel{3.15}
\end{eqnarray}
Hence, the complex shear $\gamma$ is a convolution of $\kappa$ with the
kernel $\D$, or, in other words, $\D$ describes the shear generated by
a point mass. This relation can be inverted: in 
Fourier space this convolution becomes a multiplication,
\[
\hat\gamma(\vc
\ell)=\pi^{-1}\hat{\mathcal D}(\vc \ell)\,\hat\kappa(\vc \ell)\quad{\rm for}
\quad \vc \ell\ne \vc 0\;,
\]
which can be inverted to yield
\begin{equation}
  \hat\kappa(\vc \ell) = \pi^{-1}\hat\gamma(\vc \ell)\,
  \hat{\mathcal D}^*(\vc \ell)
  \quad\hbox{for}\quad
  \vc \ell\ne\vc 0\;,
\elabel{5.3}
\end{equation}
where the Fourier transform of $\D$ is\footnote{The form of
$\hat{\mathcal D}$ can be obtained most easily by using the relations
between the surface mass density and the shear components in terms of
the deflection potential $\psi$, given in (IN-18). Fourier
transforming those immediately yields
$\hat\kappa=-|\vc\ell|^2\hat\psi/2$,
$\hat\gamma_1=-(\ell_1^2-\ell_2^2)\hat\psi/2$, 
$\hat\gamma_2=-\ell_1\ell_2\hat\psi$. Eliminating $\hat\psi$ from the
foregoing relations, the expression for $\hat{\mathcal D}$ is obtained.}
\be
  \hat{\mathcal D}(\vc \ell) = \pi
  \frac{\left(\ell_1^2-\ell_2^2+2{\rm i}\ell_1\ell_2\right)}
  {|\vc \ell|^2} \;;
\elabel{Dhat}
\ee
note that this implies that $\hat\D(\vc\ell) \hat\D^*(\vc\ell)=\pi^2$,
which has been used in obtaining (\ref{eq:5.3}). It is obvious that
$\hat\D$ is undefined for $\vc\ell=\vc 0$, which has been indicated in
the foregoing equations. Fourier
back-transformation of (\ref{eq:5.3}) then yields
\begin{eqnarray} 
  \kappa(\vc\theta) - \kappa_0 &=& 
  \frac{1}{\pi}\int_{\Real^2}\d^2\theta'\,
  {\mathcal D}^*(\vc\theta-\vc\theta')\,
  \gamma(\vc\theta')  \nonumber\\
  &=& \frac{1}{\pi}\int_{\Real^2}\d^2\theta'\,
  \Re\left[{\mathcal D}^*(\vc\theta-\vc\theta')\,
  \gamma(\vc\theta')\right]  \;.
\elabel{5.4}
\end{eqnarray}
Note that the constant $\kappa_0$ occurs since the $\vc\ell=\vc
0$-mode is undetermined. Physically, this is related to the fact that
a uniform surface mass density yields no shear. Furthermore, it is
obvious (physically, though not so easily seen mathematically) that
$\kappa$ must be real; for this reason, the imaginary part of the
integral should be zero, and taking the real-part only [as in the
second line of (\ref{eq:5.4})] makes no difference. However, in
practice this is different, since noisy data, when inserted into the
inversion formula, will produce a non-zero imaginary part. What
(\ref{eq:5.4}) shows is that if $\gamma$ can be measured, $\kappa$ can
be determined.

Before looking at this in more detail, we briefly mention some 
difficulties with the inversion formula as given above:
\bi
\item
Since $\gamma$ can at best be estimated at discrete points (galaxy
images), smoothing is required. One might be tempted to replace the
integral in (\ref{eq:5.4}) by a discrete sum over galaxy positions,
but as shown by Kaiser \& Squires (1993), the resulting mass density
estimator has infinite noise (due to the $\theta^{-2}$-behavior of the
kernel $\D$). 
\item
It is 
not the shear $\gamma$, but the reduced shear $g$ that can be
determined from the galaxy ellipticities; hence, one needs to obtain a
mass density estimator in terms of $g$. In the case of `weak' weak
lensing, i.e., where $\kappa\ll 1$ and $|\gamma|\ll 1$, then
$\gamma\approx g$. 
\item
The 
integral in (\ref{eq:5.4}) extends over $\Real^2$, whereas data are
available only on a finite field; therefore, it needs to be seen
whether modifications allow the construction of an estimator for the
surface mass density from 
finite-field shear data.
\item
To get absolute values for the surface mass density,
the additive constant $\kappa_0$ is of course a nuisance. As will be
explained soon, this indeed is the largest problem in mass
reconstructions, and is the {\it mass-sheet degeneracy} discussed in
Sect.\ 2.5 of IN.
\ei

\subsection{\llabel{WL-5.2}Improvements and generalizations}
\subsubsection{Smoothing.} Smoothing of data is needed 
to get a shear field from discrete
data points. Consider first the case that we transform (\ref{eq:5.4})
into a sum over galaxy images (ignoring the constant $\kappa_0$ for a
moment, and also assuming the weak lensing case, $\kappa\ll 1$, so
that the expectation value of $\eps$ is the shear $\gamma$), 
\be
\kappa_{\rm disc}(\vc\theta)={1\over n\, \pi}\sum_i
\Re\eck{\D(\vc\theta-\vc\theta_i)\,\eps_i} \;, 
\elabel{KSdisc}
\ee
where the sum extends over all galaxy images at positions
$\vc\theta_i$ and complex ellipticity $\eps_i$, and $n$ is the number
density of background galaxies. As shown by Kaiser \& Squires (1993),
the variance of this estimator for $\kappa$ diverges. However, one can
smooth this estimator, using a weight function $W(\Delta\theta)$
(assumed to be normalized to unity), to
obtain
\be
\kappa_{\rm smooth}(\vc\theta)=\int
\d^2\theta'\;W(|\vc\theta-\vc\theta'|)\,\kappa_{\rm
disc}(\vc\theta')\;,
\elabel{KSsmooth}
\ee 
which now has a finite variance. One might expect that, since (i)
smoothing can be represented by a convolution, (ii) the relation
between $\kappa$ and $\gamma$ is a convolution, and (iii) convolution
operations are transitive, it does not matter whether the shear field
is smoothed first and inserted into (\ref{eq:5.4}), or one uses
(\ref{eq:KSsmooth}) directly. This statement is true if the smoothing
of the shear is performed as 
\be
\gamma_{\rm smooth; 1}(\vc\theta)={1\over n}\sum_i
W(|\vc\theta-\vc\theta_i|) \,\eps_i\;.
\elabel{gamsm1}
\ee
If this expression is inserted into (\ref{eq:5.4}), one indeed
recovers the estimate (\ref{eq:KSsmooth}). However, this is not a
particularly good method for smoothing, as can be seen as follows: the
background galaxy positions will at least have Poisson noise; in fact,
since the angular correlation function even of faint galaxies is
non-zero, local number density fluctuations will be larger than
predicted from a Poisson distribution. However, in the estimator
(\ref{eq:KSdisc}) and in the smoothing procedure (\ref{eq:gamsm1}),
these local variations of the number density are not taken into
account. A much better way (Seitz \& Schneider 1995) to smooth the 
shear is given by
\be
\gamma_{\rm smooth; 2}(\vc\theta)=\eck{\sum_i
W(|\vc\theta-\vc\theta_i|)}^{-1} \;
\sum_i
W(|\vc\theta-\vc\theta_i|) \,\eps_i\;,
\elabel{gamsm2}
\ee
which takes these local number density fluctuations into account.
Lombardi \& Schneider (2001) have shown that the expectation value
of the smoothed shear estimate (\ref{eq:gamsm2}) is not exactly the
shear smoothed by the kernel $W$, but the deviation (i.e., the bias)
is very small provided the effective number of galaxy images inside
the smoothing function $W$ is substantially larger than unity, which
will always be the case for realistic applications. Lombardi \&
Schneider (2002)  then have demonstrated that the variance of
(\ref{eq:gamsm2}) is indeed substantially reduced compared to that of
(\ref{eq:gamsm1}), in agreement with the finding of Seitz \& Schneider
(1995).

When smoothed with a Gaussian kernel of angular scale
$\theta_{\rm s}$, the 
covariance of the resulting mass map is finite, and given by (Lombardi
\& Bertin 1998; van Waerbeke 2000)
\be
{\rm Cov}\rund{\kappa(\vc\theta),\kappa(\vc\theta')}
={\sigma_\eps^2\over 4\pi\theta_{\rm s}^2n}
\exp\rund{-{|\vc\theta-\vc\theta'|^2\over 2\theta_{\rm s}^2}}\;.
\elabel{Covkappa}
\ee
Thus, the larger the smoothing scale, the less noisy is the
corresponding mass map; on the other hand, the more are features
washed out. Choosing the appropriate smoothing scale is not easy; we
shall come back to this issue in Sect.\ \ref{sc:WL-5.3} below.

\subsubsection{The non-linear case, $g\ne\gamma$.}
Noting that the reduced shear $g=\gamma/(1-\kappa)$ can be estimated
from the ellipticity of images (assuming that we avoid the potentially
critical inner region of the cluster, where $|g|>1$; indeed, this case
can also be taken into account, at the price of somewhat increased
complexity), one can write:
\begin{equation}
  \kappa(\vc\theta) - \kappa_0 = 
  \frac{1}{\pi}\,\int_{\Real^2}\d^2\theta'\,
  \left[1-\kappa(\vc\theta')\right]\,\Re\left[
    \D^*(\vc\theta-\vc\theta')\,
    g(\vc\theta')
  \right]\;;
\elabel{5.7}
\end{equation}
this integral equation for $\kappa$ can be solved by iteration, and it
converges quickly (Seitz \& Schneider 1995). Note that in this case, the
undetermined constant $\kappa_0$ no longer corresponds to adding a
uniform mass sheet. What the arbitrary value of $\kappa_0$ corresponds
to can be seen as follows: 
The transformation
\begin{eqnarray}
  \kappa(\vc\theta)\to\kappa'(\vc\theta) &=&
  \lambda\kappa(\vc\theta)+(1-\lambda)
  \quad\hbox{or}\nonumber\\
  \left[1-\kappa'(\vc\theta)\right] &=&
  \lambda\left[1-\kappa(\vc\theta)\right]
\elabel{5.8}
\end{eqnarray}
changes the shear $\gamma\to \gamma'=\lambda\gamma$, and thus leaves
$g$ invariant; this is the mass-sheet degeneracy!  It can be broken if
magnification information can be obtained, since $\A\to\A'=\lambda
\A$, so that
\[
\mu \to \mu'=\lambda^{-2}\mu \;.
\]
Magnification information can be obtained from the number counts of
images (Broadhurst et al.\ 1995), owing to the magnification bias,
provided the unlensed number density is sufficiently well known. 
In principle, the mass sheet degeneracy can 
also be broken if  redshift information of the source
galaxies is available and if the sources are widely distributed in
redshift; this can be seen as follows: let
\be
Z(z_{\rm s})={ D_{\rm ds}/D_{\rm s} \over \lim_{z_{\rm s}\to\infty}
D_{\rm ds}/D_{\rm s}} \,{\rm H}(z_{\rm s}-z_{\rm d})
\elabel{bigZdef}
\ee
(H being the Heaviside step function)
be the ratio of the lens strength of a source at $z_{\rm s}$ to that
of a fiducial source at infinite redshift (see
Fig.\ts\ref{fig:Zfunction}); then, if $\kappa_\infty$ 
and $\gamma_\infty$ denote the surface mass density and shear for such
a fiducial source, the reduced shear for a source at $z_{\rm s}$ is
\be
g={Z \gamma_\infty \over 1-Z\kappa_\infty}\;,
\ee
and there is no global transformation of $\kappa_\infty$ that leaves
$g$ invariant for sources at all redshifts, showing the validity of
the above statement. However, even in this case the mass-sheet
degeneracy is only mildly broken (see Bradac et al.\ 2004). In
particular, only those regions in the cluster where the non-linearity
(i.e., the difference between $\gamma$ and $g$) is noticibly can
contribute to the degeneracy breaking, that is, the region near the
critical curves where $|g|\sim 1$.

\begin{figure}
\bmi{7.7}
\includegraphics[width=7.69cm]{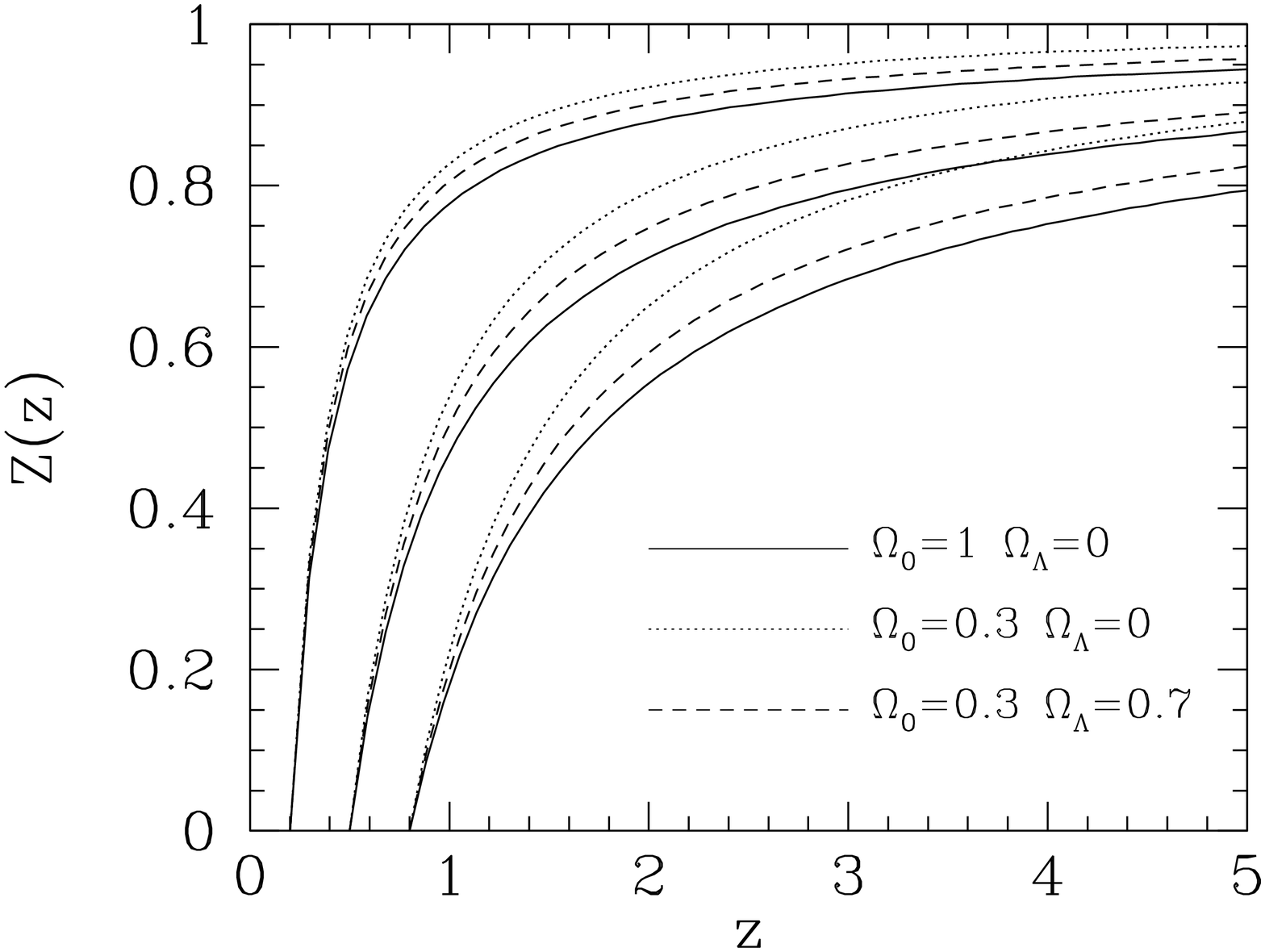}
\emi
\bmi{4.0}
\caption{The redshift weight function $Z(z_{\rm s})$, defined in
(\ref{eq:bigZdef}), for three different values of the lens redshift
$z_{\rm d}=0.2$, 0.5, and 0.8, and three different geometries of the
Universe, as indicated in the labels (here, $\Omega_{\rm m}$ is
denoted as $\Omega_0$). Asymptotically for $z_{\rm s}\to \infty$, all
curves tend to $Z=1$ (from Bartelmann \& Schneider 2001)}
\flabel{Zfunction}
\emi
\end{figure}

In the non-linear case ($\gamma\ne g$) the reduced shear needs to be
obtained from smoothing the galaxy ellipticities in the first
place. Since the relation between $g$ and $\kappa$ is non-linear, the
`transitivity of convolutions' no longer applies; one thus cannot
start from a discretization of an integral over image ellipticities
and smooth the resulting mass map later. We also note that the
accuracy with which the (reduced) shear is estimated can be improved
provided redshift estimates of individual source galaxies are
available (see Fig.\ts\ref{fig:shearaccuracy}). In
particular for high-redshift clusters, redshift information on
individual source galaxies becomes highly valuable. This can be
understood by considering a high-redshift lens, where an appreciable
fraction of faint `source' galaxies are located in front of the lens,
and thus do not contribute to the lensing signal. However, they do
contribute to the noise of the measurement. Redshift information
allows the elimination of these foreground galaxies in the shear
estimate and thus the reduction of noise.   

\begin{figure}
\bc
\includegraphics[width=8cm]{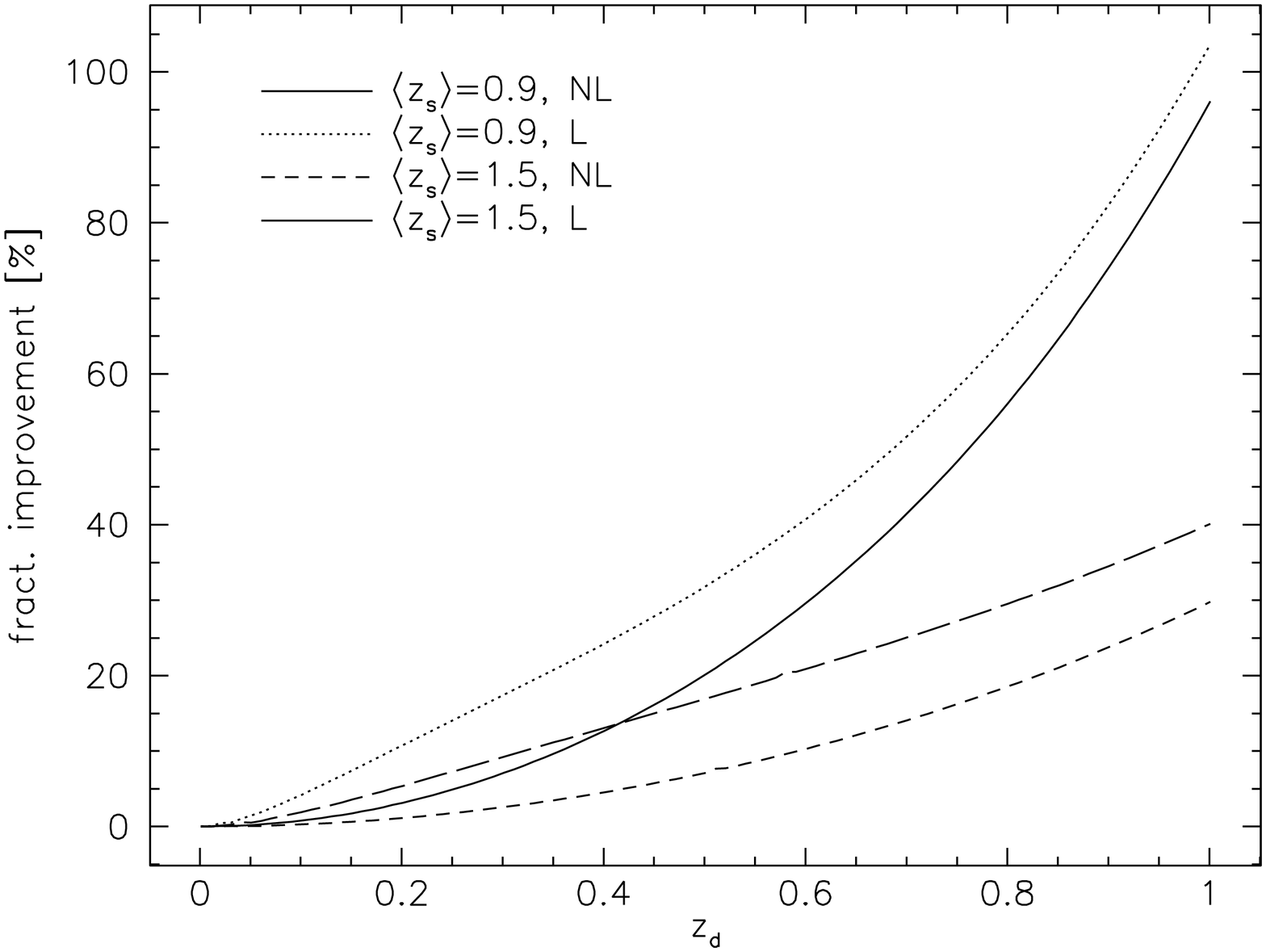}
\ec
\caption{The fractional gain in accuracy of the shear estimate when
using redshift information of individual source galaxies, relative to
the case where only the redshift distribution of the population is
known, plotted as a function of the lens redshift. It is assumed that
the sources have a broad redshift distribution, with a mean of
$\ave{z_{\rm s}}=0.9$ (solid and dotted curves) or $\ave{z_{\rm
s}}=1.5$ (short-dashed and long-dashed curves). The gain of accuracy
also depends on the lens strength; the dotted and long-dashed curves
assume local lens parameters of $\gamma_\infty=0.3=\kappa_\infty$,
whereas the solid and short-dashed curves assume only very weak
lensing, here approximated by $\gamma_\infty=0=\kappa_\infty$. One
sees that the gain is dramatic once the lens redshift becomes
comparable to the mean redshift of the source galaxies and is
therefore of great importance for high-redshift clusters (from
Bartelmann \& Schneider 2001)}
\flabel{shearaccuracy}
\end{figure}

\subsubsection{Finite-field mass reconstruction.} 
In order to obtain a mass map from a finite data field, one starts
from the relation (Kaiser 1995) 
\begin{equation}
  \nabla\kappa = \left(\begin{array}{c}
    \gamma_{1,1}+\gamma_{2,2} \\
    \gamma_{2,1}-\gamma_{1,2} \\
  \end{array}\right) \equiv \vc u_\gamma(\vc\theta)\;,
\elabel{5.10}
\end{equation}
which is a {\it local} relation between shear and surface mass
density; it can easily be derived from the definitions of $\kappa$
and $\gamma$ in terms of $\psi_{,ij}$. 
A similar relation can be obtained in terms of reduced shear,
\be
  \nabla K(\vc\theta) = \frac{-1}{1-g_1^2-g_2^2}\,
  \left(\begin{array}{cc}
    1-g_1 & -g_2 \\
    -g_2 & 1+g_1 \\
  \end{array}\right)\,
  \left(\begin{array}{c}
    g_{1,1}+g_{2,2} \\
    g_{2,1}-g_{1,2} \\
  \end{array}\right) \equiv\vc u_g(\vc\theta)\;,
\elabel{5.11aa}
\ee
where
\begin{equation}
  K(\vc\theta) \equiv \ln[1-\kappa(\vc\theta)]
\elabel{5.12}
\end{equation}
is a non-linear function of $\kappa$. Based on these local relations,
finite-field inversion relations can be derived, and several of them
appeared in the literature right after the foregoing equations have
been published. For example, it is possible to obtain finite-field
mass maps from line integrations (Schneider 1995; for other methods,
see Squires \& Kaiser 1996). Of all these finite-field methods, one
can be identified as optimal, by the following reasoning: in the case
of noise-free data, the imaginary part of (\ref{eq:5.4}) should
vanish. Since one is always dealing with noisy data (at least coming
from the finite intrinsic ellipticity distribution of the sources), in
real life the imaginary part of (\ref{eq:5.4}) will not be zero. But
being solely a noise component, one can choose that finite-field
inversion which yields a zero imaginary component when averaged over
the data field (Seitz \& Schneider 1996). One way of deriving this
mass map is obtained by a further differentiation of (\ref{eq:5.10});
this then yields a von Neumann boundary-value problem on the data
field $\cal U$ (Seitz \& Schneider 2001),
\begin{equation}
  \nabla^2 \kappa =\nabla \cdot \vc u_\gamma
  \quad\hbox{with}\quad
  \vc n\cdot \nabla \kappa = \vc n\cdot \vc u_\gamma
  \quad\hbox{on}\quad
  \partial{\mathcal U}\;,
\elabel{5.17}
\end{equation}
where $\vc n$ is the outward-directed normal on the boundary
$\partial{\mathcal U}$ of ${\mathcal U}$.  The analogous equation
holds for $K$ in terms of $g$ and $\vc u_g$,
\be
\nabla^2 K=\nabla \cdot \vc u_g
  \quad\hbox{with}\quad
  \vc n\cdot \nabla K = \vc n\cdot \vc u_g
  \quad\hbox{on}\quad
  \partial{\mathcal U}\;.
\elabel{5.17b}
\end{equation}
Note that (\ref{eq:5.17}) determines the solution $\kappa$ only up to
an additive constant, and (\ref{eq:5.17b}) determines $K$ only up to
an additive constant, i.e., $(1-\kappa)$ up to a multiplicative
factor. Hence, in both cases we recover the mass-sheet degeneracies
for the linear and non-linear case, respectively.
The numerical solution of
these equations is fast, using overrelaxation (see Press et al.\
1992).  In fact, the foregoing formulation of the problem is
equivalent (Lombardi \& Bertin 1998) to the minimization of the action
\be
  A=\int_{\mathcal U}\d^2\theta\; |\nabla\kappa(\vc\theta)-\vc
   u_\gamma(\vc\theta)|^2 \;,
\elabel{5.17a}
\end{equation}
from which the von Neumann problem can be derived as the Euler
equation of the variational principle $\delta A=0$. Furthermore,
Lombardi \& Bertin (1998) have shown that the solution of
(\ref{eq:5.17}) is `optimal', in that for this estimator the variance 
of $\kappa$ is minimized. 

Since (\ref{eq:5.17}) provides a linear relation between the shear and
the surface mass density, one expects that it can also be written in
the form 
\be
\kappa(\vc\theta)=\int_{\cal U}\d^2\theta'\;\vc
H(\vc\theta;\vc\theta')\cdot \vc u_\gamma(\vc\theta')\;,
\elabel{H-greens}
\ee
where the vector field $\vc H(\vc\theta;\vc\theta')$ is the Green's
function of the von Neumann problem (\ref{eq:5.17}). Accordingly, 
\be
K(\vc\theta)=\int_{\cal U}\d^2\theta'\;\vc
H(\vc\theta;\vc\theta')\cdot \vc u_g(\vc\theta')\;.
\elabel{H-greens1}
\ee
Seitz \& Schneider (1996) gave explicit expression for $\vc H$ in the
case of a circular and rectangular data field. 

One might ask how important the changes in the resulting mass maps are
compared to the Kaiser--Squires formula applied to a finite data
field. For that we note that applying (\ref{eq:5.4}) or (\ref{eq:5.7})
to a finite data field is equivalent to setting the shear outside the
data field to zero. Hence, the resulting mass distribution will be
such as to yield a zero shear outside the data field, despite the
fact that we have no indication from data that the shear indeed is
zero there. This induces features in the mass map, in form of a
pillow-like overall mass distribution. The amplitude of this feature
depends on the strength of the lens, its location inside the data
field, and in particular the size of the data field. Whereas for large
data fields this amplitude is small compared to the noise amplitude of
the mass map, it is nevertheless a systematic that can easily be
avoided, and should be avoided, by using the finite-field inversions,
which cause hardly any additional technical problems.

Various tests have been conducted in the literature as to the accuracy
of the various inversions. For those, one generates artificial shear
data from a known mass distribution, and compares the 
mass maps reconstructed with the various methods with the original
(e.g., Seitz \& Schneider 1996, 2001; Squires \& Kaiser 1996). One of
the surprising results of such comparisons is that in some cases, the
Kaiser \& Squires original reconstruction faired better than the
explicit finite-field inversions, although it is known to yield
systematics. The explanation for this apparent paradox is, however,
easy: the mass models used in these test consisted of one or more
localized mass peaks well inside the data field, so the shear outside
the data field is very small. Noting that the KS formula applied to a
finite data field is equivalent to setting $\gamma=0$ outside the
data field, this methods provides `information' to the reconstruction
process which is not really there, but for the mass models used in the
numerical tests is in fact close to the truth. Of course, by adding
this nearly correct `information' to the mass reconstruction, the
noise can be lowered relative to the finite-field reconstructions
where no assumptions about the shear field outside the data field is
made. 

\subsubsection{Constraints on the geometry of the Universe from weak lensing
  mass reconstructions.} 
The strength of the lensing signal depends, for a given lens redshift, on the
redshift of the sources, through the function $Z(z_{\rm s})$
(\ref{eq:bigZdef}). Suppose that the surface mass density of a cluster was
well known, and that the redshifts of background sources can be
determined. Then, by comparing the measured shear signal from sources at a
given redshift $z_{\rm s}$ with the one expected from the mass distribution,
the value of $Z(z_{\rm s})$ can be determined. Since $Z(z)$ depends on the
geometry of the Universe, parameterized through $\Omega_{\rm m}$ and
$\Omega_\Lambda$, these cosmological parameters can in principle be
determined. A similar strategy for strong lensing clusters was described at
the end of Sect.\ts\ref{sc:WL-4}.

Of course, the surface mass density of the cluster cannot assumed to
be known, but needs to be reconstructed from the weak lensing data
itself. Consider for a moment only the amplitude of the surface mass
density, assuming that its shape is obtained from the
reconstruction. Changing the function $Z(z)$ by a multiplicative
factor would be equivalent to changing the surface mass density
$\Sigma$ of the cluster by the inverse of this factor, and hence such a
constant factor in $Z$ is unobservable due to the mass-sheet
degeneracy. Hence, not the amplitude of 
the function $Z(z)$ shown in Fig.\ts\ref{fig:Zfunction} is important
here, but its shape.

Lombardi \& Bertin (1999) have suggested a method to perform cluster
mass reconstructions and at the same time determine the cosmological
parameters by minimizing the difference between 
the shear predicted from the reconstructed
mass profile and the observed image ellipticities, where the former
depends on the functional form of $Z(z)$. A nice and simple way to
illustrate such a method was given in Gautret et al.\ (2000), called
the `triplet method'. Consider three background galaxies which have a
small separation on the sky, and assume to know the three source
redshifts. Because of their closeness, one might assume that they all
experience the same tidal field and surface mass density from the
cluster. In that case, the shear of the three galaxies is described by
five parameters, the two components of $\gamma_\infty$, $\kappa$, and
$\Omega_{\rm m}$ and $\Omega_\Lambda$. From the six observables (two
components of three galaxy ellipticities), one can minimize the
difference between the predicted shear and the observed ellipticities
with respect to these five parameters, and in particular obtain an
estimate for the cosmological parameters. Repeating this process for a
large number of triplets of background galaxies, the accuracy on the
$\Omega$'s can be improved, and results from a large number of
clusters can be combined.

This procedure is probably too simple to be applied in practice; in
particular, it treats $\kappa_\infty$ and $\gamma_\infty$ for each triplet as
independent numbers, whereas the mass profile of the cluster is described by a
single scalar function. However, it nicely illustrates the principle. Lombardi
\& Bertin (1999) have used a single density profile $\kappa_\infty
(\vc\theta)$ of the cluster, but assumed that the mass-sheet degeneracy is
broken by some other means. Jain \& Taylor (2003) suggested a similar
technique for employing the lensing strength as a function of redshifts and
cosmological parameters to infer constraints on the latter.
Clearly, more work is needed in order to turn these
useful ideas into a practically applicable method.

\subsection{\llabel{WL-5.3}Inverse methods}
In addition to these `direct' methods for determining $\kappa$,
inverse methods have been developed, such as a maximum-likelihood fit
(Bartelmann et al.\ 1996; Squires \& Kaiser 1996) to the data.  There
are a number of reasons why these are in principle preferable to the
direct method discussed above. First, in the direct methods, the
smoothing scale is set arbitrarily, and in general kept constant. It
would be useful to obtain an objective way how this scale should be
chosen, and perhaps, that the smoothing scale be a function of
position: e.g., in regions with larger number densities of sources,
the smoothing scale could be reduced. Second, the direct methods do
not allow additional input coming from complementary observations; for
example, if both shear and magnification information are available,
the latter could not be incorporated into the mass reconstruction. The
same is true for clusters where strong lensing constraints are known.

\subsubsection{The shear likelihood function.}
In the inverse methods, one tries to fit a (very general) lens model
to the observational data, such that the data agree within the
estimated errors with the model. In the maximum-likelihood methods,
one parameterizes the lens by the deflection potential $\psi$ on a grid
and then minimizes the regularized log-likelihood
\begin{equation}
 -\ln {\cal L} = \sum_{i=1}^{N_{\rm g}}
 \frac{|\epsilon_i-g\rund{\vc\theta_i,\{\psi_n\}}|^2}
{\sigma_i^2\rund{\vc\theta_i,\{\psi_n\}}} 
+ 2 \ln\sigma_i(\vc\theta_i,\{\psi_n\})
+\lambda_{\rm e} S(\{\psi_n\}) \;,
 \elabel{5.35}
\end{equation}
where $\sigma_i\approx \sigma_\eps\rund{1-\abs{g(\vc\theta_i,\{\psi_n\})}^2}$
[see eq.\ (\ref{eq:noisesti}) for the case $|g|<1$ that was assumed here],
with respect to these gridded $\psi$-values; this specific form of the
likelihood assumes that the intrinsic ellipticity distribution follows a
Gaussian with width $\sigma_\eps$.\footnote{This specific form (\ref{eq:5.35})
  of the likelihood function assumes that the sheared ellipticity probability
  distribution follows a two-dimensional Gaussian with mean $g$ and dispersion
  $\sigma$; note that this assumption is not valid in general, not even when
  the intrinsic ellipticity distribution is Gaussian (see Geiger \& Schneider
  1999 for an illustration of this fact).  The exact form of the lensed
  ellipticity distribution follows from the intrinsic distribution $p_{\rm s}
  (\eps^{(\rm s)})$ and the transformation law (\ref{eq:4.6}) between
  intrinsic and lensed ellipticity, $p(\eps)=p_{\rm s}\rund{\eps^{(\rm
      s)}(\eps;g)} \det\rund{\partial \eps^{(\rm s)}/\partial \eps}$.
  However, in many cases the Gaussian approximation underlying (\ref{eq:5.35})
  is sufficient and convenient for analytical considerations.}  In order to
avoid overfitting, one needs a regularization term $S$; entropy regularization
(Seitz et al.\ 1998) seems very well suited (see Bridle et al. 1998; Marshall
et al.\ 2002 for alternative regularizations). The entropy term $S$ gets large
if the mass distribution has a lot of structure; hence, in minimizing
(\ref{eq:5.35}) one tries to match the data as closely as permitted by the
entropic term (Narayan \& Nityananda 1986). As a result, one obtains a model
as smooth as compatible with the data, but where structure shows up where the
data require it. The parameter $\lambda_{\rm e}$ is a Langrangean multiplier
which sets the relative weight of the likelihood function and the
regularization; it should be chosen such that the $\chi^2$ per galaxy image is
about unity, i.e.,
\[
\sum_{i=1}^{N_{\rm g}}
 \frac{|\epsilon_i-g\rund{\vc\theta_i,\{\psi_n\}}|^2}
{\sigma_i^2\rund{\vc\theta_i,\{\psi_n\}}} \approx N_{\rm g} \;,
\]
since then the deviation of the observed galaxy ellipticities from
their expectation value $g$ is as large as expected from the
ellipticity dispersion. This choice of the regularization parameter
$\lambda_{\rm e}$ then fixes the effective smoothing used for the
reconstruction. 

Strong lensing constraints can be incorporated into the inverse method
by adding a term to the log-likelihood function which forces the
minimum to satisfy these strong constraints nearly precisely. E.g., if
a pair of multiple images at $\vc\theta_1$ and $\vc\theta_2$ is
identified, one could add the term 
\[
\lambda_{\rm s}\abs{\vc\beta(\vc\theta_1)-\vc\beta(\vc\theta_2)}^2
=\lambda_{\rm s}\abs{\eck{\vc\theta_1
-\vc\alpha(\vc\theta_1)}- \eck{\vc\theta_2-\vc\alpha(\vc\theta_2)}}^2 
\]
to the log-likelihood; by turning up the parameter $\lambda_{\rm s}$,
its minimum is guaranteed to correspond to a solution where the
multiple image constraint is satisfied. Note that the form of this
`source-plane minimization' is simplified -- see Sect.\ts 4.6 of SL --
but in the current context this approach suffices. 

\subsubsection{Magnification likelihood.} 
Similarly, when accurate number counts of faint background galaxies
are available, the magnification information can be incorporated into
the log-likelihood function. If the number counts behave (locally) as
a power law, $n_0(>S)\propto S^{-\alpha}$, the expected number of
galaxies on the data field $\U$ then is
\be
\ave{N}=n_0\int_\U \d^2\theta\;|\mu(\vc\theta)|^{\alpha-1}\;;
\elabel{numbexpct}
\ee
see (\ref{eq:4.43}). 
The likelihood of observing $N$ galaxies at the positions
$\vc\theta_i$ can then be factorized into a term that yields the
probability of observing $N$ galaxies when the expected number is
$\ave{N}$, and one that the $N$ galaxies are at their observed
locations. Since the probability for a galaxy to be at $\vc\theta_i$
is proportional to the expected number density there,
$n=n_0\,\mu^{\alpha-1}$, the likelihood function becomes (Seitz et
al.\ 1998) 
\be
{\cal L}_\mu=P_N\rund{\ave{N}}\prod_{i=1}^N |\mu(\vc\theta_i)|^{\alpha-1}\;,
\elabel{magnilike}
\ee
with the first factor yielding the Poisson probability. Note that this
expression assumes that the background galaxies are unclustered on the
sky; in reality, where (even faint) galaxies cluster,  
this factorization does not strictly apply.

It should be pointed out that the deflection potential $\psi$, and not
the surface mass density $\kappa$, should be used as variable on the
grid, for two reasons: first, shear and $\kappa$ depend locally on
$\psi$, and are thus readily calculated by finite differencing from $\psi$,
whereas the relation between $\gamma$ and $\kappa$ is non-local and
requires summation over all gridpoints, which is of course more time
consuming.  Second, and more important, the surface mass density on a
finite field {\it does not} determine $\gamma$ on this field, since
mass outside the field contributes to $\gamma$ as well. In fact, one
can show (Schneider \& Bartelmann 1997) that the shear inside a circle
is fully determined by the mass distribution inside the circle and the
multipole moments of the mass distribution outside the circle; in
principle, the latter can thus be determined from the shear
measurement.

Despite these reasons, some authors prefer to construct inverse methods in
which the surface mass density on a grid serves as variables (e.g., Bridle et
al.\ 1998; Marshall et al.\ 2002). The fact that the mass density on a finite
field does not describe the shear in this field is accounted for in these
methods by choosing a reconstruction grid that is larger than the data field
and by allowing the surface mass density in this outer region to vary as
well. Whereas the larger numerical grid requires a larger numerical effort, in
addition to the non-local relation between $\kappa$ and $\gamma$, this is of
lesser importance, provided the numerical resources are available. Worse,
however, is the view that the mass distribution outside the data field
obtained by this method has any physical significance! It has not. This mass
distribution is solely one of infinitely many that can approximately generate
the shear in the data field from mass outside the data field. The fact that
numerical tests show that one can indeed recover some of the mass distribution
outside the data field is again a fluke, since these models are usually chosen
such that all mass distribution outside the field in contained in a boundary
region around the data field which is part of the numerical grid -- and hence,
the necessary `external' shear must be generated by a mass distribution in
this boundary zone which by construction is where it is. In real life,
however, there is no constraint on where the `external' shear contribution
comes from.

\subsection{\llabel{WL-5.4}Parameterized mass models}
Whereas the parameter-free mass maps obtained through one of the
methods discussed above provide a direct view of the mass distribution
of a cluster, their quantitative interpretation is not
straightforward. Peaks in the surface mass density can indicate the
presence of a mass concentration, or else be a peak caused by the
ellipticity noise of the galaxies. Since the estimated values for
$\kappa$ at different locations $\vc\theta$ are correlated [see eq.\
(\ref{eq:Covkappa})], it is hard to imagine `error bars' attached to
each point.  Therefore, it is often preferable to use parameterized
mass models to fit the observed data; for example, fitting shear
(and/or magnification) data to an NFW mass profile (see IN, Sect.\
6.2) yields the virial mass $M_{200}$ of the cluster and its
concentration index $c$. There are basically two methods which have
been used to obtain such parameterized models. The first one, assuming
a spherical mass model, orders the tangential component of the
observed image ellipticities into radial bins and fits a parameterized
shear profile through these bins, by minimizing a corresponding
$\chi^2$-function. One of the disadvantages of this method is that the
result of the fitting process can depend on the selected binning, but
this can be largely avoided by choosing the bins fine enough.  This
then essentially corresponds to minimizing the first term in
(\ref{eq:5.35}).

Alternatively, a likelihood method can be used, in which the
log-likelihood function (\ref{eq:5.35}) -- without the regularization
term -- is minimized,  with the values of the potential on the grid
$\{\psi_n\}$ replaced by a set of parameters which describe
the mass profile. Schneider et al.\ (2000) have used this
likelihood method to investigate with which accuracy the model
parameters of a mass profile can be obtained, using both the shear
information as well as magnification information from number counts
depletion. One of the surprising findings of this study was that the
slope of the fitted mass profile is highly degenerate if only shear
information is used; indeed, the mass-sheet degeneracy strikes again
and causes even
fairly different mass profiles to have very similar reduced shear
profiles, as is illustrated for a simple example in Fig.\
\ref{fig:SKE1}. In Fig.\ \ref{fig:SKE2}, the resulting degeneracy of
the profile slope is seen. This degeneracy can be broken if number
count information is used in addition. As seen in the middle panel of
Fig.\ \ref{fig:SKE1}, the magnification profiles of the four models
displayed are quite different and thus the number counts sensitive to
the profile slope. Indeed, the confidence regions in the parameter
fits, shown in Fig.\ts\ref{fig:SKE2},  
obtained from the magnification information are highly inclined
relative to those from the shear measurements, implying that the
combination of both methods yields much better constraints on the
model parameters. Of course, as mentioned before, the mass-sheet
degeneracy can also be broken if redshift information of individual
background galaxies is available. 

\begin{figure}[htb]
\bmi{5}
\includegraphics[bb=36 68 336 764, clip, width=4.7cm]{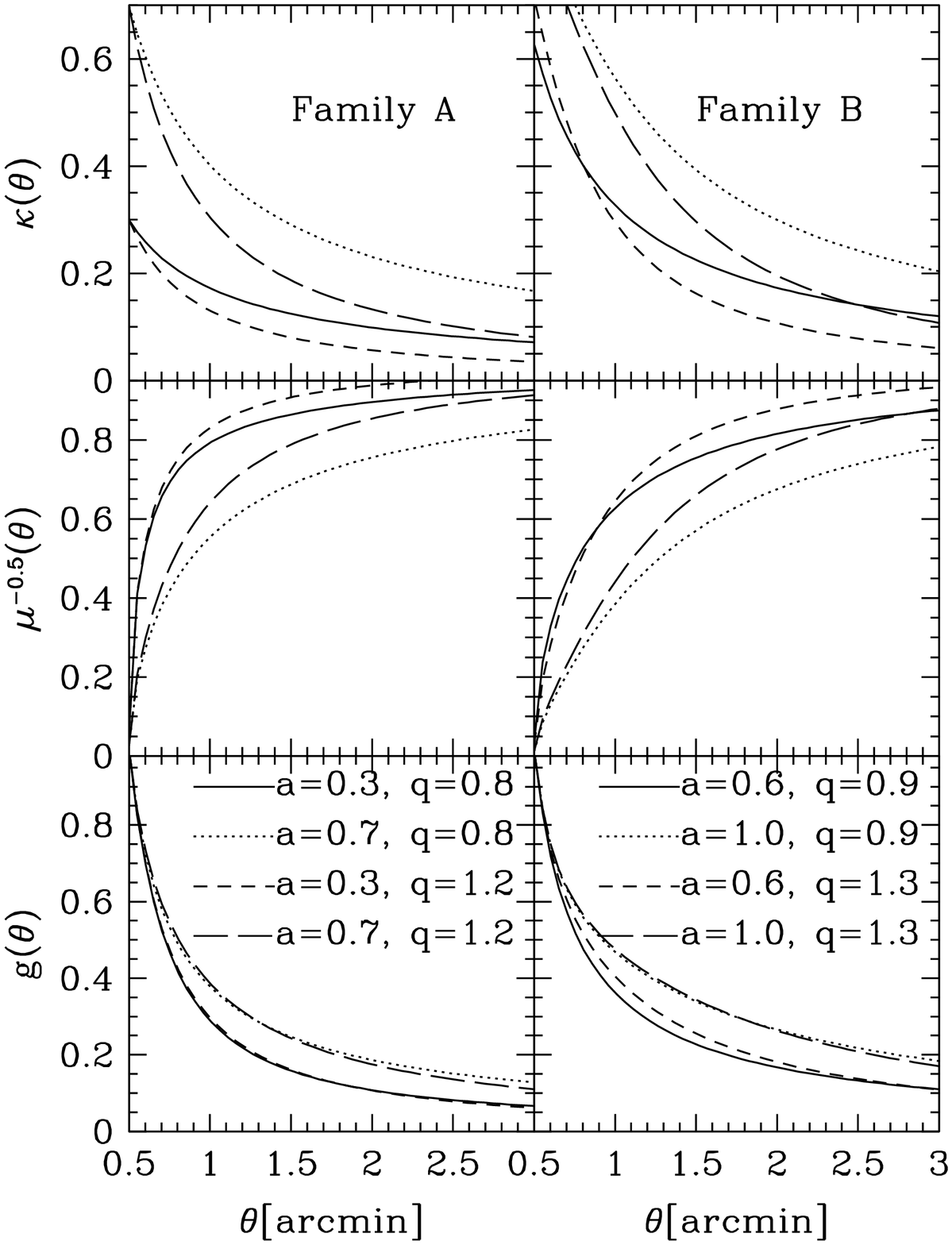}
\emi
\bmi{6.7}
\caption{The Einstein radius of a spherical mass distribution
was assumed to be $\theta_{\rm E}=0\arcminf5$, and the density profile
outside the Einstein radius was assumed to follow a power law,
$\kappa(\theta)= a(\theta/\theta_{\rm E})^{-q}$; an SIS would have
$a=1/2$ and $q=1$. The figure displays for four combinations of model
parameters the surface mass density $\kappa(\theta)$, the function
$\mu^{-1/2}$, which would be the depletion factor for source counts of
slope $\beta=1/2$, 
and the reduced shear
$g(\theta)$. As can be seen, whereas the density profiles of the four
models are quite different, the reduced shear profiles are pairwise almost
fully 
degenerate. This is due to the mass-sheet degeneracy; it implies that
it will be difficult to determine the slope $q$ of the profiles from
shear measurements alone, unless much larger fields around the cluster
are used (from Schneider et al.\ 2000)}
\flabel{SKE1}
\emi
\end{figure}

\begin{figure}[htb]
\bc
\includegraphics[width=6.8cm]{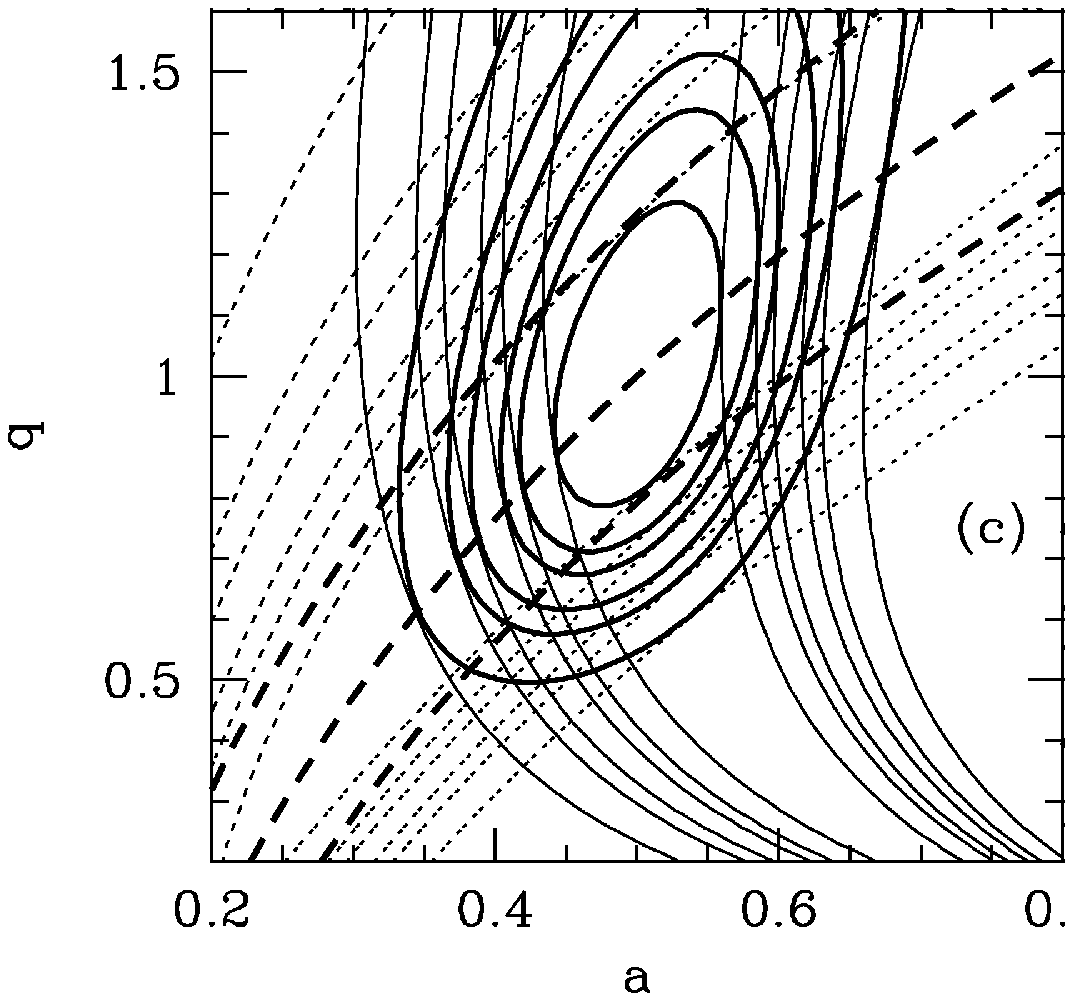}
\ec
\caption{For the power-law models of Fig.\ts\ref{fig:SKE1}, confidence
regions in the slope $q$ and amplitude $a$ are drawn, as derived from
the shear (thin solid contours), the magnification (dotted) and their
combination (thick solid). A number density of $30/\rm arcmin^2$ for
shear measurements and $120/\rm arcmin^2$ for number counts was
assumed. Thick dashed curves show models with constant total number of
galaxies in the field, demonstrating that most of the constraint from
magnification is due to the total counts, with little information
about the detailed profile. It was assumed here that the unlensed number
density of background galaxies is perfectly known; the fact that most of the
magnification information comes from the total number of galaxies in the field
implies that any uncertainty in the unlensed number density will quickly
remove most of the magnification information (from Schneider et al.\ 2000)}
\flabel{SKE2}
\end{figure}

However, in order for the magnification information to yield
significant constraints on the mass parameters, one needs to know the
unlensed number density $n_0$ of sources quite accurately. In fact,
even an uncertainty of less than $\sim 10\%$ in the value of $n_0$
renders the magnification information in relation to the shear
information essentially useless (in the frame of parameterized
models). Note that an accurate determination of $n_0$ is difficult to
achieve: since $n_0$ corresponds to the unlensed number density of
faint galaxies at the same flux limit as used for the actual data
field, one requires an accurate photometric calibration. A flux
calibration uncertainty of $0.1\,{\rm mag}$ corresponds to an
uncertainty in $n_0$ of about $\sim 5\%$ for a slope of $\alpha=0.5$,
and such uncertainties are likely at the very faint flux limits needed
to achieve a high number density of sources. In addition, the presence
of bright cluster galaxies renders the detection and accurate brightness
measurement of background galaxies difficult and requires masking of
regions around them. Nevertheless, in cases
where only magnification information is available, it can provide
information on the mass profile by itself. Such a situation can occur
for observing conditions with seeing above $\sim 1''$, when the shear
method is challenged by the smallness of faint galaxies.

The result shown in Fig.\ \ref{fig:SKE2} implies that the shape of the
mass profile cannot be very well determined from the shear method,
owing to the mass sheet degeneracy. This result extends to more
general mass profiles than power-law models; e.g., King \& Schneider
(2001) considered NFW models with their two parameters $c$ and
$r_{200}$. A fairly strong degeneracy between these two parameters was
found. Furthermore, the mass-sheet degeneracy renders it surprisingly
difficult to distinguish an isothermal mass model from an NFW
profile. The ability to distinguish these two families of models
increases with a larger field-of-view of the observations. This
expectation was indeed verified in King et al.\ (\cite{KCS-1689}) where the
wide-field imaging data of the cluster A~1689 were analyzed with the
likelihood method. Although the field size is larger than $30'$,
so that the shear profile up to $\sim 15'$ from the cluster center can be
measured, an NFW profile is preferred with less than 90\% confidence
over a power-law mass model. The determination of the mass profiles is
likely to improve when strong lensing constraints are taken into account
as well. 

The likelihood method for obtaining the parameters of a mass model is
robust in the sense that the result is only slightly affected by
substructure, as has been shown by King et al.\ (2001) using
numerically generated cluster models. However, if a `wrong'
parameterization of the mass distribution is chosen, the interpretation
of the resulting best-fit model must proceed carefully, and the resulting
physical parameters, such as the total mass, may be biased. The principal
problems with parameterized models are the same as for lens galaxies in strong
lensing: unless the parameters have a well-defined physical meaning, one
does not learn much, even if they are determined with good accuracy (see
Sect.\ts 4.7 of SL).

\subsection{\llabel{WL-5.5}Problems of weak lensing cluster mass
  reconstruction and mass determination} 
In this section, some of the major problems of determining the
mass profile of clusters from weak lensing techniques are
summarized. The finite ellipticity dispersion of galaxies generates a noise
which provides a fundamental limit to the accuracy of all shear
measurements. We will mention a number of additional issues here.

\subsubsection{Number 1: The mass-sheet degeneracy.}
As mentioned several times, the major problem is the mass-sheet
degeneracy, which implies that there is always one arbitrary constant
that is undetermined from the shear data. Number count depletion can
in principle lift this degeneracy, but this magnification effect has
been observed in only a few clusters yet, and as mentioned above, this
method has its own problems. Employing redshift information of
individual source galaxies can also break this degeneracy (Bradac et
al.\ 2004). Note that the mass-sheet degeneracy causes quite different
mass profiles to have very similar reduced shear profiles.

\subsubsection{Source redshift distribution.}
Since the critical surface mass density $\Sigma_{\rm cr}$ depends on the
source redshift, a quantitative interpretation of the weak lensing mass
reconstruction requires the knowledge of the redshift distribution of the
galaxy sample used for the shear measurements. Those are typically so faint
(and numerous) that it is infeasible to obtain individual spectroscopic
redshifts for them. There are several ways to deal with this issue: probably
the best is to obtain multi-color photometry of the fields and employ
photometric redshift techniques (e.g. Connolly et al.\ 1995; Ben\'\i tez
2000; Bolzonella et al.\ 2000).  In order for them to be accurate, the
number of bands needs to be fairly large; in addition, since much of the
background galaxy population is situated at redshifts above unity, one
requires near-IR images, as optical photometry alone cannot be used for
photometric redshifts above $z\gtrsim 1.3$ (where the $4000\,$\AA-break is
redshifted out of the optical window). The problem with near-IR photometry is,
however, that currently near-IR cameras have a substantially smaller
field-of-view than optical cameras; in addition, due to the much higher sky
brightness for ground-based near-IR observations, they extend to brighter flux
limits (or smaller galaxy number densities) than optical images, for the same
observing time. Nevertheless, upcoming wide-field near-IR cameras, such as the
VISTA project on Paranal or WIRCAM at the CFHT, will bring great progress in
this direction.

The alternative to individual
redshift estimates of background galaxies is to use the redshift
distribution obtained through spectroscopic (or detailed photometric
redshift) surveys in other fields, and identify this with the faint
background galaxy population at the same magnitude. In this way, the
redshift distribution of the galaxies can be estimated. The issues
that need to be considered here is that neither the targets for a
spectroscopic survey, nor the galaxy population from which the shear
is estimated, are strictly magnitude selected. Very small galaxies,
for example, cannot be used for a shear estimate (or are heavily
downweighted) owing to their large smearing corrections from the
PSF. Similarly, for low-surface brightness galaxies it is much harder
to determine a spectroscopic redshift. Hence, in these redshift
identifications, care needs to be excersized. 

For cluster mass reconstructions, the physical mass scale is obtained
from the average $\beta:=\ave{D_{\rm ds}/D_{\rm s}}$ over all source
galaxies.  This average is fairly insensitive to the detailed redshift
distribution, as long as the mean source redshift is substantially
larger than the lens redshift. This is typically the case for
low-redshift ($z\lesssim 0.3$) clusters. However, for higher-redshift
lenses, determining $\beta$ requires a good knowledge of the galaxy
redshift distribution.

\subsubsection{Contamination of the source sample.}
Next on the list is the contamination of the galaxy sample from which
the shear is measured by cluster galaxies; a fraction of the faint
galaxies will be foreground objects or faint cluster members. Whereas
the foreground population is automatically taken into account in the
normal lensing analysis (i.e., in determining $\beta$), the cluster
members constitute an additional population of galaxies which is not
included in the statistical redshift distribution. The galaxy sample
used for the shear measurement is usually chosen as to be
substantially fainter than the brighter cluster member galaxies;
however, the abundance of dwarf galaxies in clusters (or equivalently,
the shape of the cluster galaxy luminosity function) is not well
known, and may vary substantially from cluster to cluster (e.g.,
Trentham \& Tully 2002, and references therein). Including
cluster members in the population from which the shear is measured
weakens the lensing signal, since they are not sheared. As a
consequence, a smaller shear is measured, and a lower cluster mass is
derived. In addition, the dwarf contamination varies as a function of
distance from the cluster center, so that the shape of the mass
distribution will be affected.  Color selection of faint galaxies can
help in the selection of background galaxies, i.e., to obtain a
cleaner set of true background galaxies. Of course, cluster dwarfs, if
not properly accounted for, will also affect the magnification method.
One method to deal with this problem is to use only galaxies redder than the
Red Cluster Sequence of the cluster galaxies in the color-magnitude
diagram, as this sequence indicates the reddest galaxies at the
corresponding redshift. 

\subsubsection{Accuracy of mass determination via weak lensing.}
Comparing the `true' mass of a cluster with that measured by weak lensing is
not trivial, as one has to define what the true mass of a cluster is. Using
clusters from numerical simulations, the mass is defined as the mass inside a
sphere of radius $r_{200}$ around the cluster center within which the
overdensity is 200 times the critical density of the universe at the redshift
considered. When comparing this mass with the projected mass 
inside a circle of radius $R=r_{200}$, one should not be surprised that the
latter is larger (Metzler et al.\ 2001), since one compares apples (the mass
inside a sphere) with oranges (the mass within a cylinder). Metzler et al.\
ascribed 
this to the mass in dark matter filaments at the intersection of which massive
clusters are located, but it is most likely mainly an effect of the mass
definitions. 

The mass-sheet degeneracy tell us there is little hope to measure the `total'
mass of a cluster without further assumptions. Therefore, one natural strategy
is to assume a parameterized mass profile and see how accurately one can
determine these parameters. The effect of ellipticity noise has already been
described in Sect.\ts\ref{sc:WL-5.4}. Using simulated clusters,
Clowe et al.\ (\cite{CdeLK}) have studied the effect of asphericity and
substructure of clusters on these mass parameters, by analyzing the shear
field obtained from independent projection of the clusters. They find that the
non-spherical mass distribution and substructure induce uncertainties in the
two parameters ($r_{\rm 200}$ and the concentration $c$) of an NWF profile
which are larger than those from the ellipticity noise under very
good observing conditions. Among different projections of the same cluster,
the value of $r_{200}$ has a spread of 10 -- 15\%, corresponding to a spread
in virial mass of $\sim 40\%$. Averaging over the different projections, they
find that there is little bias in the mass determination, except for clusters
with very large ellipticity. 

\subsubsection{Lensing by the large-scale structure.}
Lensing by foreground and background density inhomogeneities (i.e., the LSS),
yields a fundamental limit to the accuracy of cluster mass estimates. Since
lensing probes the projected density, these foreground and background
inhomogeneities are present in the lensing signal. Hoekstra
(\cite{HoekNFW}) has 
investigated this effect in the determination of the parameters of an NFW mass
profile; we shall return to this issue in Sect.\ts\ref{sc:WL-9.3} below when
we consider lensing by the large-scale structure. In principle, the foreground
and background contributions can be eliminated if the individual redshifts of
the source galaxies are known, since in this case a three-dimensional mass
reconstruction becomes possible (see Sect.\ts\ref{sc:WL-7.6}); however, the
resulting cluster mass map will be very noisy.

\subsection{\llabel{WL-5.6}Results}
After the first detection of a coherent alignment of galaxy images in
two clusters by Tyson et al.\ (1990) and the development of the Kaiser
\& Squires (1993) mass reconstruction method, the cluster MS\ 1224+20
was the first for which a mass map was obtained (Fahlman et al.\
1994). This investigation of the X-ray selected cluster yielded a mass
map centered on the X-ray centroid of the cluster, but also a
surprisingly high $M/L$-ratio of $\sim 800\, h$ (here and in following
we quote mass-to-light ratios always in Solar units). This high $M/L$ ratio
has later been confirmed in an independent analysis by Fischer
(1999). This mass estimate is in strong conflict with that obtained
from a virial analysis (Carlberg 1994);
however, it is known that this cluster has a very complex structure,
is not relaxed, and most likely a superposition of galaxy
concentrations in redshift.

Since this pioneering work, mass reconstructions of many clusters have
been performed; see Mellier (1999) and Sect.\ 5.4 of BS. Here, only a
few recent results shall be mentioned, followed by a summary.

\subsubsection{Wide-field mass reconstructions.}
\begin{figure}[tb]
\bmi{7.5}
\includegraphics[width=7.3cm]{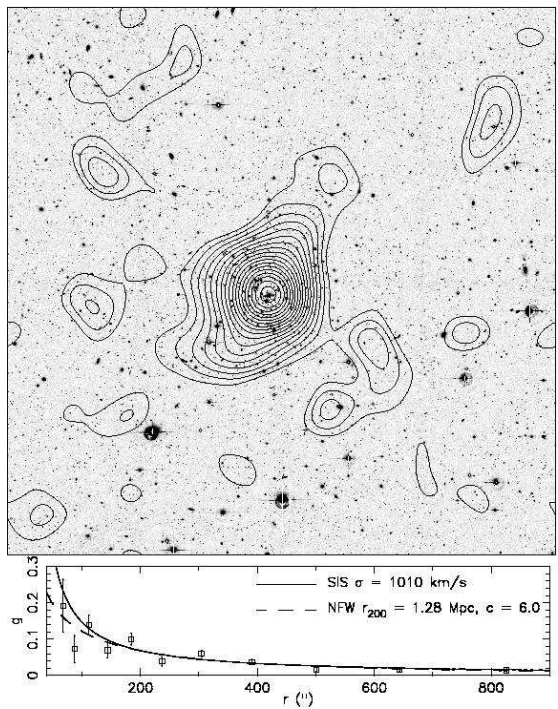}
\emi
\bmi{4.2}
\caption{Contours show the mass reconstruction of the cluster A1689,
obtained from 
data taken with the WFI at the ESO/MPG 2.2m telescope. The image is
$\sim 33'$ on a side, corresponding to $\sim 4.3\,h^{-1}\,{\rm Mpc}$
at the cluster redshift of $z_{\rm d}=0.18$. In the lower panel, the
reduced shear profile is shown, together with the best fitting SIS and
NFW models. The mass reconstruction has been smoothed by a
$1\arcminf15$ Gaussian, and contour spacing is $\Delta\kappa=0.01$. No
corrections have been applied to account for contamination of the
lensing signal by cluster dwarf galaxies -- that would increase the
mass of the best fit models by $\sim 25\%$ (taken from Clowe \&
Schneider 2001)
}
\flabel{A1689WFI}
\emi
\end{figure}
The advent of large mosaic CCD cameras provides an opportunity to map
large regions around clusters to be used for a mass reconstruction,
and thus to measure the shear profile out to the virial radius of
clusters. These large-scale observations offer the best promise to
investigate the outer slope of the mass profile, and in particular
distinguish between isothermal distributions and those following the NWF
profile. Fig.\ \ref{fig:A1689WFI} shows an example of such a mass
reconstruction, that of the cluster Abell 1689 with $z_{\rm
d}=0.182$. A significant shear is observed out to the virial
radius. The mass peak is centered on the brightest cluster galaxy, and
the overall lens signal is significant at the 13.4-$\sigma$ level. The
shear signal is fit with two models, as shown in the lower panel of
Fig.\ \ref{fig:A1689WFI}; the NWF profile yields a better fit than an
SIS profile. Two more clusters observed with the WFI by Clowe \&
Schneider (2002) yield similar results, i.e., a detection of the
lensing signal out to the virial radius, and a preference for an NWF
mass profile, although in one of the two cases this preference is
marginal. The lensing signal of such rich clusters could be
contaminated by faint cluster member galaxies; correcting for this
effect would increase the estimate of the lensing strength, but
requires multi-color imaging for source selection.

The cluster A1689 is (one of) the strongest lensing clusters known (see
Fig.\ \ref{fig:A1689ACS}); in fact, it is strong enough so that a weak
lensing signal can be significantly detected from near-IR images (King
et al.\ \cite{King-IR}) despite the fact that the usable number density of
(background) galaxies is only $\sim 3\,{\rm arcmin}^{-2}$. The
estimate of its velocity dispersion from weak lensing yields an
Einstein radius well below the distance of the giant arcs from the
cluster center. Hence, in this cluster we see a discrepancy between
the strong and weak lensing results, which cannot be easily explained
by redshift differences between the arc sources and the mean redshift
of the faint galaxies used for the weak lensing analysis. On the other
hand, A1689 is known to be not a relaxed cluster, due to the redshift
distribution of its member galaxies. This may explain the fact that
the weak lensing mass estimates is also lower than that obtained from
X-ray studies.

\subsubsection{Filaments between clusters.}
One of the predictions of CDM models for structure formation is that clusters
of galaxies are located at the intersection points of filaments formed by the
dark matter distribution. In particular, this implies that a physical pair of
clusters should be connected by a bridge or filament of (dark) matter, and
weak lensing mass reconstructions can in principle be used to search for them.
In the investigation of the $z=0.42$ supercluster MS0302+17, Kaiser et al.\ 
(1998) found an indication of a possible filament connecting two of the three
clusters, with the caveat (as pointed out by the authors) that the filament
lies just along the boundary of two CCD chips; in fact, an indepedent analysis
of this supercluster (Gavazzi et al.\ 2004) failed to confirm this filament.
Gray et al.\ (2002) saw evidence for a filament connecting the two clusters
A901A/901B in their mass reconstruction of the A901/902 supercluster field.
Another potential filament has been found in the wide-field mass
reconstruction of the field containing the pair of clusters A222/223 (Dietrich
et al.\ 2004). Spectroscopy shows that there are also galaxies at the same
redshift as the two clusters present in the `filament' (Dietrich et al.\
2002). 

\begin{figure}
\bc
\includegraphics[width=10cm,angle=-90]{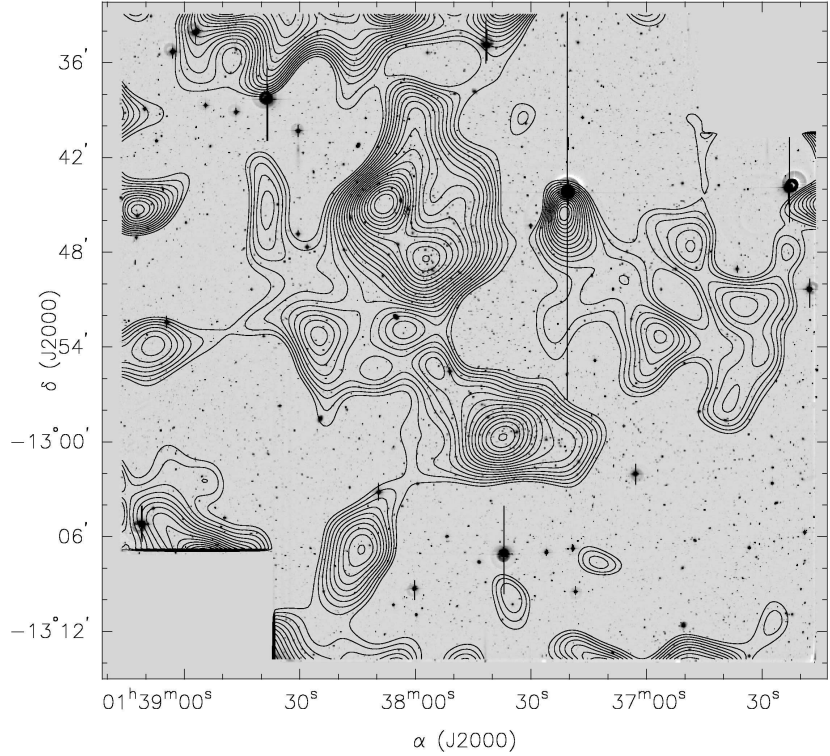}
\ec
\caption{A deep R-band image of the cluster pair Abell 222/223, obtained from
  two different pointings with the WFI@ESO/MPG\ 2.2m, with contours
showing the reconstructed $\kappa$-map. 
The two clusters are in
  the region where the pointings overlap and thus deep imaging is available
  there. Both clusters are obviously detected in the mass map, with A223 (the
  Northern one) clearly split up into two subclusters. The mass reconstruction
  shows a connection between the two clusters which can be interpreted as a
  filament; galaxies at the clusters' redshift are present in this
  inter-cluster region. A further mass concentration is seen about $13'$ to
  the South-East of A222, which is significant at the $3.5\sigma$ level and
  where a clear concentration of galaxies is visible. A
  possible red cluster sequence indicates a substantially higher redshift for
  this cluster, compared to $z\approx 0.21$ of the double cluster (from
  Dietrich et al.\ 2004)} 
\flabel{A222-223}
\end{figure}

One of the problems related to the unambiguous detection of filaments is the
difficulty to define what a `filament' is, i.e. to device a statistics to
quantify the presence of a mass bridge. The eye easily picks up a pattern and
identifies it as a `filament', but quantifying such a pattern turns out to be
very difficult, as shown by Dietrich et al.\ (2004). Because of that, it is
difficult to distinguish between noise in the mass maps, the `elliptical'
extension of two clusters pointing towards each other, and a true filament.
However, this problem is not specific to the weak investigation: even if the
true projected mass distribution of a pair of clusters were known
(e.g., from a cluster pair in numerical simulations), it is not
straightforward to define what a filament would be.

\subsubsection{Correlation between mass and light.}
Mass reconstructions on wide-fields, particularly those covering
supercluster regions, are ideally suited to investigate the relation
between mass and galaxy light. For example, a smoothed light map of
the color-selected early-type galaxies can be correlated with the
reconstructed $\kappa$-map; alternatively, assuming that light traces
mass, the expected shear map can be predicted from the early-type
galaxies and compared to the observed shear, with the mass-to-light
ratio being the essential fit parameter. Such studies have been
carried out on the aforementioned supercluster fields, as well as on
blank fields (Wilson et al.\ \cite{Wils01empty}). These studies yield
very consistent results, in that the mass of clusters is very well
traced by the distribution of early-type galaxies, but the
mass-to-light ratio seems to vary between different fields, with $\sim
400 h$ (in solar units) for the 0302 supercluster (Gavazzi et al.\
2004), $\sim 200 h$ for the A901/902 supercluster (Gray et al.\ 2002),
and $\sim 300 h$ for empty fields (Wilson et al.\
\cite{Wils01empty}) in the rest-frame B-band. When one looks in more
detail at these 
supercluster fields, interesting additional complications appear. The
three clusters in the 0302 field, as well as the three clusters in the
A901/902 field (A901 is indeed a pair of clusters) have quite
different properties. In terms of number density of color-selected
galaxies, A901a and A902 dominate the field, whereas only A901b seems
to be detected in X-rays. Considering early-type galaxies' luminosity,
A901a is the most prominent of the three clusters. In contrast to
this, A902 seems to be most massive as judged from the weak lensing
reconstruction. Similar differences between the three clusters in the
0302 field are also seen. It therefore appears that the mass-to-light
properties of clusters cover quite a range.

\subsubsection{Cluster mass reconstructions from space.}
The exquisite image quality that can be achieved with the HST -- imaging
without the blurring effects of atmospheric seeing -- suggests that such data
would be ideal for weak lensing studies. This is indeed partly true: from
space, the shape of smaller galaxy images can be measured than from the ground
where the size of the seeing disk limits the image size of galaxies that can be
used for ellipticity measurements in practice. Fig.\ \ref{fig:0939} shows an
HST image of the cluster A851 ($z_{\rm d}=0.41$), together with a mass
reconstruction. The agreement between the mass distribution and the angular
distribution of bright cluster galaxies is striking.
A detailed X-ray observation of this cluster with XMM-Newton (De
Filippis et al.\ 2003) finds two extended X-ray components coinciding
with the two maxima of the bright galaxy distribution, and thus of the
mass map shown in Fig.\ \ref{fig:0939}, in addition to several compact
X-ray sources inside the HST field. Clearly, this cluster is a
dynamically young system, as also seen by the inhomogeneities of the
X-ray temperature and metallicity of the intracluster gas.

\begin{figure}
\bmi{6.0}
\includegraphics[width=5.9cm]{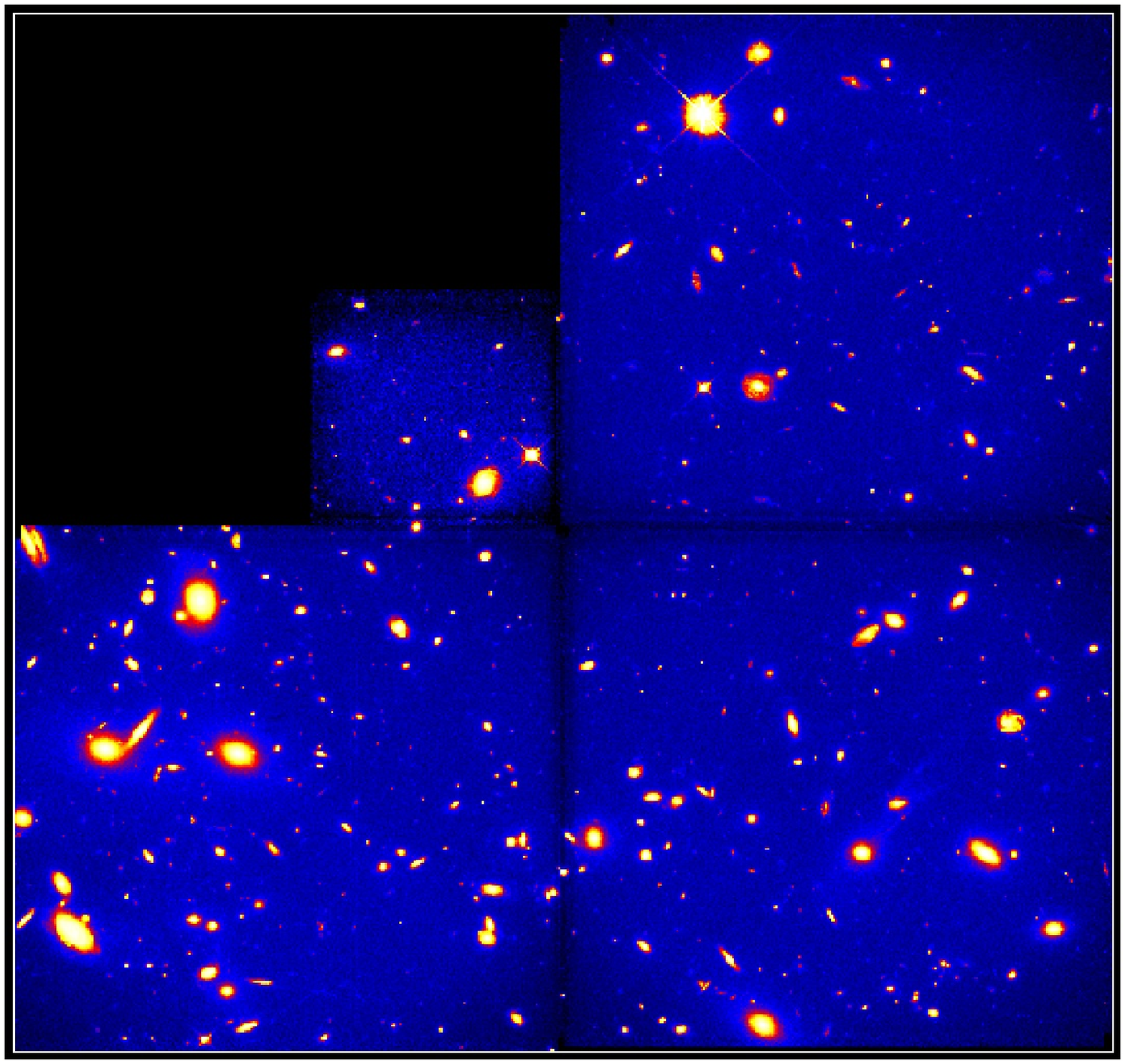}
\emi
\bmi{5.7}
\includegraphics[width=5.6cm]{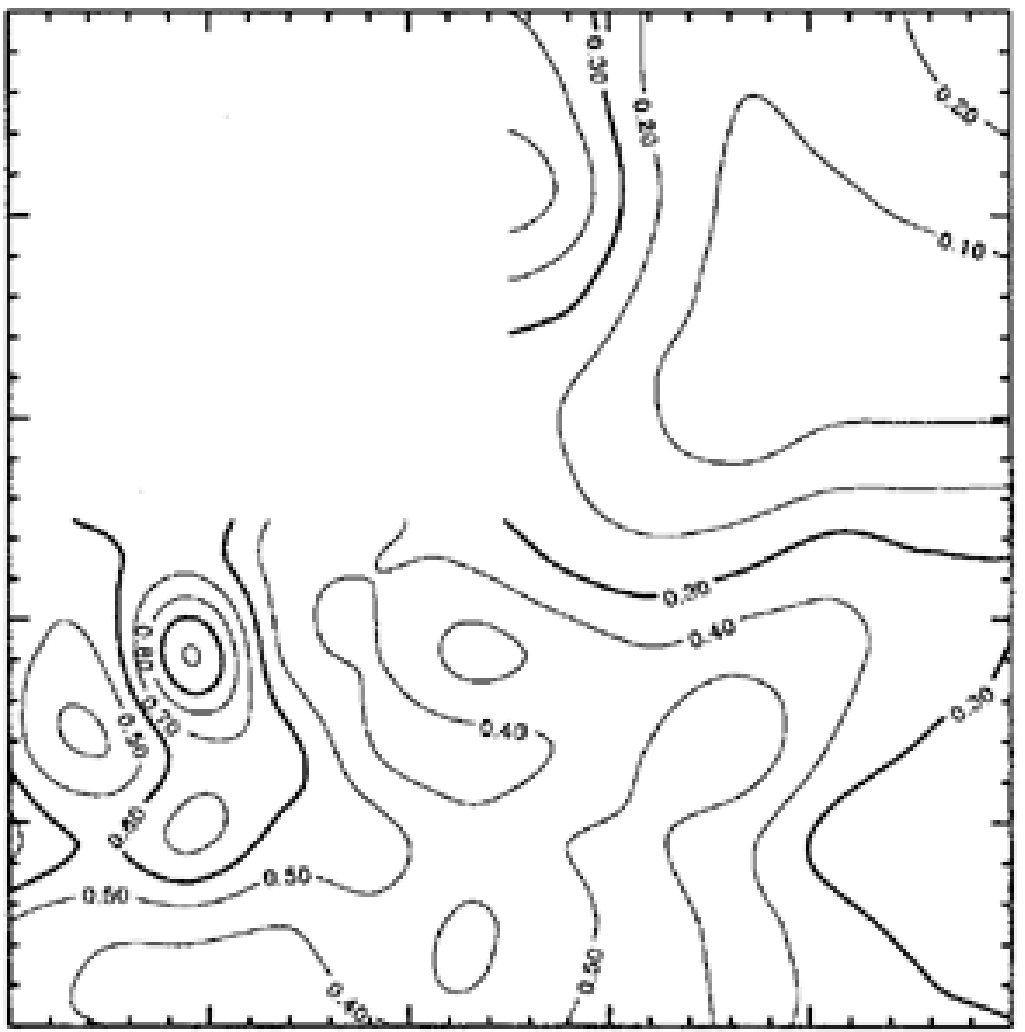}
\emi
\caption{The left panel shows an WFPC2@HST image of the cluster
Cl0939+4713 ($=$Abell 851; taken from Seitz et al.\ 1996; the field is
about $2\arcminf5$ on a side), whereas the
right panel shows a mass reconstruction obtained by Geiger \&
Schneider (1999); this was obtained using the entropy-regularized
maximum likelihood method of Seitz et al.\ (1998). One notices the
increased spatial resolution of the resulting mass map near the center
of the cluster, which this method yields `automatically' in those
regions where the shear signal is large. Indeed, this mass map
predicts that the cluster is critical in the central part, in
agreement with the finding of Trager et al.\ (1997) that strong
lensing features (multiple images plus an arc) of sources with $z\sim
4$ are seen there. The strong correlation between the distribution of
mass and that of the bright cluster galaxies is obvious: Not only does
the peak of the mass distribution coincide with the light center of
the cluster, but also a secondary maximum in the surface mass density
corresponds to a galaxy concentration (seen in the lower middle), as
well as a pronounced minimum on the left where hardly any bright
galaxies are visible}
\flabel{0939}
\end{figure}

The drawback of cluster weak lensing studies with the HST is the small
field-of-view of its WFPC2 camera, which precludes imaging of large
regions around the cluster center. To compensate for this, one can use
multiple pointings to tile a cluster. For example, Hoekstra and
collaborators have observed three X-ray selected clusters with HST
mosaics; the results from this survey are summarized in Hoekstra et
al.\ (\cite{Hoek-2053}). One example is shown in Fig.\
\ref{fig:1054-HST}, the high-redshift cluster MS1054$-$03 at $z_{\rm
d}=0.83$. Also in this cluster one detects clear substructure, here
consisting of three mass peaks, which is matched by the distribution
of bright cluster galaxies. The shape of the mass maps indicates that
this cluster is not relaxed, but perhaps in a later stage of merging,
a view also supported by its hot X-ray temperature. In fact,
new observations with Chandra and XMM-Newton of MS\ts 1054 have shown
that this cluster has a much lower temperature than measured earlier
with ASCA (Gioia et al.\ 2004). Only two of the three components seen
in the galaxy distribution and the mass reconstruction are seen in
X-rays, with the central weak lensing component being the dominant
X-ray source. The newly determined X-ray temperature is consistent
with the velocity dispersion of cluster galaxies.

\begin{figure}
\bmi{7}
\includegraphics[width=6.9cm]{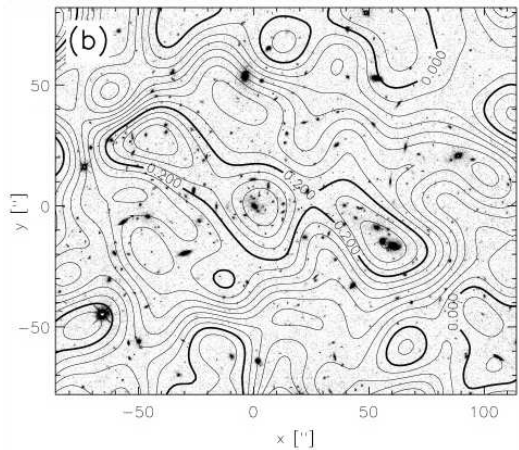}
\emi
\bmi{4.7}
\caption{Mass reconstruction (contours) of the inner part of the high-redshift
($z_{\rm d}=0.83$) cluster MS1054$-$03, based on a mosaic of six
pointings obtained with the WFPC2@HST (from Hoekstra et al.\
\cite{Hoek-1054}). The splitting of the cluster core into three subcomponents, 
also previously seen from ground-based images by Clowe et al.\ (2000),
shows that this cluster is not yet relaxed}
\flabel{1054-HST}
\emi
\end{figure}

\subsubsection{Magnification effects.}
As mentioned in Sect.\ts\ref{sc:WL-2.4}, the magnification of a lens
can also be used to reconstruct its surface mass density (Broadhurst
et al.\ 1995). Provided a population of background source galaxies is
identified whose number count slope $\alpha$ -- see (\ref{eq:4.43}) -
differs significantly from unity, local counts of these sources can be
turned into an estimator of the local magnification. If the lens is
weak, (\ref{eq:mu-simple}) provides a relation between the local
number counts and the local surface mass density. If the lens is not
weak, this relation no longer suffices, but one needs to use the full
expression
\be
|\mu|^{-1}=\abs{ (1-\kappa)^2-|\gamma|^2}\;,
\elabel{magfi1}
\ee
where we have written absolute values to account for the fact that the sign of
the magnification cannot be observed. There are two obvious difficulties with
(\ref{eq:magfi1}): the first comes from the sign ambiguities, namely whether
$\mu$ is 
positive or negative, and whether $\kappa<1$ or $>1$. Assuming that we are in
the region of the cluster where $\mu>0$ and $\kappa<1$ (that is, outside the
outer critical curve), then (\ref{eq:magfi1}) can be rewritten as
\be
\kappa=1-\sqrt{\mu^{-1}+|\gamma|^2}\;,
\elabel{magfi2}
\ee
which shows the second difficulty: in order to estimate $\kappa$ from $\mu$,
one needs to know the shear magnitude $|\gamma|$. 

There are various ways to deal with this second problem. Consider first the
case that
the (reduced) shear is also observed, in which case one better writes
\be
\kappa=1-\eck{\mu\rund{1-|g|^2}}^{-1/2}\; ;
\elabel{magfi3}
\ee
but of course, if shear measurements are available, they should be combined
with magnification observations in a more optimized way. A second method,
using magnification only, is based on the fact that $\gamma$ depends linearly
on $\kappa$ (ignoring finite-field problems here), and so (\ref{eq:magfi2})
can be turned into a quadratic equation for the $\kappa$ field (Dye \& Taylor
1998). From numerical models of clusters, van Kampen (1998) claimed that the
shear in these clusters approximately follows on average a relation of the
form $|\gamma|=(1-c)\sqrt{\kappa/c}$, with $c\sim 0.7$; however, there is (as
expected) large scatter around this mean relation which by itself has little
theoretical justification. Fig.\ts\ref{fig:0024magni} shows the mass
reconstruction of the cluster Cl\ts 0024+17 using galaxy number counts and the
two reconstruction methods just mentioned.

\begin{figure}
\bmi{6}
\includegraphics[width=5.8cm]{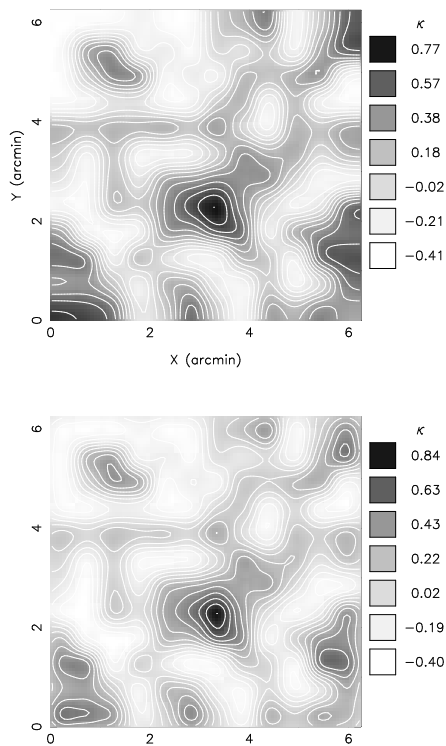}
\emi
\bmi{5.7}
\caption{Mass reconstruction of the cluster Cl\ts 0024+17 from the
  magnification method. The two different reconstructions are based on two
  different ways to turn the magnification signal -- number count depletions
  -- into a surface mass density mass, as described in the text: in the upper
  panel, a local relation between surface mass density and shear magnitude has
  been used, whereas in the lower panel, the magnification was transformed
  into a $\kappa$ map using the (non-local) quadratic dependence of the
  inverse magnification on the surface mass density field. Overall,
  these two reconstructions agree very well. To account for the presence of
  bright foreground galaxies, the data field had to be masked before local
  number densities of background galaxies were estimated -- the mask is shown
  in Fig.\ts\ref{fig:0024-mask} (from Dye et al.\ 2002)}
\flabel{0024magni}
\emi
\end{figure}

Magnification effects have been observed for a few clusters, most
noticibly Cl\ts 0024+17 (Fort et al.\ 1997; R\"ognvaldsson et al.\
2001; Dye et al.\ 2002) and A1689 (Taylor et al.\ 1998; Dye et al.\
2001). We shall describe some of the results obtained for Cl\ts
0024+17 as an example (Dye et al.\ 2002).  Since the cluster galaxies
generate a local overdensity of galaxy counts, they need to be removed
first, which can be done based on a color and magnitude
criterion. Comparison with extensive spectroscopy of this cluster
(Czoske et al.\ 2001) shows that this selection is very effective for
the brighter objects. For the fainter galaxies -- those from which the
lensing signal is actually measured -- a statistical subtraction of
foreground and cluster galaxies needs to be performed, which is done
by subtracting galaxies according to the field luminosity function
with $z<z_{\rm d}$ and cluster galaxies according to the cluster
luminosity function. The latter is based on the assumption that the
luminosity distribution of cluster galaxies is independent from the
distance to the cluster center.  Next, the field of the cluster needs
to be masked for bright objects, near which the photometry of fainter
galaxies becomes inaccurate or impossible; Fig.\ts\ref{fig:0024-mask}
shows the masked data field.  The number density of sources is then
determined from the unmasked area.  The resulting mass reconstruction
is shown in Fig.\ts\ref{fig:0024magni}.
\begin{figure}
\bmi{6}
\includegraphics[bb=4 4 359 371, clip, width=5.9cm]{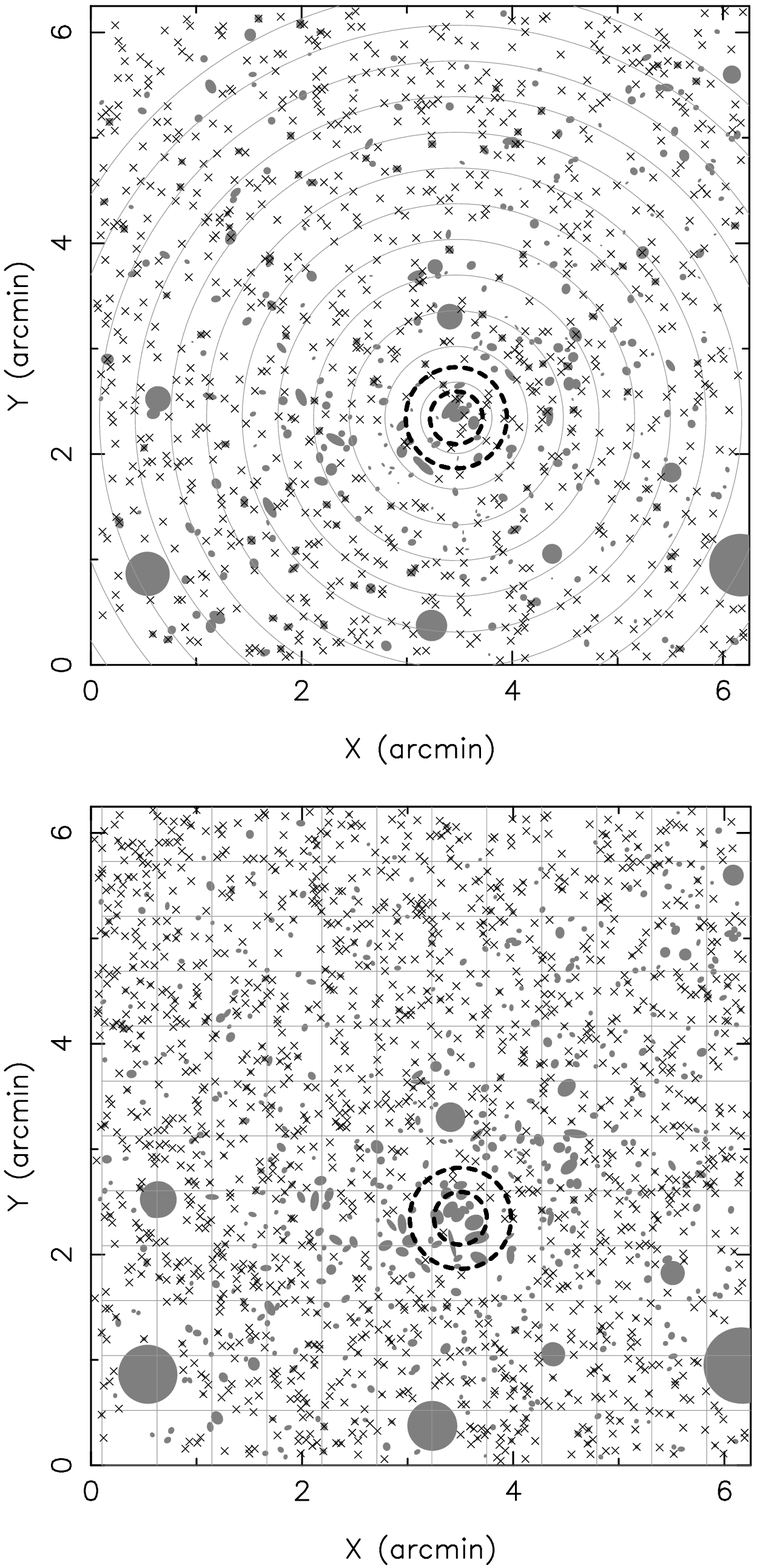}
\emi
\bmi{5.7}
\caption{The mask of the data field of the cluster Cl\ts 0024+17 (grey
  circles) and the location of putative background objects (crosses). The
  inner dashed circle shows the critical curve of the cluster as derived from
  the multiply imaged arc system (from Dye et al.\ 2002)}
\flabel{0024-mask}
\emi
\end{figure}
The results confirm the earlier finding from strong lensing (see
Sect.\ts\ref{sc:WL-4.4}) that the mass in the inner part of this cluster is
larger by a factor $\sim 3$ than estimated from its X-ray emission (Soucail
et al.\ 2000).

\subsubsection{Magnification and shear method compared.}
It is interesting to consider the relative merits of shear and
magnification methods for weak lensing studies. The number of clusters
that have been investigated with either method are quite different,
with less than a handful for which the magnification effect has been
seen. The reason for this is multifold. First, the shear method does
not need external calibration, as it is based on the assumption of
random source ellipticity; in contrast to this, the magnification
method requires the number counts of unlensed sources. Whereas this
can be obtained from the same dataset, provided it covers a
sufficiently large area, this self-calibration removes one of the
strongest appeals of the magnification effect, namely its potential to
break the mass-sheet degeneracy. Second, the magnification method is
affected by the angular correlation of galaxies, as clearly
demonstrated by Athreya et al.\ (2002) in their study of the cluster
MS\ts 1008$-$1224, where the background number counts revealed the
presence of a background cluster which, if not cut out of the data,
would contaminate the resulting mass profile substantially.  Third,
the removal of foreground galaxies, and more seriously, of faint
cluster members introduces an uncertainty in the results which is
difficult to control. Finally, the number count method yields a lower
lensing signal-to-noise than the shear method: If we consider $N_\gamma$ and
$N_\mu$ galaxies in a given patch of the sky, such that for the former
ones the ellipticities have been measured, and for the latter ones
accurate photometry is available and the galaxies are above the
photometric completeness brightness, the signal-to-noise ratio from
the shear -- see (\ref{eq:noisesti}) -- and number count methods are
\be
\rund{\rm S\over N}_\gamma={|\gamma|\over \sigma_\eps}\sqrt{N_\gamma} \;
; \;
\rund{\rm S\over N}_\mu=2\kappa |\alpha -1| \sqrt{N_\mu}\;,
\elabel{shearmagni}
\ee
where we employed (\ref{eq:mu-simple}) in the latter case and assumed that the
source galaxy positions are uncorrelated. The ratio of these
two S/N values is
\be
{\rm (S/N)_\gamma\over (S/N)_\mu}={|\gamma|\over\kappa}\;
{1\over 2\sigma_\eps |1-\alpha|}\,\sqrt{N_\gamma\over N_\mu}\;.
\elabel{S-N-comp}
\ee
For an isothermal mass profile, the first of these factors is unity. With
$\sigma_\eps \approx 0.4$ and $\alpha\approx 0.75$ for R-band counts, the
second factor is $\sim 5$. The final factor depends on the quality of the
data: in good seeing conditions, this ratio is of order unity. However, when
the seeing is bad, the photometric completeness level can be considerably
fainter than the magnitude for which the shape of galaxies can be measured
reliably. Therefore, for data with relatively bad seeing, the magnification
effect may provide a competitive means to extract weak lensing information.
Having said all of this, the magnification method will keep its position as an
alternative to shear measurements, in particular for future multi-color
datasets where the separation of foreground and cluster galaxies from the
background population can be made more cleanly.

\subsubsection{Summary.}
The mass reconstruction of clusters using weak lensing has by now become
routine; quite a few cameras at excellent sites yield data with sub-arcsecond
image quality to enable this kind of work. Overall, the reconstructions have
shown that the projected mass distribution is quite similar to that of the
projected galaxy distribution and the shape of the X-ray emission, at least
for clusters that appear relaxed. There is no strong evidence for a
discrepancy between the mass obtained from weak lensing and that from X-rays,
again with exceptions like for Cl0024+16 mentioned above (which most likely is
not a single cluster). The weak lensing mass profiles are considered more
reliable than the ones obtained from X-ray studies, since they do not rely on
symmetry or equilibrium assumptions. On the other hand, they contain
contributions from foreground and background mass inhomogeneities, and are
affected by the mass-sheet degeneracy. What is still lacking is a combined
analysis of clusters, making use of weak lensing, X-ray, Sunyaev--Zeldovich,
and galaxy dynamics measurements, although promising first attempts have been
published (e.g., Zaroubi et al.\ 1998, 2001; Reblinsky 2000; Dor\'e et al.\
2001; Marshall et al.\ 2003).

\subsection{\llabel{WL-5.7}Aperture mass and other aperture measures}
In the weak lensing regime, $\kappa\ll 1$, the mass-sheet degeneracy
corresponds to adding a uniform surface mass density $\kappa_0$.
However, one can define quantities in terms of the surface mass
density which are invariant under this transformation. In addition,
several of these quantities can be determined directly in terms of the
locally measured shear. In this section we shall present the basic properties
of the aperture measures, whereas in the following section we shall
demonstrate how the aperture mass can be used to find mass concentrations
based solely on their weak lensing properties.

\subsubsection{Aperture mass.}
Let
$U\rund{|\vc\theta|}$ be
a compensated weight (or filter) function, meaning 
$\int \d\theta\;\theta\,U(\theta)=0$,
then the {\it aperture mass}
\begin{equation}
  M_{\rm ap}(\vc\theta_0) = \int\d^2\theta\,\kappa(\vc\theta)\,
  U(|\vc\theta-\vc\theta_0|)
\elabel{5.22}
\end{equation}
is independent of $\kappa_0$, as can be easily seen. For example, if $U$ has
the shape of a Mexican hat, $M_{\rm ap}$ will have a maximum if the filter
center is centered on a mass concentration. The important point to notice is
that $M_{\rm ap}$ can be written directly in terms of the shear (Kaiser et
al.\ 1994; Schneider 1996)
\begin{equation} 
  M_{\rm ap}(\vc\theta_0) = \int\d^2\theta\,{Q(|\vc\theta|)}\,
  \gamma_{\rm t}(\vc\theta;\vc\theta_0)  \;,
\elabel{5.28}
\end{equation}
where we have defined the {\em tangential component\/}
$\gamma_{\rm t}$ of the shear relative to the point
$\vc\theta_0$ [cf. eq.\ \ref{eq:sheart+c}], and 
\be
Q(\theta)={2\over \theta^2}\int_0^\theta\d\theta'\;\theta'\,
U(\theta') - U(\theta)\;.
\elabel{apertQ}
\ee
These relations can be derived from (\ref{eq:5.10}), by rewriting the
partial derivatives in polar coordinates and subsequent integration by parts
(see Schneider \& Bartelmann 1997); it can also be derived directly from the
Kaiser \& Squires inversion formula (\ref{eq:5.4}), as shown in Schneider
(1996). Perhaps easiest is the following derivation (Squires \& Kaiser
1996): We first rewrite
(\ref{eq:5.22}) as
\bea
M_{\rm ap}&=&2\pi\int_0^{\theta_{\rm u}}
\d\vt\;\vt\,U(\vt)\,\ave{\kappa(\vt)} \nonumber \\
&=&2\pi\eck{X(\vt)\,\ave{\kappa(\vt)}}_0^{\theta_{\rm u}}
-2\pi\int_0^{\theta_{\rm u}} \d\vt\;X(\vt)\,{\d\ave{\kappa}\over
\d\vt}\;,
\elabel{MapN-1}
\eea
where $\theta_{\rm u}$ is the radius of the aperture, and we have
defined
\[
X(\theta)=\int_0^{\theta}
\d\vt\;\vt\,U(\vt)\;.
\]
This definition and the compensated nature of $U$ implies that the
boundary terms in (\ref{eq:MapN-1}) vanish. Making use of
(\ref{eq:mass-rela4}), one finds that
\[
{\d\ave{\kappa}\over \d\vt}
={\d\bar{\kappa}\over \d\vt}- {\d\ave{\gamma_{\rm t}}\over \d\vt}
=-{2\over \vt}\ave{\gamma_{\rm t}}-{\d\ave{\gamma_{\rm t}}\over
\d\vt}\;,
\]
where we used (\ref{eq:mass-rela3}) and (\ref{eq:mass-rela4}) to
obtain $\d\bar\kappa/\d\vt=-2\ave{\gamma_{\rm t}}/\vt$. Inserting the
foregoing equation into (\ref{eq:MapN-1}), one obtains
\bea
M_{\rm ap}&=&2\pi\int_0^{\theta_{\rm u}}
\d\vt\;\vt\,{2 X(\vt)\over \vt^2}\ave{\gamma_{\rm t}(\vt)}  \nonumber \\
&+&2\pi\eck{X(\vt)\,\ave{\gamma_{\rm t}(\vt)}}_0^{\theta_{\rm u}}
-2\pi\int_0^{\theta_{\rm u}}
\d\vt\;{\d X\over \d\vt}\,\ave{\gamma_{\rm t}(\vt)}\;.
\eea
The boundary term again vanishes, and one sees that the last equation
has the form of (\ref{eq:5.28}), with the weight function $Q=2
X/\vt^2-U$, reproducing (\ref{eq:apertQ}). 

We shall now consider a few properties of the aperture mass, which
follow directly from (\ref{eq:apertQ}).
\bi
\item
If $U$ has finite support, then $Q$ has finite support, which is due to the
compensated nature of $U$. This implies
that the aperture mass can be calculated on a finite data field, i.e., from
the shear in the same circle where $U\ne 0$.
\item If $U(\theta)={\rm const.}$ for $0\le \theta\le\theta_{\rm in}$, then
  $Q(\theta)=0$ for the same interval, as is see directly from
  (\ref{eq:apertQ}).  Therefore, the strong lensing regime (where $\gamma$
  deviates appreciably from $g$) can be avoided by properly choosing $U$ (and
  $Q$).
\item 
If $U(\theta)=(\pi \theta_{\rm in}^2)^{-1}$ for
$0\le\theta\le \theta_{\rm in}$, 
$U(\theta)=-[\pi (\theta_{\rm out}^2-
\theta_{\rm in}^2)]^{-1}$ for $\theta_{\rm in} < \theta\le \theta_{\rm
out}$, and $U=0$ for $\theta>\theta_{\rm out}$,
then 
$Q(\theta)=\theta_{\rm out}^2\,\theta^{-2}\left[\pi
(\theta_{\rm out}^2-\vartheta_{\rm in}^2)\right]^{-1}$ for
$\theta_{\rm in}\le\theta\le\theta_{\rm out}$, 
and $Q(\theta)=0$
otherwise. For this special choice of $U$, 
\be
M_{\rm ap}=\bar\kappa(\theta_{\rm in})-\bar\kappa(\theta_{\rm
in},\theta_{\rm out})\;,
\ee
the mean mass density inside $\theta_{\rm in}$ minus the mean density
in the annulus $\theta_{\rm in}\le\theta\le\theta_{\rm out}$ (Kaiser
1995). 
Since the latter is non-negative, this yields a lower limit to
$\bar\kappa(\theta_{\rm in})$, and thus to $M(\theta_{\rm in})$.
\ei
The aperture mass can be generalized to the case where the weight
function $U$ is constant on curves other than circles, e.g., on
ellipses, in the sense that the corresponding expressions can be
rewritten directly in terms of the shear on a finite region (see
Squires \& Kaiser 1996 for the case where $U$ is constant on a set of
self-similar curves, and Schneider \& Bartelmann 1997 for a general
set of nested curves). In general, $M_{\rm ap}$ is not a particularly
good measure for the total mass of a cluster -- since it employs a
compensated filter -- but it has been specifically designed that way to
be immune against the mass-sheet degeneracy. However, $M_{\rm ap}$ is
a very convenient measure for mass concentrations (see
Sect.\ts\ref{sc:WL-5.8}) and, as shown above, 
yields a robust lower limit on cluster masses.

\subsubsection{Aperture multipoles.}
The aperture method can also be used to calculate multipoles of the mass
distribution: define the multipoles
\be
Q^{(n)}:=\int\d^2\theta\;\abs{\vc\theta}^n\,U(|\vc\theta|)\,{\rm e}^{n{\rm
    i}\vp}\,\kappa(\vc\theta)\;,
\ee
then the $Q^{(n)}$ can again be expressed as an integral over the
shear. Here, $U$ is a radial weight function for which certain
restrictions apply (see Schneider \& Bartelmann 1997 for details), but is not
required to be compensated for $n>0$. A
few cases of interest are: a weight function $U$ which is non-zero
only within an annulus $\theta_{\rm in}\le \theta\le \theta_{\rm
out}$ and which continuously goes to zero as $\theta\to\theta_{\rm
in,out}$; in this case, the shear is required only within the same
annulus. Likewise, if $U$ is constant for $0\le\theta\le\theta_{\rm
in}$ and then decreases smoothly to zero at $\theta_{\rm out}$, only
the shear within the annulus is required to calculate the multipoles.
Aperture multipoles can be used to calculate the multipole moments of mass
concentrations like clusters directly from the shear, i.e., without obtaining
first a mass map, which allows a more direct quantification of signal-to-noise
properties. 

\subsubsection{The cross aperture.}
We have seen that the Kaiser \& Squires inversion, given by the first
expression in (\ref{eq:5.4}), must yield a real result; the imaginary part of
the integral in (\ref{eq:5.4}) vanishes in the absence of noise. Suppose one
would multiply the complex shear by ${\rm i}={\rm e}^{2{\rm i}\pi/4}$; this
would transform the real part of the integral into the imaginary part and the
imaginary part into the negative of the real part. Geometrically,
multiplication by this phase factor corresponds to rotating the shear at every
point by $45^\circ$. Hence, if all shears are rotated by $\pi/4$, the
real part of the Kaiser \& Squires inversion formula (\ref{eq:5.4}) yields
zero. This $45$-degree test has been suggested by A.\ Stebbins; it can
be used on real data to test whether typical features in the mass map are
significant, as those should have larger amplitude that spurious features
obtained from the mass reconstruction in which the shear has been rotated by
$\pi/4$ (the corresponding `mass map' then yields a good indication of the
typical noise present in the real mass map). 

One can define in analogy to (\ref{eq:5.28}) the cross aperture by replacing
the tangential component of the shear by its cross component. According to the
$45$-degree test, the resulting cross aperture should be exactly zero. Hence,
if we define for $\vc\theta_0=\vc 0$
\bea
M&:=&M_{\rm ap}+{\rm i} M_\perp\!=\!\int\! \d^2\theta\;Q(|\vc\theta|)\eck{
\gamma_{\rm t}(\vc\theta)+{\rm i}\gamma_\times(\vc\theta)} \nonumber \\
&=&-\int\!\d^2\theta\,Q(|\vc\theta|)\,\gamma(\vc\theta)\,{\rm e}^{-2{\rm
    i}\phi}\;, 
\elabel{aperture}
\eea
where $\phi$ is the polar angle of $\vc\theta$ as in (\ref{eq:sheart+c}), then
$M$ is expected to be purely real. We shall make use of this definition and
the interpretation of $M$ in later sections.

\subsection{\llabel{WL-5.8}Mass detection of clusters}
\subsubsection{Motivation.}
If a weak lensing mass reconstruction of a cluster has been performed and a
mass peak is seen, it can also be quantified by applying the aperture mass
statistics to it: placing the center of the aperture on the mass peak, and
choosing the radius of the aperture to match the extent of the mass peak will
give a significant positive value of $M_{\rm ap}$. Now consider to observe a
random field in the sky, and to determine the shear in this field. Then, one
can place apertures on this field and determine $M_{\rm ap}$ at each point. If
$M_{\rm ap}$ attains a significant positive value at some point, it then
corresponds to a point around which the shear is tangentially oriented. Such
shear patterns are generated by mass peaks according to (\ref{eq:5.22}) --
hence, a significant peak in the $M_{\rm ap}$-map corresponds to a mass
concentration (which can, in principle t least, be a mass
concentration just in 
two-dimensional projection, not necessarily in 3D).
Hence, the aperture mass statistics allows us to search for
mass concentrations on blank fields, using weak lensing methods (Schneider
1996). From the estimate (\ref{eq:4.55}), we see that the detectable mass
concentrations have to have typical cluster masses.

The reason why this method is interesting is obvious: As discussed in Sect.\
6 of IN, the abundance of clusters as a function of mass and redshift is an
important cosmological probe. Cosmological simulations are able to predict
the abundance of massive halos for a given choice of cosmological
parameters. To compare these predictions with observations, cluster samples
are analyzed. However, clusters are usually detected either as an overdensity
in the galaxy number counts (possibly in connection with color information, to
employ the red cluster sequence -- see Gladders \& Yee 2000), or from extended
X-ray sources. In both cases, one makes use of the luminous properties of the
clusters, and cosmologists find it much more difficult to predict those, as
the physics of the baryonic component of the matter is much harder to handle
than the dark matter. Hence, a method for cluster detection that is
independent of their luminosity would provide a clean probe of cosmology. From
what was said above, the aperture mass provides such method (Schneider 1996). 

To illustrate this point, we show in Fig.\ \ref{fig:JSW} the projected
mass and the corresponding shear field as it results from studying the
propagation of light rays through a numerically generated cosmological
matter distribution (Jain et al.\ 2000; we shall return to such
simulations in Sect.\ts\ref{sc:WL-6.6}). From the comparison of these
two panels, one sees that for each large mass concentration there is a
tangential shear pattern centered on the mass peak. Thus, a systematic
search for such shear patterns can reveal the presence and abundance
of peaks in the mass map.

\begin{figure}[t]
\bmi{6}
\includegraphics[width=5.7cm]{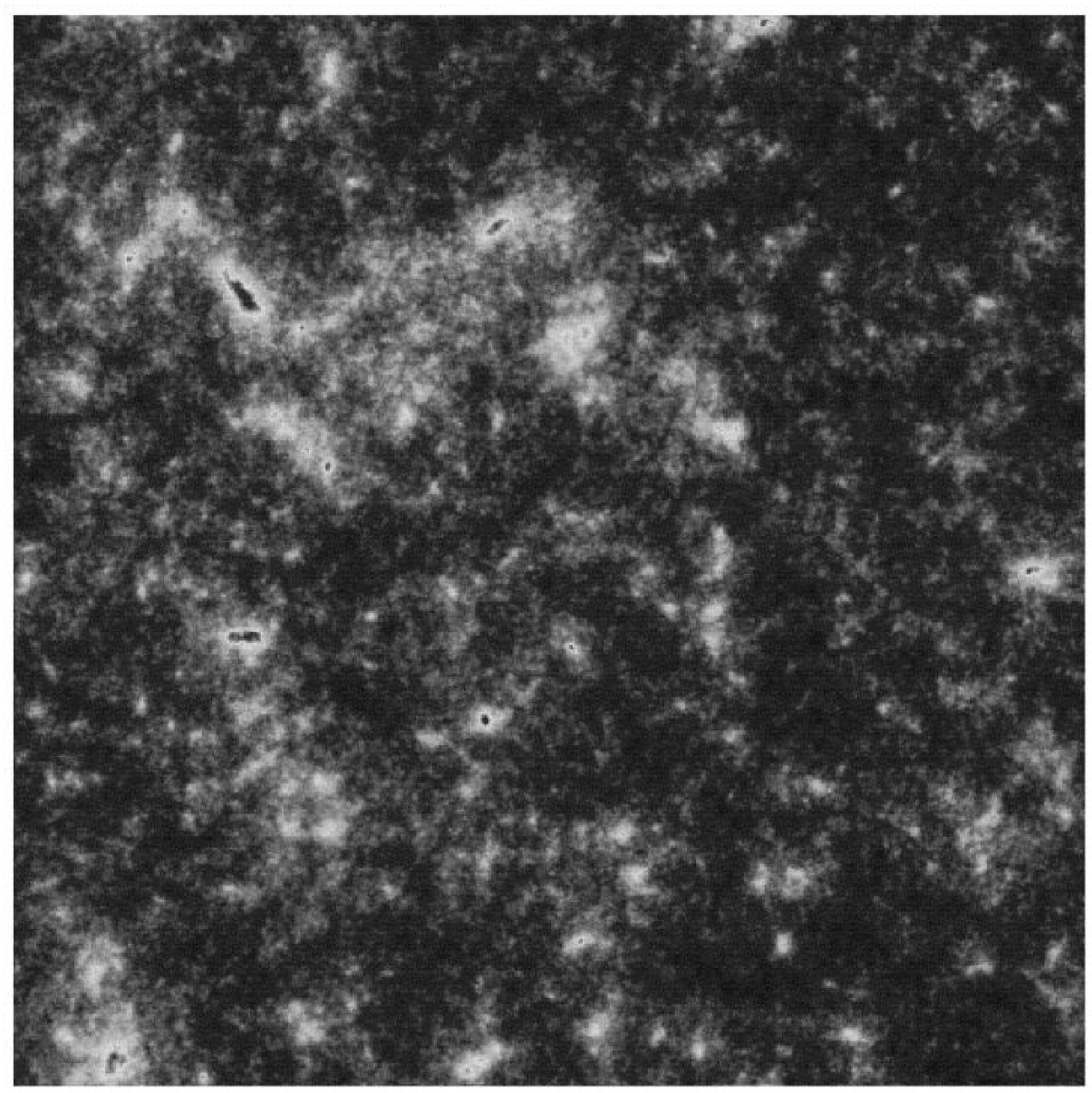}
\emi\bmi{5.7}
\includegraphics[width=5.7cm]{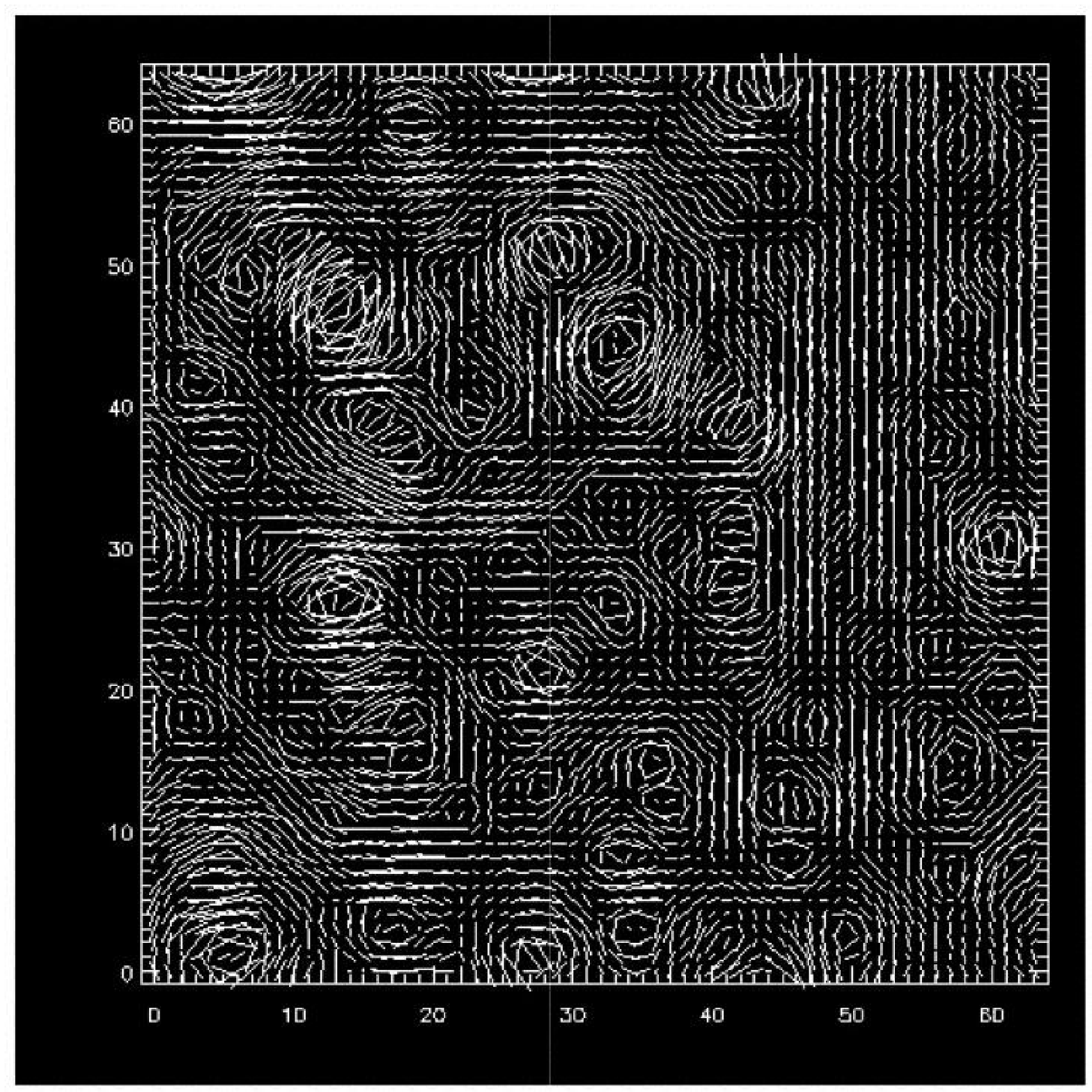}
\emi

\bmi{6}
\includegraphics[width=5.7cm]{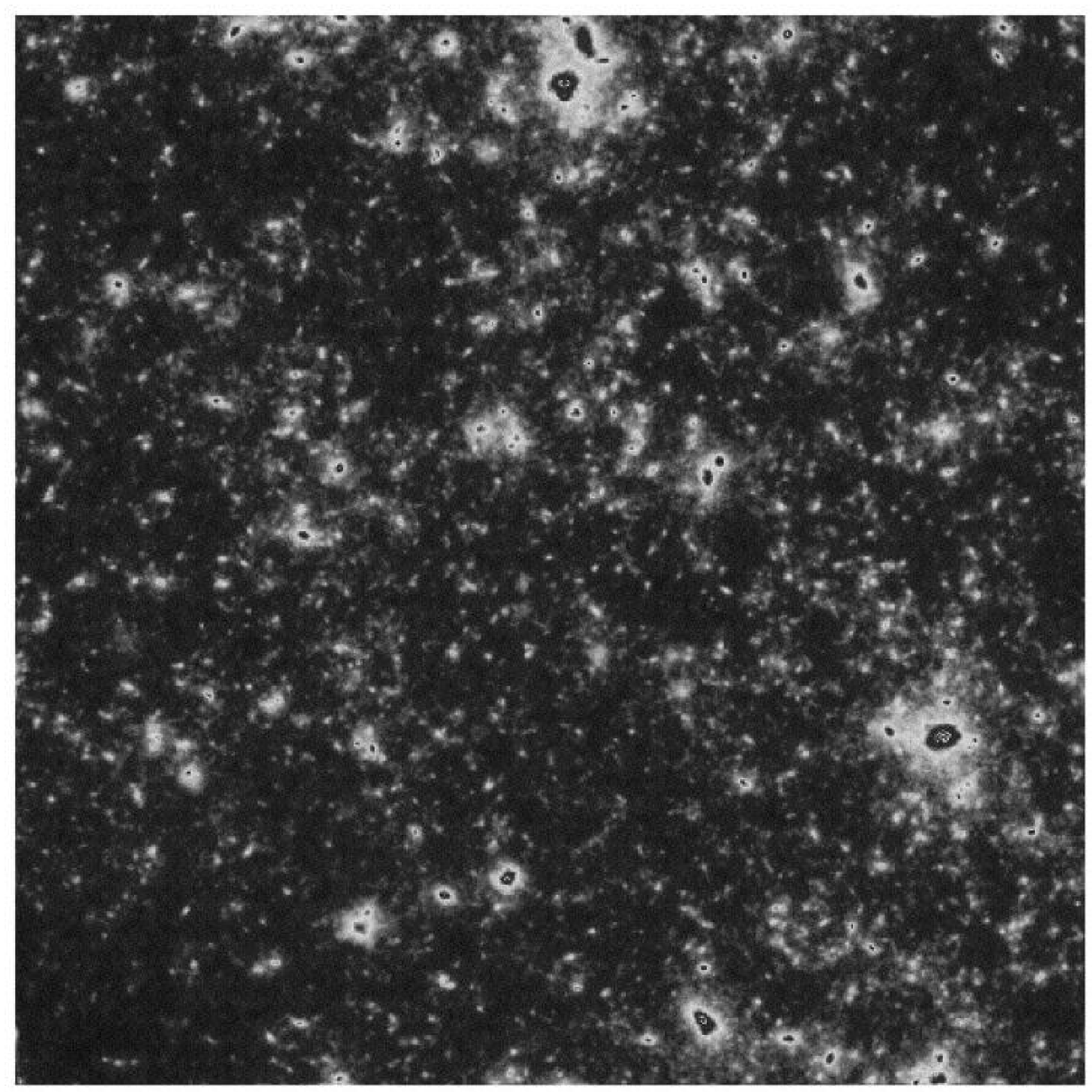}
\emi\bmi{5.7}
\includegraphics[width=5.7cm]{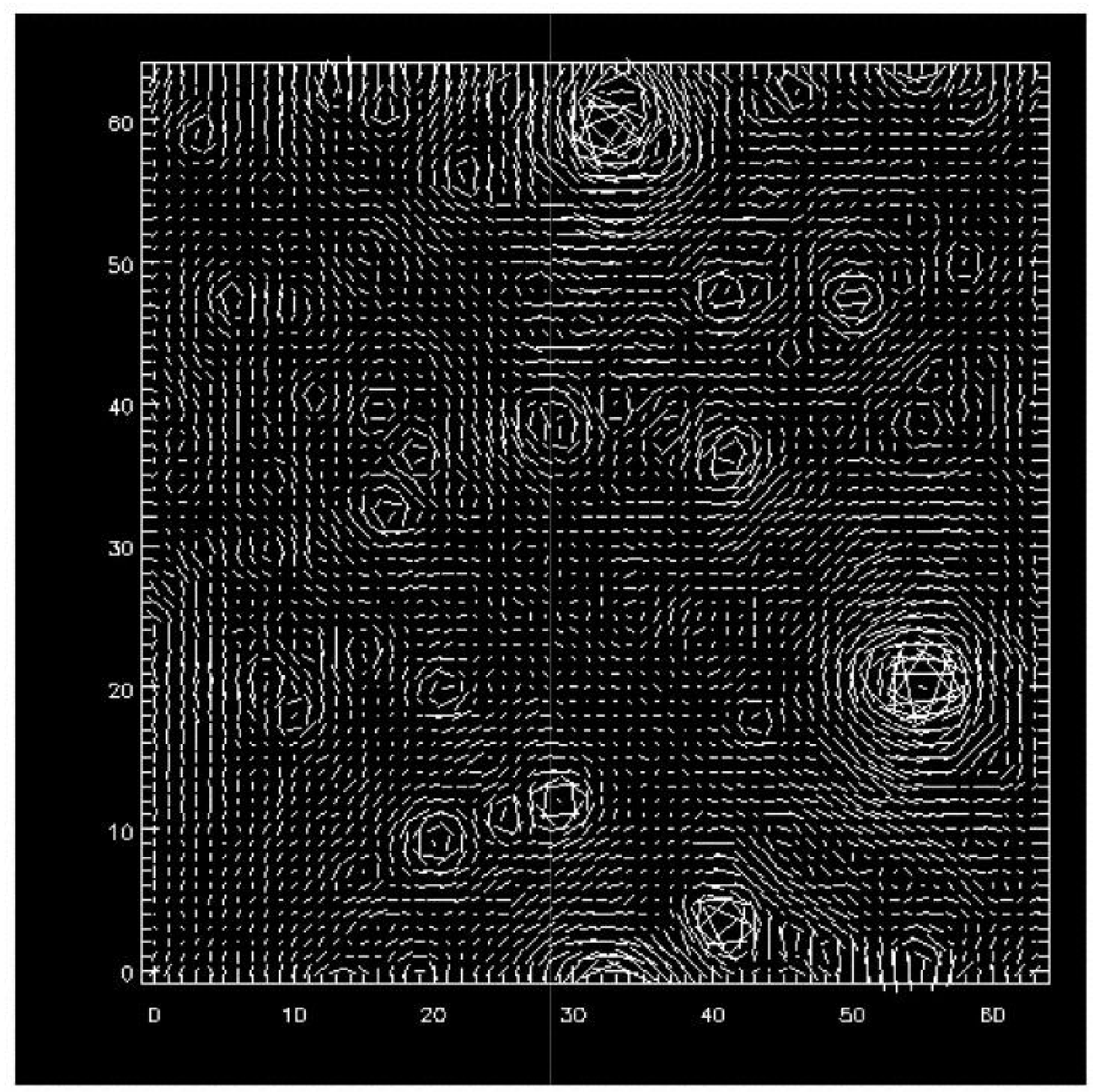}
\emi
\caption{Projected mass distribution of the large-scale structure
  (left), and the corresponding shear field (right), where the length and
  orientation of the sticks indicate the magnitude and direction of the local
  shear. The top panels correspond to an Einstein--de Sitter model of the
  Universe, whereas the bottom panels are for a low density open model. The
  size of the field is one degree on the side, and the background galaxies are
  assumed to all lie at the redshift $z_{\rm s}=1$. Note that each mass
  concentration seen in the left-hand panels generates a circular shear
  pattern at this position; this form the basic picture of the detection of
  mass concentrations from a weak lensing observation (from Jain et al.\ 
  2000)} \flabel{JSW}
\end{figure}
 
\subsubsection{The method.}
The search for mass concentrations can thus be carried out by calculating the
aperture mass on a gid over the data field and to identify significant peaks. 
A practical estimator for $M_{\rm ap}$ is obtained by replacing the
integral in (\ref{eq:5.28}) by a finite sum over image ellipticities:
\be
\hat M_{\rm ap}(\vc\theta_0)={1\over n}
\sum_i \eps_{{\rm t}i}(\vc\theta_0)\,Q(|\vc\theta_i-\vc\theta_0|)\;,
\elabel{hatMap}
\ee
where $n$ is the mean number density of galaxy images, and
$\eps_{{\rm t}i}(\vc\theta_0)$ is the ellipticity component of a galaxy at
$\vc\theta_i$ tangent to the center $\vc\theta_0$ of the aperture.
This estimator has easy-to-quantify signal-to-noise properties.
In the absence of a lensing signal, $\ave{\hat M_{\rm ap}}\equiv 0$, and
the dispersion of $\hat M_{\rm ap}(\vc\theta_0)$ is
\be
\sigma^2(\vc\theta_0)={\sigma_\eps^2\over 2n^2}
\sum_i Q^2(|\vc\theta_i-\vc\theta_0|)\;;
\elabel{59}
\ee
hence, the signal-to-noise of $\hat M_{\rm ap}(\vc\theta_0)$ is
\be
{\rm S\over N}={\sqrt{2}\over \sigma_\eps}\,
{\sum_i \eps_{{\rm t}i}(\vc\theta_0)\,Q(|\vc\theta_i-\vc\theta_0|)
\over
\sqrt{\sum_i Q^2(|\vc\theta_i-\vc\theta_0|)}} \;.
\elabel{MapSN}
\ee
The noise depends on $\vc\theta_0$, as the image number density can vary of
data field. The size (or radius) of the aperture shall be adapted to the mass
concentrations excepted: too small aperture radii miss most of the lensing
signal of real mass concentrations, but is more susceptible to noise peaks,
whereas too large aperture radii include regions of very low signal which may
be swamped again by noise. In addition, the shape of the filter function $Q$
can be adapted to the expected mass profiles of mass concentrations; e.g., one
can design filters which are particularly sensitive to NFW-like density
profiles. In order not to prejudice the findings of a survey, it may be
advantageous to use a `generic' filter function, e.g., of the form 
\be
U(\vt)={9\over \pi\theta^2}\rund{1-{\vt^2\over \theta^2}}\rund{{1\over
    3}-{\vt^2\over \theta^2}}\; ;\;\;
Q(\vt)={6\over \pi\theta^2}\,{\vt^2\over \theta^2}\rund{1-{\vt^2\over
    \theta^2}}\;. 
\elabel{U+Qgeneric}
\ee
The relation between the two expressions for $M_{\rm ap}$ given by
(\ref{eq:5.22}) and (\ref{eq:5.28}) is only valid if the aperture lies fully
inside the data field. If it does not, i.e., if the aperture crosses the
boundary of the data field, these two expressions are no longer equivalent;
nevertheless, the estimator (\ref{eq:hatMap}) still measures a tangential
shear alignment around the aperture center and thus signifies the presence of
a mass concentration. 

There are superior estimates of the significance of a detected mass
peak than using the signal-to-noise ratio (\ref{eq:MapSN}). One
consists in bootstrapping; there one calculates $M_{\rm ap}$ at a given
point (where $N$ galaxies are in the aperture) many times by randomly
drawing -- with replacement -- $N$ galaxies and tests how often is
signal negative. The fraction of cases with negative values
corresponds to the error level of having a positive detection of
$M_{\rm ap}$. Alternatively, one can conduct another Monte-Carlo
experiment, by randomizing all galaxy image orientations and
calculating $M_{\rm ap}$ from these randomized samples, and ask in
which fraction of realizations is the value of $M_{\rm ap}$ larger
than the measured value? As the randomized galaxies should show no
lensing signal, this fraction is again the probability of getting a
value as large as that measured from random galaxy orientations. 
In fact, from the central limit theorem one expects that the
probability distribution of $M_{\rm ap}$ from randomizing the image
orientations will be a Gaussian of zero mean, and its dispersion can
be calculated directly from (\ref{eq:hatMap}) to be
\be
\sigma^2(\vc\theta_0)
={1\over 2n^2}\sum_i |\eps_i|^2\,Q^2(|\vc\theta_i-\vc\theta_0|) \;,
\ee
which is similar to (\ref{eq:59}), but accounts for the moduli of
the ellipticity of the individual galaxy images.

Both of the aforementioned methods take the true ellipticity
distribution of galaxy images into account, and should yield very
similar results for the significance.  Highly significant peaks
signify the presence of a mass concentration, detected solely on the
basis of its mass, and therefore, it is a very promising search method
for clusters.

There is nothing special about the weight function
(\ref{eq:U+Qgeneric}), except mathematical simplicity. It is therefore
not clear whether these filter functions are most efficient to detect
cluster-mass matter concentrations. In fact, as shown in Schneider
(1996), the largest S/N is obtained if the filter function $U$ follows
the true mass profile of the lens or, equivalently, if $Q$ follows its
radial shear profile. Hennawi \& Spergel (2003) and Schirmer (2004)
tested a large range of filter functions, including
(\ref{eq:U+Qgeneric}), Gaussians, and those approximating an NFW
profile. Based on numerical ray-tracing simulations, Hennawi \&
Spergel conclude that the `truncated' NFW filter is most efficient for
cluster detections; the same conclusion has been achieved by Schirmer
(2004) based on wide-field imaging data. 

Furthermore, Hennawi \& Spergel have complemented their cluster search
by a `tomographic' component, assuming that the source galaxies have
(photometric) redshift estimates available. Since the lens strength is
a function of source redshift, the expected behaviour of the aperture
mass signal as a function of estimated source redshift can be used as
an additional search criteria. They shown that this additional
information increases the sensitivity of weak lensing to find mass
concentrations, in particular for higher-redshift ones; in fact, the
cluster search by Wittman et al.\ (described below) has employed the
use of redshift information. As an additional bonus, this method also
provides an estimate of the lens redshift.

\subsubsection{Results.}
In the past few years, a number of clusters and/or cluster candidates have
been detected by the weak lensing method, and a few of them shall be discussed
here. The right-hand panel of Fig.\ \ref{fig:VLT-cluster} shows the mass
reconstruction of one of the 50 FORS1@VLT fields observed in the course of a
cosmic shear survey (see Sect.\ts\ref{sc:WL-7.1}). This reconstruction shows
an obvious mass peak, indicated by a circle. The left panel shows the optical
image, and it is obvious that the location of the mass peak coincides with a
concentration of bright galaxies -- this certainly is a cluster, detected by
its weak lensing signal. However, no follow-up observations have been
conducted yet to measure its redshift.

\begin{figure}

\includegraphics[width=11.7cm]{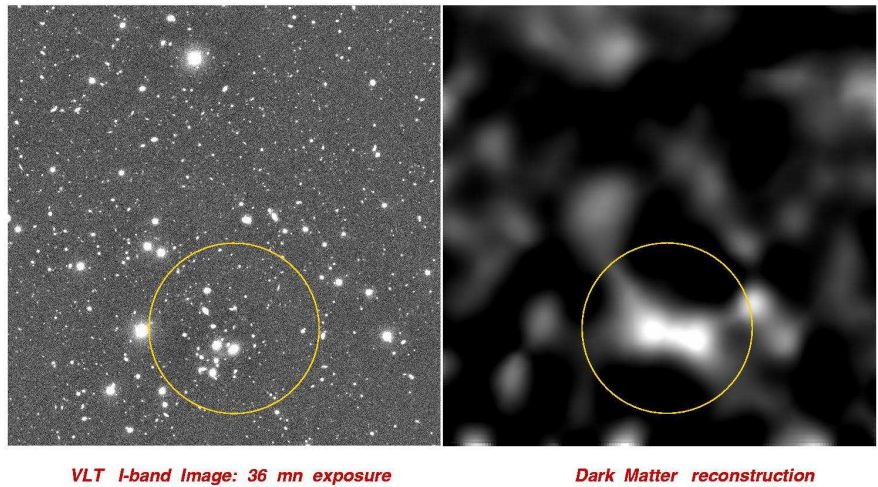}
\caption{A cosmic shear survey was carried out with the FORS1 instrument on
  the VLT (see Maoli et al.\ 2001 and Sect.\ \ref{sc:WL-7.1} below). The left
  panel shows one of the 50 fields observed in the course of this survey,
  whereas the right panel shows a weak-lensing mass reconstruction of this
  field. Obviously, a strong mass peak is detected in this reconstruction,
  indicated by the circle. At the same position, one finds a strong
  overdensity of relatively bright galaxies on the VLT image; therefore, this
  mass peak corresponds to a cluster of galaxies. A reanalysis of all 50 VLT
  fields (Hetterscheidt 2003) yielded no further significant cluster
  candidate; however, with a field size of only $\sim 6\arcminf 5$, detecting
  clusters in them is difficult unless these are positioned close to the
  field centers}
\flabel{VLT-cluster}
\end{figure}

Wittman et al.\ (2001, 2003) reported on the discovery of two clusters
from their wide-field weak lensing survey; one of them is shown in
Fig.\ \ref{fig:Wittcluster} and discussed here. First, a peak in their
mass reconstruction was identified which has a significance of
4.5$\sigma$. The location of the mass peak is identified with a
concentration of red elliptical galaxies, with the two centers
separated by about $1'$ (which is about the accuracy with which the
centers of mass concentrations are expected to be determined from mass
reconstructions). Follow-up spectroscopy confirmed the galaxy
concentration to be a cluster at redshift $z_{\rm d}=0.28$, with a
velocity dispersion of $\sigma_v\sim 600\, {\rm km/s}$. Since
multi-color photometry data are available, photometric redshift
estimates of the faint galaxy population have been obtained, and the
tangential shear around the mass peak has been investigated as a
function of this estimated redshift. The lens signal rises as the
redshift increases, as expected due to the lensing efficiency factor
$D_{\rm ds}/D_{\rm s}$. In fact, from the source redshift dependence
of the lens signal, the lens redshift can be estimated, and yields a
result within $\sim 0.03$ of the spectroscopically measured $z_{\rm
d}$. Hence, in this case not only can the presence of a cluster be
inferred from weak lensing, but at the same time a cluster redshift
has been obtained from lensing observations alone. This is one example
of using source redshift information to investigate the redshift
structure of the lensing matter distribution; we shall return to a
more general discussion of this issue in Sect.\ts\ref{sc:WL-7.6}.
 
\begin{figure}
\bmi{6.7}
\includegraphics[width=6.5cm]{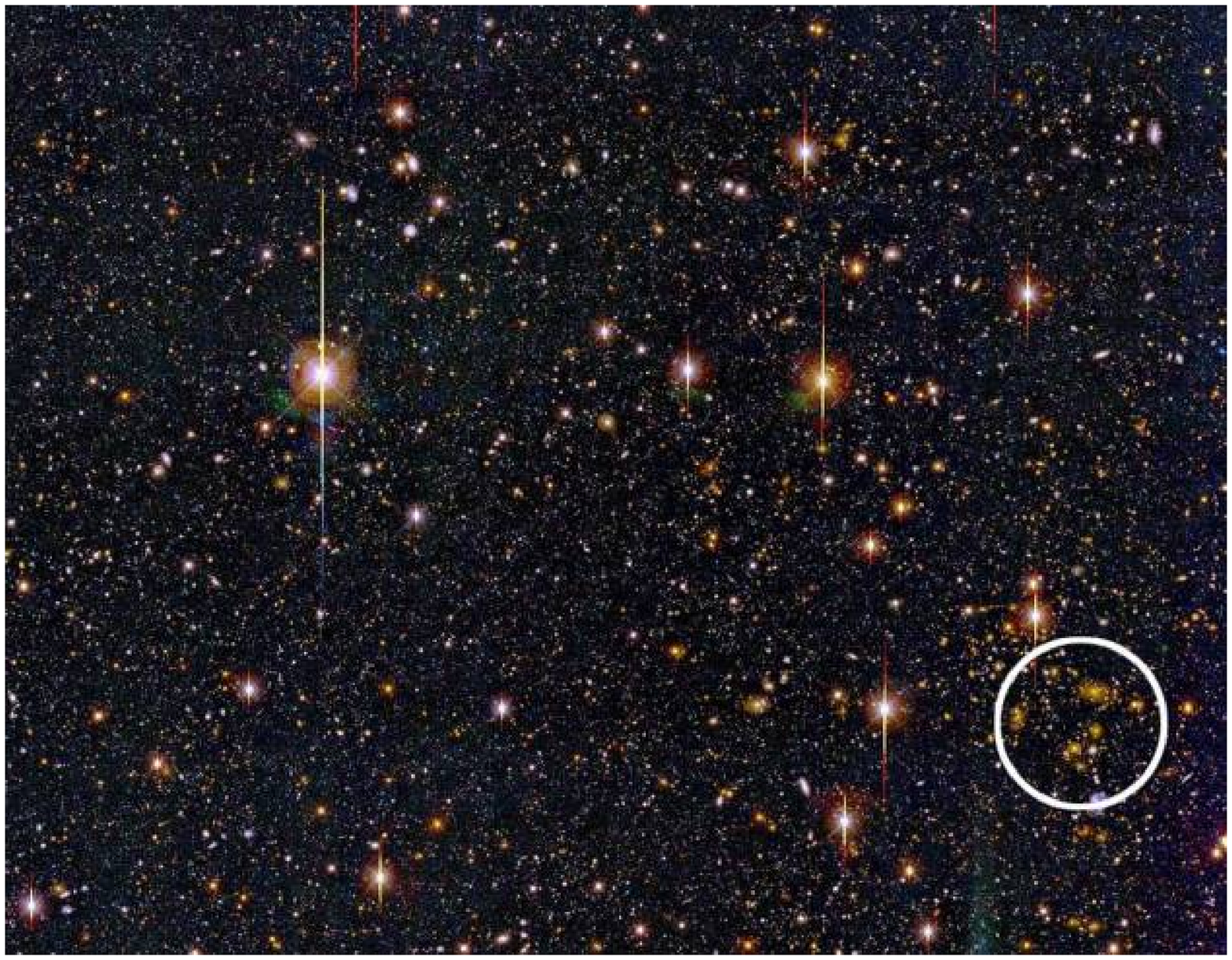}
\emi
\bmi{5.0}
\includegraphics[width=5.0cm]{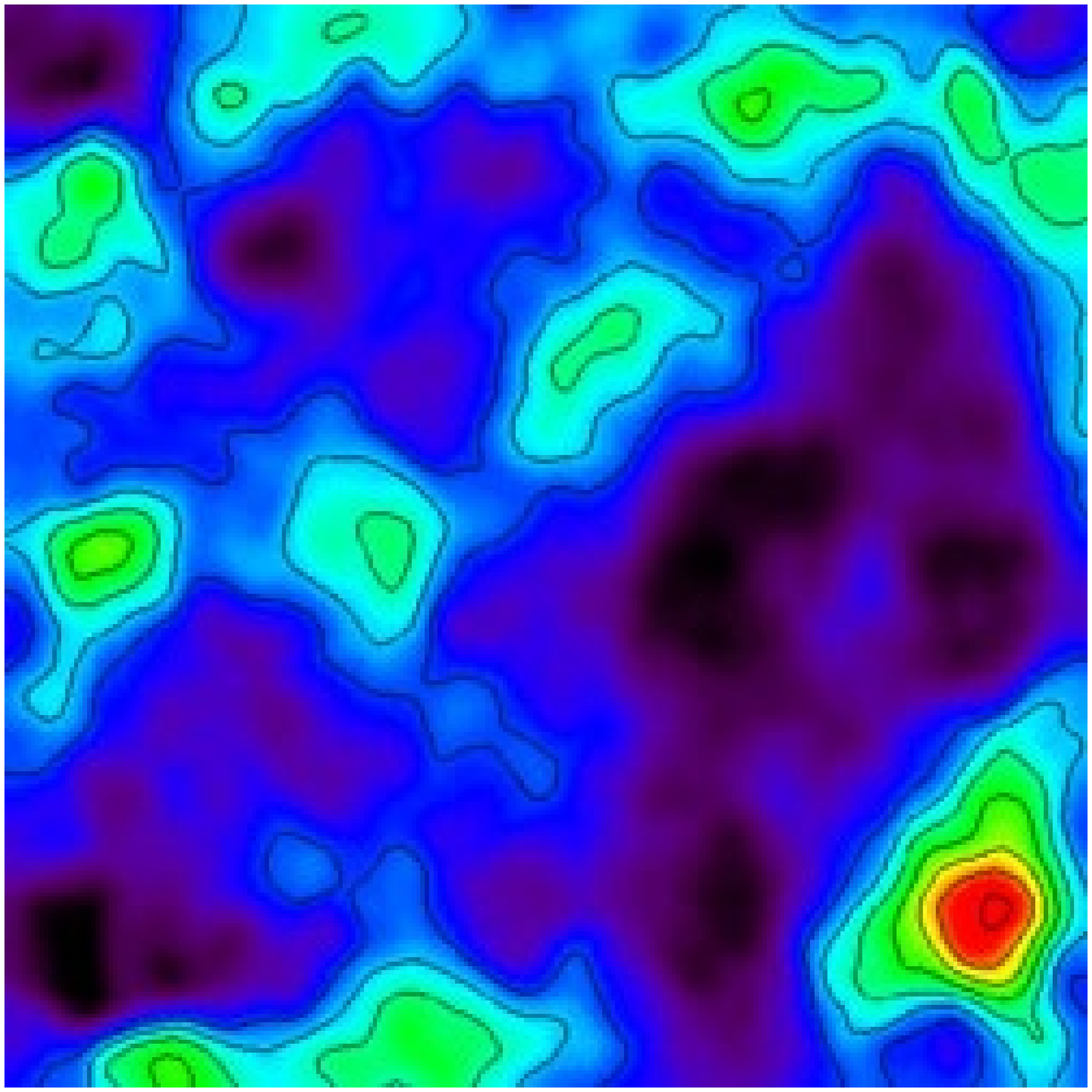}
\emi
\caption{
Left: BTC image of a blank field, right: mass reconstruction, showing
the presence of a (mass-selected) cluster near the lower right corner
-- spectroscopically verified 
to be at $z=0.276$ (from Wittman et al.\ 2001)
}
\flabel{Wittcluster}
\end{figure}

In a wide-field imaging weak lensing survey of galaxy clusters, Dahle
et al.\ (2003) detected three significant mass peaks away from the
clusters that were targeted. One of these cases is illustrated in
Fig.\ \ref{fig:Dahle-cluster}, showing the mass reconstruction in the
field of the cluster A\ 1705. The mass peak South-West of the cluster
coincides with a galaxy concentration at $z\sim 0.55$, as estimated
from their color, and an arc is seen near the brightest galaxy of this
cluster. A further cluster was detected in the wide-field image of the
A222/223 double cluster field (Dietrich et al.\ 2004) which coincides
with an overdensity of galaxies.  Hence, by now of order ten cluster-mass
matter concentrations have been discovered by weak lensing techniques
and verified as genuine clusters from optical photometry and, for some
of them, spectroscopy.
 
\begin{figure}
\bmi{6.5}
\includegraphics[width=6.3cm]{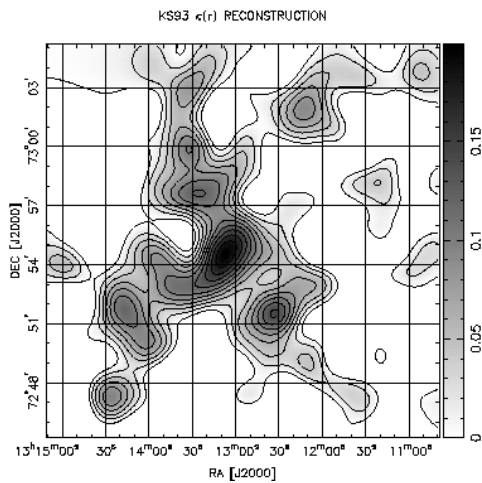}
\emi
\bmi{5.2}
\caption{Shown is the mass reconstruction of the field containing the
cluster A1705, located near the center of this field. The  peak
$\sim 4'$ to the North-East of A1705 appears to be associated with galaxies at
the same redshift as A1705. However, the peak $\sim 4'$ South-West of
A1705 seems to be associated with galaxies at considerably larger
redshift, at $z\sim 0.55\pm 0.05$, as determined from the $V-I$ colors
of the corresponding galaxy concentration. Indeed, an arc curving
around the central galaxy of this newly detected cluster candidate is
observed (from Dahle et al.\ 2003)}
\flabel{Dahle-cluster}
\emi
\end{figure}

Miyazaki et al.\ (2002) used a $2.1\, {\rm deg}^2$ deep image taken with the
Suprime-Cam wide-field imager on Subaru to search for mass peaks. They
compared their peak statistics with both, the expected peak statistics from a
noise field created by intrinsic galaxy ellipticities (Jain \& van Waerbeke
2000) as well as from N-body simulations, and found a broader distribution in
the actual data. They interpret this as statistical evidence for the presence
of mass peaks; however, their interpreation of the significant dips in the
mass map as evidence for voids cannot hold, as the density contrast of voids
is too small (since the fractional density contrast $\delta> -1$) to be
detectable with weak lensing. They find a number density of $>5\sigma$ peaks
of about $5\,{\rm deg}^{-2}$, well in agreement with predictions from Kruse \&
Schneider (1999) and Reblinsky et al.\ (1999). Schirmer (2004) investigated
about $16\,{\rm deg}^2$ of images taken with the WFI@ESO/MPG\ 2.2m, and
detected 100 $>4\sigma$-peaks, again in good agreement with theoretical
expectations. 

\subsubsection{Dark clusters?}
In addition, however, this method has the potential to discover mass
concentrations with very large mass-to-light ratio, i.e., clusters which are
very faint optically and which would be missed in more conventional surveys
for clusters. Two potential `dark clusters' have been reported in the
literature.\footnote{A third case reported in Miralles et al.\
(2002) has in the meantime been considerably weakened (Erben et al.\ 2003).}   
Umetsu \& Futamase (2000), using the WFPC2 onboard HST detected a
highly significant ($4.5\sigma$) mass concentration $1\arcminf 7$ away from
the cluster Cl\ts 1604+4304, also without an apparent overdensity of
associated galaxies. 

In the course of a wide-field weak lensing analysis of the cluster A\
1942, Erben et al.\ (2000) detected a mass peak which, using the
aperture mass statistics introduced previously, has been shown to be
highly significant ($\sim 4.7\sigma$ on the V-band image), with the
significance being obtained from the randomization and bootstrapping
techniques described above. An additional I-band image confirmed the
presence of a mass peak at the same location as on the V-band image,
though with somewhat lower significance. No concentration of galaxies
is seen near the location of the mass peak, which indicates that it
either is a very dark mass concentration, or a cluster at a fairly
high redshift (which, however, would imply an enormous mass for it),
or, after all, a statistical fluke. It is important to note that the
signal in $M_{\rm ap}$ comes from a range of radii (see Fig.\
\ref{fig:DarkClump.shear}); it is not dominated by a few highly
flattened galaxies which happen to have a fortitious orientation.
Gray et al.\ (\cite{Gray-DC-IR}) have used near-IR images to search
for a galaxy concentration in this direction, without finding an
obvious candidate. Therefore, at present it is unclear whether the
`dark clump' is indeed a very unusual cluster. A low-significance
X-ray source near its position, as obtained in a ROSAT observation of
A\ 1942, certainly needs confirmation by the more sensitive X-ray
observatory XMM.\footnote{Judging from the results of several proposal
submissions, people on X-ray TACs seem not to care too much about dark
cluster candidates.}  Of course, if there are really dark clusters,
their confirmation by methods other than weak lensing would be
extremely difficult; but even if we are dealing with a statistical
fluke, it would be very important to find the cause for it. An HST
mosaic observation of this field has been conducted; a first analysis
of these data was able to confirm the findings of Erben et al., in the
sense that the shear signal from galaxies seen in both, the HST images
and the ground-based data, have a significant tangential alignment
(von der Linden 2004). However, contrary to expectations if this was
truly a lensing mass signal, there is hardly any tangential alignment
from fainter galaxies, although they are expected to be located at
higher redshift and thus should show a stronger shear signal. However,
as a word of caution, the PSF anisotropy of WFPC2 cannot be controlled
from stars on the image, owing to the small field-of-view, and no
stellar cluster has been observed with the filter with which the dark
clump observations were conducted, so that the PSF anisotropy cannot
be accurately inferred from such calibration images. The existence of
dark clusters would be highly unexpected in view of our current
understanding of structure formation and galaxy evolution, and would
require revisions of these models.

\begin{figure} 
\bmi{7.0}
\includegraphics[width=6.8cm]{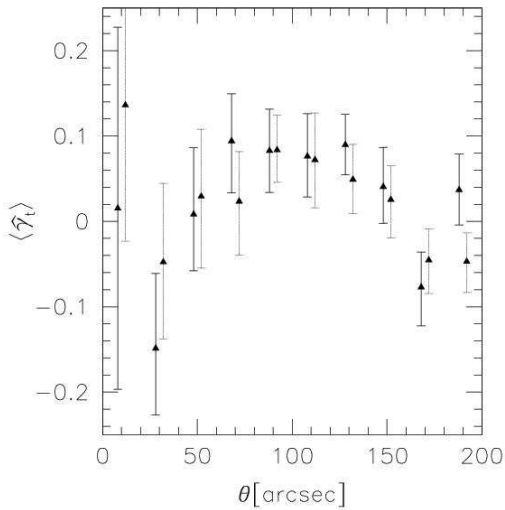}
\emi
\bmi{4.7}
\caption{Tangential shear profile from both (V- and I-band) images
around the `dark cluster' 
  candidate near the cluster A1942. For each angular scales, two points (and
  corresponding error bars) are plotted, which are derived from two different
  images of the field in the V- and I-band. It can be seen that the tangential
  shear signal extends over quite a range in radius (from Erben et al.\ 2000)} 
\flabel{DarkClump.shear}
\emi
\end{figure}
 
The search for clusters by weak lensing will certainly continue, due
to the novel properties of the cluster samples obtained that way. The
observational data required are the same as those used for cosmic
shear studies, and several very wide-field surveys are currently
conducted, as will be described in Sect.\ \ref{sc:WL-7}. Hence, we can
expect to have a sizable sample of shear-selected clusters in the near
future. The search for mass concentrations by weak lensing techniques is
affected by foreground and background inhomogeneities, which impose
fundamental limits on the reliability and completeness of such searches; we
shall return to this issue in Sect.\ts\ref{sc:WL-9.3}.

\subsubsection{Expectations.}
Kruse \& Schneider (1999) have calculated the expected number density
of lensing-detected clusters, using the aperture-mass method, for
different cosmological parameters; these have been verified in
numerical simulations of the large-scale structure by Reblinsky et
al.\ (1999). Depending on the cosmological model, a few clusters per
deg$^2$ should be detected at about the $5\sigma$ level. The
dependence of the expected number density of detectable mass peaks on
the cosmological parameters can be used as a cosmological probe; in
particular, Bartelmann et al.\ (\cite{Bart-396-21}) and Weinberg \&
Kamionkowski (\cite{WeinKami}) demonstrate that the observed abundance
of weak lensing clusters can probe the equation-of-state of the dark
energy.  Bartelmann et al.\ (2001) argued that the abundance of weak
lensing detected clusters strongly depends on their mass profile, with
an order-of-magnitude difference between NFW profiles and isothermal
spheres. Weinberg \& Kamionkowski (2002) argued, based on the
spherical collapse model of cluster formation, that a considerable
fraction of such detections are expected to be due to non-virialized
mass concentrations, which would then be considerably weaker X-ray
emitters and may be candidates for the `dark clusters'.

\section{\llabel{WL-6}Cosmic shear -- lensing by the LSS}
Up to now we have considered the lensing effect of localized mass
concentrations, like galaxies and clusters. In addition to that, light
bundles propagating through the Universe are continuously deflected
and distorted by the gravitational field of the inhomogeneous mass
distribution, the large-scale structure (LSS) of the cosmic matter
field. This distortion of light bundles causes shape and size distortions of
images of distant galaxies, and therefore, the statistics of the
distortions reflect the statistical properties of the LSS (Gunn 1967;
Blandford et al.\ 1991; Miralda-Escud\'e 1991; Kaiser 1992).

{\em Cosmic shear} deals with the investigation of this connection,
from the measurement of the correlated image distortions to the
inference of cosmological information from this distortion
statistics. As we shall see, cosmic shear has become a very important
tool in observational cosmology. From a technical point-of-view, it is
quite challenging, first because the distortions are indeed very weak
and therefore difficult to measure, and second, in contrast to
`ordinary' lensing, here the light deflection does not occur in a
`lens plane' but by a 3-D matter distribution, implying the need for a
different description of the lensing optics. We start by looking at
the description of light propagating through the Universe, and then
consider the second-order statistical properties of the cosmic shear
which reflect the second-order statistical properties of the cosmic
matter field, i.e., the power spectrum. Observational results from
cosmic shear surveys are presented in Sect.\ \ref{sc:WL-7}, whereas
higher-order statistical properties of the shear field will be treated
in Sect.\ \ref{sc:WL-9}.

\subsection{\llabel{WL-6.1}Light propagation in an inhomogeneous Universe}
In this brief, but rather technical section, we outline the derivation of the
lensing effects of the three-dimensional mass distribution between the faint
background galaxy population and us; the reader is referred to Bartelmann \&
Schneider (2001) for a more detailed discussion. The final result of this
consideration has a very simple interpretation: in the lowest-order
approximation, the 3-D cosmological mass distribution can be considered, for
sources at a single redshift $z_{\rm s}$, as an effective surface mass density
$\kappa$, just like in ordinary lensing. The resulting $\kappa$ is obtained as
a line-of-sight integral of the density contrast $\Delta\rho$, weighted by the
usual geometrical factor entering the lens equations.

The laws of light propagation follow from Einstein's General
Relativity; according to it, light propagates along the
null-geodesics of the space-time metric. As shown in SEF (see also
Seitz et al.\ 1994), one can
derive from General Relativity that 
the governing equation for the propagation of thin light bundles
through an arbitrary space-time is the equation of geodesic deviation,
\begin{equation}
  \frac{\d^2\vc\xi}{\d\lambda^2} = {\mathcal T}\,\vc\xi\;,
\elabel{6.1}
\end{equation}
where $\vc\xi$ is the separation vector of two neighboring light rays,
$\lambda$ the affine parameter along the central ray of the bundle,
and ${\mathcal T}$ is the {\em optical tidal matrix\/} which describes
the influence of space-time curvature on the propagation of light.
${\mathcal T}$ can be expressed directly in terms of the Riemann
curvature tensor. 

For the case of a weakly inhomogeneous Universe, the tidal matrix can
be explicitly calculated in terms of the peculiar Newtonian potential.  For
that, we write the slightly perturbed metric of the Universe in the
form
\begin{equation}
  \d s^2 = a^2(\tau)\left[
    \left(1+\frac{2\Phi}{c^2}\right)\,c^2\d\tau^2-
    \left(1-\frac{2\Phi}{c^2}\right)
    \left(\d w^2+f_K^2(w)\d\omega^2\right)
  \right]\;,
\elabel{3.38}
\end{equation}
where $w$ is the comoving radial distance, $a=(1+z)^{-1}$ the scale factor,
normalized to unity today, $\tau$ is the conformal time, related to the cosmic
time $t$ through $\d t=a\;\d \tau$, $f_K(w)$ is the comoving angular diameter
distance, which equals $w$ in a spatially flat model, and $\Phi(\vc x,w)$
denotes the Newtonian peculiar gravitational potential which depends on the
comoving position vector $\vc x$ and cosmic time, here expressed in terms of
the comoving distance $w$ (see Sect.\ts 4 of IN for a more detailed
description of the various cosmological terms).
In this metric, the tidal matrix ${\cal T}$ can be calculated in terms of the
Newtonian potential $\Phi$, and correspondingly,
the equation of geodesic deviation (\ref{eq:6.1})
yields the evolution equation for the comoving separation vector $\vc
x(\vc\theta,w)$ between a ray separated by an angle $\vc\theta$ at the
observer from a fiducial ray
\begin{figure}
\bc
\includegraphics[width=8.5cm]{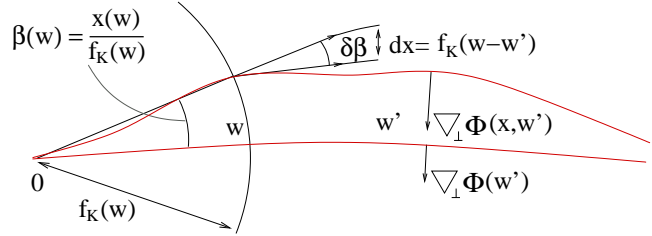}
\ec
\caption{Illustration of the evolution of the separation 
between two light rays in a curved space-time (source: T. Schrabback)}
\flabel{Light-propa}
\end{figure}
\begin{equation} 
\frac{\d^2\vc x}{\d w^2} +
K\,\vc x = -\frac{2}{c^2} \eck{ \nabla_\perp\Phi\rund{\vc
x(\vc\theta,w),w} -\nabla_\perp\Phi^{(0)}\rund{w}} \;,
\elabel{6.12}
\end{equation}
where $K=(H_0/c)^2\,(\Omega_{\rm m}+\Omega_\Lambda-1)$ is the spatial
curvature of the Universe, $\nabla_\perp=(\partial/\partial
x_1,\partial/\partial x_2)$ is the transverse {\it comoving} gradient
operator, and $\Phi^{(0)}(w)$ is the potential along the fiducial
ray.\footnote{In some of the literature, this transport equation is written
  without the term accounting for the potential along the fiducial ray. The
  idea behind this is to compare a light ray in the inhomogeneous universe
  with one in the homogeneous, unperturbed universe. Apart from the conceptual
  difficulty, this `first-order expansion' is not justified, as the light rays
  in an inhomogeneous universe can deviate quite significantly from straight
  rays in the homogeneous reference universe -- much more than the lenght
  scale of typical density fluctuations. These difficulties are all avoided if
  one starts from the exact equation of geodesic deviation, as done here.} The
formal solution of this transport equation is obtained by the method of Green's
function, to yield
\begin{equation}
  \vc x(\vc\theta,w) = f_K(w)\vc\theta - 
  \frac{2}{c^2}\!\int_0^w\!\!\! \d w'\,f_K(w-w')\! 
  \eck{ \nabla_\perp\Phi\rund{\vc
x(\vc\theta,w'),w'} -\nabla_\perp\Phi^{(0)}\rund{w'}} .
\elabel{6.15}
\end{equation}
A source at comoving distance $w$ with comoving separation $\vc x$ from the
fiducial light ray would be seen, in the absence of lensing, at the angular
separation $\vc \beta=\vc x/f_K(w)$ from the fiducial ray (this statement is
nothing but the definition of the comoving angular diameter distance).  Hence,
$\vc\beta$ is the unlensed angular position in the `comoving source plane' at
distance $w$, where the origin of this source plane is given by the
intersection point with the fiducial ray.
Therefore, in analogy with standard lens theory, we define the
Jacobian matrix
\be
\A(\vc\theta,w)={\partial\vc \beta\over\partial\vc\theta}
={1\over f_K(w)}{\partial\vc x\over\partial\vc\theta}\;,
\ee
and obtain from (\ref{eq:6.15})
\be
\A_{ij}(\vc\theta,w)=\delta_{ij}-\frac{2}{c^2}\int_0^w \!\! \d w'\,
{f_K(w-w') f_K(w')\over f_K(w)}\,\Phi_{,ik}\rund{\vc
x(\vc\theta,w'),w'}\,\A_{kj}(\vc\theta,w') \;,
\elabel{fullJacobi}
\ee
which describes the locally linearized mapping introduced by LSS
lensing. To derive (\ref{eq:fullJacobi}), we noted that
$\nabla_\perp\Phi^{(0)}$ does not depend on $\vc\theta$, and used the
chain rule in the derivative of $\Phi$.  This equation still is exact
in the limit of validity of the weak-field metric.  Next, we expand
$\A$ in powers of $\Phi$, and truncate the series after the linear
term:
\be
\A_{ij}(\vc\theta,w)=\delta_{ij}-\frac{2}{c^2}\int_0^w \!\! \d w'\,
{f_K(w-w') f_K(w')\over f_K(w)}\,\Phi_{,ij}\rund{f_K(w')\vc\theta,w'} \;.
\ee
Hence, to linear order, the distortion can be obtained by integrating along
the unperturbed ray $\vc x=f_K(w)\,\vc\theta$; this is also called the Born
approximation. Corrections to the Born approximation are necessarily of order
$\Phi^2$. Throughout this article, we will employ the Born approximation;
later, we will comment on its accuracy.  If we now
define the deflection potential
\be
\psi(\vc\theta,w):=\frac{2}{c^2}\int_0^w \!\! \d w'\,
{f_K(w-w') \over f_K(w)\;f_K(w')}\,\Phi\rund{f_K(w')\vc\theta,w'} \;,
\ee
then $\A_{ij}=\delta_{ij}-\psi_{,ij}$, just as in ordinary lens theory.
{\em In this approximation, lensing by the 3-D matter
distribution can be treated as an equivalent lens plane with
deflection potential $\psi$, mass density $\kappa=\nabla^2\psi/2$, and
shear $\gamma=(\psi_{,11}-\psi_{,22})/2+{\rm i}\psi_{,12}$. }

\subsection{\llabel{WL-6.2}Cosmic shear: the principle}
\subsubsection{The effective surface mass density.}
Next, we relate $\kappa$ to fractional density contrast $\delta$ of
matter fluctuations in the Universe; this is done in a number of
steps:
\ben
\item
To obtain $\kappa=\nabla^2 \psi/2$, 
take the 2-D Laplacian of $\psi$, and add the term $\Phi_{,33}$ in the
resulting integrand; this latter term vanishes in the line-of-sight
integration, as can be seen by integration by parts.
\item
We make use of the 3-D Poisson equation in comoving coordinates
\be
\nabla^2\Phi={3 H_0^2\Omega_{\rm m}\over 2 a}\delta
\ee
to obtain
\be
\kappa(\vc\theta,w)=\frac{3H_0^2\Omega_{\rm m}}{2c^2}\,
  \int_0^w\,\d w'\,\frac{f_K(w')f_K(w-w')}{f_K(w)}\,
  \frac{\delta\rund{f_K(w')\vc\theta,w'}}{a(w')}\;.
\elabel{6.21}
\end{equation}
Note that $\kappa$ is proportional to $\Omega_{\rm m}$, since lensing is
sensitive to $\Delta\rho\propto \Omega_{\rm m}\,\delta$, not just to
the density contrast $\delta=\Delta\rho/\bar\rho$ itself.
\item
For a redshift distribution of sources with $p_z(z)\,\d z=p_w(w)\,\d w$,
the effective surface mass density becomes
\bea
\kappa(\vc\theta)&=&\int\d w\;p_w(w)\,\kappa(\vc\theta,w) \nonumber \\
&=&
\frac{3H_0^2\Omega_{\rm m}}{2c^2}\,
  \int_0^{w_{\rm h}}\d w\;g(w)\,f_K(w)\,
\frac{\delta\rund{f_K(w)\vc\theta,w}}{a(w)} \;,
\elabel{5.11}
\eea
with
\be
g(w)=\int_w^{w_{\rm h}}\d w'\;p_w(w'){f_K(w'-w)\over f_K(w')}\;,
\elabel{g-fact}
\ee
which is the source-redshift weighted lens efficiency factor
$D_{\rm ds}/D_{\rm s}$ for a
density fluctuation at distance $w$, and
$w_{\rm h}$ is the comoving horizon distance, obtained from $w(a)$ by letting
$a\to 0$. 
\een
The expression (\ref{eq:6.21}) for the effective surface mass density
can be interpreted in a very simple way. Consider a redshift interval
of width $\d z$ around $z$, corresponding to the proper radial
distance interval $\d D_{\rm prop}= |c\,\d t|= H^{-1}(z)  (1+z)^{-1}\,c\,\d
z$. The surface mass density in this interval
is $\Delta\rho\,\d D_{\rm prop}$, where only the density contrast
$\Delta \rho=\rho-\bar\rho$ acts as a lens (the `lensing effect' of
the mean matter density of the Universe is accounted for by the
relations between angular diameter distance and redshift; see
Schneider \& Weiss 1988a). Dividing
this surface mass density by the corresponding critical surface
mass density, and integrating along the line-of-sight to the sources,
one finds
\be
\kappa=\int_0^{z_{\rm s}} \d z\;{4\pi G\over c^2}\,
{D^{\rm ang}_{\rm d} D^{\rm ang}_{\rm ds}\over D^{\rm ang}_{\rm s}}
\,{\d D_{\rm prop}\over \d z}\,
\Delta\rho\;.
\ee
This expression is equivalent to (\ref{eq:6.21}), as can be easily
shown (by the way, this is a good excersize for practicing the use of
cosmological quantities like redshift, distances etc.).

\subsubsection{Limber's equation.}
The density field $\delta$ is assumed to be a realization of a random
field. It is the properties of the random field that cosmologists can hope to
predict, and not a specific realization of it. In particular, the second-order
statistical properties of the density field are described in terms of the
power spectrum (see IN, Sect.\ 6.1). We shall therefore look at the relation
between the quantities relevant for lensing and the power spectrum
$P_\delta(k)$ of the matter distribution in the Universe. The basis of this
relation is formed by Limber's equation.  If $\delta$ is a homogeneous and
isotropic 3-D random field, then the projections
\be
g_i(\vc\theta)=\int\d w\;q_i(w)\,\delta\rund{f_K(w)\vc\theta,w}
\ee
also are (2-D) homogeneous and isotropic random fields, where the
$q_i$ are weight functions. In particular, the correlation function
\be
C_{12}=\ave{g_1(\vc\vp_1)\,g_2(\vc\vp_2)}\equiv
C_{12}(|\vc\vp_1-\vc\vp_2|) 
\ee
depends only on the modulus of the separation vector. The original
form of the Limber (1953) equation relates $C_{12}$ to the correlation
function of $\delta$ which is a line-of-sight
projection. Alternatively, one can consider the Fourier-space analogy
of this relation: The power spectrum $P_{12}(\ell)$ -- the Fourier
transform of $C_{12}(\theta)$ -- depends linearly on $P_\delta(k)$
(Kaiser 1992, 1998), 
\be
P_{12}(\ell)=\int\d w \;{q_1(w)\, q_2(w)\over f_K^2(w)}\,
P_\delta\rund{{\ell\over f_K(w)},w}  \;,
\elabel{limber}
\ee
if the largest-scale structures in $\delta$ are much smaller than the
effective range $\Delta w$ of the projection.
Hence, we obtain the (very reasonable) result that the 2-D power at
angular scale $1/\ell$ is obtained from the 3-D power at
length scale $f_K(w)\,(1/\ell)$, integrated over $w$. 

Comparing
(\ref{eq:5.11}) with (\ref{eq:limber}), one sees that 
$\kappa(\vc\theta)$ is such a projection of $\delta$ with
the weights
$q_1(w)=q_2(w)= (3/2)(H_0/c)^2\Omega_{\rm m}g(w)f_K(w)/a(w)$, so that
\begin{equation}
  P_\kappa(\ell) = \frac{9H_0^4\Omega_{\rm m}^2}{4c^4}\,
  \int_0^{w_{\rm h}}\d w\,\frac{g^2(w)}{a^2(w)}\,
  P_\delta\left(\frac{\ell}{f_K(w)},w\right)\;.
\elabel{6.25}
\end{equation}
The power spectrum $P_\kappa$, if observable, can therefore be used to
constrain the 3-D power spectrum $P_\delta$. For a number of
cosmological models, the power spectrum $P_\kappa(\ell)$ is plotted in
Fig.\ \ref{fig:powerspects}. Predictions of $P_\kappa$ are plotted
both for assuming linear growth of the density structure (see Sect.\
6.1 of IN), as well as the prescription of the fully nonlinear power
spectrum as given by the fitting formulae of Peacock \& Dodds (1996).
From this figure one infers that the nonlinear evolution of the
density fluctuations becomes dominant for values of $\ell\gtrsim 200$,
corresponding to an angular scale of about $30'$; the precise values
depend on the cosmological model and the redshift distribution of the
sources. Furthermore, the dimensionless power spectrum
$\ell^2\,P_\kappa(\ell)$, that is, the power per logarithmic bin,
peaks at around $\ell \sim 10^4$, corresponding to an angular scale of
$\sim 1'$, again somewhat depending on the source redshift
distribution. Third, one notices that the shape and amplitude of
$P_\kappa$ depends on the values of the cosmological parameters;
therefore, by measuring the power spectrum, or quantities directly
related to it, one can constrain the values of the cosmological
parameters. We consider next appropriate statistical measures of the
cosmic shear which are directly and simply related to the power
spectrum $P_\kappa$.

\begin{figure}
\includegraphics[width=11.7cm]{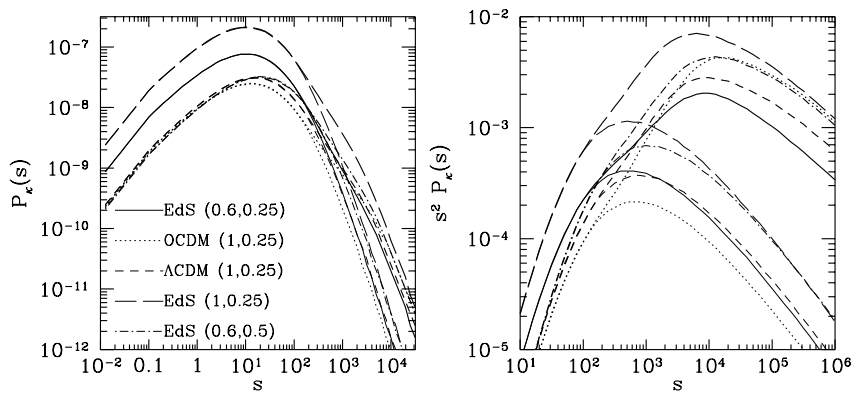}
\caption{The power spectrum $P_\kappa(\ell)$ (left panel) and its
dimensionless form $\ell^2\,P_\kappa(\ell)$ (right panel) for several
cosmological models (where here, $\ell$ is denoted by $s$).
Specifically, EdS denotes an $\Omega_{\rm m}=1$, $\Omega_\Lambda=0$
Einstein-de Sitter model, OCDM an open $\Omega_{\rm m}=0.3$,
$\Omega_\Lambda=0$ Universe, and $\Lambda$CDM a flat, low-density
$\Omega_{\rm m}=0.3$, $\Omega_\Lambda=0.7$ model.  Numbers in
parenthesis indicate $(\Gamma_{\rm spect},\sigma_8)$, where
$\Gamma_{\rm spect}$ is the shape parameter of the power spectrum (see
IN, Sect.\ 6.1) and $\sigma_8$ is the power-spectrum normalization.
For these power
spectra, the mean redshift of the galaxy distribution was assumed to
be $\ave {z_{\rm s}}=1.5$.  Thin curves show the power spectra
assuming linear evolution of the density fluctuations in the Universe,
and thick curves use the fully non-linear evolution, according to the
prescription of Peacock \& Dodds (1996).  For angular scales below
$\sim 30'$, corresponding to $\ell\ge 200$, the non-linear evolution of
the power spectrum becomes very important (from Schneider et al.\
1998a)}
\flabel{powerspects}
\end{figure}

\subsection{\llabel{WL-6.3}Second-order cosmic shear measures}
We will now turn to statistical quantities of the cosmic shear field
which are quadratic in the shear, i.e., to second-order shear
statistics. Higher-order statistical properties, which already have
been detected in cosmic shear surveys, will be considered in Sect.\
\ref{sc:WL-9}.  As we shall see, all second-order statistics of the
cosmic shear yield (filtered) information about, and are fully
described in terms of $P_\kappa$.  The most-often used second-order
statistics are:
\bi
\item
The two-point correlation function(s) of the shear, $\xi_{\pm}(\theta)$,
\item
the shear dispersion in a (circular) aperture,
$\ave{|\bar\gamma|^2}(\theta)$, and 
\item
the aperture mass dispersion, $\ave{M_{\rm ap}^2}(\theta)$.
\ei
Those will be discussed next, and their relation to $P_\kappa(\ell)$
shown. As a preparation, consider the Fourier transform of $\kappa$, 
\be
\hat\kappa(\vc\ell)=\int\d^2\theta\,{\rm e}^{{\rm
i}\vc\ell\cdot\vc\theta}\,\kappa(\vc\theta)\;;
\ee
then,
\be
\ave{\hat\kappa(\vc\ell)\hat\kappa^{*}(\vc\ell')}
=(2\pi)^2\,\delta_{\rm D}(\vc\ell-\vc\ell')\,P_\kappa(\ell)\;,
\ee
which provides another definition of the power spectrum $P_\kappa$ [compare
  with eq.\ (123) of IN]. The Fourier transform of the shear is
\be
\hat\gamma(\vc\ell)=\rund{\ell_1^2-\ell_2^2+2{\rm i}\ell_1\ell_2
\over \abs{\vc\ell}^2}\hat\kappa(\vc\ell)
={\rm e}^{2{\rm i}\beta}\,\hat\kappa(\vc\ell) \;,
\elabel{N10}
\ee
where $\beta$ is the polar angle of the vector $\vc\ell$; this follows
directly from (\ref{eq:5.3}) and (\ref{eq:Dhat}). 
Eq.\ (\ref{eq:N10}) implies that
\be
\ave{\hat\gamma(\vc\ell)\hat\gamma^*(\vc\ell')}
=(2\pi)^2\,\delta_{\rm D}(\vc\ell-\vc\ell')\,P_\kappa(\ell).
\ee
Hence, the power spectrum of the shear is the same as that of the
surface mass density. 

\subsubsection{Shear correlation functions.}
Consider a pair of points (i.e., galaxy images); their separation direction
$\vp$ (i.e. the polar angle of the separation vector $\vc\theta$) 
is used to define the tangential and cross-component of the
shear at these positions {\em for this pair},
$\gamma_{\rm t}=-\Re\rund{\gamma\,{\rm e}^{-2{\rm i}\vp}}$, 
$\gamma_{\times}=-\Im\rund{\gamma\,{\rm e}^{-2{\rm i}\vp}}$, 
as in (\ref{eq:sheart+c}). Using these two shear components, one can
then define the correlation functions $\ave{\gamma_{\rm t}\gamma_{\rm
t}}(\theta)$ and $\ave{\gamma_\times 
\gamma_\times}(\theta)$, as well as the mixed correlator. However, it
turns out to be more convenient to define the following combinations, 
\be
\xi_\pm(\theta)=\ave{\gamma_{\rm t}\gamma_{\rm t}}(\theta)
\pm\ave{\gamma_\times 
\gamma_\times}(\theta)\;,\;\;
\xi_\times(\theta)=\ave{\gamma_{\rm t}\gamma_{\times}}(\theta)
\;.\nonumber 
\ee
Due to parity symmetry, $\xi_\times(\theta)$ is expected to vanish, since
under such a transformation, $\gamma_{\rm t}\to \gamma_{\rm t}$, but
$\gamma_\times\to -\gamma_\times$.
Next we relate the shear correlation functions to the power spectrum
$P_\kappa$:
Using the definition of $\xi_\pm$, replacing $\gamma$ in terms of
$\hat\gamma$, and making use of relation between $\hat\gamma$ and
$\hat\kappa$, one finds (e.g., Kaiser 1992)
\be
\xi_+(\theta)=\int_0^\infty
{\d\ell\,\ell\over 2\pi}\,{\rm J}_0(\ell\theta)\,
P_\kappa(\ell)\;; \;\;  
\xi_-(\theta)=\int_0^\infty
{\d\ell\,\ell\over 2\pi}\,{\rm J}_4(\ell\theta)\,
P_\kappa(\ell)\;,
\elabel{xi-power}
\ee
where ${\rm J}_n(x)$ is the n-th order Bessel function of first kind.
$\xi_\pm$ can be measured as follows: on a data field, select all
pairs of faint galaxies with separation within $\Delta\theta$ of
$\theta$ and then
take the average $\ave{\eps_{{\rm t}i}\,\eps_{{\rm t}j}}$ over all
these pairs; since $\eps_i=\eps_i^{(\rm s)}+\gamma(\vc\theta_i)$, the
expectation 
value of $\ave{\eps_{{\rm t}i}\,\eps_{{\rm t}j}}$ is 
$\ave{\gamma_{\rm t}\gamma_{\rm t} }(\theta)$, provided source
ellipticities are uncorrelated. 
Similarly, the correlation for the cross-components is obtained.
It is obvious how to generalize this estimator in the presence of a weight
factor for the individual galaxies, as it results from the image analysis
described in Sect.\ \ref{sc:WL-3.5}. 

\subsubsection{The shear dispersion.}
Consider a circular aperture of radius $\theta$; the mean shear in this
aperture is $\bar\gamma$. Averaging over many such apertures, one
defines the shear dispersion
$\ave{|\bar\gamma|^2}(\theta)$.
It is related to the power spectrum through
\be
\ave{\abs{\bar\gamma}^2}(\theta)={1\over 2\pi}\int\d\ell\,\ell\,
P_\kappa(\ell)\,W_{\rm TH}(\ell\theta)\;,\;\;
{\rm where} \;\;
W_{\rm TH}(\eta)={4 {\rm J}_1^2(\eta)\over \eta^2}
\elabel{sheardisp}
\ee
is the top-hat filter function (see, e.g., Kaiser 1992).
A practical unbiased estimator of the mean shear in the aperture is 
$\hat{\bar{\gamma}}=N^{-1}\sum_{i=1}^N \eps_i$,
where $N$ is the number of galaxies in the aperture. However, the square of
this expression is {\em not} an unbiased estimator of
$\ave{\abs{\bar\gamma}^2}$, since the diagonal terms of the resulting double
sum yield additional terms, since ${\rm E}\rund{\eps_i
  \eps^*_i}=|\gamma(\vc\theta_i)|^2 +\sigma_\eps^2$. An unbiased estimate for
the shear dispersion is obtained by omitting the diagonal terms, 
\be
\widehat{\ave{\abs{\bar\gamma}^2}}={1\over N(N-1)}\sum_{i\ne j}^N
\eps_i\,\eps^*_j\;.
\elabel{shear.disp.est}
\ee
This expression is then averaged over many aperture placed on the data
field. Again, the generalization to allow for weighting of galaxy images is
obvious. Note in particular that this estimator is not positive
semi-definite. 

\subsubsection{The aperture mass.}
Consider a circular aperture of radius $\theta$; for a point inside the
aperture, define the tangential and cross-components of the shear relative
to the center of the aperture (as before);
then define
\be
M_{\rm ap}(\theta)=\int\d^2\vt\;Q(|\vc\vt|)\,\gamma_{\rm t}(\vc\vt)\;,
\ee
where $Q$ is a weight function with support $\vt\in[0,\theta]$.
If we  use the function $Q$ given in (\ref{eq:U+Qgeneric}),
the dispersion of $M_{\rm ap}(\theta)$ is related to
power spectrum by (Schneider et al.\ 1998a)
\be
\ave{M_{\rm ap}^2}(\theta)={1\over
2\pi}\int_0^\infty\d\ell\;\ell\,P_\kappa(\ell)\, W_{\rm ap}(\theta\ell)\;,\;\;
{\rm with} \;\;
W_{\rm ap,1}(\eta):={576{\rm J}_4^2(\eta)\over \eta^4}\;.
\elabel{N26a}
\ee
Crittenden et al.\ (2002) suggested a different pair $U$ and $Q$ of filter
functions, 
\be
U(\vt)={1\over 2\pi\,\theta^2}\eck{1-\rund{\vt^2\over 2\theta^2}}\,
\exp\rund{-{\vt^2 \over 2\theta^2}}\; ;\;\;
Q(\vt)={\vt^2\over 4\pi\theta^4}\exp\rund{-{\vt^2\over 2\theta^2}}\;.
\elabel{U+Q.CNPT}
\ee
These function have the disadvantage of not having finite support; however,
due to the very strong fall-off for $\vt\gg\theta$, for many practical
purposes the support can be considered effectively as finite. 
This little drawback is
compensated by the convenient analytic properties of these filter functions,
as we shall see later. For example, the relation of the corresponding aperture
mass dispersion is again given by the first of eqs.\ (\ref{eq:N26a}), but the
filter function simplifies to 
\be
W_{\rm ap,2}(\eta)={\eta^4\over 4}\,{\rm e}^{-\eta^2} \;.
\ee
Whereas the filter functions which relate the power spectrum to the
shear correlation functions, i.e., the Bessel function appearing in
(\ref{eq:xi-power}), and to the shear dispersion, given by $W_{\rm
TH}$, are quite broad filters, implying that these statistics at a
given angular scale depend on the power spectrum over a wide range of
$\ell$, the two filter function $W_{\rm ap,1,2}$ are very localized
and thus the aperture mass dispersion yields highly localized
information about the power spectrum (see Bartelmann \& Schneider
1999, who showed that replacing the filter function $W$ by a
delta-`function' causes an error of only $\sim 10\%$). Hence, the
shape of $\ave{M_{\rm ap}^2}(\theta)$ directly reflects the shape of
the power spectrum as can also be seen in Fig.\ \ref{fig:map+gammasq}
below.

\subsubsection{Interrelations.}
These various 2-point statistics all depend linearly on the power
spectrum $P_\kappa$; therefore, one should not be too surprised that
they are all related to each other (Crittenden et al.\ 2002). The
surprise perhaps is that these interrelations are quite simple. 
First, the relations between $\xi_\pm$ and $P_\kappa$ can be inverted,
making use of the orthonormality relation of Bessel functions:
\be
P_\kappa(\ell)=2\pi\int_0^\infty\d\theta\,\theta\,\xi_+(\theta)\,{\rm
J}_0(\ell\theta)=2\pi\int_0^\infty\d\theta\,\theta\,\xi_-(\theta)\,{\rm
J}_4(\ell\theta) \;.
\elabel{Pfromxi}
\ee
Next, we 
take one of these and plug them into the relation (\ref{eq:xi-power}) 
between the other
correlation function and $P_\kappa$, to find:
\be
\xi_+(\theta)=\xi_-(\theta)
+\int_\theta^\infty{\d\vt\over\vt}\xi_-(\vt)\rund{4-12{\theta^2\over
\vt^2}}\;;
\ee
\be
\xi_-(\theta)=\xi_+(\theta)
+\int_0^\theta{\d\vt\,\vt\over\theta^2}\xi_+(\vt)
\rund{4-12{\vt^2\over \theta^2}}\,.
\ee
These equations show that the two shear correlation functions are not
independent of each other, the reason for that being that the shear
(which itself is a two-component quantity) is derived from a single
scalar field, namely the deflection potential $\psi$. We shall return
to this issue further below.
Using (\ref{eq:Pfromxi}) in the equation for the shear dispersion, one
finds 
\[
\ave{\abs{\bar\gamma}^2}(\theta)=
\int_0^{2\theta}{\d\vt\,\vt\over  \theta^2}
\;\xi_+(\vt)\,S_+\rund{\vt\over\theta}
=
\int_0^\infty{\d\vt\,\vt\over  \theta^2}
\;\xi_-(\vt)\,S_-\rund{\vt\over\theta} \;,
\]
where the $S_\pm$ are simple functions, given explicitly in Schneider et al.\
(2002a) and plotted in Fig.\ts\ref{fig:T+Splots}.
\begin{figure}
\bmi{6}
\includegraphics[width=5.8cm]{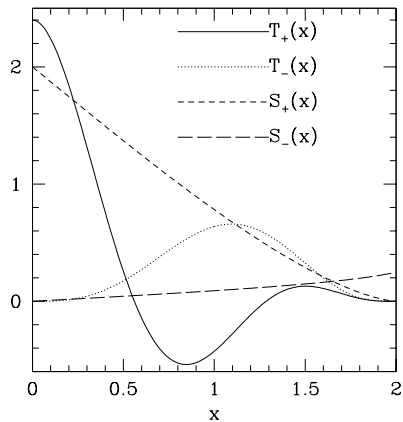}
\emi
\bmi{5.7}
\caption{The function $S_\pm(x)$ and $T_\pm(x)$ which relate the shear and
aperture mass dispersion to the correlation functions. Note that $S_-$
does not vanish for $x>2$, as is the case for the other three functions
(from Schneider et al.\ 2002a)}
\flabel{T+Splots}
\emi
\end{figure}
Finally, 
the same procedure for the aperture mass dispersion lets us write 
\be
\ave{M_{\rm ap}^2}(\theta)=\int_0^{2\theta}{\d\vt\,\vt\over  \theta^2}
\;\xi_+(\vt)\,T_+\rund{\vt\over\theta} 
=
\int_0^{2\theta}{\d\vt\,\vt\over  \theta^2}
\;\xi_-(\vt)\,T_-\rund{\vt\over\theta} \;,
\elabel{Mapfromxi}
\ee
again with analytically known functions $T_\pm$, given for the filter function
(\ref{eq:U+Qgeneric}) in Schneider et al.\ (2002a), and for the filter
function (\ref{eq:U+Q.CNPT}) in Jarvis et al.\ (2003b).  Hence, all these
2-point statistics can be evaluated from the correlation functions
$\xi_\pm(\theta)$, which is of particular interest, since they can be measured
best: Real data fields contain holes and gaps (like CCD defects; brights
stars; nearby galaxies, etc.) which makes the placing of apertures difficult;
however, the evaluation of the correlation functions is not affected by gaps,
as one uses all pairs of galaxy images with a given angular separation.
Furthermore, it should be noted that the aperture mass dispersion at angular
scale $\theta$ can be calculated from $\xi_\pm(\vt)$ in the finite range
$\vt\in[0,2\theta]$, and the shear dispersion can be calculated from
$\xi_+$ on $\vt\in[0,2\theta]$, but not from $\xi_-$ on a finite
interval; this is due to the fact that $\xi_-$ on small scales does
not contain the information of the power spectrum on large scales,
because of the filter function ${\rm J}_4$ in (\ref{eq:xi-power}).

We also note that from a cosmic shear survey, the power spectrum
$P_\kappa$ can be determined directly, as has been investigated by
Kaiser (1998), Seljak (1998) and Hu \& White (2001). This is {\it not}
done by applying (\ref{eq:Pfromxi}), as these relations would require
the determination of the correlation function for all separation, but
by more sophisticated methods. A simple example (though not optimal)
is to consider the measured shear field on the square; Fourier
transforming it and binning modes in $|\vc \ell|$ then yields an
estimate of the power spectrum, once the power from the intrinsic
ellipticity dispersion is subtracted. Better methods aim at minimizing
the variance of the reconstructed power spectrum (Seljak 1998; Hu \&
White 2001). As mentioned before, the aperture mass dispersion is a
filtered version of the power spectrum with such a narrow filter, that
it contains essentially the same information as $P_\kappa$ over the
corresponding angular scale and at $\ell\sim 5/\theta$, provided
$P_\kappa$ has no sharp features.

\subsection{\llabel{WL-6.4}Cosmic shear and cosmology}
\subsubsection{Why cosmology from cosmic shear?}
Before continuing, it is worth to pause for a second and ask the question why
one tries to investigate cosmological questions by using cosmic shear -- since
it is widely assumed that the CMB will measure `all' cosmological quantities
with high accuracy. Partial answers to this question are: 
\bi
\item
Cosmic shear measures the mass distribution at much lower redshifts
($z\lesssim 1$) and at smaller physical scales [$R\sim 0.3\,h^{-1}\,
(\theta/1')\,{\rm Mpc}$] than the CMB; indeed, it is the only way to
map out the dark matter distribution directly without any assumptions
about the relation between dark and baryonic matter.
\item
Cosmic shear measures the non-linearly evolved mass
distribution and its associated power spectrum $P_\delta(k)$; hence, in
combination with the CMB it allows us to study the evolution of the
power spectrum and in particular, provide a very powerful test of the
gravitational instability paradigm for structure growth.
\item
As was demonstrated by the recent results from the WMAP satellite
(Bennett et al.\ 2003), the strongest constraints are derived when
combining CMB measurements (constraining the power spectrum on large
spatial scales) with measurements on substantially smaller scales, to
break parameter degeneracies remaining from the CMB results alone (see
Sper\-gel et al.\ 2003). Hu \& Tegmark (1999) have explicitly
demonstrated how much the accuracy of estimates of cosmological
parameters is improved when the CMB results from missions like WMAP
and later Planck is complemented by cosmic shear measurements (see
Fig.\ \ref{fig:hu+tegmark}). In fact, as we shall see later,
combinations of CMB anisotropy measurements have already been combined
with cosmic shear measurements (see Fig.\ \ref{fig:lens+CMB}) and lead
to substantially improved constraints on the cosmological parameters.
\item
It provides a fully independent way to probe the cosmological model;
given the revolutionary claims coming from the CMB, SN\ Ia, and the
LSS of the galaxy distribution, namely that more than 95\% of the
contents in the Universe is in a form that we have not the slightest idea
about what it is (the names `dark matter' and `dark energy' reflect our
ignorance about their physical nature), an
additional independent verification of these 
claims is certainly welcome. 
\item
For a foreseeable future, astronomical observations will provide the
only possibility to probe the dark energy empirically. The equation of
state of the dark energy can be probed best at relatively low
redshifts, that is with SN\ts Ia and cosmic shear observations,
whereas CMB anisotropy measurements are relatively insensitive to the
properties of the dark energy, as the latter was subdominant at the
epoch of recombination.
\item
As we have seen in Sect.\ \ref{sc:WL-5.8}, cosmic shear studies provide a new
and highly valuable search method for cluster-scale matter
concentrations.
\ei

\begin{figure}
\bc
\includegraphics[width=8cm]{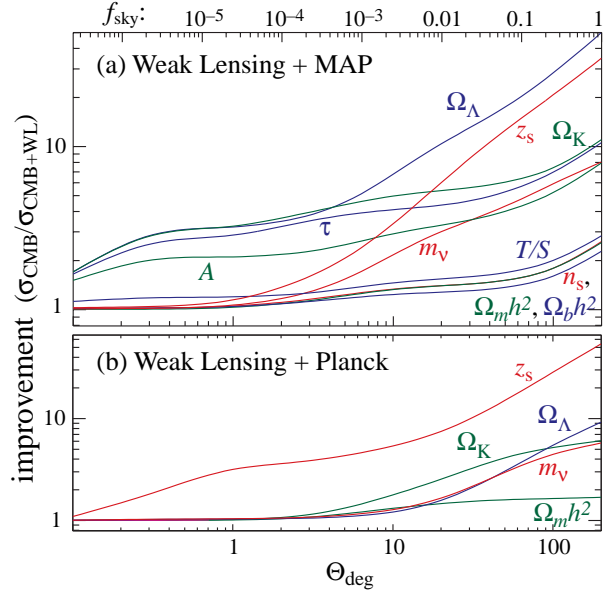}
\ec
\caption{The improvement of the accuracy of cosmological parameters
when supplementing CMB data from WMAP (upper panel) and the Planck
satellite (lower panel) by a cosmic shear survey of solid angle
$\theta^2\pi$. The accuracies are significantly improved, certainly when
combined with WMAP, but even in combination with Planck, the
accuracies of the density parameters can be increased, when using
next-generation cosmic shear surveys with hundreds of square degrees
(from Hu \& Tegmark 1999)}
\flabel{hu+tegmark}
\end{figure}

\subsubsection{Expectations.}
The cosmic shear signal depends on the cosmological model,
parameterized by $\Omega_{\rm m}$, $\Omega_\Lambda$, and the shape
parameter $\Gamma_{\rm spect}$ of the power spectrum, the
normalization of the power spectrum, usually expressed in terms of
$\sigma_8$, and the redshift distribution of the sources.  By
measuring $\xi_\pm$ over a significant range of angular scales one can
derive constraints on these parameters.  To first order, the amplitude
of the cosmic shear signal depends on the combination $\sim \sigma_8\,
\Omega_{\rm m}^{0.5}$, very similar to the cluster abundance.
Furthermore, the cosmic shear signal shows a strong dependence on the source
redshift distribution. These dependencies are easily understood qualitatively:
A higher normalization $\sigma_8$ increases $P_\delta$ on all scales, thus
increasing $P_\kappa$. The increase with $\Omega_{\rm m}$ is mainly due to the
prefactor in (\ref{eq:6.25}), i.e. due to the fact that the light deflection
depends on $\Delta\rho$, not just merely on $\delta=\Delta\rho/\bar\rho$, as
most other cosmological probes. Finally, increasing the redshift of sources
has two effects: first, the lens efficiency $D_{\rm ds}/D_{\rm s}=f_K(w_{\rm
  s}-w)/f_K(w_{\rm s})$ at given distance $w$ increases as the sources are
moved further away, and second, a larger source redshift implies a longer ray
path through the inhomogeneous matter distribution. 

In Fig.\ \ref{fig:map+gammasq} the predictions of the shear dispersion
and the aperture mass dispersion are shown as a function of angular
scale, for several cosmological models. The dependencies of the power
spectrum $P_\kappa$ on cosmological parameters and $\ell$ is reflected
in these cosmic shear measures. In particular, the narrow filter
function which relates the aperture mass dispersion to the power
spectrum implies that the peak in $\ell^2 P_\kappa(\ell)$ at around
$\ell\sim 10^4$ (see Fig.\ts\ref{fig:powerspects}) translates into a
peak of $\ave{M_{\rm ap}^2}$ at around $\theta\sim 1'$. The non-linear
evolution of the power spectrum is dominating the cosmic shear result
for scales below $\sim 30'$; the fact that the non-linear prediction
approach the linear ones at somewhat smaller scales for the shear
dispersion $\ave{|\bar\gamma|^2}$ is due to the fact that this
statistics corresponds to a broad-band filter $W_{\rm TH}$
(\ref{eq:sheardisp}) of $P_\kappa$ which includes the whole range of
small $\ell$ values, which are less affected by non-linear evolution.

\begin{figure}
\includegraphics[width=11.7cm]{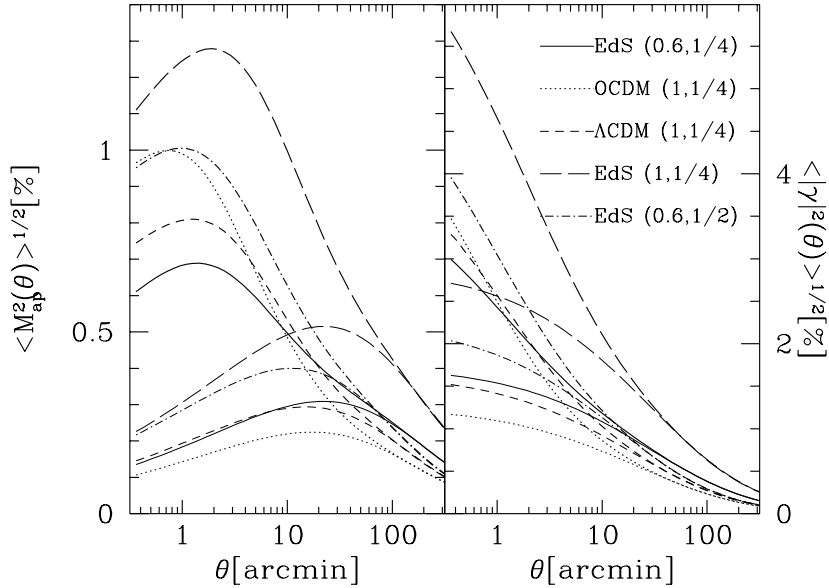}

\caption{The square root of the aperture mass dispersion (left) and of
the shear dispersion (right), for the same cosmological models as were used
for Fig.\ \ref{fig:powerspects}, again with results from assuming linear growth
of structure in the Universe shown as thin curves, whereas the fully
non-linear evolution was taken into account for the thick curves.  One sees
that the aperture 
mass signal is considerably smaller than that of the shear dispersion;
this is due to the fact that the filter function $W_{\rm ap}$ is much
narrower than $W_{\rm TH}$; hence, at a given angular scale,
$\ave{M_{\rm ap}^2}$ samples less power than
$\ave{|\bar\gamma|^2}$. However, this also implies that the aperture
mass dispersion provides much more localized information about the
power spectrum than the shear dispersion and is therefore a more
useful statistics to consider. Other advantages of $\ave{M_{\rm
ap}^2}$ will be described further below. For scales below $\sim 30'$,
the non-linear evolution of the power spectrum becomes very important
(from Schneider et al.\ 1998a)
}
\flabel{map+gammasq}
\end{figure}

\subsubsection{Deriving constraints.}
From the measured correlation functions $\xi_\pm(\theta)$ (or any
other measure of the cosmic shear, but we will concentrate on the
statistics which is most easily obtained from real data), obtaining
constraints on cosmological parameters can proceed through maximizing
the likelihood ${\cal L}(p|\xi^{\rm obs})$, which yields the
probability for the set of cosmological parameters being $p$, given
the observed correlation function $\xi^{\rm obs}$. This likelihood is
given by the probability $P(\xi^{\rm obs}|p)$ that the observed
correlation function is $\xi^{\rm obs}$, given the parameters $p$. For
a given set of parameters $p$, the correlation function $\xi(p)$ is
predicted. If one assumes that the observed correlations $\xi^{\rm
obs}$ are drawn from a (multi-variate) Gaussian probability
distribution, then
\[
P(\xi^{\rm obs}|p)={1\over (2\pi)^{n/2} \sqrt{\det \rm
    Cov}}\,\exp\rund{-\chi^2(p,\xi^{\rm obs})\over 2}\;,
\]
with
\be
\chi^2(p,\xi^{\rm obs})=\sum_{ij}\rund{\xi_i(p)-\xi_i^{\rm obs}}\,{\rm
Cov}^{-1}_{ij} \,\rund{\xi_j(p)-\xi_j^{\rm obs}}\;.
\elabel{chisqxi}
\ee
Here, the $\xi_i=\xi(\theta_i)$ are the values of the correlation
function(s) (i.e., either $\xi_\pm$, or using both) in angular bins,
$n$ is the number of angular bins in case either one of the $\xi_\pm$
is used, or if both are combined, twice the number of angular bins,
and ${\rm Cov}_{ij}^{-1}$ is the inverse of the covariance matrix, which is
defined as
\be
{\rm Cov}_{ij}=\ave{\eck{\xi_i(p)-\xi^{\rm obs}_i}\eck{\xi_j(p)-\xi^{\rm
      obs}_j}} \;,
\ee
where the average is over multiple realizations of the cosmic shear
survey under consideration. ${\rm Cov}_{ij}$ can be determined either
from the $\xi_\pm$ itself, from simulations, or estimated from the
data in terms of the $\xi_\pm^{\rm obs}$ (see Schneider et al.\
2002b; Kilbinger \& Schneider 2004, Simon et al.\ 2004).
Nevertheless, the calculation of the covariance is fairly
cumbersome, and most authors have used approximate methods to derive
it, such as the field-to-field variations of the measured
correlation. In fact, this latter approach may be more accurate than
using the analytic expressions of the covariance in terms of the
correlation function, which are obtained by assuming that the shear
field is Gaussian, so that the four-point correlation function can be
factorized as products of two-point correlators.
As it turns out, $\xi_+ (\theta)$ is strongly correlated across
angular bins, much less so for $\xi_-(\theta)$; this is due to the
fact that the filter function that describes $\xi$ in terms of the
power spectrum $P_\kappa$ is much broader for $\xi_+$ (namely ${\rm
J}_0$) than ${\rm J}_4$ which applies for $\xi_-$.  

The accuracy with which $\xi_\pm$ can be measured, and thus the
covariance matrix, depends on the number density of galaxies (that is,
depth and quality of the images), the total solid angle covered by the
survey, and its geometrical arrangement (compact survey vs. widely
separated pointings). The accuracy is determined by a combination of
the intrinsic ellipticity dispersion and the cosmic (or sampling)
variance.  The likelihood function then becomes
\be {\cal L}(p|\xi^{\rm obs})={1\over (2\pi)^{n/2} \sqrt{\det \rm
Cov}}\,\exp\rund{-\chi^2(p,\xi^{\rm obs})\over 2} \; P_{\rm
prior}(p)\;, \elabel{likelihood} 
\ee 
where $P_{\rm prior}(p)$ contains
prior information (or prejudice) about the parameters to be
determined. For example, the redshift distribution of the sources (at
given apparent magnitude) is fairly well known from spectroscopic
redshift surveys, and so the prior probability for $z_{\rm s}$ would
be chosen to be a fairly narrow function which describes this prior
knowledge on the redshifts. One often assumes that all but a few
parameters are known precisely, and thus considers a restricted space
of parameters; this is equivalent to replacing the prior for those
parameters which are fixed by a delta-`function'. If $m$ parameters
are assumed to be undetermined, but one is mainly interested in the
confidence contours of $m'<m$ parameters, then the likelihood function
is integrated over the remaining $m-m'$ parameters; this is called
marginalization and yields the likelihood function for these $m'$
parameters.

There are two principal contributions to the `noise' of cosmic shear
measurements. One is the contribution coming from the finite intrinsic
ellipticity dispersion of the source galaxies, the other due to the
finite data fields of any survey. This latter effect implies that only
a {\it typical} part of the sky is mapped, whose properties will in
general deviate from the {\it average} properties of such a region in
the sky for a given cosmology. This effect is called cosmic
variance, or sample variance.
Whereas the noise from intrinsic ellipticity dispersions
dominates at small angular scales, at scales beyond a few arcminutes
the cosmic variance is always the dominating effect (e.g., Kaiser
1998; White \& Hu 2000).

Of course, all of what was said above can be carried over to the other
second-order shear statistics, with their respective covariance
matrices. The first cosmic shear measurements were made in terms of
the shear dispersion and compared to theoretical prediction from a
range of cosmological models. As is true for the correlation
functions, the shear dispersion is strongly correlated between
different angular scales. This is much less the case for the aperture
mass dispersion, where the correlation quickly falls off once the
angular scales differ by more than a factor $\sim 1.5$ (see Schneider
et al.\ 2002b).  Even less correlated is the power spectrum
itself. These properties are of large interest if the results from a
cosmic shear survey are displayed as a curve with error bars; for the
aperture mass dispersion and the power spectrum estimates, these
errors are largely uncorrelated. However, for deriving cosmological
constraints, the correlation function $\xi_\pm$ are most useful since
they contain all second-order information in the data, in addition of
being the primary observable.

\subsection{\llabel{WL-6.5}E-modes, B-modes}
In the derivation of the lensing properties of the LSS, we ended up
with an equivalent surface mass density.  In particular, this implied
that $\A$ is a symmetric matrix, and that the shear can be obtained in
terms of $\kappa$ or $\psi$. Now, the shear is a 2-component quantity,
whereas both $\kappa$ and $\psi$ are scalar fields. This then
obviously implies that
the two shear components are not independent of each other!

Recall that (\ref{eq:5.10}) yields a relation between the gradient of
$\kappa$ and the first derivatives of the shear components; in
particular, (\ref{eq:5.10}) implies that $\nabla\times \vc
u_\gamma\equiv 0$, yielding a local constraint relation between the
second derivatives of the shear components. The validity of this constraint
equation guarantees 
that the imaginary part of (\ref{eq:5.4}) vanishes.  This constraint
is also present at the level of 2-point statistics, since one expects
from (\ref{eq:Pfromxi}) that
\[
\int_0^\infty\d\theta\,\theta\,
\xi_+(\theta){\rm J}_0(\theta\ell)
=
\int_0^\infty\d\theta\,\theta\,
\xi_-(\theta){\rm J}_4(\theta\ell)\;.
\]
Hence, the two correlation functions $\xi_\pm$ are not independent.
The observed shear field is not guaranteed to satisfy these relations,
due to noise, remaining systematics, or other effects. Therefore,
searching for deviations from this relation allows a check for these
effects.  However, there might also be a `shear' component present
that is not due to lensing (by a single equivalent thin matter sheet
$\kappa$). Shear components which satisfy the foregoing relations are called
E-modes; those which don't are B-modes -- these names are exported
from the polarization of the CMB, which has the same mathematical
properties as the shear field, namely that of a polar.

\begin{figure}
\bmi{6}
\includegraphics[width=5.8cm]{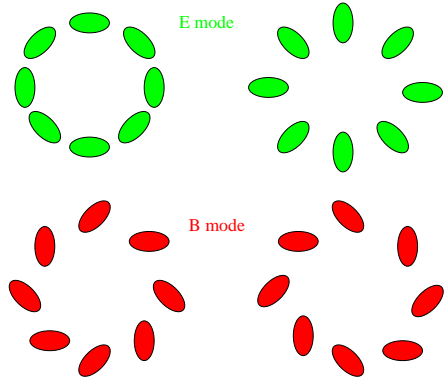}
\emi
\bmi{5.7}
\caption{Sketch of the distinction between E- and B-modes of the shear. The
  upper row shows a typical E-mode shear pattern coming from a mass
  overdensity (left) or underdensity (right), yielding tangential and radial
  alignment of the shear, respectively. The lower row shows a B-mode pattern,
  which is obtained from the E-mode pattern by rotating all shears by
  $45^\circ$. Those cannot be produced from gravitational lensing (from van
  Waerbeke \& Mellier 2003)}
\flabel{EBmodescetch}
\emi
\end{figure}

The best way to separate these modes locally is provided by the
aperture measures: 
$\ave{M^2_{\rm ap}(\theta)}$ is sensitive {\it only} to E-modes.
If one defines in analogy -- recall (\ref{eq:aperture})
\be
M_{\perp}(\theta)=\int\d^2\vt\;Q(|\vc\vt|)\,\gamma_{\times}(\vc\vt)\;,
\elabel{Mapperp}
\ee
then $\ave{M^2_\perp(\theta)}$ is sensitive {\it only} to B-modes.  In
fact, one can show that for a pure E-mode shear field, $M_\perp\equiv
0$, and for a pure B-mode field, $M_{\rm ap}\equiv 0$. Furthermore, in
general (that is, even if a B-mode is present), $\ave{M_{\rm ap}}=0$,
since $\ave{\kappa}=0$, and $\ave{M_\perp}=0$, owing to parity
invariance: a non-zero mean value of $M_\perp$ would introduce a net
orientation into the shear field. Using the same argument, one finds
that $\ave{M_{\rm ap}^m M_\perp^n}=0$ for $n$ odd (Schneider 2003).

\subsubsection{E/B-mode decomposition of a shear field.}
There are a number of (equivalent) ways to decompose a shear field into its
two modes. One is provided by the Kaiser \& Squires mass reconstruction
(\ref{eq:5.4}), which yields, for a general shear field, a complex surface
mass density $\kappa=\kappa^{\rm E} + {\rm i}\kappa^{\rm B}$. Another
separation is obtained by considering the vector field $\vc u_\gamma(\vc
\theta)$ (\ref{eq:5.10}) obtained from the first derivatives of the shear
components. This vector will in general not be a gradient field; its gradient
component corresponds to the E-mode field, the remaining one to the
B-mode. Hence one defines
\be
\nabla^2\kappa^{\rm E}=\nabla\cdot\vc u_\gamma \quad; \quad
\nabla^2\kappa^{\rm B}=\nabla\times\vc u_\gamma\; .
\ee
In full analogy with the `lensing-only' case (i.e., a pure E-mode), one
defines the (complex) potential $\psi(\vc\theta)=\psi^{\rm E}(\vc\theta)+{\rm
  i}\psi^{\rm B}(\theta)$ by the Poisson equation $\nabla^2\psi=2\kappa$, and
the shear is obtained in terms of the complex $\psi$ in the usual way, 
\bea
\gamma&=&\gamma_1+{\rm i}\gamma_2=\rund{\psi_{,11}-\psi_{,22}}/2+{\rm
  i}\psi_{,12} \nonumber \\
&=&\eck{{1\over 2}\rund{\psi^{\rm E}_{,11}-\psi^{\rm E}_{,22}}
-\psi^{\rm B}_{,12}} +{\rm i}\eck{\psi^{\rm E}_{,12}
+{1\over 2}\rund{\psi^{\rm B}_{,11}-\psi^{\rm B}_{,22}}} \;.
\eea
On the level of second-order statistics, one considers the Fourier transforms
of the E- and B-mode convergence, and defines the two power spectra $P_{\rm
  E}$, $P_{\rm B}$, and the 
cross-power spectrum $P_{\rm EB}$ by
\bea
\ave{\hat\kappa^{\rm E}(\vc\ell)\hat\kappa^{\rm E*}(\vc\ell')}
&=&(2\pi)^2\,\delta_{\rm D}(\vc\ell-\vc\ell')\,P_{\rm E}(\ell)\;,
\nonumber \\
\ave{\hat\kappa^{\rm B}(\vc\ell)\hat\kappa^{\rm B*}(\vc\ell')}
&=&(2\pi)^2\,\delta_{\rm D}(\vc\ell-\vc\ell')\,P_{\rm B}(\ell)\;,
\\
\ave{\hat\kappa^{\rm E}(\vc\ell)\hat\kappa^{\rm B*}(\vc\ell')}
&=&(2\pi)^2\,\delta_{\rm D}(\vc\ell-\vc\ell')\,P_{\rm EB}(\ell)\;.
\nonumber
\eea
From what was said above, the cross power $P_{\rm EB}$ vanishes for
parity-symmetric shear fields, and we shall henceforth ignore it.
The shear correlation functions now depend on the power spectra of both modes,
and are given as (Crittenden et al.\ 2002; Schneider et al.\ 2002a)
\bea
\xi_+(\theta)
&=&\int_0^\infty
{\d\ell\,\ell\over 2\pi}\,{\rm J}_0(\ell\theta)
\eck{P_{\rm E}(\ell)+P_{\rm B}(\ell)}\;,
\nonumber \\
\xi_-(\theta) &=&
\int_0^\infty{\d\ell\,\ell\over 2\pi}\,{\rm J}_4(\ell\theta)
\eck{P_{\rm E}(\ell)-P_{\rm B}(\ell)}\;.
\nonumber
\eea
Hence, in the presence of B-modes, the $\xi_-$ correlation function cannot be
obtained from $\xi_+$, as was the case for a pure E-mode shear field. The
inverse relation (\ref{eq:Pfromxi}) now gets modified to 
\bea
P_{\rm E}(\ell)&=&\pi\int_0^\infty\d\theta\,\theta\,
\eck{\xi_+(\theta){\rm J}_0(\ell\theta)
+\xi_-(\theta){\rm J}_4(\ell\theta)}\;,
\nonumber\\
P_{\rm B}(\ell)&=&\pi\int_0^\infty\d\theta\,\theta\,
\eck{\xi_+(\theta){\rm J}_0(\ell\theta)
-\xi_-(\theta){\rm J}_4(\ell\theta)}\;.
\elabel{N15}
\eea
Hence, the two power spectra can be obtained from the shear
correlation functions. However, due to the infinite range of
integration, one would need to measure the correlation functions over
all angular scales to apply the previous equations for calculating the
power spectra. Much more convenient for the E/B-mode decomposition is
the use of the aperture measures, since one can show that
\bea
\ave{M_{\rm ap}^2}(\theta)&=&
{1\over 2\pi}\int_0^\infty\d\ell\;\ell\,P_{\rm E}(\ell) \, 
W_{\rm ap}(\theta\ell)\;,
\nonumber \\
\ave{M_{\perp}^2}(\theta)&=&
{1\over 2\pi}\int_0^\infty\d\ell\;\ell\,P_{\rm B}(\ell) \, 
W_{\rm ap}(\theta\ell)\;,
\elabel{N28}
\eea
so that these two-point statistics clearly separate E- and B-modes. In
addition, as mentioned before, they provide a highly localized measure
of the corresponding power spectra, since the filter function $W_{\rm
ap}(\eta)$ involved is very narrow. As was true for the E-mode only
case, the aperture measures can be expressed as finite integrals over
the correlation functions,
\bea
\ave{M_{\rm ap}^2}(\theta)\!\! 
&=&\!\!{1\over
2}\int{\d\vt\,\vt\over\theta^2}
\eck{\xi_+(\vt)\,T_+\!\!\rund{\vt\over\theta}+\xi_-(\vt)\,T_-\!\!\rund{\vt\over\theta}}
\;,\nonumber \\
\ave{M_{\perp}^2}(\theta)\!\!&=&\!\!{1\over
2}\int{\d\vt\,\vt\over\theta^2}
\eck{\xi_+(\vt)\,T_+\!\!\rund{\vt\over\theta}-\xi_-(\vt)\,T_-\!\!\rund{\vt\over\theta}}
\;,
\elabel{N29}
\eea
where the two functions $T_\pm$ are the same as in
(\ref{eq:Mapfromxi}) and have been given explicitly in Schneider et
al.\ (2002a) for the weight function $Q$ given in
(\ref{eq:U+Qgeneric}), and in Jarvis et al.\ (2003) for the weight
function (\ref{eq:U+Q.CNPT}). Hence, the relations (\ref{eq:N29})
remove the necessity to calculate the aperture measures by placing
apertures on the data field which, owing to gaps and holes, would make
this an inaccurate and biased determination. Instead, obtaining the
correlation functions from the data is all that is needed.

The relations given above have been applied to recent cosmic shear surveys,
and significant B-modes have been discovered (see Sect.\ \ref{sc:WL-7}); the
question now is what are they due to? As mentioned before, the noise,
which contributes to both E- and B-modes in similar strengths, could
be underestimated, the cosmic variance which also determines the error bars on
the aperture measures and which depends on fourth-order statistical properties
of the shear field could also be underestimated, 
there could be remaining systematic effects, or
B-modes could indeed be present. There are two
possibilities known to generate a B-mode through lensing: The
first-order in $\Phi$ (or `Born') approximation may not be strictly
valid, but as shown by ray-tracing simulations through cosmic matter
fields (e.g., Jain et al.\ 2000), the resulting B-modes are expected to
be very small. Clustering of sources also yields a finite B-mode
(Schneider et al.\ 2002a), but again, this effect is much smaller than
the observed amplitude of the B-modes (see
Fig.\ts\ref{fig:EB-correl}).

\begin{figure}
\bc
\includegraphics[width=7.8cm]{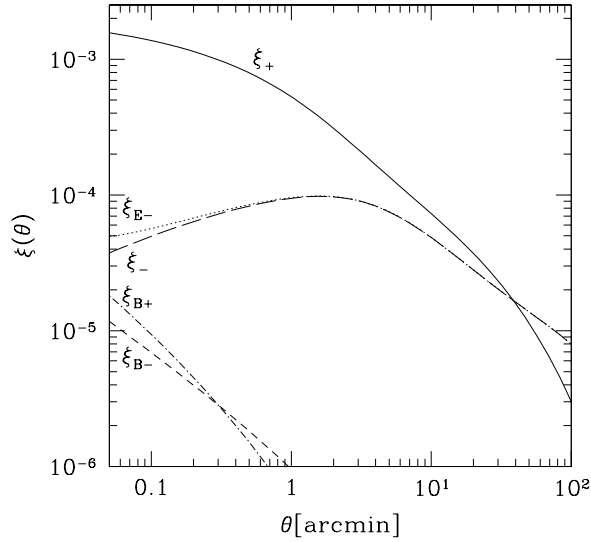}
\ec
\caption{The correlation functions $\xi_\pm(\theta)$ for a
$\Lambda$CDM model with $\Gamma_{\rm spect}=0.21$ and $\sigma_8=1$,
and a source population with mean redshift of $\ave{z_{\rm
s}}=1.5$. Also plotted are the corresponding correlation functions
that arise separately from the E- and B-modes, with the $\xi_{{\rm E}+}$
mode curve coinciding within the line thickness with $\xi_+$. In this
calculation, the clustering of the faint galaxy population was taken
into account, and they give rise to a very small B-mode contribution, as can
be seen from the $\xi_{{\rm B}\pm}$ curves. The smallness of the B-mode due to
intrinsic source clustering renders this effect not viable to explain the
B-modes observed in some of the cosmic shear surveys (from Schneider et al.\
2002a)} 
\flabel{EB-correl}
\end{figure}

\subsubsection{Intrinsic alignment of source galaxies.}
Currently the best guess for the generation of a finite B-mode are
intrinsic correlations of galaxy ellipticities.  Such intrinsic
alignments of galaxy ellipticities can be caused by tidal
gravitational fields during galaxy formation, owing to tidal
interactions between galaxies, or between galaxies and clusters.
Predictions of the alignment of the projected ellipticity of the
galaxy mass can be made analytically (e.g. in the frame of tidal
torque theory) or from numerical simulations; however, the predictions
from various groups differ by large factors (e.g., Croft \& Metzler
2000; Crittenden et al.\ 2001; Heavens et al.\ 2000; Jing 2002) which
means that the process is not well understood at present. For example,
the results of these studies depend on whether one assumes that the
light of a galaxy is aligned with the dark matter distribution, or
aligned with the angular momentum vector of the dark halo. This is
related to the question of whether the orientation of the galaxy light
(which is the issue of relevance here) is the same as that of the
mass.

If intrinsic alignments play a role, then
\be
\xi_+=\ave{\eps_i\,\eps^*_j}=\ave{\eps_i^{(\rm s)}\,\eps^{(\rm s)*}_j}
+\xi_+^{\rm lens}\;,
\elabel{xip-intr1}
\ee
and measured correlations $\xi_\pm$ contain both components, the
intrinsic correlation and the shear.  Of course, there is no reason
why intrinsic correlations should have only a B-mode. If a B-mode
contribution is generated through this process, then the measured
E-mode is most likely also contaminated by intrinsic alignments.
Given that intrinsic alignments yield ellipticity correlations only
for spatially close sources (i.e., close in 3-D, not merely in
projection), it is clear that the deeper a cosmic shear survey is, and
thus the broader the redshift distribution of source galaxies, the
smaller is the relative amplitude of an intrinsic signal. Most of the
theoretical investigations on the strength of intrinsic alignments
predict that the deep cosmic shear surveys (say, with mean source
redshifts of $\ave{z_{\rm s}}\sim 1$) are affected at a $\sim 10\%$
level, but that shallower cosmic shear surveys are more strongly
affected; for them, the intrinsic alignment can be of same order or
even larger than the lensing signal.

However, the intrinsic signal can be separated from the lensing signal
if redshift information of the sources is available, owing to the fact
that $\ave{\eps_i^{(\rm s)}\,\eps^{(\rm s)*}_j}$ will be non-zero only
if the two galaxies are at essentially the same redshift. Hence, if
$z$-information is available (e.g., photometric redshifts), then
galaxy pairs which are likely to have similar redshifts are to be
avoided in estimating the cosmic shear signal (King \& Schneider 2002;
Heymans \& Heavens 2002, Takada \& White 2004). This will change the
expectation value of the shear correlation function, but in a
controllable way, as the redshifts are assumed to be known. Indeed,
using (photometric) redshifts, one can simultaneously determine the
intrinsic and the lensing signal, essentially providing a cosmic shear
tomography (King \& Schneider 2003). This again is accomplished by
employing the fact that the intrinsic correlation can only come from
galaxies very close in redshift. Hence, in the presence of intrinsic
alignments, the redshift dependent correlation functions
$\xi_\pm(z_1,z_2;\theta)$ between galaxies with estimated redshifts
$z_i$ are expected to show a strong peak over the range
$|z_1-z_2|\lesssim \Delta z$, where $\Delta z$ is the typical
uncertainty in photometric redshifts. It is this peak that allows one
to identify and subtract the intrinsic signal from the correlation
functions.  An efficient method to calculate the covariance of the
redshift-dependent correlation functions has been developed by Simon
et al.\ (2004), where the improvement in the constraints on
cosmological parameters from redshift information has been studied,
confirming the earlier results by Hu (1999) which were based on
considerations of the power spectrum.

Brown et al.\ (2002) obtained a measurement of the intrinsic
ellipticity correlation from the Super-COSMOS photographic plate data,
where the galaxies are at too low a redshift for cosmic shear playing
any role. Heymans et al.\ (2003) used the COMBO-17 data set (that will
be described in Sect.\ts\ref{sc:WL-7.3} below) for which accurate
photometric redshifts are available to measure the intrinsic
alignment. The results from both studies is that the models predicting
a large intrinsic amplitude can safely be ruled out. Nevertheless,
intrinsic alignment affects cosmic shear measurements, at about the
2\% level for a survey with the depth of the VIRMOS-DESCART survey,
and somewhat more for the slightly shallower COMBO-17 survey. Hence,
to obtain precision measurements of cosmic shear, very important for
constraining the equation of state of dark energy, these physically
close pairs of galaxies need to be identified in the survey, making
accurate photometric redshifts mendatory.

\subsubsection{Correlation between intrinsic ellipticity and shear.}
The relation (\ref{eq:xip-intr1}) above implicitly assumes that the shear is
uncorrelated with the intrinsic shape of a neighboring galaxy. However, as
pointed out by Hirata \& Seljak (2004), this is not necessarily the
case. Hence consider galaxies at two significantly different redshifts
$z_i<z_j$. For them, the first term in (\ref{eq:xip-intr1}) vanishes. However,
making use of $\eps=\eps^{(\rm s)}+\gamma$, one finds
\be
\ave{\eps_i\eps_j^*}=\ave{\eps_i^{(\rm s)}\gamma_j^*} +\xi_+^{\rm lens}\;,
\elabel{xip-intr2}
\ee
where the first term on the right-hand side describes the correlation between
the intrinsic ellipticity of the lower-redshift galaxy with the shear along
the l.o.s. to the higher-redshift one. The correlation can in principle be
non-zero: if the intrinsic alignment of the light of a galaxy is determined by
the large-scale tidal gravitational field, then this tidal field at the
redshift $z_i$ causes both, an alignment of the nearer galaxy and a
contribution to the shear of the more distant one (see
Fig.\ts\ref{fig:HiraSelj}). This alignment effect can 
therefore not be removed by considering only pairs at different redshifts. 

\begin{figure}
\bc
\includegraphics[width=7cm]{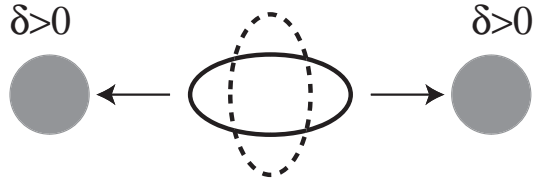}
\ec
\caption{A tidal gravitational field, for example caused by two matter
  concentrations, can produce an alignment of a galaxy situated at the same
  redshift (indicated by the solid ellipse), as well as contributing to the
  shear towards a more distant galaxy (as indicated by the dashed ellipse) (from
  Hirata \& Seljak 2004)}
\flabel{HiraSelj}
\end{figure}

The importance of this effect depends on the nature of the alignment of
galaxies relative to an external tidal field. If the alignment is linear in
the tidal field strength, then this effect can be a serious contaminant of the
cosmic shear signal, in particular for relatively shallow surveys (where the
mean source redshift is small); in particular, this effect can yield much
larger contaminations than the intrinsic alignment given by the first term in
(\ref{eq:xip-intr1}).  As can be seen from Fig.\ts\ref{fig:HiraSelj}, the
resulting contribution is negative, hence decreases the lensing signal.  If,
however, the intrinsic alignment depends quadratically on the tidal field, as
is suggested by tidal torque theory, than this effect is negligible. Whether
or not this effect is relevant needs to be checked from observations. Assuming
that the matter density field is represented approximately by the galaxy
distribution, the latter can be used to estimate the tidal gravitational
field, in particular its direction. Alternatively, since the correlation
between the intrinsic alignment and the shear towards more distant galaxies
has a different redshift dependence than the lensing shear signal, these two
contributions can be disentangled from the $z$-dependence of the signal.

It should be noted that the use of photometric redshifts also permits
to study the cosmic shear measures as a function of source redshift;
hence, one can probe various redshift projections $P_\kappa(\ell)$ of
the underlying power spectrum $P_\delta(k;z)$ separately. This is due
to the fact that the cosmic shear signal from different populations of
galaxies (i.e., with different redshift distributions) lead to
different weight functions $g(w)$ [see (\ref{eq:g-fact})], and thus to
different weighting in the projection (\ref{eq:6.25}) of the power
spectrum. Not surprisingly, uncertainties of cosmological parameters
are thereby reduced (Hu 1999; Simon et al.\ 2004). Also, as shown by
Taylor (2001), Hu \& 
Keeton (2002) and Bacon \& Taylor (2003), in principle the
three-dimensional mass distribution $\delta(\vec x)$ can be
reconstructed if the redshifts of the source galaxies are known (see
Sect.\ts\ref{sc:WL-7.6}).

\subsection{\llabel{WL-6.6}Predictions; ray-tracing simulations}
The power spectrum of the convergence $P_\kappa$ can be calculated
from the power spectrum of the cosmological matter distribution
$P_\delta$, using (\ref{eq:6.25}); the latter in turn is determined by
the cosmological model. However, since the non-linear evolution of the
power spectrum is essential for making accurate quantitative
predictions for the shear properties, there is no analytic method
known how to calculate the necessary non-linear $P_\delta$. As was
mentioned in Sect.\ 6.1 of IN, fairly accurate fitting formulae exist
which yield a closed-form expression for $P_\delta$ and which can be
used to obtain $P_\kappa$ (see, e.g., Jain \& Seljak 1997).
Nevertheless, there are a number of reasons why this purely analytic
approach should at least be supplemented by numerical simulations.
\bi
\item
First, the fitting formulae for $P_\delta$ (Peacock \& Dodds 1996;
Smith et al.\ 2003) have of course only a finite accuracy, and are
likely to be insufficient for comparison with results from the ongoing
cosmic shear surveys which are expected to yield very accurate
measurements, owing to their large solid angle.
\item
A second reason why simulations are needed is to test whether the
various approximations that enter the foregoing analytical treatment
are in fact accurate enough. To recall them, we employed the Born
approximation, i.e., neglected terms of higher order than linear in
the Newtonian potential when deriving the convergence, and we assumed
that the shear everywhere is small, so that the difference between
shear and reduced shear can be neglected, at least on average. This,
however, is not guaranteed: regions in the sky with large shear are
most likely also those regions where the convergence is particularly
large, and therefore, there one expects a correlation between $\gamma$
and $\kappa$, which can affect the dispersion of
$g=\gamma/(1-\kappa)$.
\item
Third, whereas fairly accurate fitting formulae exist for
the power spectrum, this is not the case for higher-order statistical
properties of the matter distribution; hence, when considering higher-order
shear statistics (Sect.\ \ref{sc:WL-9}), numerical simulations will most
likely be the only way to obtain accurate predictions.
\item
The covariance of the shear correlations (and all other second-order
shear measures) depends on fourth-order statistics of the shear field,
for which hardly any useful analytical approximations are
available. The analytical covariance estimates are all based on the
Gaussian assumption for the fourth-order correlators.
Therefore, simulations are invaluable for the calculation
of these covariances, which can be derived for arbitrary survey
geometries. 
\ei

\def\hMpc{h^{-1}\,{\rm Mpc}}
\subsubsection{Ray-tracing simulations: The principle.}
The simulations proceed by following light rays through the
inhomogeneous matter distribution in the Universe. The latter is
generated by cosmological simulations of structure evolution. Those
start at an early epoch by generating a realization of a Gaussian
random field with a power spectrum according to the cosmological model
considered, and follow the evolution of the density and velocity field
of the matter using Newtonian gravity in an expanding Universe. The
mass distribution is represented by discrete particles whose evolution
in time is followed.  A finite volume of the Universe is simulated
this way, typically a box of comoving side-length $L$, for which
periodic boundary conditions are applied. This allows one to use Fast
Fourier Transforms (FFT) to evaluate the gravitational potential and
forces from the density distribution. The box size $L$ should be
chosen such that the box contains a representative part of the real Universe,
and must therefore be larger than the largest scales on which
structure is expected, according to the power spectrum; a reasonable
choice is $L\gtrsim 100 \hMpc$. The number of grid points and the
number of particles that can be distributed in this volume is limited
by computer memory; modern simulations work typically with $256^3$
points and the same number of particles, though larger simulations
have also been carried out; this immediately yields the
size of grid cells, of order $0.5\hMpc$. This comoving length, if
located at a redshift of $z\sim 0.3$ (which is about the most relevant
for cosmic shear), subtends an angle of roughly $2'$ on the sky. The
finite number of particles yields the mass resolution of the
simulations, which is typically $\sim 10^{10} h^{-1}M_\odot$,
depending on cosmological parameters.

In order to obtain higher spatial resolution, force calculations are
split up into near-field and far-field forces. The gravitational force
due to the distant matter distribution is obtained by grid-based FFT
methods, whereas the force from nearby masses is calculated from
summing up the forces of individual particles; such simulations yield
considerably higher resolution of the resulting mass
distribution. Since the matter in these simulations is represented by
massive particles, these can undergo strong interactions, leading to
(unphysical) large orbital deflections. In order to avoid these
unphysical strong collisions, the force between pairs of particles is
modified at short distances, typically comparable to the mean
separation of two particles in the simulation. This softening length
defines the minimum length scale on which the results from numerical
simulations can be considered reliable.  Cosmological simulations
consider either the dark matter only or, more recently, the
hydrodynamics effects of baryons have been incorporated as well.

The outcome of such simulations, as far as they are relevant here, are
the 3-dimensional positions of the matter particles at different
(output) times or redshifts. In order to study the light propagation
through this simulated mass distribution, one employs multiple
lens-plane theory. First, the volume between us and sources at some
redshift $z_{\rm s}$ is filled with boxes from the cosmological
simulations. That is, the comoving distance $w_{\rm s}=w(z_{\rm s})$
is split up into $n$ intervals of length $L$, and the mass
distribution at an output time close to $t_i=t(w=(i-1/2)L)$ is
considered to be placed at this distance. In this way, one has a light
cone covered by cubes containing representative matter
distributions. Since the mass distributions at the different times
$t_i$ are not independent of each other, but one is an evolved version
of the earlier one, the resulting mass distribution is highly
correlated over distances much larger than $L$. This can be avoided by
making use of the statistical homogeneity and isotropy of the mass
distribution: each box can be translated by an arbitrary
two-dimensional vector, employing the periodicity of the mass
distribution, and rotated by an arbitrary angle; furthermore, the
three different projections of the box can be used for its
orientation. In this way -- a kind of recycling of numerical results --
the worst correlations are removed.  

Alternatively, one can combine
the outputs from several simulations with different realizations of
the initial conditions. In this case, one can use simulation boxes of
different spatial extent, to match the comoving size of a big light
cone as a function of redshift. That is, for a given light-cone size,
only relatively small boxes are needed at low redshifts, and bigger
ones at higher redshift (see White \& Hu 2000).

Second, the mass in each of these boxes is projected along the
line-of-sight, yielding a surface mass density at the appropriate
comoving distance $w_i=(i-1/2)L$. Each of these surface mass densities
can now be considered a lens plane, and the propagation of light can
be followed from one lens plane to the next; the corresponding theory
was worked out in detail by Blandford \& Narayan (1986; see also
Chap.\ 9 of SEF), but applied as early as 1970 by Refsdal (1970) for a
cosmological model consisting of point masses only (see also Schneider
\& Weiss 1988a,b).  Important to note is that the surface mass density
$\Sigma$ in each lens plane is the projection of
$\Delta\rho=\rho-\bar\rho$ of a box, so that for each lens plane,
$\ave{\Sigma}=0$.  As has been shown in Seitz et al.\
(1994), this multiple lens-plane approach presents a well-defined
discretization of the full 3-dimensional propagation equations.  Light
bundles are deflected and distorted in each lens plane and thus
represented as piecewise straight rays. The resulting Jacobi matrix
$\A$ is then obtained as a sum of products of the tidal matrices in
the individual lens planes, yielding a discretized version of the form
(\ref{eq:fullJacobi}) for $\A$. The result of such simulations is then
the matrix $\A(\vc\theta)$ on a predefined angular grid, as well as
the positions $\vc\beta(\vc\theta)$ in the source plane. The latter
will not be needed here, but have been used in studies of multiple
images caused by the LSS (see Wambsganss et al.\ 1998).

One needs special care in applying the foregoing prescription; in particular,
in the smoothing process to obtain a mass distribution from the discrete
particles; Jain et al.\ (2000) contains a detailed discussion on these
issues.\footnote{For other recent ray-tracing simulations related to cosmic
  shear, see e.g.\ Barber et al.\ (2000); Hamana \& Mellier (2001); Premadi et
  al.\ (2001); Taruya et 
  al.\ (2002); Fluke et al.\ (2002); Barber (2002); Vale \& White (2003).}
The finite spatial resolution in the simulations translates into a
redshift-dependent angular resolution, which degrades for the low redshift
lens planes; on the other hand, those have a small impact on the light
propagation due to the large value of $\Sigma_{\rm cr}$ for them [see eq.\ 
(10) of IN]. The discreteness of particles gives rise to a shot-noise term in
the mass distribution, yielding increased power on small angular scales.

\subsubsection{Results from ray-tracing simulations.}
We shall summarize here some of the results from ray-tracing
simulation:
\bi
\item
Whereas the Jacobi matrix in this multi-deflection situation is no
longer symmetric, the contribution from the asymmetry is very small. The power
spectrum of the asymmetric part of $\A$ is at least three orders of
magnitude smaller than the power spectrum $P_\kappa$, for sources at
$z_{\rm s}=1$ (Jain et al.\ 2000). This result is in accord
with analytical expectations (e.g., Bernardeau et al.\ 1997; Schneider
et al.\ 1998a), i.e., that terms quadratic in the
Newtonian potential are considerably smaller than first-order terms,
and supports the validity of the Born approximation. Furthermore, this
result suggests that a simpler method for predicting cosmic shear
distributions from numerical simulations may be legitimate, namely to
project the mass distribution of all lens planes along the grid of
angular positions, with the respective weighting factors, according to
(\ref{eq:6.21}), i.e., employing the Born approximation. 
Of course this simplified method is computationally
much faster than the full ray-tracing.
\item
The power spectra obtained reproduce the ones derived using
(\ref{eq:6.25}), over the range of wavevectors which are only mildly
affected by resolution and discreteness effects. This provides an
additional check on the accuracy of the fitting formulae for the
non-linear power spectrum. 
\item
The simulation results give the full two-dimensional shear map, and
thus can be used to study properties other than the second-order ones,
e.g., higher-order statistics, or the occurrence of circular shear
patters indicating the presence of strong mass concentrations.  An
example of such maps is shown in Fig.\ \ref{fig:JSW}. These shear maps
can be used to simulate real surveys, e.g, including the holes in the
data field resulting from masking or complicated survey geometries,
and thus to determine the accuracy with which the power spectra can be
determined from such surveys. Note that in order to quantify the error
(or covariance matrix) of any second-order statistics, one needs to
know the fourth-order statistics, which in general cannot be obtained
analytically when outside the linear (Gaussian) regime.  
Simulations are also used to obtain good survey strategies.  
\ei

\subsubsection{Higher-order correction terms.}
Up to now we have considered the lowest-order approximation of the
Jacobi matrix (\ref{eq:fullJacobi}) and have argued that this provides
a sufficiently accurate description. Higher-order terms in $\Phi$ were
neglected since we argued that, because the Newtonian potential is very
small, these should play no important role. However, this argument is
not fully correct since, whereas the potential certainly is small, its
derivatives are not necessarily so. Of course, proper ray-tracing
simulation take these higher-order terms automatically into account.

We can consider the terms quadratic in $\Phi$ when expanding
(\ref{eq:fullJacobi}) to higher order. There are two such terms, one
containing the product of second-order derivatives of $\Phi$, the
other a product of first derivatives of $\Phi$ and its third
derivatives. The former is due to lens-lens coupling: The shear and
surface mass densities from different redshifts (or lens planes, in
the discretized approximation) do not simply add, but multi lens plane
theory shows that the tidal matices from different lens planes get
multiplied. The latter term comes from dropping the Born approximation
and couples the deflection of a light ray (first derivative of $\Phi$)
with the change of the tidal matrix with regards to the position
(third derivatives of $\Phi$). These terms are explitly given in the
appendix of Schneider et al.\ (1998a), in Bernardeau et al.\
(1997) and in Cooray \& Hu (2002) and found to be indeed small, providing
corrections of at most a few percent. Furthermore, Hamana (2001) has shown
that the magnification bias caused by the foreground matter inhomogeneities on
the selection of background galaxies has no practical effect on second-order
cosmic shear statistics. 

Another effect that affects the power spectrum $P_\kappa$ is the
difference between shear and reduced shear, the latter being the
observable. Since the correlation function of the reduced shear is the
correlation function of the shear plus a term containing a product of
two shears and one surface mass density, this correction depends
linearly on the third-order statistical properties of the projected
mass $\kappa$. Also this correction turns out to be very small;
moreover, it does not give rise to any B-mode contribution (Schneider
et al.\ 2002a). 

\section{\llabel{WL-7}Large-scale structure lensing: results}

After the theory of cosmic shear was considered in some detail in the previous
section, we shall summarize here the observational results that have been
obtained so far. In fact, as we will see, progress has been incredibly fast
over the past $\sim$four years, with the first detections reported in 2000,
and much larger surveys being available by now, with even larger ones ongoing
or planned. Already by now, cosmic shear is one of the pillars on which our
cosmological model rests.

The predictions discussed in the previous section have shown that the rms
value of cosmic shear is of the order of $\sim 2\%$ on angular scales of $\sim
1'$, and considerably smaller on larger scales. These small values make the
measurements of cosmic shear particularly challenging, as the observational
and instrumental effects described in Sect.\ \ref{sc:WL-3} are expected to be
larger than the cosmic shear signal, and thus have to be understood and
removed with great precision. For example, the PSF anisotropy of nearly all
wide-field cameras is considerably larger than a few percent and thus needs to
be corrected for. But, as also discussed in Sect.\ \ref{sc:WL-3}, methods have
been developed and thoroughly tested which are able to do so.

\subsection{\llabel{WL-7.1}Early detections of cosmic shear}

Whereas the theory of cosmic shear was worked out in the early 1990's
(Blandford et al.\ 1991; Miralda-Escud\'e 1991; Kaiser 1992), it took
until the year 2000 before this effect was first
discovered.\footnote{An early heroic attempt by Mould et al.\ (1994)
to detect cosmic shear on a single $\sim 9'\times 9'$ field only
yielded an upper limit, and the putative detection of a shear signal
by Schneider et al.\ (1998b; see also Fort et al.\ 1996) in three
$2'\times 2'$ fields is, due to the very small sky area, of no
cosmological relevance.}  The reason for this evolution must be seen
by a combination of instrumental developments, i.e.\ the wide-field
CCD mosaic cameras, and the image analysis software, like IMCAT (the
software package encoding the KSB method discussed in Sect.\
\ref{sc:WL-3.5}), with which shapes of galaxies can be accurately
corrected for PSF effects. Then in March 2000, four groups
independently announced their first discoveries of cosmic shear (Bacon
et al.\ 2000; Kaiser et al.\ 2000; van Waerbeke et al.\ 2000, Wittman
et al.\ 2000). In these surveys, of the order of $10^5$ galaxy images
have been analyzed, covering about 1\ deg$^2$. Later that year, Maoli
et al.\ (2001) reported a significant cosmic shear measurement from
50 widely separated FORS1@VLT images, each of size $\sim
6\arcminf5\times 6\arcminf5$, which also agreed with the earlier
results.  The fact that the results from four independent teams
agreed within the respective error bars immediately lend credit to
this new window of observational cosmology. This is also due to the
fact that 4 different telescopes, 5 different cameras (the UH8K and
CFH12K at CFHT, the $8'\times 16'$-imager on WHT, the BTC at the
4m-CTIO telescope and FORS1 at the VLT), independent data reduction
tools and at least two different image analysis methods have been
used. These early results are displayed in Fig.\
\ref{fig:SC-early-results}, where the (equivalent) shear dispersion is
plotted as a function of effective circular aperture radius, together
with the predictions for several cosmological models. It is
immediately clear that a high-normalization Einstein-de Sitter model
can already be excluded from these early results, but the other three
models displayed are equally valid approximations to the data.

\begin{figure}
\includegraphics[width=11.7cm]{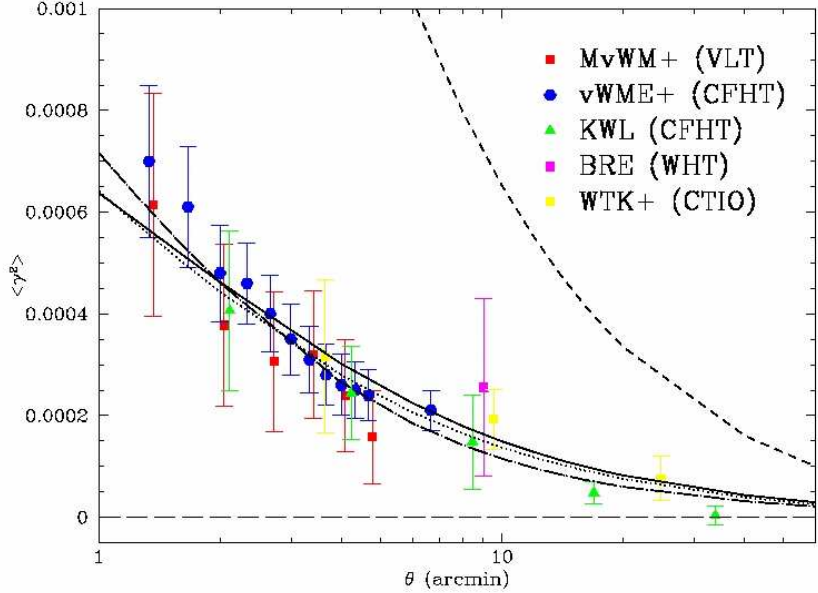}
\caption{Shear dispersion as a function of equivalent circular aperture radius
  as obtained from the first five measurements of cosmic shear (MvWM+: Maoli
  et al.\ 2001; vWME+: van Waerbeke et al.\ 2000; KWL: Kaiser, Wilson \&
  Luppino 2000; BRE: Bacon, Refregier \& Ellis 2000; WTK: Wittman et al.\
  2000). The data points within each team are not statistically independent,
  due to the fairly strong covariance of the shear dispersion on different
  angular scales, but points from different teams are independent (see
text). The 
  error bars contain the noise from the intrinsic ellipticity dispersion
  and, for some of the groups, also an estimate of cosmic variance. The four
  curves are predictions from four cosmological models; the upper-most one
  corresponds to an Einstein-de Sitter Universe with normalization
  $\sigma_8=1$, and can clearly be excluded by the data. The other three models
  are cluster normalized -- see Sect.\ 4.4 of IN -- and all provide
equally good fits
  to these early data (courtesy: Y.\ Mellier)}
\flabel{SC-early-results}
\end{figure}

Maoli et al.\ (2001) considered the constraints one obtains by combining the
results from these five surveys, in terms of the normalization parameter
$\sigma_8$ of the power spectrum.  The confidence contours in the $\Omega_{\rm
m}-\sigma_8$-plane are shown in Fig.\ \ref{fig:Maoli-constraints}. There is
clearly a degeneracy between these two parameters from the data sets
considered, roughly tracing $\sigma_8\sim 0.59\Omega_{\rm m}^{-0.47}$;
although the best fitting model is defined by 
$\Omega_{\rm m}=0.26$, $\sigma_8=1.1$, it cannot be significantly
distinguished from, e.g., a $\Omega_{\rm m}=1$, $\sigma_8=0.62$ model since
the error bars displayed in Fig.\ \ref{fig:SC-early-results} are too large and
the range of angular scales over which the shear was measured is too small. In
Fig.\ \ref{fig:Maoli-constraints}, the solid curve displays the normalization
as obtained from the abundance of massive clusters, which is seen to follow
pretty much the valley of degeneracy from the cosmic shear analysis. This fact
should not come as a surprise, since the cluster abundance probes the power
spectrum on a comoving scale of about $8\hMpc$, which is comparable to the
median scale probed by the cosmic shear measurements. However, the predictions
of the cluster abundance rely on the assumption that the initial density field
was Gaussian, whereas the cosmic shear prediction is independent of this
assumption, which therefore can be tested by comparing the results from both
methods. 

\begin{figure}
\bmi{7}
\includegraphics[bb= 109 340 485 702, clip,width=6.8cm,angle=0]{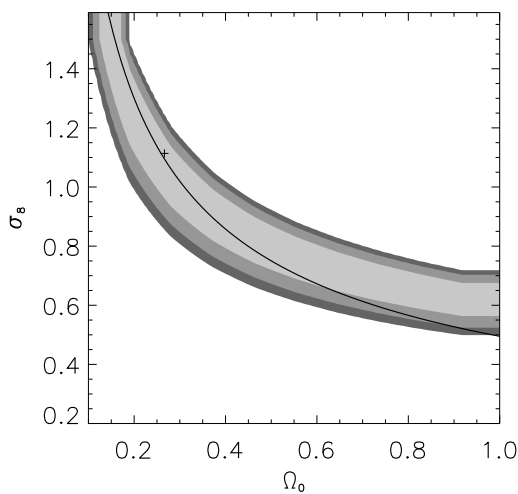}
\emi
\bmi{4.7}
\caption{Constraints on $\Omega_{\rm m}$ and $\sigma_8$ from the five
surveys shown in Fig.\ts\ref{fig:SC-early-results}; shown are 1, 2 and
3-$\sigma$ confidence regions. The cross denotes the best-fitting
model, but as can be seen, these two parameters are highly degenerate
with the data used. The solid curve displays the constraint from
cluster normalization (from Maoli et al.\ 2001)}
\flabel{Maoli-constraints} 
\emi
\end{figure}

\subsection{\llabel{WL-7.2}Integrity of the results}
As mentioned before, the cosmic shear effects are smaller than many
observational effects (like an anisotropic PSF) that could mimic a
shear; it is therefore necessary to exclude as much as possible such
systematics from the data. The early results described above were
therefore accompanied by quite a large number of tests; they should be
applied to all cosmic shear surveys as a sanity check.  A few of those
shall be mentioned here.

\subsubsection{Stellar ellipticity fits.}
The ellipticity of stellar objects should be well fitted by a
low-order function, so one is able to predict the PSF anisotropy at
galaxy locations. After subtracting this low-order fit from the
measured stellar ellipticities, there should be no coherent spatial
structure remaining, and the ellipticity dispersion of the corrected
ellipticities should be considerably smaller that the original ones,
essentially compatible with measurement noise. 

\subsubsection{Correlation of PSF anisotropy with corrected galaxy
ellipticities.} 
After correcting for the anisotropy of
the PSF, there should remain no correlation between the corrected
galaxy ellipticities and the ellipticity of the PSF. This correlation
can be measured by considering $\ave{\eps\,\eps^*}$, where $\eps$ is
the corrected galaxy ellipticity (\ref{eq:KSB-shearest}), and $\eps^*$
the uncorrected stellar ellipticity (i.e., the PSF anisotropy). Bacon et
al.\ (2000) found that for fairly low signal-to-noise galaxy images,
this correlation was significantly different from zero, but for
galaxies with high S/N (only those entered their cosmic shear
analysis), no significant correlation remained. The same was found in
van Waerbeke et al.\ (2000), except that the average $\ave{\eps_1}$ was
slightly negative, but independent of $\eps^*_1$. The level of
$\ave{\eps_1}$ was much smaller than the estimated cosmic shear, and
does not affect the latter by more than $10\%$.

\subsubsection{Spatial dependence of mean galaxy ellipticity.}
When a cosmic shear survey consist of many uncorrelated fields, the mean
galaxy ellipticity at a given position on the CCD chips should be zero, due to
the assumed statistical isotropy of the shear field. If, on the other hand,
the shear averaged over many fields shows a dependence on the chip position,
most likely optical distortions and/or PSF effects have not been properly
accounted for. 

\subsubsection{Parity invariance.}
The two-point correlation function
$\xi_\times(\theta)=\ave{\gamma_{\rm t}\gamma_\times}(\theta)$ is
expected to vanish for a density distribution that is parity
symmetric. More generally, every astrophysical cause for a `shear'
signal (such as intrinsic galaxy alignments, or higher-order lensing
effects) is expected to be invariant under parity transformation. A
significant cross-correlation $\xi_\times$ would therefore indicate
systematic effects in the observations and/or data analysis.

\subsection{\llabel{WL-7.3}Recent cosmic shear surveys}
Relatively soon after the announcement of the first cosmic shear detections,
additional results were published. These newer surveys can roughly be
classified as follows: deep surveys, shallower, but much wider surveys, and
special surveys, such as obtained with the Hubble Space Telescope. We shall
mention examples of each of these classes here, without providing a complete
list. 

\begin{figure}
\bmi{6.0}
\includegraphics[width=5.7cm,angle=0]{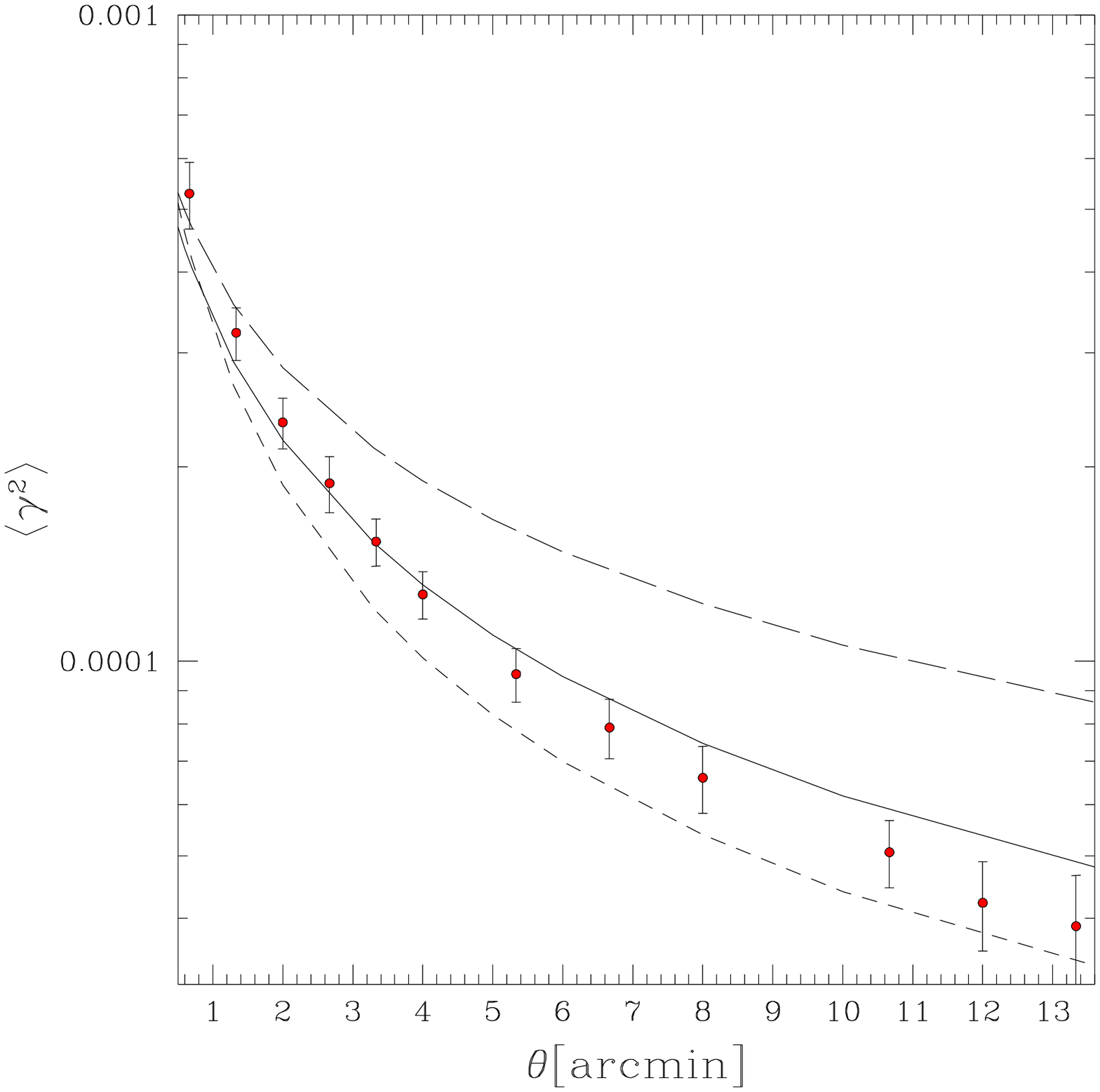}
\emi\bmi{5.7}
\includegraphics[width=5.7cm,angle=0]{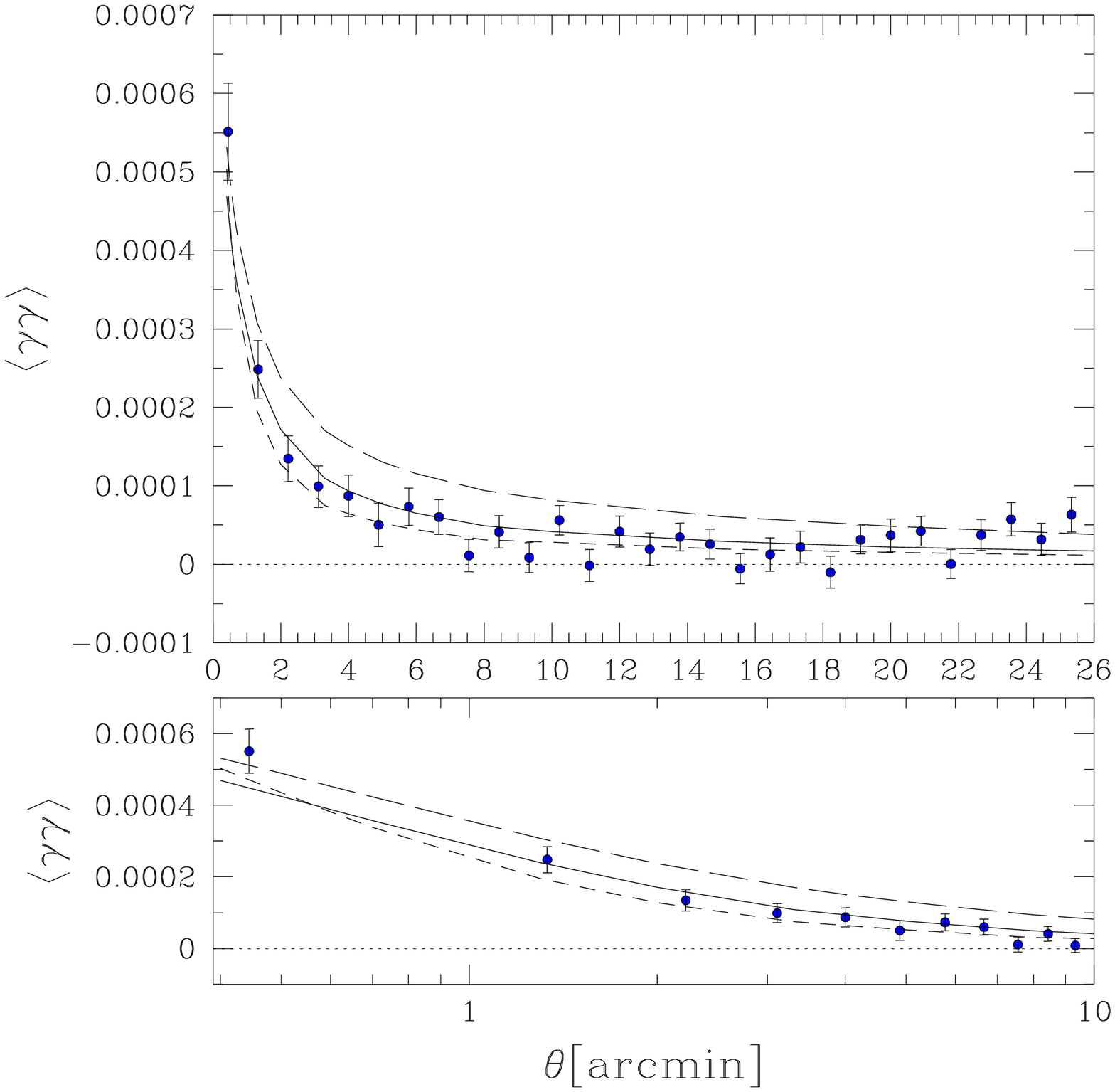}
\emi

\caption{The shear dispersion as a function of aperture radius (left) and the
  shear correlation function $\xi_+(\theta)$ (right) as measured from the
  VIRMOS-DESCART survey (van Waerbeke et al.\ 2001). The lower panel on the
  right shows an enlargement with logarithmic axis of the larger figure. The
  error bars were calculated from simulations in which  the galaxy images have
  been randomized in orientation. The curves show predictions from three
  different cosmological models, corresponding to $(\Omega_{\rm
  m},\Omega_\Lambda,\sigma_8)= (0.3,0,0.9)$ (open model, short-dashed curves),
  $(0.3,0.7,0.9)$ (low-density flat model, solid curves), and $(1,0,0.6)$
  (Einstein-de Sitter Universe, long-dashed curves). In all cases, the shape
  parameter of the power spectrum was set to $\Gamma_{\rm spect}=0.21$. The
  redshift distribution of the sources was assumed to follow the law
  (\ref{eq:z-distri}), with $\alpha=2$, $\beta=1.5$ and $z_0=0.8$,
  corresponding to a mean redshift of $\bar z\approx 1.2$
}
\flabel{vW01-gamxi}
\end{figure}

\subsubsection{Deep surveys.}
Currently the largest of the deep surveys from which cosmic shear
results have been published 
is the VIRMOS-DESCART survey,
carried out with the CFH12K camera at the CFHT; this camera covers about
$45'\times 30'$ in one exposure. The exposure time of the images, taken in the
I-band, is one hour. The survey covers four fields of $2^\circ \times 2^\circ$
each, of which roughly $8.5\,{\rm deg}^2$ have been used for a weak lensing
analysis up to now (van Waerbeke et al.\ 2001, 2002). About 20\% of the area
is masked out, to account for diffraction spikes, image defects, bright and
large foreground objects etc. The number density of galaxy images used for the
cosmic shear analysis is about $17\,{\rm arcmin}^{-2}$.  A small part of this
survey was used for the early cosmic shear detection (van Waerbeke et al.\ 
2000). Compared to the earlier results, the error bars on the shear
measurements are greatly reduced, owing to the much better statistics. We show
in Fig.\ \ref{fig:vW01-gamxi} the shear dispersion and the correlation
function as measured from this survey.  Furthermore, this survey yielded the
first detection of a significant $\ave{M_{\rm ap}^2}$-signal; we shall come
back to this later. In order to compare the measured shear signal with
cosmological predictions, one needs to assume a redshift distribution for the
galaxies; a frequently used parameterization for this is
\be
p(z)=N\,\rund{z\over z_0}^\alpha \exp\eck{-\rund{z\over z_0}^\beta}\;,
\elabel{z-distri}
\ee
where $\alpha$ and $\beta$ determine the shape of the redshift distribution,
$z_0$ the characteristic redshift, and $N$ is a normalization factor, chosen
such as $\int \d z\, p(z)=1$. 

Another example of a deep survey is the Suprime-Cam
survey (Hamana et al.\ 2003), a $2.1\,{\rm deg}^2$ survey taken with
the wide-field camera Suprime-Cam (with a $34'\times 27'$
field-of-view) at the 8.2-m Subaru telescope. With an exposure time of
$30\,{\rm min}$, the data is considerably deeper than the
VIRMOS-DESCART survey, due to the much larger aperture of the
telescope. After cuts in the object catalog, the resulting number
density of objects used for the weak lensing analysis is $\approx
30\,{\rm arcmin}^{-2}$. Fig.\ \ref{fig:SupCam-PSF} shows how small the
PSF anisotropy is, and that the correction with a fifth-order
polynomial over the whole field-of-view in fact reduces the remaining
stellar ellipticities considerably. This survey has detected a
significant cosmic shear signal, as measured by the shear
correlation functions and the aperture mass dispersion, over angular
scales $2'\lesssim \theta\lesssim 40'$. The shear signal increases as
fainter galaxies are used in the analysis, as expected, since fainter
galaxies are expected to be at larger mean redshift and thus show a
stronger shear signal.

\begin{figure}
\bmi{7}
\includegraphics[width=6.9cm]{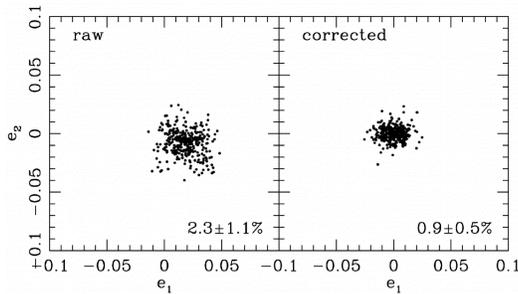}
\emi
\bmi{4.7}
\caption{
Stellar ellipticities before and after correction for PSF anisotropies in the
Suprime-Cam survey. Numbers give mean and dispersion of stellar ellipticities
$|\chi|$ (from Hamana et al.\ 2003)}
\flabel{SupCam-PSF}
\emi
\end{figure}

Bacon et al.\ (2003) combine images taken at the Keck\ts II telescope and the
WHT. For the former, 173 fields were used, each having a f.o.v. of $2'\times
8'$; and the data from WHT were obtained from 20 different fields, covering
about 1\ts deg$^2$ in total. The large number of fields minimizes the sample
variance of this particular survey, and the two instruments used allowed a
cross-check of instrumental systematics. 

\subsubsection{Very wide surveys.}
Within a given observing time, instead of mapping a sky region to fairly deep
magnitudes, one can also map larger regions with smaller exposure time; since
most of the surveys have been carried out with goals in addition to cosmic
shear, the survey strategy will depend on these other considerations. We shall
mention two very wide surveys here.

Hoekstra et al.\ (\cite{hoek595}; also Hoekstra et al.\ \cite{hoek55})
used the Red Cluster Sequence (RCS) survey, a survey designed to
obtain a large sample of galaxy clusters using color selection
techniques (Gladders \& Yee 2000). The cosmic shear analysis is based
on $53\,{\rm deg}^2$ of $R_C$-band data, spread over 13 patches on the
sky and observed with two different instruments, the CFH12K@CFHT for
Northern fields, and the Mosaic\ II camera at the CTIO 4m telescope in
the South. The integration times are 900\ s and 1200\ s,
respectively. The shear dispersion as measured with the two
instruments are in satisfactory agreement and thus can be safely
combined. Owing to the shallower magnitude, the detected shear is
smaller than in the deeper surveys mentioned above: on a scale of
$2.5\,{\rm arcmin}$, the shear dispersion is $\ave{|\bar\gamma|^2}\sim
4\times 10^{-5}$ in the RCS survey, compared to $\sim 2\times 10^{-4}$
in the deeper VIRMOS-DESCART survey (see Fig.\ \ref{fig:vW01-gamxi}),
in accordance with expectations.

Jarvis et al.\ (2003) presented a cosmic shear survey of $75\, {\rm
deg}^2$, taken with the BTC camera and the Mosaic\ II camera on the
CTIO 4m telescope, with about half the data taken with each
instrument. The survey covers 12 fields, each with sidelength of $\sim
2.5^\circ$. For each pointing, three exposures of $5\, {\rm min}$ were
taken, making the depth of this survey comparable to the RCS. A total
of $\sim 2\times 10^6$ galaxies with $R\le 23$ were used for the shear
analysis.  Since this survey has some peculiar properties which are
very educational, it will be discussed in somewhat more detail. The
first point to notice is the large pixel size of the BTC, of
$0\arcsecf 43$ per pixel -- for comparison, the CFH12K has $\sim
0\arcsecf 20$ per pixel. With a median seeing of $1\arcsecf 05$, the
PSF is slightly undersampled with the BTC. Second, the PSF anisotropy
on the BTC is very large, as shown in Fig.\ \ref{fig:Jarvis-PSF} -- a
large fraction of the exposures has stellar images with ellipticities
higher than 10\%. Obviously, this renders the image analysis and the
correction for PSF effects challenging. As shown on the right-hand
part of Fig.\ \ref{fig:Jarvis-PSF}, this challenge is indeed met. This
fact is very nicely illustrated in Fig.\ \ref{fig:Jarvis-gscorr},
where the corrected stellar ellipticities are shown as a function of
the PSF anisotropy; in essence, the correction reduces the PSF
anisotropy by nearly a factor of 300!

\begin{figure}
\bmi{6}
\includegraphics[width=5.7cm]{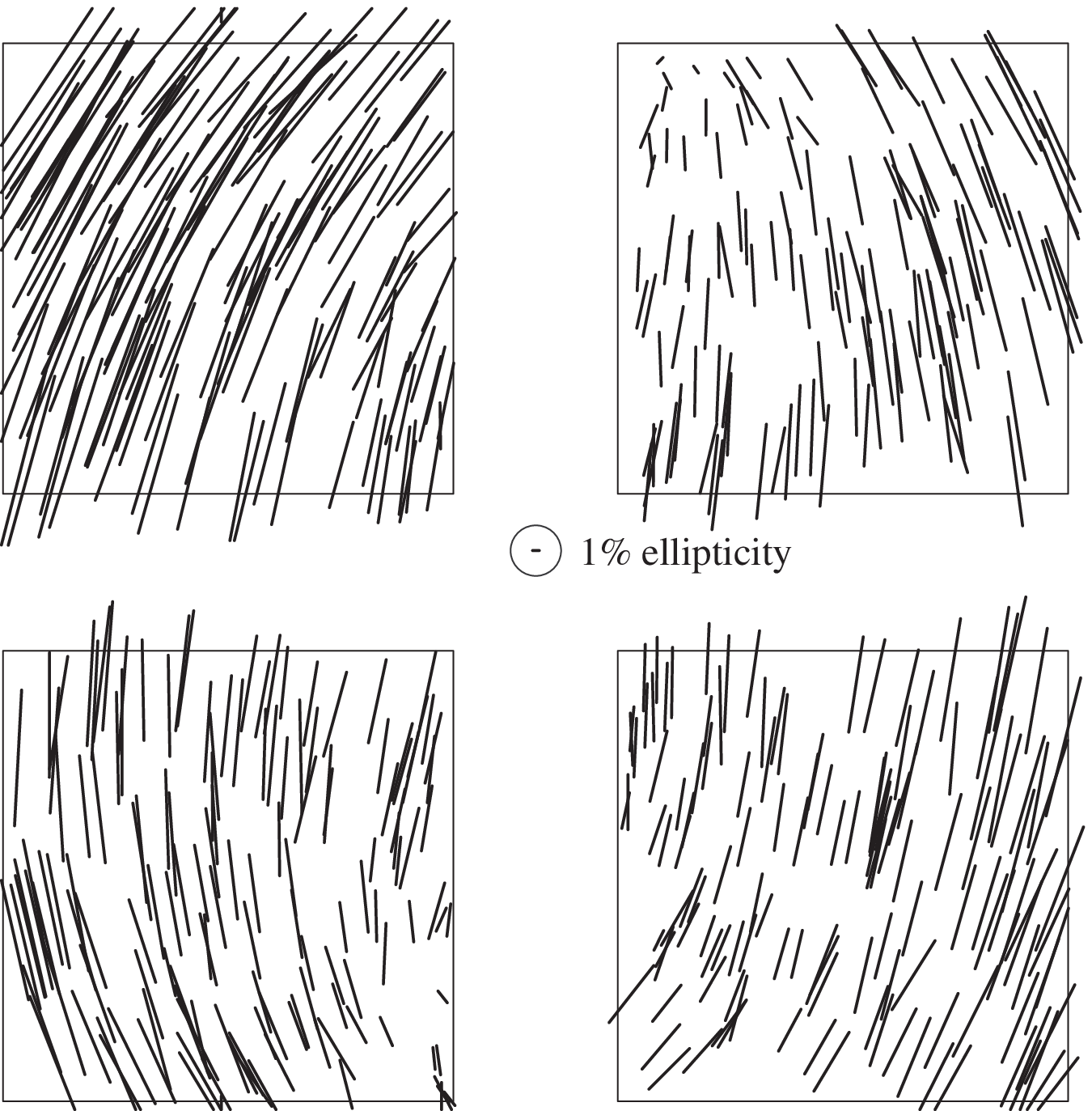}
\emi
\bmi{5.7}
\includegraphics[width=5.7cm]{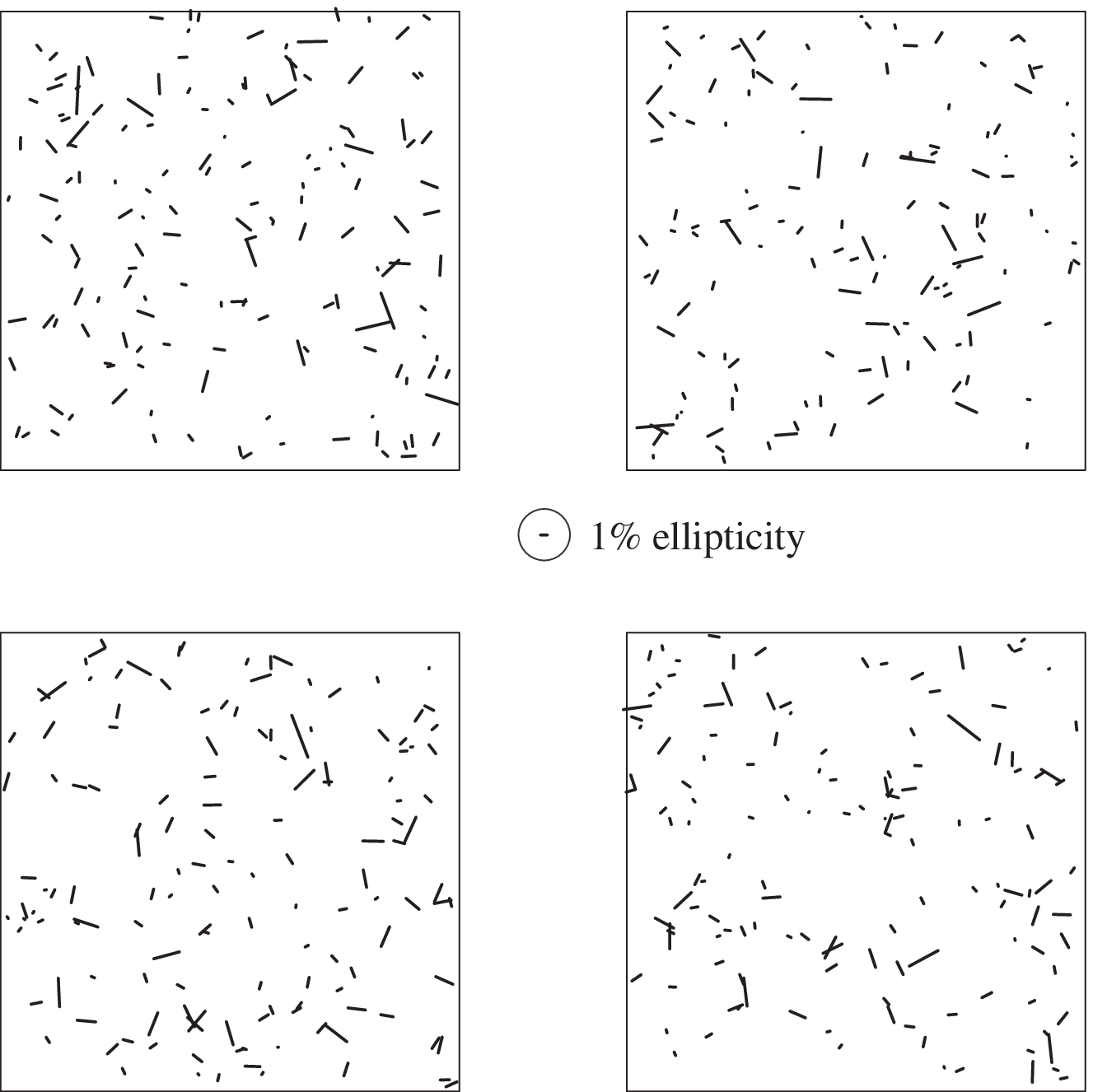}
\emi
\caption{On the left-hand side, the raw ellipticities of stars are shown for
  the four CCDs of the BTC instrument; for reference, a 1\% ellipticity is
  indicated. After correcting for the PSF anisotropy, the remaining stellar
  ellipticities (shown on the right) are of order 1--2\%, and essentially
  uncorrelated with position on the chip, i.e., they are compatible with
  measurement noise (from Jarvis et al.\ 2003)
}
\flabel{Jarvis-PSF}
\end{figure}
 
The third point to notice is that the image analysis for this survey
has not been carried out with IMCAT (as for most of the other
surveys), but by a different image analysis method described in
Bernstein \& Jarvis (2002). In this respect, this survey is
independent of all the others described in this section; it is
important to have more than one image analysis tool to check potential
systematics of either one.

\begin{figure}
\includegraphics[width=11.7cm]{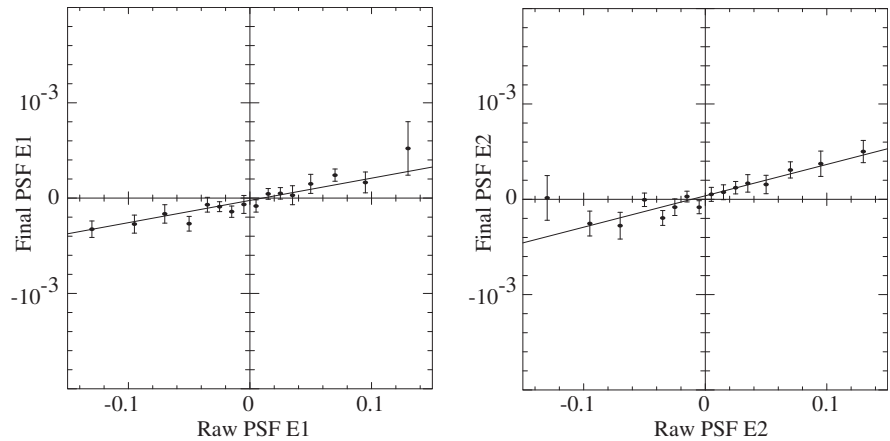}
\caption{These two plots show the two components of the stellar ellipticities
  as measured on the data ($x$-axis) and after correction, from the
Jarvis et al.\ (2003) survey. The slope of the
  straight line is about 1/300, meaning that the strong PSF anisotropy can be
  corrected for up to this very small residual. The final PSF anisotropy is
  well below $5\times 10^{-4}$. This figure, together with Fig.\
  \ref{fig:Jarvis-PSF}, demonstrates how well the procedures for PSF
  corrections work (from Jarvis et al.\ 2003)}
\flabel{Jarvis-gscorr}
\end{figure}

One of the amazing results from the CTIO cosmic shear survey is that the shear
dispersion can be measured with about a $3\sigma$ significance on each of the
12 fields. Hence, this provides a shear dispersion measurement on scales
larger than 1\ degree (the radius of a circle with area of the mean area of
the 12 fields of $\sim 6.2\,{\rm deg}^2$); the shear dispersion on these
scales is $\ave{|\bar\gamma|^2}=0.0012\pm 0.0003$. 

\subsubsection{Special surveys.}
There are a number of cosmic shear surveys which cover a much smaller
total area than the ones mentioned above, and are thus not competitive
in terms of statistical accuracy, but which have some special
properties which give them an important complementary role. One
example are surveys carried out with the Hubble Space Telescope. Since
for them the PSF is much smaller than for ground-based observations,
PSF corrections in measuring galaxy ellipticities are expected to be
correspondingly smaller. The drawback of HST observations is that its
cameras, at least before the installment of the ACS, have a small
field-of-view, less than $1\,{\rm arcmin}^2$ for the STIS CCD, and
about $5\, {\rm arcmin}^2$ for WFPC2. This implies that the total area
covered by HST surveys are smaller than those achievable from the
ground, and that the number of stars per field are very small, so that
PSF measurements are typically not possible on those frames which are
used for a cosmic shear analysis. Hence, the PSF needs to be measured
on different frames, e.g., taken on star clusters, and one needs to
assume (this assumption can be tested, of course) that the PSF is
fairly stable in time. In fact, this is not really true, as the
telescopes moves in and out the Earth's shadow every orbit, thereby
changing its temperature and thus changing its length (an effect
called breathing).  A further potential problem of HST observations is
that the WFPC2 has a pixel scale of $0\arcsecf 1$ and thus
substantially undersamples the PSF; this is likely to be a serious
problem for very faint objects whose size is not much larger than the
PSF size.

Cosmic shear surveys from two instruments onboard HST have been
reported in the literature so far. One of the surveys uses archival
data from the Medium Deep Survey, a mostly parallel survey carried out
with the WFPC2. Refregier et al.\ (2002) used 271 WFPC2
pointings observed in the I-band, selected such that each of them is
separated from the others by at least $10'$ to have statistically
independent fields. They detected a shear dispersion on the scale of
the WFC-chips (which is equivalent to a scale $\theta\sim 0\arcminf
72$) of $\ave{|\bar\gamma|^2} \sim 3.5\times 10^{-4}$, which is a
3.8$\sigma$ detection. The measurement accuracy is lower than that,
owing to cosmic variance and uncertainties in the redshift
distribution of the sources.  
H\"ammerle et al.\ (2002) used archival
parallel data taken with STIS; from the 121 fields which are deep
enough, have multiple exposures, and are at sufficiently high galactic
latitude, they obtained a shear dispersion of
$\ave{|\bar\gamma|^2}\sim 15\times 10^{-4}$ on an effective scale of
$\sim 30''$, a mere 1.5$\sigma$ detection. This low significance is
due to the small total area covered by this survey. On the other hand,
since the pixel scale of STIS is half of that of WFPC2, the
undersampling problem is much less in this case.
A larger set of STIS parallel observations were analyzed with
respect to cosmic shear by Rhodes et al.\ (2004) and Miralles et al.\
(2003). Whereas Rhodes et al.\ obtained a significant ($\sim
5\sigma$) detection on an angular scale of $\sim 30''$, 
Miralles et al.\ concluded that the degradation of the STIS CCD in
orbit regarding the charge transfer efficiency prevents a solid
measurement of weak lensing. The discrepancies between these two
works, which are based to a large degree on the same data set, is
unclear at present. Personally I consider this discrepancy as a
warning sign that weak lensing measurement based on small
fields-of-view, and correspondingly too few stars to control the PSF
on the science exposures, need to be regarded with extreme caution. 

The new ACS onboard HST offers better prospects for cosmic shear
measurements, since it has a substantially larger field-of-view. A
first result was derived by Schrabback (2004), again based on parallel
data. He found that the PSF is not stable in time, but that the
anisotropy pattern changes among only a few characteristic
patterns. He used those as templates, and the (typically a dozen)
stars in the science frames to select a linear combination of these
templates for the PSF correction of individual frames, thereby
obtaining a solid detection of cosmic shear from the early ACS data. 

A further survey that should be mentioned here is the one conducted on
COMBO17 fields (Brown et al.\ 2003). COMBO17 is a one square degree
survey, split over four fields, taken with the WFI at the ESO/MPG 2.2m
telescope on La Silla, in 5 broad-band and 12 medium-band filters. In
essence, therefore, this multi-band survey produces low-resolution
spectra of the objects and thus permits to determine very accurate
photometric redshifts of the galaxies taken for the shear
analysis. Therefore, for the analysis of Brown et al., the redshift
distribution of the galaxies is assumed to be very well known and not
a source of uncertainty in translating the cosmic shear measurement
into a constraint on cosmological parameters. We shall return to this
aspect in Sect.\ts\ref{sc:WL-7.6}. The data set was reanalyzed by Heymans et
al.\ (2004) where special care has been taken to identify and remove the
signal coming from intrinsic alignment of galaxy shapes.

\subsection{\llabel{WL-7.4}Detection of B-modes}
The recent cosmic shear surveys have measured the aperture mass
dispersion $\ave{M_{\rm ap}^2(\theta)}$, as well as its counterpart
$\ave{M_\perp^2(\theta)}$ for the B-modes (see Sect.\
\ref{sc:WL-6.5}). These aperture measures are obtained in terms of the
directly measured shear correlation functions, using the relations
(\ref{eq:N29}). As an example, we show in Fig.\ \ref{fig:RCS-aperture}
the aperture measures as obtained from the Red Cluster Sequence survey
(Hoekstra et al.\ \cite{hoek595}). A significant measurement of
$\ave{M_{\rm ap}^2(\theta)}$ is obtained over quite a range of angular
scales, with a peak around a few arcminutes, as predicted from CDM
power spectra (see Fig.\ \ref{fig:map+gammasq}). In addition to that,
however, a significant detection of $\ave{M_\perp^2(\theta)}$
signifies the presence of B-modes. As discussed in Sect.\
\ref{sc:WL-6.5}, those cannot be due to cosmic shear. The only
plausible explanation for them, apart from systematics in the
observations and data analysis, is an intrinsic alignment of
galaxies. If this is the cause of the B-modes, then one would expect
that the relative contribution of the B-mode signal decreases as
higher-redshift galaxies are used for the shear measurement. In fact,
this expectation is satisfied, as shown in Fig.\
\ref{fig:RCS-aperture}, where the galaxy sample is split into a bright
and faint part, and the relative amplitude of the B-mode signal is
smaller for the fainter (and thus presumably more distant) sample.

\begin{figure}
\bc
\includegraphics[width=10cm]{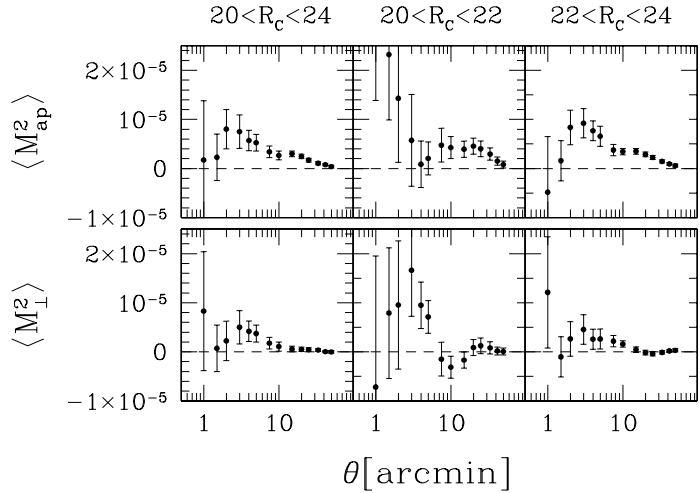}
\ec
\caption{The aperture mass dispersion $\ave{M_{\rm ap}^2(\theta)}$
(top panels) and the cross aperture dispersion
$\ave{M_\perp^2(\theta)}$ (bottom panels) from the RCS survey
(Hoekstra et al.\ \cite{hoek595}). In the left panels, all galaxies
with apparent magnitude $20\le R_C\le 24$ are used, the middle and
right panels show the same statistics for the brighter and fainter
subsamples of background galaxies, respectively. Error bars in the
former are larger, owing to the smaller number of bright galaxies}
\flabel{RCS-aperture}
\end{figure}

Similar detections of a B-mode signal have been obtained by the other
surveys.  For example, van Waerbeke et al.\ (2001) reported a
significant B-mode signal on angular scales of a few arcminutes. In
the reanalysis of the VIRMOS-DESCART data, van Waerbeke et al.\ (2002)
reported that the B-mode on these scales was caused by the polynomial
PSF anisotropy fit: the third-order function (fitted for each chip
individually) has its largest amplitude near the boundary of the chips
and is least well constrained there, unless one finds stars close to
these edges. If a second-order polynomial fit is used, the B-modes on
a few arcminute scales disappear. Van Waerbeke et al.\ (2002)
calculate the aperture statistics from the uncorrected stellar
ellipticities in their survey and found that the `E- and B-modes' of
the PSF anisotropy have very similar amplitude and shape (as a
function of $\theta$). This similarity is unlikely to change in the
course of the PSF correction procedure. Thus, they argue, that if the
B-mode is due to systematics in the data analysis, a systematic error
of very similar amplitude will also affect the E-mode. Jarvis et al.\
(2003) found a significant B-mode signal on angular scales below $\sim
30'$; hence, despite their detection of an E-mode signal over a large
range of angular scales $1'\lesssim \theta\lesssim 100'$, one suspects
that part of this signal might be due to non-lensing effects.

Given our lack of understanding about the origin of the B-mode signal,
and the associated likelihood that any effect causing a B-mode signal
also contributes a non-lensing part to the E-mode signal, one needs a
prescription on how to use the detected E-mode signal for a
cosmological analysis. Depending on what one believes the B-modes are
due to, this prescription varies. For example, if the B-mode is due to
a residual systematic, one would add its signal in quadrature to the
error bars of the E-mode signal, as done in van Waerbeke et al.\
(2002). On the other hand, if the B-mode signal is due to intrinsic
alignments of galaxies, as is at least suggested for the RCS survey
from Fig.\ts\ref{fig:RCS-aperture} owing to its dependence on galaxy
magnitudes, then it could be more reasonable to subtract the B-mode
signal from the E-mode signal, if one assumes that intrinsic
alignments produce similar amplitudes of both modes [which is far from
clear, however; Mackey et al.\ (2002) find that the E-mode signal from
intrinsic alignments is expected to be $\sim 3.5$ times higher than
the corresponding B-mode signal].

Owing to the small size of the fields observed with the early HST instruments,
no 
E/B-mode decomposition can be carried out from these surveys -- the largest
size of these fields is smaller than the angular scale at which the aperture
mass dispersion is expected to peak (see Fig.\ \ref{fig:map+gammasq}). 
However, future cosmic shear studies carried out with ACS images will most
likely be able to detect, or set upper bounds on the presence of B-modes.

In fact, it is most likely that (most of) the B-mode signal seen in
the cosmic shear surveys is due to remaining systematics. Hoekstra
(2004) investigated the PSF anisotropy of the CFH12k camera using
fields with a high number density of stars. Randomly selecting about
100 stars per CCD, which is the typical number observed in high
galactic latitute fields, he fitted a second-order polynomial to these
stars representing the PSF anisotropy. Correcting with this model all
the stars in the field, the remaining stellar ellipticities carry
substantial E- and B-mode signals, essentially on all angular scales,
but peaking at about the size of a CCD.  A substantially smaller
residual is obtained if the ellipticities of stars in one of the
fields is corrected by a more detailed model of the PSF anisotropy as
measured from a different field; this improvement indicates that the
PSF anisotropy pattern in the data set used by Hoekstra is fairly
stable between different exposures. This, however, is not necessarily
the case in other datasets. Nevertheless, if one assumes that the PSF
anisotropy is a superposition of two effects, one from the properties
of the telescope and instrument itself, the other from the specific
observation procedure (e.g., tracking, wind shake, etc.), and further
assuming that the latter one affects mainly the large-scale properties
of the anisotropy pattern, then a superposition of a PSF model
(obtained from a dense stellar field and describing the small-scale
properties of the anisotropy pattern) plus a low-order polynomial can
be a better representation of the PSF anisotropy. This indeed was
verified in the tests made by Hoekstra (2004).  In their reanalysis of
the VIRMOS-DESCART survey, van Waerbeke et al. (2004) have fitted the
PSF anisotropy with a rational function, instead of a polynomial. This
functional form was suggested by the study of Hoekstra (2004). When
correcting the galaxy ellipticities with this new PSF model,
essentially no more B-modes in the VIRMOS-DESCART survey are
detected. Further studies on PSF anisotropy corrections need to be
conducted; possibly the optimal way of dealing with them will be
instrument-specific.

\subsection{\llabel{WL-7.5} Cosmological constraints}
The measured cosmic shear signal can be translated into constraints on
cosmological parameters, by comparing the measurements with
theoretical predictions. In Sect.\ \ref{sc:WL-6.4} we have outlined
how such a comparison can be made; there, we have concentrated on the
shear correlation functions as the primary observables. However, the
detection of significant B-modes in the shear field makes the aperture
measures the `better' statistics to compare with
predictions. They can be calculated from the shear correlation
functions, as shown in (\ref{eq:N29}).  Calculating a likelihood
function from the aperture mass dispersion proceeds in the same way as
outlined in Sect.\ \ref{sc:WL-6.4} for the correlation functions.

We have argued in Sect.\ \ref{sc:WL-6.3} that $\ave{M_{\rm
ap}^2(\theta)}$ provides very localized information about the power
spectrum $P_\kappa(\ell)$ and is thus a very useful
statistic. One therefore might expect that the aperture mass
dispersion as calculated from the shear correlation functions contains
essentially all the second-order statistical information of the
survey. This is not true, however; one needs to recall that the shear
correlation function $\xi_+$ is a low-pass filter of the power
spectrum, and thus contains information of $P_\kappa$ on angular
scales larger than the survey size. This information is no longer
contained in the aperture mass dispersion, owing to its localized
associated filter. Therefore, in order to keep this long-range
information in the comparison with theoretical predictions, it is
useful to complement the estimates of $\ave{M_{\rm ap}^2(\theta)}$
with either the shear dispersion, or the correlation function $\xi_+$,
at a scale which is not much smaller than the largest scale at which
$\ave{M_{\rm ap}^2(\theta)}$ is measured. Note, however, that this
step implicitly assumes that on these large angular scales, the shear
signal is essentially free of B-mode contributions. If this assumption
is not true, and cannot be justified from the survey data, then this
additional constraint should probably be dropped.

\begin{figure}
\bmi{6.2}
\includegraphics[width=6.2cm]{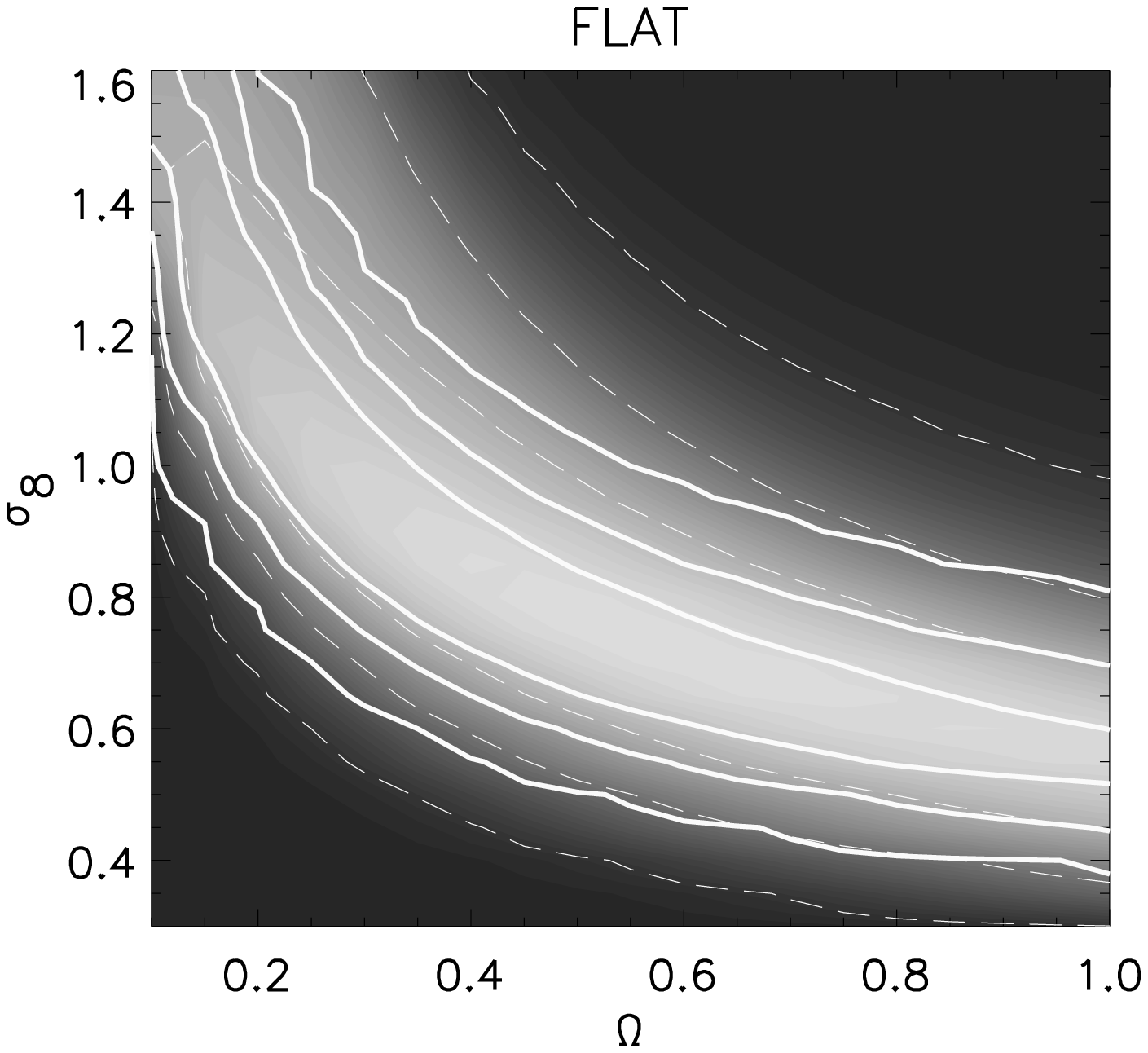}
\emi
\bmi{5.5}
\includegraphics[width=5.5cm]{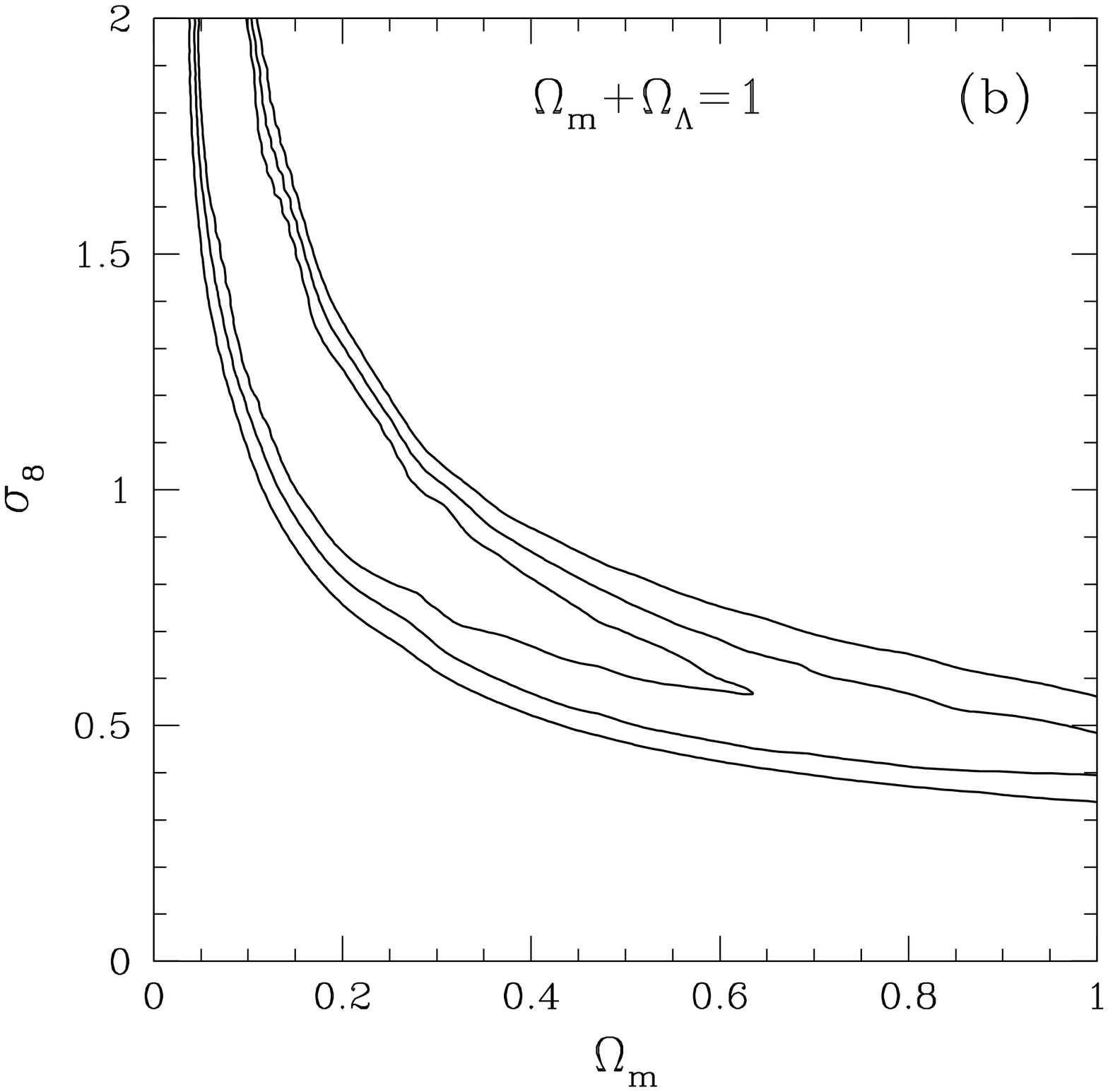}
\emi
\caption{Constraints on $\Omega_{\rm m}$ and $\sigma_8$ from two cosmic shear
  surveys. Left: The VIRMOS-DESCART survey (van Waerbeke et al.\ 2002). The
  grey-scale and dashed contours show the 68\%, 95\% and 99.9\% confidence
  regions with a marginalization over the range $\Gamma_{\rm
    spect}\in[0.05,0.7]$, and mean galaxy redshift in the range $\bar z_{\rm
    s}\in[0.50,1.34]$, whereas the solid contours show the same confidence
  regions with the stronger priors $\Gamma_{\rm spect}\in [0.1,0.4]$ and $\bar
  z_{\rm s}\in [0.8, 1.1]$. Right: The RCS survey (Hoekstra et al.\ 2002),
  showing the 1, 2, and 3$\sigma$ confidence regions for a prior $\Gamma_{\rm
    spect}\in [0.05,0.5]$ and mean redshift $\bar z_{\rm s}\in [0.54,0.66] $.
  In both cases, a flat Universe has been assumed} 
\flabel{Omega-sigma8}
\end{figure}

The various constraints on parameters that have been derived from the cosmic
shear surveys differ in the amount of prior information that has been used. As
an example, we consider the analysis of van Waerbeke et al.\ (2002). These
authors have considered a model with four free parameters: $\Omega_{\rm m}$,
the normalization $\sigma_8$, the shape parameter $\Gamma_{\rm spect}$ and the
characteristic redshift $z_{\rm s}$ (or, equivalently, mean redshift $\bar
z_{\rm s}$) of their galaxy sample, assuming a flat Universe, i.e.,
$\Omega_\Lambda=1-\Omega_{\rm m}$. They have used a flat prior for
$\Gamma_{\rm spect}$ and $\bar z_{\rm s}$ in a fairly wide interval over which
they marginalized the likelihood function (see Fig.\
\ref{fig:Omega-sigma8}). Depending on the width of these intervals, the
confidence regions are more or less wide. It should be noted that the
confidence contours close if $\Gamma_{\rm spect}$ and $\bar z_{\rm s}$ are
assumed to be known (see van Waerbeke et al.\ 2001), but when these two
parameters are kept free, $\Omega_{\rm m}$ and $\sigma_8$ are degenerate.

The right panel of Fig.\ \ref{fig:Omega-sigma8} shows the
corresponding constraints as obtained from the RCS survey. Since this
survey is shallower and only extends to magnitudes where spectroscopic
surveys provide information on their redshift distribution, the range
of $\bar z_{\rm s}$ over which the likelihood is marginalized is
smaller than for the VIRMOS-DESCART survey.  Correspondingly, the
confidence region is slightly smaller in the case. Even smaller
confidence regions are obtained if external information is used:
Hoekstra et al.\ (\cite{hoek595}) considered Gaussian priors with
$\Omega_{\rm m}+\Omega_\Lambda=1.02\pm 0.06$, as follows from pre-WMAP
CMB results, $\Gamma_{\rm spect}=0.21 \pm 0.03$, as follows from the
2dF galaxy redshift survey, and $\bar z_{\rm s}=0.59\pm 0.02$, for
which the width of the valley of maximum likelihood narrows
considerably. Jarvis et al.\ (2003) used for their estimate of
cosmological parameters the aperture mass dispersion at three angular
scales plus the shear dispersion at $\theta=100'$, and they considered
alternatively the E-mode signal, and the E-mode signal $\pm$ the
B-mode signal, to arrive at constraints on the $\Omega_{\rm
m}$--$\sigma_8$ parameter plane. Since the CTIO survey samples a
larger angular scale than the other surveys (data at small angular
scales are discarded owing to the large B-mode signal there), the
results are much less sensitive to $\Gamma_{\rm spect}$; furthermore,
for the same reason the Jarvis et al.\ results are much less sensitive
to the fit of the non-linear power spectrum according to Peacock \&
Dodds (1996) which van Waerbeke et al\ (2002) found to be not accurate
enough for some cosmological models. In fact, if instead of the
Peacock \& Dodds fitting formula, the fit by Smith et al\ (2003) is
used to describe the non-linear power spectrum, the resulting best
estimate of $\sigma_8$ is decreased by 8\% for the RCS survey (as
quoted in Jarvis et al.\ 2003).

For the RCS and the CTIO surveys, the covariance matrix was obtained from
field-to-field variations, i.e., ${\rm
  Cov}_{ij}=\ave{(d_i-\mu_i)(d_j-\mu_j)}$, where $\mu_i$ is the mean of the
observable $d_i$ (e.g., the aperture mass dispersion at a specific angular
scale) over the independent patches of the survey, and angular brackets denote
the average over all independent patches. The estimate of the covariance
matrix for the VIRMOS-DESCART survey is slightly different, as it has only
four independent patches.

To summarize the results from these surveys, each of them found that a
combination of parameters of the form $\sigma_8\Omega_{\rm m}^\alpha$ is
determined best from the data, with $\alpha\sim 0.55$, where the exact
value of $\alpha$ depends on the survey depth. If we consider the specific
case of $\Omega_{\rm m}=0.3$ which is close to the concordance value that was
recently confirmed by WMAP, then the VIRMOS-DESCART survey yields
$\sigma_8=0.94\pm0.12$, the RCS survey has 
$\sigma_8=0.91^{+0.05}_{-0.12}$, which improves to
$\sigma_8=0.86^{+0.04}_{-0.05}$ if the stronger (Gaussian) priors mentioned
above are used, and the CTIO survey yields $\sigma_8=0.71^{+0.12}_{-0.16}$,
here as 2$\sigma$ limits. Whereas these results are marginally in mutual
agreement, the CTIO value for $\sigma_8$ is lower than the other two. The
higher values are also supported by results from the WFPC2 survey by Refregier
et al.\ (2002), who find $\sigma_8=0.94\pm 0.17$, Bacon et al.\ (2003) with
$\sigma_8=0.97\pm0.13$, and the earlier surveys discussed in Sect.\
\ref{sc:WL-7.1}. The only survey supporting the low value of the CTIO survey
is COMBO17 (Brown et al.\ 2002; see also the reanalysis of this dataset by
Heymans et al.\ 2004). Most likely, these remaining discrepancies
will be clarified in the near future; see discussion below. It should also be
noted that at least for some of the surveys, a large part of the uncertainty
comes from the unknown redshift distribution of the galaxies; this situation
will most likely improve, as efficient spectrographs with large multiplex
capability become available at 10m-class telescopes, which will in the near
future deliver large galaxy redshift surveys at very faint magnitudes. Those
can be used to much better constrain the redshift distribution of the source
galaxies in cosmic shear surveys.

\subsection{\llabel{WL-7.6}3-D lensing}
As mentioned several times before, using individual source redshift
information, as will become available in future multi-color wide-field
surveys, can improve the cosmological constraints obtained from weak
lensing.  In this section we shall therefore summarize some of the
work that has been published on this so-called 3-D lensing.

\subsubsection{Three-dimensional matter distribution.}
Provided the redshifts of individual source galaxies are known (or estimated
from their multiple colors), one can derive the 3-D matter distribution, not
only its projection. The principle of this method can be most easily
illustrated in the case of a flat Universe, for which the surface mass density
$\kappa(\vc\theta,w)$ for sources at comoving distance $w$ becomes -- see
(\ref{eq:5.11}) 
\be
\kappa(\vc\theta,w)={3 H_0^2\Omega_{\rm m}\over 2 c^2}
\int_0^w \d w'\;{w' (w-w')\over w}\,{\delta(w'\vc\theta,w')\over a(w')}\;.
\elabel{3-D-1}
\ee
Multiplying this expression by $w$ and differentiating twice yields
\[
{\d^2\over \d w^2}\rund{w\,\kappa(\vc\theta,w)}=
{3 H_0^2\Omega_{\rm m}\over 2 c^2}\,
{w\over a(w)}\;\delta(w\vc\theta,w)\;,
\]
which therefore allows one to obtain the three-dimensional density contrast
$\delta$ in terms of the surface mass densities $\kappa$ at different source
redshifts. As we have seen in Sect.\ts\ref{sc:WL-5}, there are several methods
how to obtain the surface mass density from the observed shear. To illustrate
the 3-D method, we use the finite-field reconstruction in the form of
(\ref{eq:H-greens}), for which one finds
\be
\delta(w\vc\theta,w)={2 c^2\over 3 H_0^2\Omega_{\rm m}}\,{a(w)\over w}
\int\d^2 \theta'\;\vc H(\vc\theta;\vc\theta')\cdot
{d^2\over \d w^2}\eck{w\,\vc u_\gamma(\vc\theta',w)}\;.
\elabel{3-D-2}
\ee
Taylor (2001) derived the foregoing result, but concentrated on the 3-D
gravitational potential instead of the mass distribution, and Bacon \& Taylor
(2003) and Hu \& Keeton (2003) discussed practical implementations of this
relation. First to note is the notorious mass-sheet degeneracy, which in the
present context implies that one can add an arbitrary function of $w$ to the
reconstructed density contrast $\delta$. This cannot be avoided, but if the
data field is sufficiently large, so that averaged over it, the density
contrast is expected to vanish, this becomes a lesser practical problem. For
such large data fields, the above mass reconstruction can be substituted in
favour of the simpler original Kaiser \& Squires (1993) method. Still
more freedom is present in the reconstruction of the gravitational
potential. The second problem is one of smoothing: owing to the noisiness of
the observed shear field, the $w$-differentiation (as well as the
$\theta$-differentiation present in the construction of the vector field $\vc
u_\gamma$) needs to be carried out on the smoothed shear field. A
discretization of the observed shear field, as also suggested by the finite
accuracy of photometric redshifts, can be optimized with respect to this
smoothing (Hu \& Keeton 2003). 

A first application of this methods was presented in Taylor et al.\ (2004) on
one of the COMBO17 fields which contains the supercluster A901/902.  The
clusters present clearly show up also in the 3-D mass map, as well as a
massive structure behind the cluster A902 at higher redshift. Already earlier,
Wittman et al.\ (2001, 2003) estimated the redshifts of clusters found in
their deep blank-field data by studying the dependence of the weak lensing
signal on the estimated source redshifts, and subsequent spectroscopy showed
that these estimates were fairly accurate.

\subsubsection{Power spectrum estimates.}
A redshift-dependent shear field can also be used to improve on the
cosmological constraints obtained from cosmic shear. Hu (1999) has pointed out
that even crude information on the source redshifts can strongly
reduce the uncertainties of cosmological parameters. In fact, the
3-D power spectrum can be constructed from redshift-dependent shear data (see,
e.g., Heavens 2003, Hu 2002, and references therein). 
For illustration purposes, one can use the $\kappa$ power spectrum for
sources at fixed 
comoving distance $w$, which reads in a flat Universe -- see (\ref{eq:6.25})
\be
P_\kappa(\ell,w)={9 H_0^4\Omega_{\rm m}^2\over 4 c^4}
\int_0^w\d w'\;{(w-w')^2\over w^2\,a^2(w')}\,
P_\delta\rund{{\ell\over w'},w'}\;.
\elabel{3-D-3}
\ee
Differentiating $w^2 P_\kappa$ three times w.r.t. $w$ then yields (Bacon et
al.\ 2004) 
\be
P_\delta(k,w)={2 c^4\over 9 H_0^4\Omega_{\rm m}^2}\,a^2(w)\,
{\d^3 \over \d w^3}\eck{w^2\,P_\kappa(w k,w)}\;.
\elabel{3-D-4}
\ee
In this way, one could obtain the three-dimensional power spectrum of
the matter. However, this method is essentially useless, since it is
both very noisy (due to the third-order derivatives) and throws away
most of the information contained in the shear field, as it makes use
only of shear correlations of galaxies having the same redshift, and
not of all the pairs at different distances. A much better approach to
construct the three-dimensional power spectrum is given, e.g., by Pen
et al.\ (2003).

In my view, the best use of three-dimensional data is to construct the
shear correlators $\xi_\pm(\theta;z_1,z_2)$, as they contain all
second-order statistical information in the data and at the same time
allow the identification and removal of a signal from intrinsic shape
correlations of galaxies (King \& Schneider 2003). From these
correlation functions, one can calculate a $\chi^2$ function as in
(\ref{eq:chisqxi}) and minimize it w.r.t. the wanted parameters. One
problem of this approach is the large size of the covariance matrix,
which now has six arguments (two angular separations and four
redshifts). However, as shown in Simon et al.\ (2004), it can be
calculated fairly efficiently, provided one assumes that the
fourth-order correlations factorize into products of two-point
correlators, i.e., Gaussian fields (if this assumption is dropped, the
covariance must be calculated from cosmological N-body simulations).

Bacon et al.\ (2004) used the COMBO17 data to derive the shape of the power
spectrum, using the redshift dependent shear correlations. They parameterize
the power spectrum in the form $P(k,z)\propto A k^\alpha {\rm e}^{-s z}$, so
that it is described by an amplitude $A$, a local slope $\alpha$ and a growth
parameter $s$ which describes how the amplitude of the power spectrum declines
towards higher redshifts. In fact, the slope $\alpha=-1.2$ was fixed to the
approximate value in $\Lambda$CDM models over the relevant range of spatial
scales and redshifts probed by the COMBO17 data (since the data used cover
only 1/2\ deg$^2$, reducing the number of free parameters by fixing $\alpha$
is useful). The evolution of the power spectrum is found with high
significance in the data. Furthermore, the authors show that the use of
redshift information improves the accuracy in the determination of $\sigma_8$
by a factor of two compared to the 2-D cosmic shear analysis of the same data
(Brown et al.\ 2003). 

The main application of future multi-waveband cosmic shear surveys will be to
derive constraints on the equation of state of dark energy, as besides lensing
there are only a few methods available to probe it, most noticibly the
magnitude-redshift relation of SN\ts Ia. Since dark energy starts to dominate
the expansion of the Universe only at relatively low redshifts, little
information about its properties is obtainable from the CMB anisotropies
alone. For that reason, quite a number of workers have considered the
constraints on the dark energy equation of state that can be derived from
future cosmic shear surveys (e.g., Huterer 2002; Hu \cite{HuDEDM};
Munshi \& Wang 2003; Hu \& Jain 2003;
Abazajian \& Dodelson 2003; Benabed \& van Waerbeke \cite{BenaLvW}; Song \&
Knox \cite{SongKnox}). The results of these are very encouraging; the
sensitivity on the dark energy properties is due to its influence on structure
growth. With (photometric) redshift information on the source galaxies, the
evolution of the dark matter distribution can be studied by weak lensing, as
shown above. Van Waerbeke \& Mellier (2003) have compared the expected
accuracy of the cosmic shear result from the ongoing CFHT Legacy Survey with
the variation of various dark energy models and shown that the CFHTLS will be
able to discriminate between some of these models, with even much better
prospects from future space-based wide-field imaging surveys (e.g., Hu \& Jain
2003).

\subsection{\llabel{WL-7.7}Discussion}
The previous sections have shown that cosmic shear research has matured;
several groups have successfully presented their
results, which is important in view of the fact that the effects one wants to
observe are small, influenced by various effects, and therefore, independent
results from different instruments, groups, and data analysis techniques are
essential in this research. We have also seen that the results from the
various groups tend to agree with each other, with a few very interesting
discrepancies remaining whose resolution will most likely teach us even more
about the accuracies of data analysis procedures. 

\subsubsection{Lessons for cosmology.}
A natural question to ask is, what has cosmic shear taught us so far
about cosmology? The most important constraint coming from the
available cosmic shear results is that on the normalization
$\sigma_8$, for which only few other accurate methods are
available. We have seen that cosmic shear prefers a value of
$\sigma_8 \approx 0.8 - 0.9$, which is slightly larger than current
estimates from the abundance of clusters, but very much in agreement
with the measurement of WMAP. The estimate from the cluster abundance
is, however, not without difficulties, since it involves several
scaling relations which need to be accurately calibrated; hence,
different authors arrive at different values for $\sigma_8$ (see,
e.g., Pierpaoli, Scott \& White 2001; Seljak 2002; Schuecker et al.\
2003). The accuracy with which $\sigma_8$ is determined from CMB data
alone is comparable to that of cosmic shear estimates; as shown in
Spergel et al.\ (2003), more accurate values of $\sigma_8$ are
obtained only if the CMB measurements are combined with measurements
on smaller spatial scales, such as from galaxy redshift surveys and
the Lyman alpha forest statistics. Thus, the $\sigma_8$-determination
from cosmic shear is certainly competitive with other measurements.
Arguably, cosmic shear sticks out in this set of smaller-scale
constraints due to the fewer physical assumptions needed for its
interpretation.

But more importantly, it provides a fully independent method to measure
cosmological parameters. Hence, at present the largest role of the cosmic
shear results is that it provides an independent approach to determining these
parameters; agreement with those obtained from the CMB, galaxy redshift
surveys and other methods are thus foremost of interest in that they provide
additional evidence for the self-consistency of our cosmological model
which, taken at face value, is a pretty implausible one: we should always keep
in mind that we are claiming that our Universe consists of 4.5\% normal
(baryonic) matter, with the rest being shared with stuff that we have given
names to (`dark matter', `dark energy'), but are pretty ignorant about what
that actually is. Insofar, cosmic shear plays an essential role in shaping our
cosmological view, and has become one of the pillars on which our standard
model rests. 

\subsubsection{Agreement, or discrepancies?} 
How to clarify the remaining discrepancies that were mentioned before
-- what are they due to? One needs to step back for a second and be
amazed that these results are in fact so well in agreement as they
are, given all the technical problems a cosmic shear survey has to
face (see Sect.\ts\ref{sc:WL-3}). Nevertheless, more investigations
concerning the accuracy of the results need to be carried out, e.g.,
to study the influence of the different schemes for PSF corrections on
the final results. For this reason, it would be very valuable if the
same data set is analyzed by two independent groups and to compare the
results in detail. Such comparative studies may be a prerequisite for
the future when much larger surveys will turn cosmic shear into a tool
for precision cosmology.

\subsubsection{Joint constraints from CMB anisotropies and cosmic shear.}
As mentioned before, the full power of the CMB anisotropy measurements
is achieved when these results are combined with constraints on
smaller spatial scales. The tightest constraints from WMAP are
obtained when it is combined with results from galaxy redshift surveys
and the statistics of the Ly$\alpha$ forest absorption lines (Spergel
et al.\ 2003). Instead of the latter, one can instead use results from
cosmic shear, as it provides a cleaner probe of the statistical
properties of the matter distribution in the Universe. As was pointed
out before (e.g., Hu \& Tegmark 1999; see Fig.\ \ref{fig:hu+tegmark}),
the combination of CMB measurements with cosmic shear results is
particularly powerful to break degeneracies that are left from using
the former alone. Contaldi et al.\ (2003) used the CMB anisotropy
results from WMAP (Bennett et al.\ 2003), supplemented by anisotropy
measurements on smaller angular scales from ground-based experiments,
and combined them with the cosmic shear aperture mass dispersion from
the RCS survey (Hoekstra et al.\ \cite{hoek595}). As is shown in Fig.\
\ref{fig:lens+CMB}, the constraints in the $\Omega_{\rm
m}$-$\sigma_8$-parameter plane are nearly mutually orthogonal for the
CMB and cosmic shear, so that the combined confidence region is
substantially smaller than each of the individual regions.

\begin{figure}
\bmi{7.0}
\includegraphics[width=6.9cm]{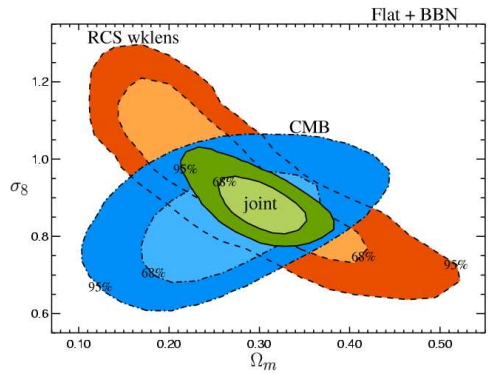}
\emi
\bmi{4.7}
\caption{The confidence region in the $\Omega_{\rm m}$-$\sigma_8$-plane 
  obtained from the two-dimensional marginalized likelihood. Shown are the
  68\% and 95\% confidence regions derived individually from the CMB and the
  RCS cosmic shear survey, as well as those obtained by combining both
  constraints (Contaldi et al.\ 2003) } 
\flabel{lens+CMB}
\emi
\end{figure}

\subsubsection{Wide vs.\ deep surveys.}
In designing future cosmic shear surveys, the survey strategy needs to
decide the effective exposure time. For a given total observing time (the
most important practical constraint), one needs to find a compromise between
depth and area. Several issues need to be considered in this respect:
\bi
\item
The lensing signal increases with redshift, and therefore with
increasing depth of a survey; it should therefore be easier to detect
a lensing signal in deep surveys. Furthermore, by splitting the galaxy
sample into subsamples according to the magnitude (and/or colors), one
can study the dependence of the lensing signal on the mean source
redshift, which is an important probe of the evolution of the matter
power spectrum, and thus of cosmology. If one wants to probe the
(dark) matter distribution at appreciable redshifts ($z\sim 0.5$), one
needs to carry out deep surveys.
\item
A wider survey is more likely to probe the linear part of the power spectrum
which is more securely predicted from cosmological models than the non-linear
part; on the other hand, measurement of the latter, when compared with precise
models (e.g., from numerical simulations), can probe the non-linear
gravitational clustering regime.
\item
Depending on the intrinsic galaxy alignment, one would prefer deeper
surveys, since the relative importance of the intrinsic signal
decreases with increasing survey depth. Very shallow surveys may in
fact be strongly affected by the intrinsic signal (e.g., Heymans \&
Heavens 2003). On the other hand, for precision measurements, as will
become available in the near future, one needs to account for the
intrinsic signal in any case, using redshift information (at least in
a statistical sense), and so shallow surveys lose this potential
disadvantage. In fact, the redshift estimates of shallower surveys are
easier to obtain than for deeper ones.
\item
In this context, one needs to compromise between area and and the
number of fliters in which exposures should be taken. Smaller area
means worse statistics, e.g., larger effects of cosmic variance, but
this has to be balanced against the additional redshift
information. Also, if a fixed observing time is used, one needs to
account for the weather, seeing and sky brightness distribution. One
should then device a strategy that the best seeing periods are used to
obtain images in the filter which is used for shape measurements, and
bright time shall be spent on the longest wavelength bands. 
\item
Fainter galaxies are smaller, and thus more strongly affected by the
point-spread function. One therefore expects that PSF corrections are on
average smaller for a shallow survey than for a deeper one. In addition, the
separation between stars and galaxies is easier for brighter (hence,
larger) objects. 
\ei
The relative weight of these arguments is still to be decided. Whereas some of
the issues could be clarified with theoretical investigations (i.e., in order
to obtain the tightest constraints on cosmological parameters, what is the
optimal choice of area and exposure time, with their product being fixed),
others (like the importance of intrinsic alignments) still remain
unclear. Since big imaging surveys will be conducted with a broad range of
scientific applications in mind, this choice will also depend on those
additional science goals.

\subsubsection{Future surveys.}
We are currently witnessing the installment of square-degree cameras
at some of the best sites, among them Megacam at the CFHT, and
OmegaCAM at the newly built VLT Survey Telescope (the 2.6m VST) on
Paranal (I present here European-biased prospects, as I am most
familiar with these projects).  Weak lensing, and in particular cosmic
shear has been one of the science drivers for these instruments, and
large surveys will be carried out with them.  Already ongoing is the
CFHT Legacy Survey, which will consist of three parts; the most
interesting one in the current context is a $\sim 160\,{\rm deg}^2$
survey with an exposure time of $\sim 1\,{\rm h}$ in each of five
optical filters. This survey will therefore yield a more than ten-fold
increase over the current VIRMOS-DESCART survey, with corresponding
reductions of the statistical and cosmic variance errors on
measurements. The multi-color nature of this survey implies that one
can obtain photometric redshift estimates at least for a part of the
galaxies which will enable the suppression of the potential
contribution to the shear signal from intrinsic alignments of
galaxies. A forecast of the expected accuracy of cosmological
parameter estimates from the CFHTLS combined with the WMAP CMB
measurements has been obtained by Tereno et al.\ (2004).  It is
expected that a substantial fraction of the VST observing time will be
spend on multi-band wide-field surveys which, if properly designed,
will be extremely useful for cosmic shear research. In order to
complement results from the CFHTLS, accounting for the fact that the
VST has smaller aperture than the CFHT (2.6m vs. 3.6m), a somewhat
shallower but wider-field survey would be most reasonable. For both of
these surveys, complementary near-IR data will become available after
about 2007, with the WirCam instrument on CFHT, and the newly build
VISTA 4m-telescope equipped with a wide-field near-IR camera on
Paranal, which will yield much better photometric redshift estimates
than the optical data alone. Furthermore, with the PanStarrs project,
a novel method for wide-field imaging and a great leap forward in the
data access rate will be achieved.

Towards the end of the decade, a new generation of cosmic shear
surveys may be started; there are two projects currently under debate
which would provide a giant leap forward in terms of survey area
and/or depth. One is a satellite project, SNAP/JDEM, originally
designed for finding and follow-up of high-redshift supernovae to
study the expansion history of the Universe and in particular to learn
about the equation of state of the dark energy. With its large CCD
array and multi-band imaging, SNAP will also be a wonderful instrument
for cosmic shear research, yielding photometric redshift estimates for
the faint background galaxies, and it is expected that the observing
time of this satellite mission will be split between these two probes
of dark energy.  The other project under discussion is the LSST, a 8m
telescope equipped with a $\sim 9\,{\rm deg}^2$ camera; such an
instrument, with an efficiency larger than a factor 40 over
Megacam@CFHT, would allow huge cosmic shear surveys, easily obtaining
a multi-band survey over all extragalactic sky (modulo the constraints
from the hemisphere).  Since studying the equation of state of dark
energy will be done most effectively with good photometric redshifts
of source galaxies, the space experiment may appear more promising,
given the fact that near-IR photometry is needed for a reliable
redshift estimate, and sufficiently deep near-IR observations over a
significant area of sky is not possible from the ground.

\section{\llabel{WL-8}The mass of, and associated with galaxies}
\subsection{\llabel{WL-8.1}Introduction}
Whereas galaxies are not massive enough to show a weak lensing signal
individually (see eq.\ \ref{eq:4.55}), the signal of many galaxies can
be superposed statistically.  Therefore, if one considers sets of
foreground (lens) and background galaxies, then in the mean, in a
foreground-background galaxy pair, the image ellipticity of the
background galaxy will be preferentially oriented in the direction
tangent to the line connecting foreground and background galaxy.  The
amplitude of this tangential alignment then yields a mean lensing
strength that depends on the redshift distributions of foreground and
background galaxies, and on the mass distribution of the former
population.  This effect is called galaxy-galaxy lensing and will be
described in Sect.\ \ref{sc:WL-8.2} below; it measures the mass
properties of galaxies, provided the lensing signal is dominated by the
galaxies themselves. This will not be the case for larger angular
separations between foreground and background galaxies, since then the
mass distribution in which the foreground galaxies are embedded (e.g.,
their host groups or clusters) starts to contribute significantly to
the shear signal. The interpretation of this signal then becomes more
difficult. On even larger scales, the foreground galaxies contribute
negligibly to the lens signal; a spatial correlation between the lens
strength and the foreground galaxy population then reveals the
correlation between light (galaxies) and mass in the Universe. This
correlated distribution of galaxies with respect to the underlying
(dark) matter in the Universe -- often called the bias of galaxies --
can be studied with weak lensing, as we shall describe in Sect.\
\ref{sc:WL-8.3} by using the shear signal, and in Sect.\
\ref{sc:WL-8.4} employing the magnification effect. It should be pointed out
here that our lack of knowledge about the relation between the spatial
distribution of galaxies and that of the underlying (dark) matter is
one of the major problems that hampers the quantitative interpretation
of galaxy redshift surveys; hence, these lensing studies can provide
highly valuable input into the conclusions drawn from these redshift
surveys regarding the statistical properties of the mass distribution
in the Universe.

\subsection{\llabel{WL-8.2}Galaxy-galaxy lensing}
\subsubsection{The average mass profile of galaxies.}
Probing the mass distribution of galaxies usually proceeds with
dynamical studies of luminous tracers. The best-known method is the
determination of the rotation curves of spiral galaxies, measuring the
rotational velocity of stars and gas as a function of distance from
the galaxy's center (see Sofue \& Rubin 2001 for a recent
review). This then yields the mass profile of the galaxy, i.e.\ $M(\le
r)\propto v_{\rm rot}^2(r)\,r$.  For elliptical galaxies, the dynamics
of stars (like velocity dispersions and higher-order moments of their
velocity distribution, as a function of $r$) is analyzed to obtain
their mass profiles; as the kinematics of stars in ellipticals is more
complicated than in spirals, their mass profiles are more difficult to
measure (e.g., Gerhard et al.\ 2001). In both cases, these dynamical
methods provided unambiguous evidence for the presence of a dark
matter halo in which the luminous galaxy is embedded; e.g., the
rotation curves of spirals are flat out to the most distant point
where they can be measured. The lack of stars or gas prevents the
measurement of the mass profile to radii beyond the luminous extent of
galaxies, that is beyond $\sim 10 h^{-1}\,{\rm kpc}$.  Other luminous
tracers that have been employed to study galaxy masses at larger radii
include globular clusters that are found at large galacto-centric
radii (Cot\'e et al.\ 2003), planetary nebulae, and satellite
galaxies. Determining the relative radial velocity distribution of the
latter with respect to their suspected host galaxy leads to estimates
of the dark matter halo out to distances of $\sim 100 h^{-1}\,{\rm
kpc}$. These studies (e.g., Zaritsky et al. 1997) have shown that the
dark matter halo extends out to at least these distances.

One of the open questions regarding the dark matter profile of galaxies is the
spatial extent of the halos. The dynamical studies mentioned above are all
compatible with the mass profile following approximately an isothermal law
($\rho\propto r^{-2}$), which has to be truncated at a finite radius
to yield a finite total mass. Over the
limited range in radii, the isothermal profile cannot easily be distinguished
from an NFW mass profile (see IN, Sect.\ 6.2), for which measurements at
larger distances are needed (the mass distribution in the central parts of
galaxies is affected by the baryons and thus not expected to follow the NFW
profile; see Sect.\ts 7 of SL). 

Weak gravitational lensing provides a possibility to study the mass profiles
of galaxies at still larger radii. Light bundles from distant background
galaxies provide the `dynamical tracers' that cannot be found physically
associated with the galaxies. Light bundles get distorted in such a way
that on average, images of background sources are oriented tangent to the
transverse direction connecting foreground (lens) and background (source)
galaxy. The first attempt to detect such a galaxy-galaxy lensing signal was
reported in Tyson et al.\ (1984), but the use of photographic plates and the
relatively poor seeing prevented a detection. Brainerd et al.\ (1996)
presented the first detection and analysis of galaxy-galaxy lensing. Since
then, quite a number of surveys have measured this effect, some of them using
millions of galaxies.

\subsubsection{Strategy.}
Consider pairs of fore- and background galaxies, with separation in a given
angular separation bin. The expected lensing signal is seen as a statistical
tangential alignment of background galaxy images with respect to foreground
galaxies. For example, 
if $\phi$ is the angle between the major axis of the background galaxy and
the connecting line, values $\pi/4\le \phi\le \pi/2$ should be slightly
more frequent than $0\le\phi\le\pi/4$ (see Fig.\ \ref{fig:BBS1}).
Using the fact that the intrinsic orientations of background galaxies are
distributed isotropically, one can show (Brainerd et al.\ 1996) that
\be
p(\phi)={2\over \pi}\eck{1-\gamma_{\rm t}\ave{1\over |\eps^{\rm s}|}
\,\cos(2\phi)} \;,
\elabel{alignmentprob}
\ee
where $\phi\in [0,\pi/2]$ and $\gamma_{\rm t}$ is the mean tangential shear in
the angular bin chosen. 
Thus, the amplitude of the $\cos$-wave yields the (average) strength of the
shear.

\begin{figure}
\bmi{7}
\includegraphics[width=6.8cm]{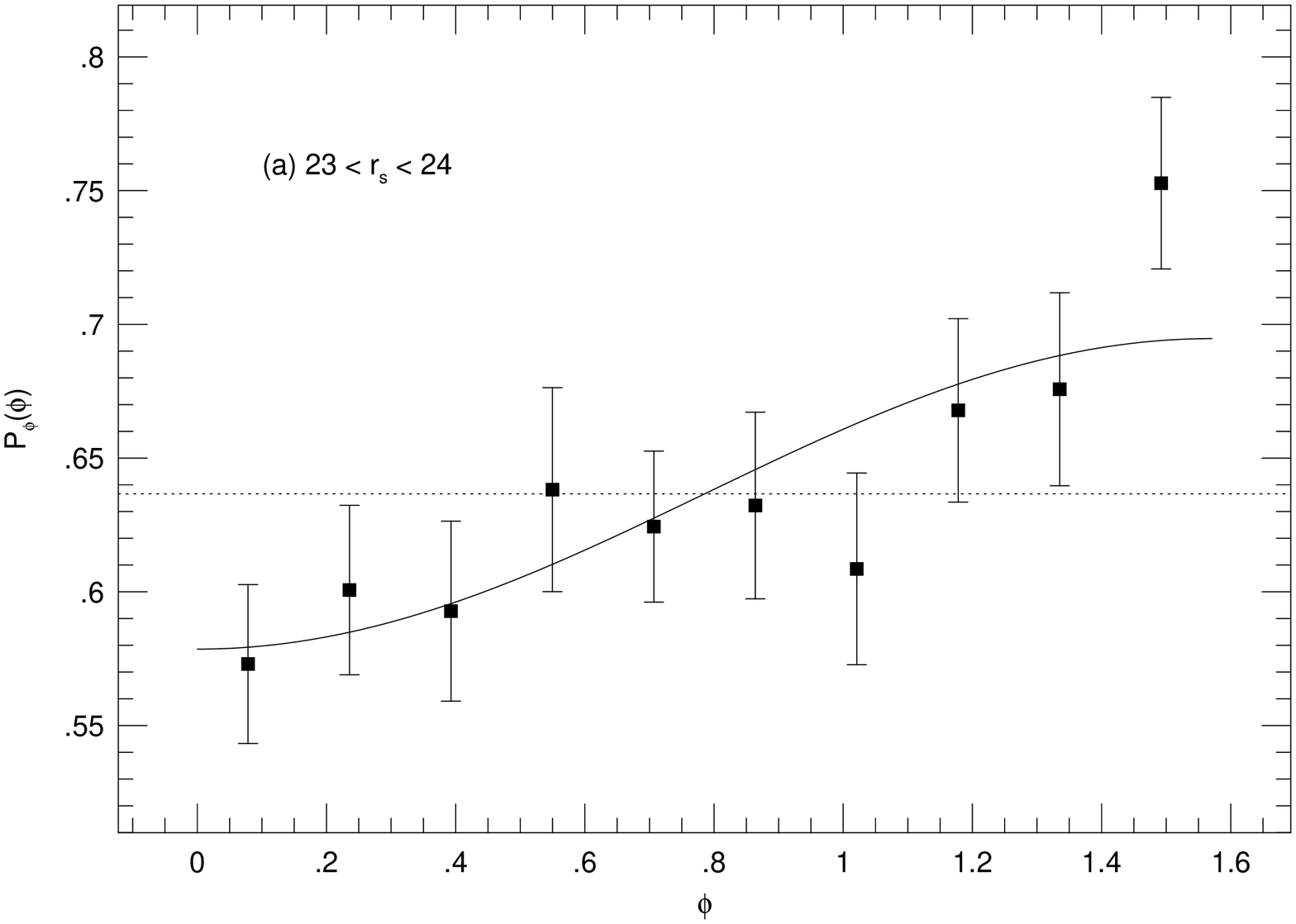}
\emi
\bmi{4.7}
\caption{ The probability distribution $p(\phi)$ of the angle $\phi$
between the major axis of the background galaxy image and the
connecting line to the foreground galaxy is plotted for the sample of
Brainerd et al.\ (1996), together with the best fit according to
(\ref{eq:alignmentprob}). The galaxy pairs have separation $5''\le
\Delta\theta\le 34''$, and are foreground-background selected by their
apparent magnitudes.  } 
\flabel{BBS1} 
\emi
\end{figure}

The mean tangential ellipticity $\ave{\eps_{{\rm t}} (\theta)}$ of
background galaxies relative to the direction towards foreground
galaxies measures the mean tangential shear at separation
$\theta$. Since the signal is averaged over many
foreground--background pairs, it measures the average mass profiles of
the foreground galaxies. For sufficiently large samples of galaxies,
the lens sample can be split into several subsamples, e.g., according
to their color and/or morphology (early-type vs. late-type galaxies),
or, if redshift estimates are available, they can be binned according
to their luminosity. Then, the mass properties can be derived for each
of the subsamples.

The distinction between foreground and background galaxies is ideally
performed using redshift information. This is indeed the case for the
galaxy-galaxy lensing studies based on the Sloan Digital Sky Survey,
for which early results have been reported by McKay et al.\ (2001);
all lens galaxies used there have spectroscopic redshifts, whereas the
source galaxies are substantially fainter than the lens galaxies so
that they can be considered as a background population. For other
surveys, the lack of redshift information requires the separation of
galaxies to be based solely on their apparent magnitudes: fainter
galaxies are on average at larger distances than brighter
ones. However, the resulting samples of `foreground' and `background'
galaxies will have (often substantial) overlap in redshift, which
needs to be accounted for statistically in the quantitative analysis
of these surveys.

\subsubsection{Quantitative analysis.}
The measurement of the galaxy-galaxy lensing signal provides the
tangential shear as a function of pair separation, $\gamma_{\rm
t}(\theta)$.  Without information about the redshifts of individual
galaxies, the separation of galaxies into a `foreground' and
`background' population has to be based on apparent magnitudes only.
In the ideal case of a huge number of foreground galaxies, one could
investigate the mass properties of `equal' galaxies, by finely binning
them according to redshift, luminosity, color, morphology
etc. However, in the real world such a fine binning has not yet been
possible, and therefore, to convert the lensing signal into physical
parameters of the lens, a parameterization of the lens population is
needed. We shall outline here how such an analysis is performed.

The first ingredient is the redshift probability distribution $p(z|m)$ of
galaxies with apparent magnitude $m$ which is assumed to be known from
redshift surveys (and/or their extrapolation to fainter magnitudes). This
probability density depends on the apparent magnitude $m$, with a broader
distribution and larger mean redshift expected for fainter $m$.  Since the
distribution of `foreground' and `background' galaxies in redshift is known
for a given survey, the probabilities $p(z|m)$ can be employed to calculate
the value of $D_{\rm ds}/D_{\rm s}$, averaged over all foreground--background
pairs (with this ratio being set to zero if $z_{\rm s}\le z_{\rm d}$). For
given physical parameters of the lenses, the shear signal is proportional to
this mean distance ratio.  

The mass profiles of galaxies are parameterized according to
their luminosity. For example, a popular parameterization is that of a
truncated isothermal sphere, where the parameters are the
line-of-sight velocity dispersion
$\sigma$ (or the equivalent circular velocity $V_{\rm c}=\sqrt{2}\sigma$) and
a truncation radius $s$ at which the $\rho\propto r^{-2}$ isothermal density
profile turns into a steeper $\rho\propto r^{-4}$ law. The velocity dispersion
is certainly dependent on the luminosity, as follows from the Tully-Fisher and
Faber-Jackson relations for late- and early-type galaxies, respectively. One
therefore assumes the scaling $\sigma=\sigma_* (L/L_*)^{\beta/2}$, where $L_*$
is a fiducial luminosity (and which conveniently can be chosen close to the
characteristic luminosity of the Schechter luminosity function). Furthermore,
the truncation scale $s$ is assumed to follow the scaling
$s=s_*(L/L_*)^\eta$. The total mass of a galaxy then is $M\propto \sigma^2 s$,
or $M=M_* (L/L_*)^{\beta+\eta}$.  

Suppose $m$ and $z$ were given; then, the luminosity of galaxy would be known,
and for given values of the parameters $\sigma_*$, $s_*$, $\beta$ and $\eta$,
the mass properties of the lens galaxy would be determined. However, since $z$
is not known, but only its probability distribution, only the probability
distribution of the lens luminosities, and therefore the mass properties, are
known. One could in principle determine the expected shear signal $\gamma_{\rm
  t}(\theta)$ for a given survey by calculating the shear signal for a given
set of redshifts $z_i$ for all lens and source galaxies, and then averaging
this signal over the $z_i$ using the redshift probability distribution
$p(z_i|m_i)$. However, this very-high dimensional integration cannot be
performed; instead, one uses a Monte-Carlo integration method (Schneider \& Rix
1997): Given the positions $\vc\theta_i$ and magnitudes $m_i$ of the galaxies,
one can draw for each of them a redshift according to $p(z_i|m_i)$, and then
calculate the shear at all positions $\vc\theta_i$ corresponding to a source
galaxy, for each set of parameters $\sigma_*$, $s_*$, $\beta$ and $\eta$. This
procedure can be repeated several times, yielding the expected shear
$\ave{\gamma_i}$ and its dispersion $\sigma_{\gamma,i}$ for each source
galaxy's position. One can then calculate the likelihood function
\be
{\cal L}=\prod_{i=1}^{N_{\rm s}}{1\over \pi
  (\sigma_\eps^2+\sigma_{\gamma,i}^2)} \,
\exp\rund{- {|\eps_i-\ave{\gamma_i}|^2 \over 
\sigma_\eps^2+\sigma_{\gamma,i}^2}}\;,
\ee
where $\sigma_\eps$ is the intrinsic ellipticity dispersion of the
galaxies. ${\cal L}$ depends on the parameters of the model, and can be
maximized with respect to them, thereby yielding estimates of 
$\sigma_*$, $s_*$, $\beta$ and $\eta$.

\subsubsection{First detection}
The galaxy-galaxy lensing effect was first found by Brainerd et al.\
(1996), on a single $9\arcminf6\times 9\arcminf6 $ field. They
considered `foreground' galaxies in the magnitude range $m\in[20,23]$,
and `background' galaxies with $m\in[23,24]$; this yielded 439
foreground and 506 background galaxies, and 3202 pairs with
$\Delta\theta\in[5'',34'']$.\footnote{The lower angular scale has been
chosen to avoid overlapping isophotes of foreground and background
galaxies, whereas the upper limit was selected since it gave the
largest signal-to-noise for the deviation of the angular distribution
shown in Fig.\ \ref{fig:BBS1} from a uniform one.}  For these pairs,
the distribution of the alignment angle $\phi$ is plotted in Fig.\
\ref{fig:BBS1}. This distribution clearly is incompatible with the
absence of a lens signal (at the 99.9\% confidence level), and thus
provides a solid detection.

They analyzed the lens signal $\gamma_{\rm t}(\theta)$ in a way similar to the
method outlined above, except that their Monte-Carlo simulations also
randomized the positions of galaxies. The resulting likelihood yields
$\sigma_*\approx 160^{+50}_{-60}\,{\rm km/s}$ (90\% confidence interval),
whereas for $s_*$ only a lower limit of $25 h^{-1}\,{\rm kpc}$ (1$\sigma$) is
obtained; the small field size, in combination with the relative insensitivity
of the lensing signal to $s_*$ once this value is larger than the mean
transverse separation of lensing galaxies, prohibited the detection of an
upper bound on the halo size. 

\subsubsection{Galaxy-galaxy lensing from the Red-Sequence Cluster Survey
  (RCS).} Several groups have published results of their galaxy-galaxy
lensing surveys since its first detection. Here we shall describe the
results of a recent wide-field imaging survey, the RCS; this survey
was already described in the context of cosmic shear in Sect.\
\ref{sc:WL-7.3}. 45.5\ square degrees of single-band imaging data were
used (Hoekstra et al.\ \cite{hoek67}). Choosing lens galaxies with
$19.5\le R_C\le 21$, and source galaxies having $21.5\le R_C \le 24$
yielded $\sim 1.2\times 10^5$ lenses with median redshift of 0.35 and
$\sim 1.5\times 10^6$ sources with median redshift of $\sim 0.53$,
yielding $\ave{D_{\rm ds}/D_{\rm s}}= 0.29\pm 0.01$ for the full
sample of lenses and sources.  Fig.\ \ref{fig:HYG1} shows the shear
signal for this survey.

\begin{figure}
\bmi{6.5}
\includegraphics[width=6.4cm]{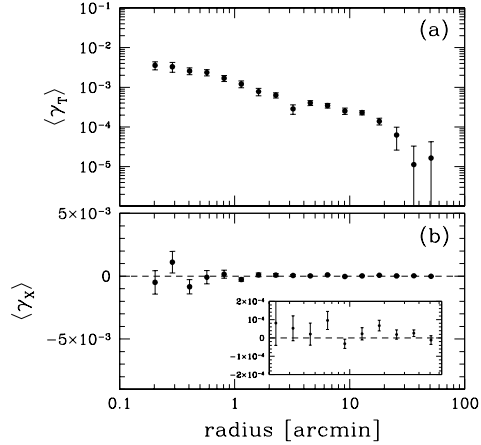}
\emi
\bmi{5.2}
\caption{
(a) Tangential shear as a function of angular separation, obtained from the
RCS survey; the shear signal is detected out to nearly one degree scale. (b)
Cross shear signal, which is expected to vanish identically in the absence of
systematic effects on the ellipticity measurements. As can be seen, the cross
signal in indeed compatible with zero. The inset expands the scale, to better
show the error bars (from Hoekstra et al.\ 2003)
}
\flabel{HYG1}
\emi
\end{figure}

The lens signal is affected by galaxies counted as lenses, but which in fact
are in the foreground. As long as they are not physically associated with lens
galaxies, this effect is accounted for in the analysis, i.e., in the value of
$\ave{D_{\rm ds}/D_{\rm s}}$. However, if fainter galaxies cluster around lens
galaxies, this produces an additional effect. Provided the orientation of the
associated faint galaxies are random with respect to the separation vector to
their bright neighbor, these physical pairs just yield a dilution of the shear
signal. The amplitude of this effect can be determined from the angular
correlation function of bright and faint galaxies, and easily corrected for.
Once this has been done, the corrected shear signal within $10''\le \theta\le
2'$ has been fitted with an SIS model, yielding a mean velocity dispersion of
the lens galaxies of $\sqrt{\ave{\sigma^2}}=128\pm 4\,{\rm km/s}$. If the
scaling relations between galaxy luminosity and velocity dispersion as
described above is employed, with $\beta=0.6$, the result is $\sigma_*=140\pm
4\,{\rm km/s}$ for $L_*=10^{10} h^{-2} L_\odot$ in the blue passband.

To interpret the shear results on larger angular scales, the SIS model
no longer suffices, and different mass models need to be employed.
Using a truncated isothermal model, the best-fitting values of the
scaling parameters $\beta=0.60\pm 0.11$ and
$\eta=0.24^{+0.26}_{-0.22}$ are obtained, when marginalizing over all
other parameters. Furthermore, $\sigma_*=137\pm 5\,{\rm km/s}$, in
very close agreement with the results from small $\theta$ and the SIS
model; this is expected, since most of the signal comes from these
smaller separations. Most interesting, the analysis also yields an
estimate of the truncation scale of $s_*=(185\pm 30)h^{-1}\,{\rm
kpc}$, providing one of only a few estimates of the scale of the dark
matter halo. Hoekstra et al.\ also performed the analysis in the frame
of an NFW mass model.

These results can then be used to calculate the mass-to-light ratio of an
$L_*$ galaxy and, using the scaling, of the galaxy population as a
whole. Considering only galaxies with $M\ge 10^{10} h^{-1} M_\odot$, the mean
mass-to-light ratio inside the virial radius of galaxy halos is about 100 in
solar units.

\subsubsection{The shape of dark matter halos.}
In the mass models considered before, the mass distribution of
galaxies was assumed to be axi-symmetric. In fact, this assumption is
not crucial, since the relation between shear and surface mass
density, $\gamma_{\rm t}(\vt)=\bar\kappa(\vt)-\kappa(\vt)$ is true for
a general mass distribution, provided $\gamma_{\rm t}$ and
$\kappa(\vt)$ are interpreted as the mean tangential shear and mean
surface mass density on a circle of radius $\vt$, and
$\bar\kappa(\vt)$ as the mean surface mass density inside this circle
(see eq.\ts\ref{eq:mass-rela4}). However, deviations from axial
symmetry are imprinted on the shear signal and can in principle be
measured. If the mass distribution is `elliptical', the shear along the
major axis (at given distance $\vt$) is larger than that along the
minor axis, and therefore, an investigation of the strength of the
shear signal relative to the orientation of the galaxy can reveal a
finite ellipticity of the mass distribution. For that, it is necessary
that the orientation of the mass distribution is (at least
approximately) known. Provided the orientation of the mass
distribution follows approximately the orientation of the luminous
part of galaxies, one can analyze the direction dependence of the
shear relative to the major axis of the light distribution 
(Natarajan \& Refregier 2000). Hoekstra et al.\ (\cite{hoek55})
have used the RCS to search for such a direction dependence; they
parameterized the lenses with a truncated isothermal profile with
ellipticity $\eps_{\rm mass}=f\eps_{\rm light}$, where $f$ is a free
parameter. The result $f=0.77\pm 0.2$ indicates first that the mass
distribution of galaxies is not round (which would be the case for
$f=0$, which is incompatible with the data), and second, that the mass
distribution is rounder than that of the light distribution, since
$f<1$. However, it must be kept in mind that the assumption of equal
orientation between light and mass is crucial for the interpretation
of $f$; misalignment causes a decrease of $f$. Note that numerical
simulations of galaxy evolution predict such a misalignment between
total mass and baryons, with an rms deviation of around $20^\circ$
(van den Bosch et al.\ 2002). Given the above result on $f$, it is therefore
not 
excluded that the flattening of halos is very similar to that of the
light. Also note that this result yields a value averaged over all
galaxies; since the lens efficiency of elliptical galaxies (at given
luminosity) is larger than that of spirals, the value of $f$ is
dominated by the contributions from early-type galaxies.

\subsubsection{Results from the Sloan Survey.}
The Sloan Digital Sky Survey (e.g., York et al.\ 2000) will map a
quarter of the sky in five photometric bands, and obtain spectra of
about one million galaxies. A large fraction of the data has already
been taken by SDSS, and parts of this data have already been released
(Abazajian et al.\ 2004). The huge amount of photometric data in
principle is ideal for weak lensing studies, as it beats down
statistical uncertainties to an unprecedented low level. However, the
site of the telescope, the relatively large pixel size of $0\arcsecf
4$, the relatively shallow exposures of about one minute and the
drift-scan mode in which data are taken (yielding excellent
flat-fielding, and thus photometric properties, somewhat at the
expense of the shape of the PSF) render the data less useful for,
e.g., cosmic shear studies: the small mean redshift of the galaxies
yields a very small expectation value of the cosmic shear, which can
easily be mimicked by residuals from PSF corrections. However,
galaxy-galaxy lensing is much less sensitive to larger-scale PSF
problems, since the component of the shear used in the analysis is not
attached to pixel directions, but to neighboring galaxies, and thus
varies rapidly with sky position. Another way of expressing this fact
is that the galaxy-galaxy lensing signal would remain unchanged if a
uniform shear would be added to the data; therefore, SDSS provides an
great opportunity for studying the mass profile of galaxies.

Fischer et al.\ (2000) reported the first results from the SDSS, and a larger
fraction of the SDSS data was subsequently used in a galaxy-galaxy lensing
study by McKay et al.\ (2001), where also the spectroscopic redshifts of the
lens galaxies were used. Their sample consists of $\sim 31000$ lens galaxies
with measured redshifts, and $\sim 3.6\times 10^6$ source galaxies selected in
the brightness range $18\le r\le 22$. For this magnitude range, the redshift
distribution of galaxies is fairly well known, leaving little calibration
uncertainty in the interpretation of the shear signal. In particular, there is
very little overlap in the redshift distribution of source and lens galaxies. 
The data set has been subjected to a large number of tests, to reveal
systematics; e.g., null results are obtained when the source galaxies are
rotated by $45^\circ$ (or, equivalently, if $\gamma_\times$ is used instead of
$\gamma_{\rm t}$), or if the lens galaxies are replaced by an equal number of
randomly distributed points relative to which the tangential shear component
is measured. Since the redshifts of the lens galaxies are known, the shear can
be measured directly in physical units, so one can determine 
\be
\Delta\Sigma_+=\bar\Sigma(\le R)-\Sigma(R)
\ee
in $M_\odot/{\rm pc}^2$ as a function of $R$ in kpc. 

\begin{figure}
\bmi{6} 
\includegraphics[width=5.8cm]{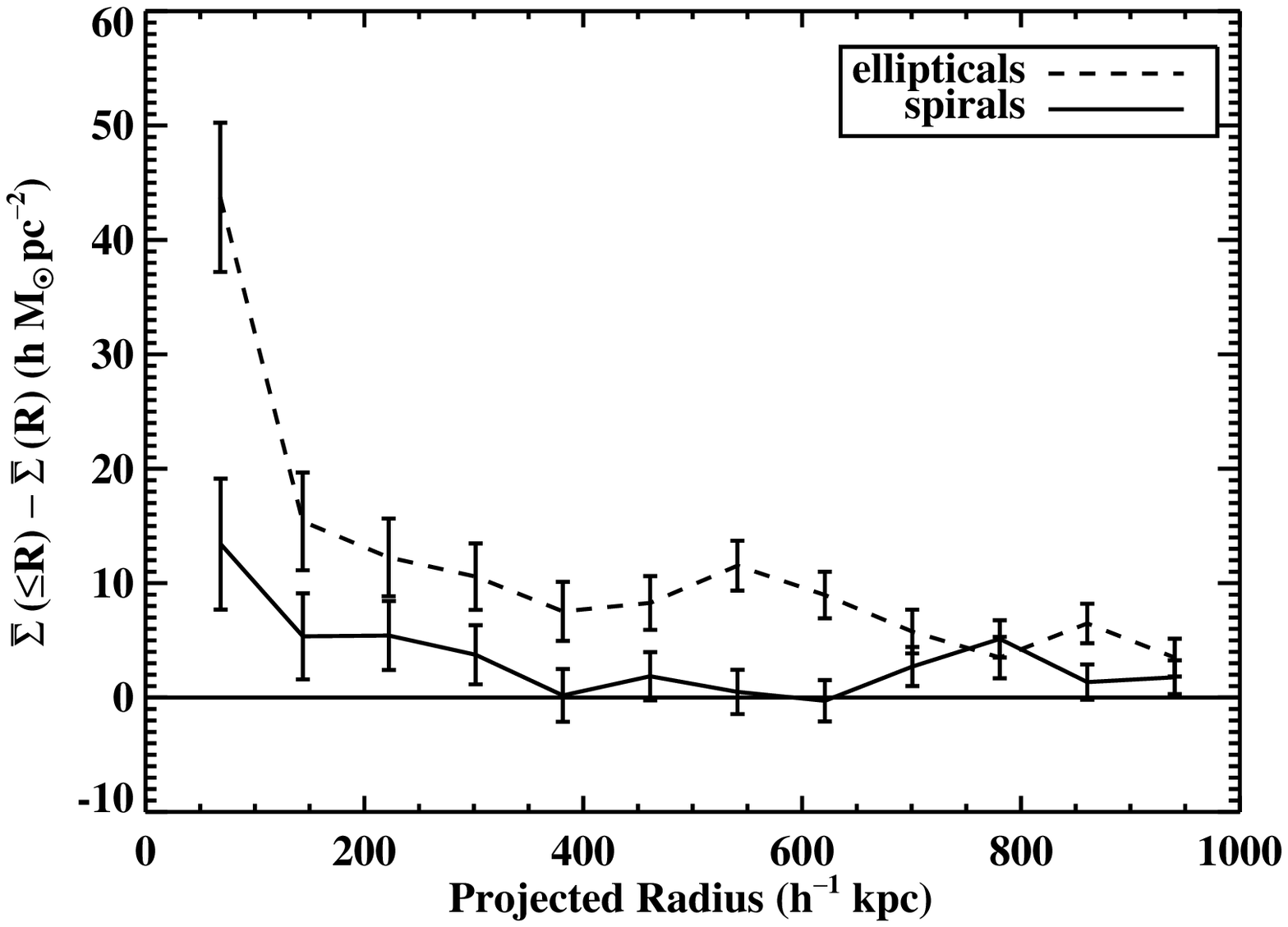}

\includegraphics[width=5.8cm]{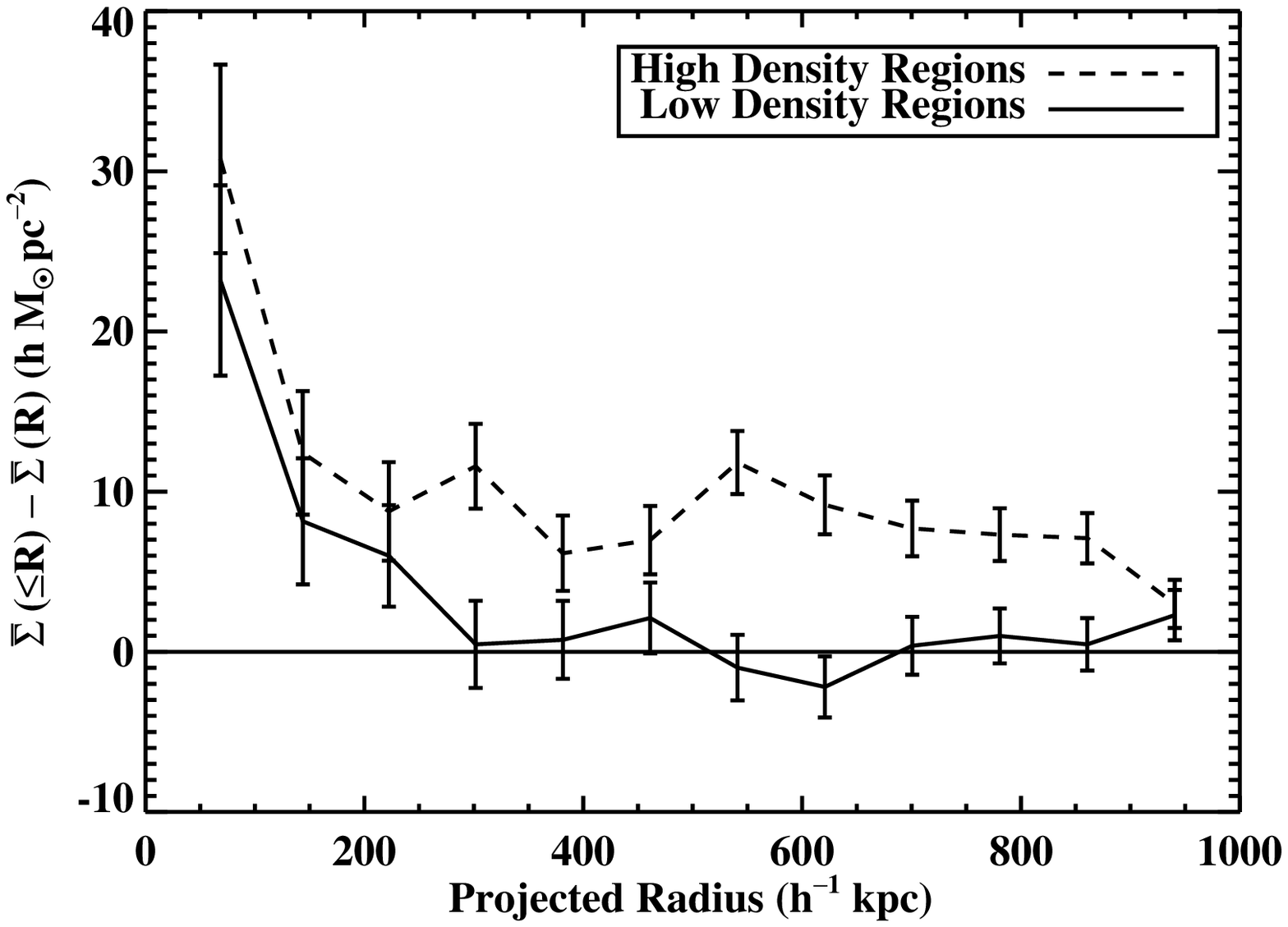}
\emi
\bmi{5.7}
\caption{
The galaxy-galaxy lensing signal from the SDSS plotted against physical radius
$R$. The lens sample has been subdivided into early- and late-type galaxies
(upper panel), and in galaxies situated in dense environments vs. those with a
smaller neighboring galaxy density (lower panel). The figure clearly shows
that the lensing signal is dominated by elliptical galaxies, and by those
located in dense environment. Owing to the morphology-density relation of
galaxies, these two results are not mutually independent. Note that the
lensing signal can be measured out to $1h^{-1}\,{\rm Mpc}$, considerably
larger than the expected size of galaxy halos; therefore, the shear at these
large separations is most likely caused by the larger-scale mass distribution
in which the galaxies are embedded
(from McKay et al.\ 2001)
}
\flabel{McKay1}
\emi
\end{figure}

Fig.\ \ref{fig:McKay1} shows the lensing result from McKay et al.\ (2001),
where the lens sample has been split according to the type of galaxy (early
vs. late type) and according to the local spatial number density of galaxies,
which is known owing to the spectroscopic redshifts. The fact that most of the
signal on small scales is due to ellipticals is expected, as they are more
massive at given luminosity than spirals. The large spatial extent of the
shear signal for ellipticals relative to that of spirals can be interpreted
either by ellipticals having a larger halo than spirals, or that ellipticals
are preferentially found in high-density environments, which contribute to the
lens signal on large scales. This latter interpretation is supported by the
lower panel in Fig.\ \ref{fig:McKay1} which shows that the signal on large
scales is entirely due to lens galaxies in dense environments. This then
implies that the galaxy-galaxy lensing signal on large scales no longer
measures the density profile of individual galaxies, but gets more and more
dominated by group and cluster halos in which these (predominantly early-type)
galaxies are embedded.

A separation of these contributions from the data themselves is not possible
at present, but can be achieved in the frame of a theoretical model.  Guzik \&
Seljak (2001) employed the halo model for the distribution of matter in the
universe (see Cooray \& Sheth 2002) to perform this separation. There, the
galaxy-galaxy lensing signal either comes from matter in the same halo in
which the galaxy is embedded, or due to other halos which are physically
associated (i.e., clustered) with the former. This latter contribution is
negligible on the scales below $\sim 1 h^{-1}\,{\rm Mpc}$ on which the SDSS
obtained a measurement. The former contribution can be split further into two
terms: the first is from the dark matter around the galaxies themselves,
whereas the second is due to the matter in groups and clusters to which the
galaxies might belong. The relative amplitude of these two terms depends on
the fraction of galaxies which are located in groups and clusters; the larger
this fraction, the more important are larger-scale halos for the shear signal.
Guzik \& Seljak estimate from the radial dependence of the SDSS signal that
about 20\% of galaxies reside in groups and clusters; on scales larger than
about $200 h^{-1}\,{\rm kpc}$ their contribution dominates. The virial mass of
an early-type $L_*$ galaxy is estimated to be $M_{200}(L_*)=(9.3\pm 2.2)\times
10^{11} h^{-1} M_\odot$, and about a factor of three smaller for late-type
galaxies (with luminosity measured in a red passband; the differences are
substantially larger for bluer passbands, owing to the sensitivity of
the luminosity to star
formation activity in late types). From the mass-to-light ratio in red
passbands, Guzik \& Seljak estimate that an $L_*$ galaxy converts about
10--15\% of its virial mass into stars. Since this fraction is close to the
baryon fraction in the universe, they conclude that most of the baryons of an
$L_*$ galaxy are transformed into stars. For more massive halos, the
mass-to-light ratio increases ($M/L\propto L^{0.4\pm 0.2}$), and therefore
their conversion of baryons into stars is smaller -- in agreement with what we
argued about clusters, where most of the baryons are present in the form of a
hot intracluster gas.

Yang et al.\ (2003) studied the cross-correlation between mass and galaxies
using numerical simulations of structure formation and semi-analytic models of
galaxy evolution. The observed dependence of the galaxy-galaxy lensing signal
on galaxy luminosity, morphological type and galaxy environment, as obtained by
McKay et al.\ (2001), is well reproduced in these simulations. The galaxy-mass
correlation is affected by satellite galaxies, i.e. galaxies not situated at
the center of their respective halo. Central galaxies can be selected by
restricting the foreground galaxy sample to relatively isolated galaxies. The
galaxy-galaxy lensing signal for such central galaxies can well be described
by an NFW mass profile, whereas this no longer is true if all galaxies are
considered. Combining the measurement with the simulation, they find that an
$L_*$-galaxy typically resides in a halo with a virial mass of $\sim 2\times
10^{12}h^{-1}M_\odot$. 

\begin{figure}
\bc
\includegraphics[width=11cm]{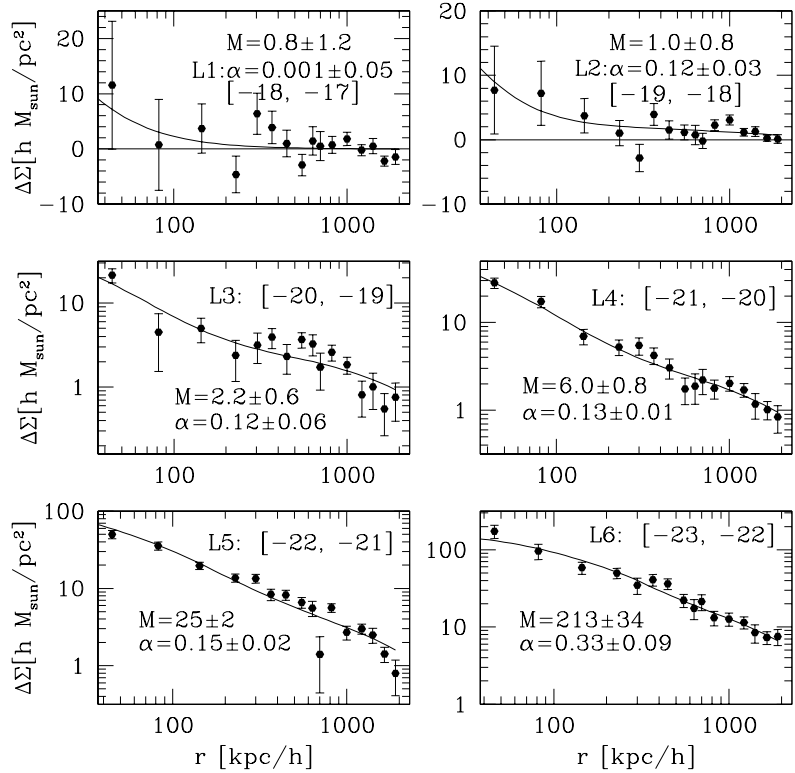}
\ec
\caption{The galaxy-galaxy lensing signal for six luminosity bins of foreground
  galaxies, as indicated by the absolute magnitude interval in each panel. The
  curves show a two-parameter model fitted to the data, based on the halo
  model, and the fit parameters are indicated: $M$ is the virial mass of the
  halo (in units of $10^{11}h^{-1}M_\odot$) in which the galaxies reside, and
  $\alpha$ is the fraction of the galaxies which are not central inside the
  halo, but satellite galaxies (from Seljak et al.\ 2004)}
\flabel{Selj-GGL-SDSS}
\end{figure}

With the SDSS progressing, larger datasets become available, allowing a more
refined analysis of galaxy-galaxy lensing (Sheldon et al.\ 2004; Seljak et
al.\ 2004). In the analysis of Seljak et al.\ (2004), more than $2.7\times
10^5$ galaxies with spectroscopic redshifts have been used as foreground
galaxies, and as background population
those fainter galaxies for which photometric redshifts have been
estimated. The resulting signal is shown in Fig.\ts\ref{fig:Selj-GGL-SDSS},
for six different bins in (foreground) galaxy luminosity.

In a further test to constrain systematic effects in the data, Hirata et al.\ 
(2004) have used spectroscopic and photometric redshifts to study the question
whether an alignment of satellite galaxies around the lens galaxies can affect
the galaxy-galaxy lensing signal from the SDSS; they obtain an upper limit of
a 15\% contamination.

The SDSS already has yielded important information about the mass properties
of galaxies; taken into account that only a part of the data of the
complete survey have been used in the studies mentioned above, an analysis of
the final survey will yield rich harvest when applied to a galaxy-galaxy
lensing analysis.

\subsubsection{Lensing by galaxies in clusters.}
As an extension of the method presented hitherto, one might use galaxy-galaxy
lensing also to specifically target the mass profile of galaxies in the inner
part of clusters. One might expect that owing to tidal stripping, their dark
matter halo has a considerably smaller spatial extent than that of the galaxy
population as a whole. The study of this effect with lensing is more
complicated than galaxy-galaxy lensing in the field, both observationally and
from theory. Observationally, the data sets that can be used need to be taken
in the inner part of massive clusters; since these are rare, a single
wide-field image usually contains at most one such cluster. Furthermore, the
number of massive galaxies projected near the center of a cluster is fairly
small.  Therefore, in order to obtain good statistics, the data of different
clusters should be combined. Since the cores of clusters are optically bright,
measuring the shape of faint background galaxies is more difficult than 
in a blank field. From the theoretical side, the lensing strength of the
cluster is much stronger than that of the individual cluster galaxies, and so
this large-scale shear contribution needs to be accounted for in the
galaxy-galaxy lensing analysis. 

Methods for performing this separation between cluster and galaxy
shear were developed by Natarajan \& Kneib (1997) and Geiger \&
Schneider (1998). Perhaps the simplest approach is provided by the
aperture mass methods, applied to the individual cluster galaxies;
there one measures the tangential shear inside an annulus around each
cluster galaxy. This measure is insensitive to the shear
contribution which is linear in the angular variable $\vc\theta$, which is a
first local approximation to the larger-scale shear
caused by the cluster. Alternatively, a mass model of the (smoothed)
cluster can be obtained, either from strong or weak lensing
constraints, or preferentially both, and subtracted from the shear
signal around galaxies to see their signal. However, once the mass
fraction in the galaxies becomes considerable, this method starts to
become biased. Geiger \& Schneider (1999) have suggested to simultaneously
perform a weak lensing mass reconstruction of the cluster and a
determination of the parameters of a conveniently parameterized mass
model of cluster galaxies (e.g., the truncated isothermal sphere);
since the maximum likelihood method for the mass reconstruction (see
Sect.\ \ref{sc:WL-5.3}) was used, the solution results from maximizing
the likelihood with respect to the mass profile parameters (the
deflection potential on a grid) and the galaxy mass parameters.

\begin{figure}
\bc
\includegraphics[width=8cm]{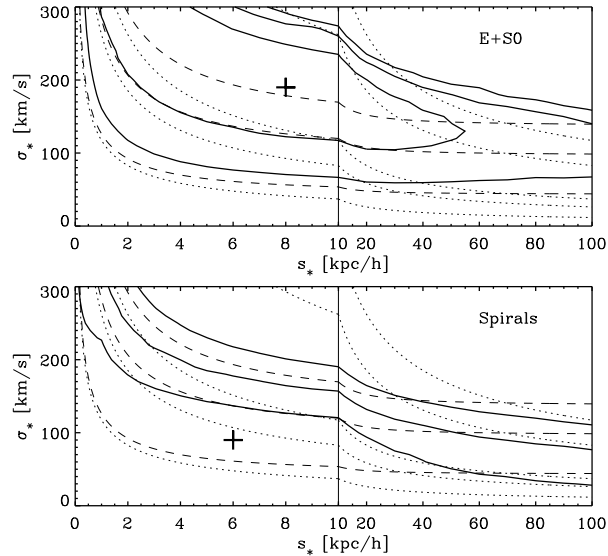}
\ec
\caption{Significance contours (solid)
for galaxy properties obtained from
galaxy-galaxy lensing of galaxies in the cluster Cl\ts 0939+4713. The
parameters are the velocity dispersion $\sigma_*$ and the halo
truncation radius $s_*$ of an $L_*$-galaxy. Based on HST data (see
Fig.\ts\ref{fig:0939}), a simultaneous reconstruction of the cluster
mass profile and the determination of the galaxy mass parameters
was performed. No significant lensing signal is seen from the 55
late-type galaxies (lower panel), but a clear detection and upper
bound to the halo size is detected for the 56 early-types. Dashed and
dotted curves connect models with the same mass inside $8h^{-1}\,{\rm
kpc}$ and total mass of an $L_*$-galaxy, respectively (from Geiger \&
Schneier 1999)}
\flabel{GeiSchn}
\end{figure}

Natarajan et al.\ (1998), by analyzing HST data of the cluster AC114,
concluded that the truncation radius of a fiducial $L_*$ galaxy in this
cluster is $\sim 15 h^{-1}\,{\rm kpc}$; similarly, Geiger \& Schneider (1999)
showed that the best-fitting truncation radius for early-type galaxies in the
cluster A851 is $\sim 10 h^{-1}\,{\rm kpc}$ (see Fig.\ts\ref{fig:GeiSchn}).
Although the uncertainties are fairly large, these results indicate that
indeed galaxies near cluster centers have a halo size considerably smaller
than the average galaxy. The sample of clusters which can be investigated
using this method will dramatically increase once the cluster sample observed
with the new ACS camera onboard HST becomes available and gets properly
analyzed.

\subsection{\llabel{WL-8.3}Galaxy biasing: shear method}
On small scales, galaxy-galaxy lensing measures the mass profile of galaxies,
whereas on intermediate scales the environment of galaxies starts to dominate
the shear signal. On even larger scale (say, beyond $\sim 1 h^{-1}\,{\rm
  Mpc}$), the host halo contribution becomes negligible. Beyond that distance,
any signal must come from the correlation of galaxy positions with the mass
distribution in the Universe. This correlation, and the related issue of
galaxy biasing (see Sect.\ 6.1 of IN), can ideally be studied with weak
lensing. In this section we shall outline how these quantities can be
determined from shear measurements, and describe some recent results. As we
shall see, this issue is intimately related to galaxy-galaxy lensing. The next
section deals with the magnification of distant sources caused by mass
overdensities correlated with galaxies and thereby causing an apparent
correlation between high-redshift sources and low-redshift galaxies; the
amplitude of this signal is again proportional to the correlation between
galaxies and the underlying dark matter.

An interesting illustration of the correlation between galaxies and
mass has been derived by Wilson et al.\ (2001). They studied 6 fields
with $30'\times 30'$ each, selected bright early-type galaxies from
their $V-I$ colors and $I$ magnitudes and measured the shear from faint
galaxies. Assuming that mass is strongly correlated with early-type
galaxies, these can be used to predict the shear field, with an
overall normalization given by the mean mass-to-light ratio of the
early-type galaxies. This correlation has indeed been found, at the
5.2-$\sigma$ significance level, and a value of $M/L\approx 300 h$ in
solar units has been obtained, assuming a flat low-density Universe. 

%
%

\subsubsection{The galaxy-mass correlation and the bias parameter.}
First, the concept of the correlation between galaxies and mass shall be
described more quantitatively. The mass density inhomogeneities are
described, as before, by the dimensionless density contrast $\delta(\vc
x,w)$. In analogy to this quantity, one defines the number density contrast
$\delta_{\rm g}(\vc x,w)$ of galaxies as 
\be
 \delta_{\rm g}(\vc x,w):={n(\vc x,w)-\bar n(w) \over \bar n(w)}\;,
\elabel{deltagal}
\ee
where $n(\vc x,w)$ is the number density of galaxies at comoving position $\vc
x$ and comoving distance $w$ (the latter providing a parameterization of cosmic
time or redshift), and $\bar n(w)$ is the mean number density of
galaxies at that 
epoch. Since the galaxy distribution is discrete, the true number density is
simply a sum of delta-functions. What is meant by $n$ is that the probability
of finding a galaxy in the volume $\d V$ situated 
at position $\vc x$ is $n(\vc x)\,\d V$.

The relation between $\delta$ and $\delta_{\rm g}$ describes the relative
distribution of galaxies and matter in the Universe. The simplest case is that
of an {\em unbiased} distribution, for which $\delta_g=\delta$; then, the
probability of finding a galaxy at any location would be just proportional to
the matter density. However, one might expect that the relation between
luminous and dark matter is more complicated. For example, galaxies are
expected to form preferentially in the high-density peaks in the early
Universe, which would imply that there are proportionally more galaxies within
mass overdensities. This led to the introduction of the concept of biasing
(e.g., Bardeen et al.\ 1986; Kaiser 1984). The simplest form of biasing,
called linear deterministic biasing, is provided by setting $\delta_{\rm
  g}=b\,\delta$, with $b$ being the bias parameter. One might suspect that the
relative bias is approximately constant on large scales, where the density
field is still in its linear evolution (i.e., on scales $\gtrsim 10
h^{-1}\,{\rm Mpc}$ today). On smaller scales, however, $b$ most likely is no
longer simply a constant. For example, the spatial distribution of galaxies in
clusters seems to deviate from the radial mass profile, and the distributions
of different galaxy types are different. Furthermore, by comparing the
clustering properties of galaxies of different types, one can determine their
relative bias, from which it is concluded that more luminous galaxies are more
strongly biased than less luminous ones, and early-type galaxies are more
strongly clustered than late-types (see Norberg et al.\ 2001 and Zehavi et
al.\ 2002 for recent results from the 2dFGRS and the SDSS). This is also
expected from theoretical models and numerical simulations which show that
more massive halos cluster more strongly (e.g., Sheth et al.\ 2001; Jing
1998).  In order to account for a possible scale dependence of the bias, one
considers the Fourier transforms of $\delta$ and $\delta_{\rm g}$ and relates
them according to
\be
\hat\delta_{\rm g}(\vc k,w)=b(|\vc k|,w)\,\hat\delta(\vc k,w)\;,
\elabel{bias}
\ee
thus accounting for a possible scale and redshift dependence of the bias.

Even this more general bias description is most likely too simple, as it is
still deterministic. Owing to the complexity of galaxy formation and
evolution, it is to be expected that the galaxy distribution is subject to
stochasticity in excess to Poisson sampling (Tegmark \& Peebles 1998;
Dekel \& Lahav 1999). 
To account for that, another
parameter is introduced, the correlation parameter $r(|\vc k|,w)$, which in
general will also depend on scale and cosmic epoch. To define it, we first
consider the correlator 
\be
\ave{\hat\delta(\vc k,w)\,\hat\delta^*_{\rm g}(\vc k',w)}
=(2\pi)^3\,\delta_{\rm D}(\vc k-\vc k')\,P_{\delta\rm g}(|\vc k|,w)\;,
\elabel{crosspower}
\ee
where the occurrence of the delta function is due to the statistical
homogeneity of the density fields, and $P_{\delta\rm g}$ denotes the
cross-power between galaxies and matter. The correlation parameter $r$ is then
defined as 
\be
r(|\vc k|,w)= { P_{\delta\rm g}(|\vc k|,w) \over  \sqrt{P_\delta(|\vc
k|,w)\,P_{\rm g}(|\vc k|,w)}} \;.
\elabel{r-def}
\ee
In the case of {\em stochastic biasing}, the definition of the bias parameter
is modified to
\be
P_{\rm g}(|\vc k|,w)=b^2(|\vc k|,w)\,P_\delta(|\vc k|,w)\;,
\elabel{bias-def}
\ee
which agrees with the definition (\ref{eq:bias}) in the case of $r\equiv 1$,
but is more general since (\ref{eq:bias-def}) no longer relates the phase of
(the Fourier transform of) $\delta_{\rm g}$ to that of $\delta$. Combining the
last two equations yields
\be
P_{\delta\rm g}(|\vc k|,w)=b(|\vc k|,w)\,r(|\vc k|,w)\,P_\delta(|\vc k|,w)\;.
\elabel{Pdeltag}
\ee
We point out again that galaxy redshift surveys are used to determine the
two-point statistics of the galaxy distribution, and therefore $P_{\rm g}$; in
order to relate there measurements to $P_\delta$, assumptions on the
properties of the bias have to be made. As we shall discuss next, weak lensing
can determine both the bias parameter and the correlation parameter.

\subsubsection{The principle.}
In order to determine $b$ and $r$, the three power spectra defined
above (or functions thereof) need to be measured. Second-order cosmic
shear measures, as discussed in Sect.\ \ref{sc:WL-6}, are proportional
to the power spectrum $P_\delta$. The correlation function of galaxies
is linearly related to $P_{\rm g}$. In particular, the
three-dimensional correlation function is just the Fourier transform
of $P_{\rm g}$, whereas the angular correlation function contains a
projection of $P_{\rm g}$ along the line-of-sight and thus follows
from Limber's equation as discussed in Sect.\
\ref{sc:WL-6.2}. Finally, the cross-power $P_{\delta\rm g}$ describes
the correlation between mass and light, and thus determines the
relation between the lensing properties of the mass distribution in the
Universe to the location of the galaxies. Galaxy-galaxy lensing on
large angular scales (where the mass profile of individual galaxies no
longer yields a significant contribution) provides one of the measures
for such a correlation. Hence, measurements of these three statistical
distributions allow a determination of $r$ and $b$.

As we shall consider projected densities, we relate the density field of
galaxies on the sky to the spatial distribution. Hence, consider a population
of (`foreground') galaxies with spatial number density $n(\vc x,w)$.  The
number density of these galaxies on the sky at $\vc\theta$ is then
$N(\vc\theta)=\int\d w\; \nu(w)\, n(f_k(w)\vc\theta,w)$, where $\nu(w)$ is the
redshift-dependent selection function, describing which fraction of the
galaxies at comoving distance $w$ are included in the sample. Foremost, this
accounts for the fact that for large distances, only the more luminous
galaxies will be in the observed galaxy sample, but $\nu$ can account also for
more subtle effects, such as spectral features entering or leaving the
photometric bands due to redshifting. The mean number density of galaxies on
the sky is $\bar N=\int\d w\;\nu(w)\,\bar n(w)$; the redshift distribution, or
more precisely, the distribution in comoving distance, of these galaxies
therefore is $p_{\rm f}(w)=\nu(w)\,\bar n(w)/\bar N$, thus relating
the selection 
function $\nu(w)$ to the redshift distribution. Using the definition
(\ref{eq:deltagal}), one then finds that
\be
N(\vc\theta)=\bar N\eck{1+\int\d w\;p_{\rm f}(w)\,\delta_{\rm
    g}(f_K(w)\vc\theta, w)}\; .
\elabel{galdensi}
\ee
We shall denote the fractional number density by $\kappa_{\rm
  g}(\vc\theta):=\eck{N(\vc\theta) - \bar N}/\bar N = \int\d w\;p_{\rm
f}(w)\,\delta_{\rm g}(f_K(w)\vc\theta, w)$.

\subsubsection{Aperture measures.}
We have seen in Sect.\ \ref{sc:WL-6.3} that the aperture mass dispersion
provides a very convenient measure of second-order cosmic shear
statistics. Therefore, it is tempting to use aperture measures also for the
determination of the bias and the mass-galaxy correlation. Define in analogy
to the definition of the aperture mass $M_{\rm ap}$ in terms of the projected
mass density the aperture counts (Schneider 1998),
\be
{\cal N}(\theta) = \int \d^2\vt\; U(|\vc\vt|)\, \kappa_{\rm g}(\vc\vt) \;,
\ee
where the integral extends over the aperture of angular radius $\theta$, and
$\vc\vt$ measures the position relative to the center of the aperture. An
unbiased estimate of the aperture counts is $\bar N^{-1}\sum_i
U(|\vc\theta_i|)$, where the $\vc\theta_i$ are the positions of the galaxies.
We now consider the dispersion of the aperture counts,
\be
\ave{{\cal N}^2(\theta)}=\int\d^2\vt\;U(|\vc\vt|)\int\d^2\vt'\;U(|\vc\vt'|)
\ave{\kappa_{\rm g}(\vc\vt)\,\kappa_{\rm g}(\vc\vt')}\;.
\elabel{ENNsq}
\ee
The correlator in the last expression is the angular two-point correlation
function $\omega(\Delta\vt)$ of the galaxies; its Fourier transform is the
angular power spectrum $P_\omega(\ell)$ of galaxies. Using the definition of
$\kappa_{\rm g}$ together with the result (\ref{eq:limber}) allows us to
express $P_\omega$ in terms of the three-dimensional power spectrum of the
galaxy distribution,
\bea
P_\omega(\ell)&=&\int \d w\;{p_{\rm f}^2(w)\over f_K^2(w)}\,b^2\rund{{\ell\over
    f_K(w)},w} \,P_\delta\rund{{\ell\over f_K(w)},w} \nonumber \\
  &=& \bar b^2\int \d w\;{p_{\rm f}^2(w)\over f_K^2(w)}
  \,P_\delta\rund{{\ell\over f_K(w)},w}
\;,
\eea
where we made use of (\ref{eq:bias-def}), and in the final step we defined the
mean bias parameter $\bar b$ which is a weighted average of the bias
parameter over the redshift distribution of the galaxies and which depends on
the angular wave number $\ell$. To simplify notation, we shall drop the bar on
$b$ and consider the bias factor as being conveniently averaged over redshift
(and later, also over spatial scale). The aperture count dispersion then
becomes 
\be
\ave{{\cal N}^2(\theta)}={1\over 2\pi}\int \d\ell\;\ell\,P_\omega(\ell)
W_{\rm ap}(\theta\ell)=2\pi\, b^2\, H_{\rm gg}(\theta)\;,
\elabel{HOEENN}
\ee
where $W_{\rm ap}$ is given in (\ref{eq:N26a}), and we have defined 
\be
  H_{\rm gg}(\theta)=\int\d w\; {p_{\rm f}^2(w)\over f_K^2(w)}\,
  {\cal P}(w,\theta)\;,
\ee
with 
\be
{\cal P}(w,\theta) = {1\over (2\pi)^2} \int\d\ell\;\ell\,
P_\delta\rund{{\ell\over f_K(w)},w}\,W_{\rm ap}(\theta\ell) \;.
\ee
Using the same notation (following Hoekstra et al.\ \cite{hoek604}), we can
write the aperture 
mass dispersion as
\be
\ave{M_{\rm ap}^2(\theta)} = {9\pi\over 2}\rund{H_0\over c}^4\Omega_{\rm
  m}^2\, H_\kappa(\theta)\;,
\elabel{HOEMap}
\ee
with
\be
H_\kappa(\theta) = \int \d w\;{g^2(w)\over a^2(w)} \,{\cal P}(w,\theta)\;,
\ee
where $g(w)$ (see eq.\ \ref{eq:g-fact}) describes the source-redshift weighted
efficiency factor of a lens at distance $w$. One therefore obtains an
expression for the bias factor,
\be
b^2={9\over 4}\rund{H_0\over c}^4 {H_\kappa(\theta) \over H_{\rm
    gg}(\theta)}\, \Omega_{\rm m}^2\,{ \ave{{\cal N}^2(\theta)} \over
\ave{M_{\rm ap}^2(\theta)}} = f_{\rm b}(\theta)\Omega_{\rm m}^2
\,{ \ave{{\cal N}^2(\theta)} \over
\ave{M_{\rm ap}^2(\theta)}} \;.
\elabel{b-value}
\ee
Note that $f_{\rm b}(\theta)$ depends, besides the aperture radius $\theta$,
on the cosmological parameters $\Omega_{\rm m}$ and $\Omega_\Lambda$, but for
a given cosmological model, it depends only weakly on the filter scale
$\theta$ and on the adopted power spectrum $P_\delta$ (van Waerbeke 1998;
Hoekstra et al.\ \cite{hoek604}). This is due to the fact that both,
$\ave{{\cal N}^2(\theta)}$ and  $\ave{M_{\rm ap}^2(\theta)}$ are linear in the
power spectrum, through the functions $H$, and in both cases they probe only a
very narrow range of $k$-values, owing to the narrow width of the filter
function $W_{\rm ap}$. Hence, the ratio 
$\ave{{\cal N}^2(\theta)} / \ave{M_{\rm ap}^2(\theta)}$ is expected to be very
close to a constant if the bias factor $b$ is scale independent.

Next we consider the correlation coefficient $r$ between the dark matter
distribution and the galaxy field. Correlating $M_{\rm ap}(\theta)$ with
${\cal N}(\theta)$ yields
\bea
\ave{M_{\rm ap}(\theta){\cal
    N}(\theta)}&=&\int\d^2\vt\;U(|\vc\vt|)\int\d^2\vt'\;U(|\vc\vt'|) 
\ave{\kappa(\vc\vt)\,\kappa_{\rm g}(\vc\vt')} 
\nonumber \\
&=&3\pi\rund{H_0\over c}^2\Omega_{\rm m}\,b\,r\,H_{\kappa\rm g}(\theta)
\;,
\elabel{MapENN}
\eea
with
\be
H_{\kappa\rm g}(\theta)=\int\d w\;{p_{\rm f}(w)\,g(w)\over a(w)\,f_K(w)}
{\cal P}(w,\theta)\;.
\ee
It should be noted that $\ave{M_{\rm ap}(\theta){\cal
    N}(\theta)}$ is a first-order statistics in the cosmic shear. It
correlates the shear signal with the location of galaxies, which are assumed
to trace the total matter distribution. As shown in Schneider (1998), the
signal-to-noise of this correlator is higher than that of $\ave{M_{\rm
    ap}^2}$, and therefore was introduced as a convenient statistics for the
detection of cosmic shear. In fact, in their original analysis of the RCS,
based on $16\,{\rm deg}^2$,
Hoekstra et al.\ (2001) obtained a significant signal for  $\ave{M_{\rm
    ap}(\theta){\cal N}(\theta)}$, but not for $\ave{M_{\rm
    ap}^2(\theta)}$.
Combining (\ref{eq:HOEENN}) and (\ref{eq:HOEMap}) with (\ref{eq:MapENN}), the
correlation coefficient $r$ can be expressed as 
\be
r={ \sqrt{H_\kappa(\theta)\,H_{\rm gg}(\theta)} \over
H_{\kappa\rm g}(\theta)}\,{\ave{M_{\rm ap}(\theta){\cal
    N}(\theta)} \over \sqrt{\ave{M_{\rm ap}^2(\theta)}\,
\ave{{\cal N}^2(\theta)}}}
=f_{\rm r}(\theta)\,{\ave{M_{\rm ap}(\theta){\cal
    N}(\theta)} \over \sqrt{\ave{M_{\rm ap}^2(\theta)}\,
\ave{{\cal N}^2(\theta)}}} \;.
\elabel{r-value}
\ee
As was the case for $f_{\rm b}$, the function $f_{\rm r}$ depends only very
weakly on the filter scale and on the adopted form of the power spectrum, so
that a variation of the (observable) final ratio with angular scale would
indicate the scale dependence of the correlation coefficient.

Whereas the two aperture measures $M_{\rm ap}$ and ${\cal N}$ can in principle
be obtained from the data field by putting down circular apertures, and the
corresponding second-order statistics can likewise be determined through
unbiased estimators defined on these apertures, this is not the method of
choice in practice, due to gaps and holes in the data field. Note that in our
discussion of cosmic shear in Sect.\ \ref{sc:WL-6.3}, we have expressed
$\ave{M_{\rm ap}^2(\theta)}$ in terms of the shear two-point correlation
functions $\xi_\pm(\theta)$ -- see (\ref{eq:Mapfromxi}) -- just for this
reason. In close analogy, ${\cal N}^2(\theta)$ can be expressed in terms of
the angular correlation function $\omega(\theta)$ of the projected galaxy
positions, as seen by (\ref{eq:ENNsq}), or more explicitly, when replacing the
power spectrum $P_\omega(\ell)$ in (\ref{eq:HOEENN}) by its Fourier transform,
which is the angular correlation function, one finds 
\be
\ave{{\cal
    N}^2(\theta)}=\int_0^{2\theta}{\d\vt\;\vt\over\theta^2}\,
    \omega(\vt)\,T_+\rund{\vt\over\theta} \;, 
\ee
where the function $T_+$ is the same as that occurring in
(\ref{eq:Mapfromxi}). 
Correspondingly, we introduce the power spectrum $P_{\rm g\kappa}(\ell)$,
which is defined as
\be
\ave{\hat\kappa(\vc\ell)\hat\kappa^*_{\rm g}(\vc\ell')}
=(2\pi)^2\delta_{\rm D}(\vc\ell-\vc\ell')\,P_{\kappa{\rm g}}(|\vc\ell|)\;.
\ee
Applying (\ref{eq:limber}), as well as the definitions of the bias and
correlation functions, this projected cross-power spectrum is related to the
3-D density contrast by
\be
P_{\kappa{\rm g}}(\ell)={3\over 2}\rund{H_0\over c}^2\Omega_{\rm m}b r
\int\d w{g(w)p_{\rm f}(w)\over a(w) f_K(w)}\,P_\delta\rund{{\ell\over
    f_K(w)},w} \;.
\ee
The angular correlation function $\ave{\kappa(\vc\vt)\kappa(\vc\vt')}$
occurring in (\ref{eq:MapENN})
can then be replaced by its Fourier transform $P_{\kappa{\rm g}}$.
On the other hand, since the Fourier transform of the surface mass density
$\kappa$ is simply related to that of the shear, one can consider the 
correlation between the
galaxy positions with the tangential shear component,
\bea
\ave{\gamma_{\rm t}(\theta)}&:=&
\ave{\kappa_{\rm g}(\vc 0)\gamma_{\rm t}(\vc\theta)} \nonumber \\
&=&-\int{\d^2\ell\over(2\pi)^2}\int{\d^2\ell'\over(2\pi)^2}
{\rm e}^{2{\rm i}(\beta'-\vp)} \exp\rund{-{\rm i}\vc\theta\cdot\vc\ell'}
\ave{\hat\kappa_{\rm g}(\vc\ell)\hat\kappa(\vc\ell')} \nonumber \\
&=&
{1\over 2\pi}\int\d\ell\;\ell\,{\rm J}_2(\theta \ell)\,P_{\kappa\rm
  g}(\ell)  \\ & \Rightarrow &
P_{\kappa\rm g}(\ell)=2\pi\int\d\theta\,\theta\,\ave{\gamma_{\rm t}(\theta)}\,
  J_2(\theta\ell) \;. \nonumber
\eea
Note that $\ave{\gamma_{\rm t}(\theta)}$ is just the galaxy-galaxy lensing
signal discussed in Sect.\ \ref{sc:WL-8.2}; this shows very clearly that
galaxy-galaxy lensing measures the correlation of mass and light in the
Universe. In terms of this mean tangential shear, the aperture mass and galaxy
number counts can be written as
\be
\ave{M_{\rm ap}(\theta){\cal N}(\theta)}=\int_0^{2\theta} {\d\vt\;\vt\over
  \theta^2} \,\ave{\gamma_{\rm t}(\vt)}\,T_2\rund{\vt\over\theta}\;,
\elabel{MapNfinal}
\ee
where the function $T_2$ is defined in a way similar to $T_\pm$ and given 
explicitly as 
\be
T_2(x)=576\int_0^\infty {\d t\over t^3}\,{\rm J}_2(xt)\,\eck{{\rm
    J}_4(t)}^2\;;
\ee
this function vanishes for $x>2$, so that the integral in
(\ref{eq:MapNfinal}) extends over a finite interval only. Hence, all
three aperture correlators can be calculated from two-point
correlation functions which can be determined from the data directly,
independent of possible gaps in the field geometry.

\subsubsection{Results from the RCS.}
Hoekstra et al.\ (\cite{hoek604}) have applied the foregoing equations to a
combination of their RCS survey and the VIRMOS-DESCART survey. The former was
used to determine $\ave{{\cal N}^2}$ and $\ave{M_{\rm ap}{\cal N}}$, the
latter for deriving $\ave{M_{\rm ap}^2}$. As pointed out by these authors,
this combination of surveys is very useful, in that the power spectrum at a
redshift around $z\sim 0.35$ can be probed; indeed, they demonstrate that the
effective redshift distribution over which the power spectrum, and thus $b$
and $r$ are probed, are well matched for all three statistics for their choice
of surveys. `Foreground' galaxies for the measurement of $\omega(\theta)$ and
$\ave{\gamma_{\rm t}(\theta)}$ are chosen to have $19.5\le R_C\le 21$,
`background' galaxies are those with $21.5\le R_C\le 24$.
\begin{figure}
\bmi{6}
\includegraphics[width=6cm]{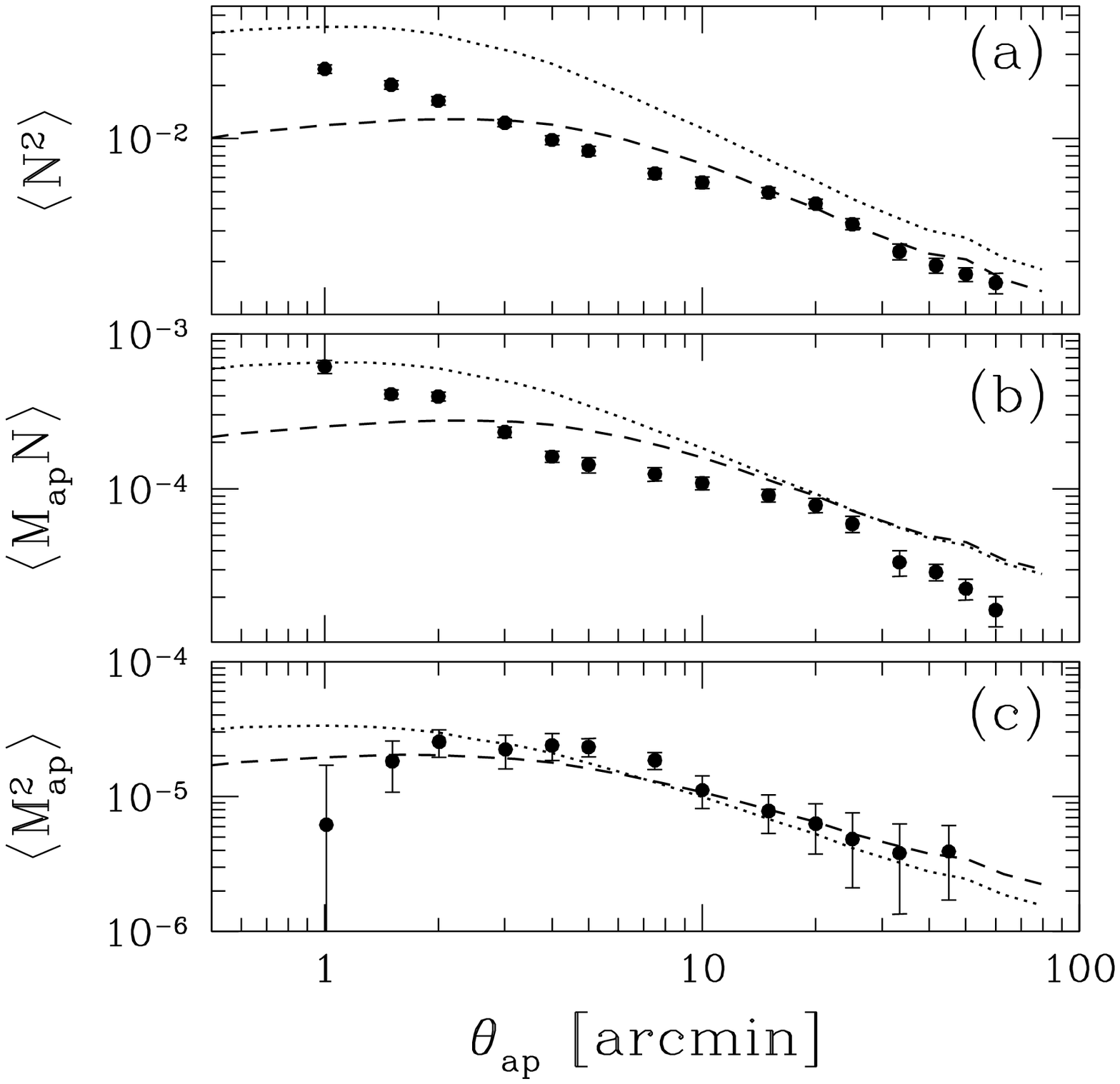}
\emi
\bmi{5.7}
\includegraphics[width=5.7cm]{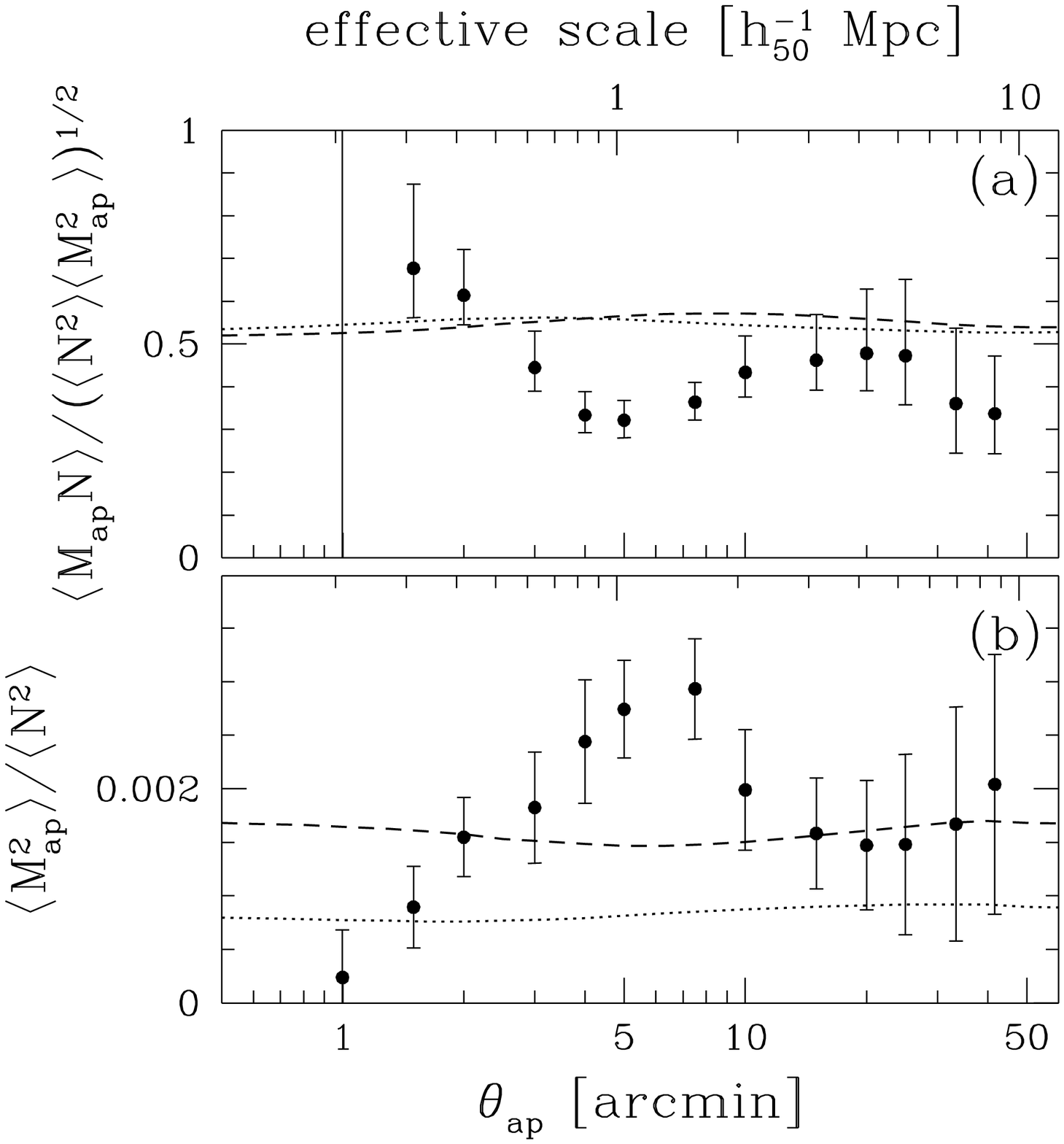}
\emi

\caption{The left figure displays the three aperture statistics as measured by
  combining the RCS and the VIRMOS-DESCART survey. Points show measured
  values, as determined from the correlation functions. The right
  panels display the ratios of the aperture statistics as they appear in
  (\ref{eq:b-value}) and (\ref{eq:r-value}). The dotted and
  dashed curves in all panels show the predictions for an OCDM and a
  $\Lambda$CDM model, 
  respectively, both with $\Omega_{\rm m}=0.3$, $\sigma_8=0.9$, and
  $\Gamma_{\rm spect}=0.21$, for the fiducial values of $b=1=r$. The fact that
  the curves in the right panels are nearly constant show the
  near-independence of $f_{\rm b}$ and $f_{\rm r}$ on the filter scale. The
  upper axis in the right panels show the effective physical scale on which
  the values of $b$ and $r$ are measured (from
  Hoekstra et al.\ \cite{hoek604})}
\flabel{Hoek-1}
\end{figure}
In Fig.\ \ref{fig:Hoek-1} the three aperture statistics are shown as a
function of angular scale, as determined from their combined survey, whereas
in the right panels, the ratios of these statistics as they appear in
(\ref{eq:b-value}) and (\ref{eq:r-value}) are displayed. Also shown are
predictions of these quantities from two cosmological models, assuming
$b=1$ and $r=1$. The fact that these model predictions are fairly constant in
the right-hand panels shows that the factors $f_{\rm b}$ and $f_{\rm r}$ are
nearly independent of the radius $\theta$ of the aperture, as mentioned
before. 

\begin{figure}
\bmi{7}
\includegraphics[width=7cm]{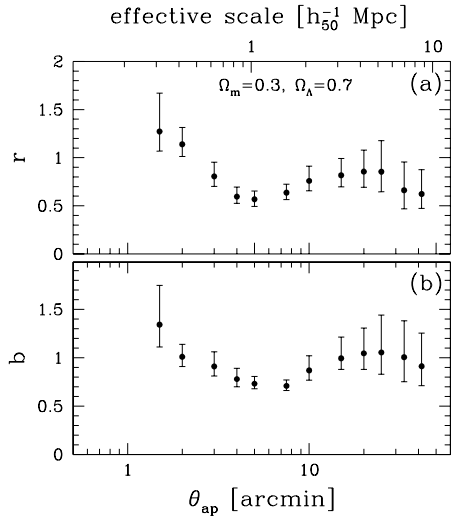}
\emi
\bmi{4.7}
\caption{The values of the bias and correlation coefficient, as determined
  from (\ref{eq:b-value}) and (\ref{eq:r-value}) and the results shown in
  Fig.\ \ref{fig:Hoek-1}; here, a $\Lambda$CDM model has been assumed for the
  cosmology dependence of the functions $f_{\rm b}$ and $f_{\rm r}$. The upper
  axis indicates the effective scale on which $b$ and $r$ are measured (from
  Hoekstra et al.\ \cite{hoek604})
}
\flabel{Hoek-2}
\emi
\end{figure}
The results for the bias and correlation factor are shown in Fig.\
\ref{fig:Hoek-2}, as a function of angular scale and effective physical
scale, corresponding to a median redshift of $z\sim 0.35$. The results
indicate that the bias factor and the galaxy-mass correlation coefficient are
compatible with a constant value on large scales,
$\gtrsim 5 h^{-1}\,{\rm Mpc}$, but on smaller scales both seem to change
with scale. The transition between these two regimes occurs at about the scale
where the density field at redshift $z\sim 0.35$ turns from linear to
non-linear evolution. In fact, in the non-linear regime one does not expect a
constant value of both coefficients, whereas in the linear regime, constant
values for them appear natural. It is evident from the figure that the error
bars are still too large to draw definite conclusions about the behavior of
$b$ and $r$ as a function of scale, but the approach to investigate the
relation between galaxies and mass is extremely promising and will certainly
yield very useful insight when applied to the next generation of cosmic shear
surveys. In particular, with larger surveys than currently available,
different cuts in the definition of foreground and background galaxies can be
used, and thus the redshift dependence of $b$ and $r$ can be
investigated. This is of course optimized if (photometric) redshift estimates
for the galaxy sample become available.

\subsubsection{Results from the SDSS.}
The large sample of galaxies with spectroscopic redshifts already available now
from the SDSS permits an accurate study of the biasing properties of these
galaxies (see the end of Sect.\ts\ref{sc:WL-8.2}). 
Two different approaches should be mentioned here: the first follows
along the line discussed above and has been published in Sheldon et al.\
(2004). In short, the galaxy-galaxy signal can be translated into the
galaxy-mass cross-correlation function $\xi_{\rm gm}$, due to the knowledge of
galaxy redshifts. The ratio of $\xi_{\rm gm}$ and the galaxy two-point
correlation function $\xi_{\rm gg}$ then depends on the ratio $r/b$.
In Fig.\ts\ref{fig:Sheldon-GGL} we show the galaxy-mass correlation as a
function of linear scale, as well as the ratio $b/r$. Note that from the SDSS
no cosmic shear measurement has been obtained yet, owing to the complex PSF
properties, and therefore $b$ and $r$ cannot be measured separately from this
data set.

\begin{figure}
\bmi{6.5}
\includegraphics[width=6.3cm]{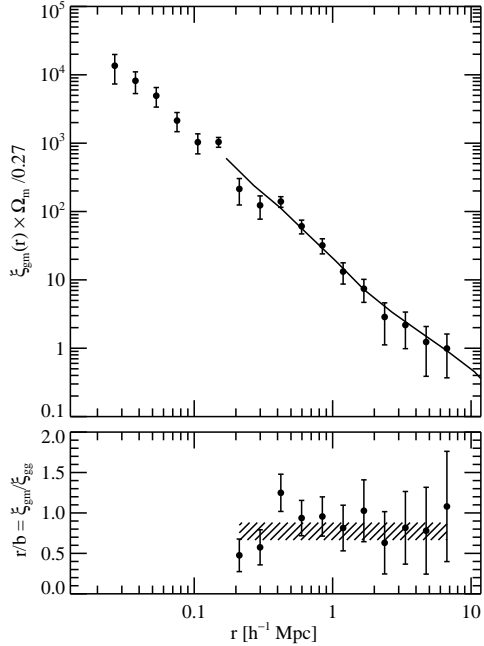}
\emi
\bmi{5.2}
\caption{The galaxy-mass cross-correlation function $\xi_{\rm gm}(r)$, as a
  function of linear scale (dots with error bars), scaled to a matter density
  parameter of $\Omega_{\rm m}=0.27$, 
as well as the two-point
  galaxy correlation function obtained from the same set of (foreground)
  galaxies (solid curve). The ratio between these two is given in the lower
  panel, which plots $b/r$ as a function of scale. Over the full range of
  scales, $\xi_{\rm gm}$ can be well approximated by a power law, $\xi_{\rm
  gm}=(r/r_0)^{-\gamma}$, with slope $\gamma=1.79\pm 0.06$ and correlation
  length $r_0=(5.4\pm 0.7)(\Omega_{\rm m}/0.27)^{-1/\gamma}h^{-1}\,{\rm
  Mpc}$. The ratio $r/b\approx (1.3\pm 0.2)(\Omega_{\rm m}/0.27)$ is
  consistent with being scale-independent}
\flabel{Sheldon-GGL}
\emi
\end{figure}

The galaxy-mass correlation function follows a power law over more
than two orders-of-magnitude in physical scale, and its slope is very
similar to the slope of the galaxy two-point correlation
function. Hence, the ratio between these two is nearly
scale-independent. When splitting the sample into blue and red, and
early- and late-type galaxies, the correlation length is larger for
the red and the early-type ones. Furthermore, as expected, the lensing
signal increases with the velocity dispersion in early-type galaxies.

An alternative approach was taken by Seljak et al.\ (2004). Their
starting point is the fact that the biasing properties of dark matter
halos is very well determined from cosmological simulations. This is
of course not true for the biasing of galaxies. The bias parameter of
galaxies with luminosity $L$ is given as
\be
b(L)=\int \d M\; p(M|L)\,b_{\rm h}(M)\;,
\ee
where $b_{\rm h}$ is the bias of halos of mass $M$ relative to the
large-scale matter distribution, and $p(M|L)$ is the probability that
a galaxy with luminosity $L$ resides in a halo of mass $M$. This
latter probability distribution is then parameterized for any
luminosity bin, by assuming that a fraction $1-\alpha$ of all galaxies
in the luminosity bin considered are at the center of their parent
halos, whereas the remaining fraction $\alpha$ are satellite
galaxies. For the central galaxies, a unique mass $M(L)$ is assigned,
whereas for the non-central ones, a mass distribution is assumed. The
values of $\alpha$ and $M$ for six luminosity bins are shown in the
various panels of Fig.\ts\ref{fig:Selj-GGL-SDSS}; they are obtained by
fitting the galaxy-galaxy lensing signal with the model just
described.  The main reason why the mass spectrum can be probed is
that the numerous low-mass galaxy halos contribute to the lensing
signal only at relatively small scales, whereas at larger scales the
higher-mass halos dominate the signal; hence, different halo masses appear
at different separations in the galaxy-galaxy lensing signal. In this
way, $b(L)$ can be determined, which depends on the non-linear mass
scale $M_*$ (see Sect.\ts 6.2 of IN). The bias parameter is a
relatively slowly varying function of galaxy luminosity for $L\lesssim
L_*$, approaching a value $\sim 0.7$ for very low-luminosity galaxies,
but quickly rises for $L>L_*$.

Seljak et al.\ combined these measurements of the bias parameter with
the clustering properties of the SDSS galaxies and the WMAP results on
the CMB anisotropy, and derived new constraints on
$\sigma_8=0.88\pm0.06$ and the bias parameter of an $L_*$-galaxy,
$b_*=0.99\pm 0.07$; furthermore, the combination of these datasets is
used to obtain new constraints on the standard cosmological
parameters. This work has opened up a new way on how to employ the
results from galaxy-galaxy lensing as a cosmological tool.

\subsection{\llabel{WL-8.4}Galaxy biasing: magnification method}
High-redshift QSOs are 
observed to be correlated on the sky
with lower-redshift galaxies and clusters. This topic has indeed an
interesting history: The detection of very close associations of high-$z$ QSOs
with low-$z$ galaxies (see Arp 1987, and references therein) has been claimed
as evidence against the cosmological interpretation of the QSO
redshifts, as the probabilities of observing such close pairs of objects which
are physically unrelated were claimed to be vanishingly small. However, these
probabilities were obtained a posteriori, and of course, any specific
configuration has a vanishingly small probability. Since the cosmological
interpretation of QSO redshifts is supported by overwhelming evidence, the
vast majority of researchers consider these associations as a statistical
fluke. 

A physical possibility to generate the association of background sources with
foreground objects is provided by the magnification bias caused by lensing:
the number counts of background sources is changed in regions where a
foreground lens yields magnifications different from unity -- see Sect.\ 5 of
IN. Thus, close to a galaxy where $\mu>1$, the number counts of bright
background QSOs can be enhanced since the slope of their counts is steeper
than unity. There have been various attempts in the literature to `explain'
the observed QSO-galaxy associations by invoking the magnification bias,
either with a smooth galaxy mass distribution or by including the effects of
microlensing; see SEF for a detailed discussion of this
effect. The bottom line, however, is that the magnification effect is by far
not large enough to account for the small (a posteriori) probabilities of the
observed individual close associations.

The topic has be revived, though in a different direction, by the finding that
high-redshift AGNs are statistically associated with low-redshift galaxies.
Fugmann (1990) provided evidence that radio-selected high-$z$ AGNs from the
1-Jansky-catalog are correlated with relatively bright (and therefore low-$z$)
galaxies taken from the Lick catalog, an analysis that later on was repeated
by Bartelmann \& Schneider (1993), using a slightly different statistics.
Different samples of foreground and background populations have been employed
in further studies, including the correlation between 1-Jansky AGN with bright
IRAS galaxies (Bartelmann \& Schneider 1994; Bartsch et al.\ 1997), high-$z$
QSOs with clusters from the Zwicky catalog of clusters (Rodrigues-Williams \&
Hogan 1994; Seitz \& Schneider 1995b), 1-Jansky AGNs with red galaxies from
the APM catalog (Ben\'\i tez \& Mart\'\i nez-Gonz\'alez 1995; see also Norman
\& Impey 2001), to mention just a few. Radio-selected AGN are considered to be
a more reliable probe since their radio flux is unaffected by extinction, an
effect which could cause a bias (if the sky shows patchy extinction, both
galaxies and QSOs would have correlated inhomogeneous distributions on the
sky) or anti-bias (if extinction is related to the lensing matter) for
flux-limited optical surveys of AGNs, and which therefore needs to be taken
into account in the correlation analysis of optically-selected AGNs. 
However, most radio source catalogs are not fully optically identified and
lack redshifts, and using incomplete radio surveys therefore can induce a
selection bias (Ben\'{\i}tez et al.\ 2001). These latter authors investigated
the correlation between two completely identified radio catalogs with the
COSMOS galaxy catalog, and found a very significant correlation signal.

The upshot of all these analyses is that there seems to be a positive
correlation between the high-$z$ sources and the low-$z$ objects, on
angular scales between $\sim 1'$ and about $1^\circ$. The
significances of these correlations are often not very large, they
typically are at the 2--3$\sigma$ level, essentially limited by the
finite number of high-redshift radio sources with a large flux (the
latter being needed for two reasons: first, only radio surveys with a
high flux threshold, such as the 1-Jansky catalog, have been
completely optically identified and redshifts determined, which is
necessary to exclude low-redshift sources which could be physically
associated with the `foreground' galaxy population, and second,
because the counts are steep only for high fluxes, needed to obtain a
high magnification bias.)  If this effect is real, it cannot be
explained by lensing caused by individual galaxies; the angular region
on which galaxies produce an appreciable magnification is just a few
arcseconds.  However, if galaxies trace the underlying (dark) matter
distribution, the latter can yield magnifications (in the same way as
it yields a shear) on larger scales. Thus, an obvious qualitative
interpretation of the observed correlation is therefore that it is due
to magnification of the large-scale matter distribution in the
Universe of which the galaxies are tracers. This view is supported by
the finding (M\'enard \& P\'eroux 2003) that there is a significant
correlation of bright QSOs with metal absorption systems in the sense
that there are relatively more bright QSOs with an aborber than
without; this effect shows the expected trend from magnification bias
caused by matter distributions associated with the absorbing material.

We therefore consider a flux-limited sample of AGNs, with distance
probability distribution $p_{\rm Q}(w)$, and a sample of galaxies with
distance distribution $p_{\rm f}(w)$. It will be assumed that the AGN sample
has been selected such that it includes only objects with redshift larger than
some threshold $z_{\rm min}$, corresponding to a minimum comoving distance
$w_{\rm min}$, which is larger than the distances of all galaxies in the
sample. We define the AGN-galaxy correlation function as 
\be
w_{\rm Qg}(\theta)={\ave{\eck{N_{\rm g}(\vc\phi)-\bar N_{\rm g}}\,
\eck{N_{\rm Q}(\vc\phi+\vc\theta)-\bar N_{\rm Q}}} \over
\bar N_{\rm g}\,\bar N_{\rm Q}} \;;
\ee
where $N_{\rm g}(\vc\phi)$ and $N_{\rm Q}(\vc\phi)$ are the observed
number densities of galaxies and AGNs, respectively. The former is
given by (\ref{eq:galdensi}). The observed number density of AGN is
affected by the magnification bias. Provided the unlensed counts can
be described (locally) as a power-law in flux, $N_{\rm Q,0}(>S)\propto
S^{-\beta}$, then from (108) of IN we find that $N_{\rm
Q}(\vc\phi)=N_{\rm Q,0}\,\mu^{\beta-1}(\vc\phi)$, where $\mu(\vc\phi)$
is the magnification in the direction $\vc\phi$. Then, if the
magnifications that are relevant are small, we can approximate
\be
\mu(\vc\phi)\approx 1+2\kappa(\vc\phi)=1+\delta\mu(\vc\phi)\;,
\ee
and the projected surface mass density $\kappa$ is given by (\ref{eq:5.11})
with $p_w$ in (\ref{eq:g-fact}) replaced by $p_{\rm Q}$. Assuming that the
magnifications do not affect the mean source counts
$\bar N_{\rm Q}$, the cross-correlation becomes
\be
w_{\rm Qg}(\theta)=2(\beta-1)\bar b(\theta)\,\bar r(\theta) \,
w_{\kappa\rm g}(\theta)\;,
\ee
where $\bar b$ and $\bar r$ are the effective bias factor of the
galaxies and the mean galaxy-mass correlation function just as in
Sect.\ts\ref{sc:WL-8.3}, and $w_{\kappa\rm g}$ is the correlation
between the projected density field $\kappa$ and the projected number
density of galaxies $\kappa_{\rm g}$, defined after
(\ref{eq:galdensi}), which is the Fourier transform of $P_{\kappa{\rm
g}}(\ell)$ defined in (\ref{eq:Pdeltag}).  Hence, a measurement of
this correlation, together with a measurement of the correlation
function of galaxies, can constrain the values of $b$ and $r$ (Dolag
\& Bartelmann 1997; M\'enard \& Bartelmann 2002).

The observed correlation between galaxies and background AGN appears
to be significantly larger than can be accounted for by the models
presented above. On scales of a few arcmin, Ben\'{\i}tez et al.\
(\cite{BSM-G}) argued that the observed signal exceeds the theoretical
expectations by a factor of a few. This discrepancy can be attributed
to either observational effects, or shortcomings of the theoretical
modelling. Obviously, selection effects can easily produce spurious
correlations, such as patchy dust obscuration or a physical
association of AGNs with the galaxies. Furthermore, the weak lensing
approximation employed above can break down on small angular
scales. Jain et al.\ (2003, see also Takada \& Hamana 2003) argued
that the simple biasing model most likely breaks down for the small
scales where the discrepancy is seen, and employed the halo model for
describing the large-scale distribution of matter and galaxies to
predict the expected correlations. For example, the strength of the
signal depends sensitively on the redshifts, magnitudes and galaxy
type.

At present, the shear method to determine the bias factor and the galaxy-mass
correlation has yielded more significant results than the magnification
method, owing to the small complete and homogeneous samples of high-redshift
AGNs. As pointed out by M\'enard \& Bartelmann (2002), the SDSS may well
change this situation shortly, as this survey will obtain $\sim 10^5$
homogeneously selected spectroscopically verified AGNs. Provided the effects
of extinction can be controlled sufficiently well, this data should provide a
precision measurement of the QSO-galaxy correlation function.

\section{\llabel{WL-9}Additional issues in cosmic shear}
\subsection{\llabel{WL-9.1}Higher-order statistics}
On the level of second-order statistics, `only' the power spectrum is
probed. If the density field was Gaussian, then the power spectrum
would fully characterize it; however, in the course of non-linear
structure evolution, non-Gaussian features of the density field are
generated, which show up correspondingly in the cosmic shear field and
which can be probed by higher-order shear statistics.  The usefulness
of these higher-order measures for cosmic shear has been pointed out
in Bernardeau et al.\ (1997), Jain \& Seljak (1997), Schneider et al.\
(1998a) and van Waerbeke et al.\ (1999); in particular, the
near-degeneracy between $\sigma_8$ and $\Omega_{\rm m}$ as found from
using second-order statistics
can be broken.
However, there are serious problems with higher-order shear
statistics, that shall be illustrated below in terms of the third-order
statistics. 

But first, we can give a simple argument why third-order statistics is able to
break the degeneracy between $\Omega_{\rm m}$ and $\sigma_8$. Consider a
density field on a scale where the inhomogeneities are just weakly
non-linear. One can then employ second-order perturbation theory for the
growth of the density contrast $\delta$. Hence, we write
$\delta=\delta^{(1)}+\delta^{(2)}+\dots$, where $\delta^{(1)}$ is the density
contrast obtained from linear perturbation theory, and $\delta^{(2)}$ is the
next-order term. This second-order term is quadratic in the linear density
field, $\delta^{(2)}\propto \rund{\delta^{(1)}}^2$. The linear density field is
proportional to $\sigma_8$, and the projected density $\kappa\propto
\Omega_{\rm m}\sigma_8$. Hence, in the linear regime, $\ave{\kappa^2}\propto
\Omega_{\rm m}^2\sigma_8^2$, where $\ave{\kappa^2}$ shall denote here any
second-order shear estimator. The lowest order contribution to the third-order
statistics is of the form
\[
\ave{\kappa^3}\propto \rund{\delta^{(1)}}^2\,\delta^{(2)}
\propto \Omega_{\rm m}^3\,\sigma_8^4\;,
\]
since the term $\rund{\delta^{(1)}}^3$ yields no contribution owing to the
assumed Gaussianity of the linear density field. Hence, a skewness statistics
of the form 
\[
\ave{\kappa^3}/\ave{\kappa^2}^2\propto \Omega_{\rm m}^{-1}
\]
will be independent of the normalization $\sigma_8$, at least in this
simplified perturbation approach. In more accurate estimates, this is not
exactly true; nevertheless, the functional dependencies of the second- and
third-order shear statistics on $\sigma_8$ and $\Omega_{\rm m}$ are different,
so that these parameters can be determined separately.

\subsubsection{The shear three-point correlation function.}
Most of the early studies on three-point statistics concentrated on the
third-order moment of the surface mass density $\kappa$ in a circular
aperture, $\ave{\kappa(\theta)}$; however, this is not a directly measureable
quantity, and therefore useful only for theoretical considerations. As for
second-order statistics, one should consider the correlation functions, which
are the quantities that can be obtained best directly from the data and which
are independent of holes and gaps in the data field.  The three-point
correlation function (3PCF) of the shear has three independent variables (e.g.
the sides of a triangle) and 8 components; as was shown in Schneider \&
Lombardi (2003), none of these eight components vanishes owing to parity
invariance (as was suspected before -- this confusion arises because little
intuition is available on the properties of the 3PCF of a polar).  This then
implies that the covariance matrix has 6 arguments and 64 components!  Of
course, this is too hard to handle efficiently, therefore one must ask which
combinations of the components of the 3PCF are most useful for studying the
dark matter distribution.  Unfortunately, this is essentially unknown yet. An
additional problem is that the predictions from theory are less well
established than for the second-order statistics.

A further complication stems from a certain degree of arbitrariness on how to
define the 8 components of the 3PCF. For the 2PCF, the vector between any pair
of points defines a natural direction with respect to which tangential and
cross components of the shear are defined; this is no longer true for three
points. On the other hand, the three points of a triangle define a set of
centers, such as the `center of mass', or the center of the in- or
circum-circle. After choosing one of these centers, one can define the two
components of the shear which are then independent of the coordinate frame.

Nevertheless, progress has been achieved. From ray-tracing simulations
through a cosmic matter distribution, the 3PCF
of the shear can be determined (Takada \& Jain \cite{TaJa1}; see also
Zaldarriaga \& Scoccimarro 2003; furthermore, the three-point cosmic shear
statistics can also be determined in the frame of the halo model, see Cooray
\& Hu 2001; Takada \& Jain \cite{TaJa2}), whereas
Schneider \& Lombardi (2003) 
have defined the `natural components' of the shear 3PCF which are
most easily related to the bispectrum of the underlying matter
distribution. Let $\gamma^{\rm c}(\vc\theta_i)= \gamma_{\rm t}+{\rm
i}\gamma_\times = -\gamma\,{\rm e}^{-2{\rm i}\zeta_i}$ be the complex shear
measured in the frame which is rotated by the angle $\zeta_i$ relative to the
Cartesian frame, so that the real and imaginary parts of $\gamma^{\rm c}$ are
the tangential and cross components of the shear relative to the chosen center
of the triangle (which has to be defined for each triplet of points
separately). Then the natural components are defined as 
\bea
\Gamma^{(0)}&=&\ave{\gamma^{\rm c}(\vc\theta_1)\,
\gamma^{\rm c}(\vc\theta_2)\,\gamma^{\rm c}(\vc\theta_3)} \;, \nonumber \\
\Gamma^{(1)}&=&\ave{\gamma^{\rm c *}(\vc\theta_1)\,
\gamma^{\rm c}(\vc\theta_2)\,\gamma^{\rm c}(\vc\theta_3)} \;,
\eea
and correspondingly for $\Gamma^{(2)}$ and $\Gamma^{(3)}$. Each of the natural
components of the 3PCF
constitutes a complex number, which depends just on the three
separations between the points. Special care is required for labelling
the points, 
and one should follow the rule that they are labeled in a counter-clock
direction around the triangle. If such a unique prescription is not
systematically applied, confusing and wrong conclusions will be obtained about
the behaviour of the shear 3PCF with respect to  parity transformations (as
the author has experienced painfully enough). In Schneider et al.\ (2004),
explicit relations are derived for the natural components of the shear 3PCF in
terms of the bispectrum (that is, the generalization of the power spectrum for
the three-point statistics) of the underlying mass distribution $\kappa$.

\subsubsection{Third-order aperture statistics.}
Alternatively, aperture measures can be defined to measure the third-order
statistics. Schneider et al.\ (1998a) calculated $\ave{M_{\rm ap}^3}(\theta)$
in the frame of the quasi-linear structure evolution model and showed it to be
a strong function of $\Omega_{\rm m}$. Van Waerbeke et al.\ (2001) calculated
the third-order aperture mass, using a fitting formula of the non-linear
evolution of the dark matter bispectrum obtained by Scoccimarro \& Couchman
(2001) and pointed out the strong sensitivity with respect to cosmological
parameters. 
Indeed, as mentioned before,
$\ave{M_{\rm ap}^3}$ is sensitive only to the E-modes of the shear field. One
might be tempted to use $\ave{M_\perp^3}(\theta)$ as a measure for
third-order B-mode statistics, but indeed, this quantity vanishes owing to
parity invariance (Schneider 2003). However, $\ave{M_\perp^2\,M_{\rm
ap}}$ is a measure for 
the B-modes at the third-order statistical level.  Jarvis et al.\ (2004) have
calculated $\ave{M_{\rm ap}^3(\theta)}$ in terms of the shear 3PCF, for the
weight function (\ref{eq:U+Q.CNPT}) in the definition of $M_{\rm ap}$.
Schneider et al.\ (2004) have shown that this relation is most easily
expressed in terms of the natural components of the shear 3PCF. On the other
hand, Jarvis et al.\ (2004) have expressed $\ave{M_{\rm ap}^3(\theta)}$ in
terms of the bispectrum of $\kappa$, and as was the case for the aperture
dispersion in relation to the power spectrum of $\kappa$, the third-order
aperture mass is a very localized measure of the bispectrum and is sensitive
essentially only to modes with three wavevectors with equal magnitudes. For
that reason, Schneider et al.\ (2004) have generalized the definition of the
third-order aperture measures, correlating the aperture mass of three
different sizes, $\ave{M_{\rm ap}(\theta_1)M_{\rm ap}(\theta_2)M_{\rm
    ap}(\theta_3)}$. This third-order statistics is again a very localized
measure of the bispectrum, but this time with wave vectors of different
magnitude 
$\ell_i\approx \pi/\theta_i$, and therefore, by considering the third-order
aperture mass for all combinations of $\theta_i$, one can probe the full
bispectrum. Therefore, the third-order aperture mass correlator with three
independent arguments (i.e., angular scales) should contain essentially the
full third-order statistical information of the $\kappa$-field, since in
contrast to the two-point 
statistics, the shear 3PCF does not contain information about long-wavelength
modes.

Furthermore, the third-order aperture statistics can be expressed directly in
terms of the shear 3PCF through a simple integration, very similar to the
relations (\ref{eq:N29}) for the two-point statistics. Finally, the other
three third-order aperture statistics (e.g., $\ave{M_\perp(\theta_1)M_{\rm
    ap}(\theta_2)M_{\rm ap}(\theta_3)}$) can as well be obtained from the
natural components of the shear 3PCF. These correlators are expected to vanish
if the shear is solely due to lensing, but intrinsic alignments of galaxies can
lead to finite correlators which include B-modes. However, as shown in
Schneider (2003), $\ave{M_{\rm ap}(\theta_1)M_{\rm
    ap}(\theta_2)M_\perp(\theta_3)}$, as well as 
$\ave{M_\perp(\theta_1)M_\perp(\theta_2)M_\perp(\theta_3)}$, are expected
to vanish even in the presence of B-modes, since these two correlators are not
invariant with respect to a parity transformation. Therefore, non-zero results
of these two correlators signify the violation of parity invariance and
therefore provide a clean check on the systematics of the data and their
analysis. 

\subsubsection{First detections.}
Bernardeau et al.\ (2002) measured for the first time a significant third-order
shear from the VIRMOS-DESCART survey, employing a suitably filtered integral
over the measured 3PCF (as defined in Bernardeau et al.\ 
2003). Pen et al.\ (\cite{Pen3PCF}) used the aperture statistics to detect a
skewness in the same data set.  The accuracy of these measurements is not
sufficient to derive strong constraints on cosmological parameters, owing to
the limited sky area available.  However, with the upcoming large cosmic shear
surveys, the 3PCF will be measured with high accuracy. Determining
the 3PCF from observed galaxy ellipticities cannot be done by
straightforwardly considering any triple of galaxies -- there are just too
many. Jarvis et al.\ (2004) and Zhang \& Pen (\cite{ZhangPen}) have developed
algorithms for calculating the 3PCF in an efficient way.

Based on the halo model for the description of the LSS, Takada \& Jain
(\cite{TaJa2}) studied the dependence of the shear 3PCF on cosmological
parameters. For relatively large triangles, the 3PCF provides a means to break
the degeneracies of cosmological parameters that are left when using the
second-order statistics only, as argued above.  For small triangles, the 3PCF
is dominated by the one-halo term, and therefore primarily probes the mass
profiles of halos.  Ho \& White (\cite{HoWhite}) show that the 3PCF on small
angular scales also contains information on the asphericity of dark matter
halos.  The full power of third-order statistics is achieved once redshift
information on the source galaxies become available, in which case the
combination of the 2PCF and 3PCF provides a sensitive probe on the
equation-of-state of the dark energy (Takada \& Jain \cite{TaJa3}).

\subsubsection{Beyond third order.}
One might be tempted to look into the properties of the fourth-order shear
statistics (though I'm sure the reader can control herself in
doing this -- but see Takada \& Jain 2002). OK, the four-point correlation
function has 16 components and depends on 5 variables, not to mention the
corresponding covariance or the redshift dependent fourth-order correlator.
One can consider correlating the aperture mass of four different angular
sizes, but in contrast to the third-order statistics, this is expected not to
contain the full information on the trispectrum (which describes
the fourth-order statistical properties of $\kappa$). Perhaps a combination of
this fourth-order aperture mass with the average of the fourth power of the
mean shear in circular apertures will carry most of the information. And how
much information on cosmological parameters does the fourth-order shear
statistics contain? And even higher orders?

Already the third-order shear statistic is not acccurately predictable from
analytic descriptions of the non-linear evolution of the matter
inhomogeneities, and the situation worsens with even higher order.\footnote{In
  the limits of small and large angular scales, analytic approximations can be
  obtained. For small scales, the highly non-linear regime is often described
  by the hierarchical ansatz and hyperextended perturbation theory (see Munshi
  \& Jain 2001 and references therein), whereas on very large scales
  second-order perturbation theory can be used. Nevertheless, the range of
  validity of these perturbation approximations and their accuracy have to be
  checked with numerical simulations.} One therefore needs to refer to
detailed ray-tracing simulations. Although they are quite time consuming, I do
not see a real bottleneck in this aspect: Once a solid and accurate
measurement of the three-point correlation function becomes available,
certainly considerable effort will be taken to compare this with numerical
simulations (in particular, since such a measurement is probably a few years
ahead, in which the computer power will increase by significant factors). If
we accept this point, then higher-order statistics can be obtained from these
simulations, and several can be `tried out' on the numerical data such that
they best distinguish between different models. For example, one can consider
the full probability distribution $p(M_{\rm ap};\theta)$ on a given data set
(Kruse \& Schneider 2000; Reblinsky et al\ 1999; Bernardeau \& Valageas 2000;
Munshi et al.\ 2004). To 
obtain this from the observational data, one needs to place apertures on the
data field which, as we have argued, is plagued with holes and gaps in the
data. However, we can place the same gaps on the simulated data fields and
therefore simulate this effect. Similarly, the numerical simulations should be
used to find good strategies for combining second- and third-order shear
statistics (and potentially higher-order ones) for an optimal distinction
between comological model parameters, and, in particular, the
equation-of-state of Dark Energy.  
Another issue one needs to consider for third- (and higher-)order cosmic shear
measures is that intrinsic clustering of sources, and the correlation between
galaxies and the dark matter distribution generating the shear shear field has
an influence on the expected signal strength (Bernardeau 1998; Hamana 2001;
Hamana et al.\ 2002).
Obviously, there are still a lot of
important studies to be done.

\subsubsection{Third-order galaxy-mass correlations.}
We have shown in Sect.\ts\ref{sc:WL-8} how galaxy-galaxy lensing can
be used to probe the correlation between galaxies and the underlying
matter distribution. With the detection of third-order shear
statistics already in currently available data sets, one might expect
that also higher-order galaxy-mass correlations can be measured from
the same data. Such correlations would then probe, on large angular
scales, the higher-order biasing parameters of galaxies, and thereby
put additional constraints on the formation and evolution of
galaxies. M\'enard et al.\ (2003) considered the correlation between
high-redshift QSOs and pairs of foreground galaxies, thus generalizing
the methods of Sect.\ts\ref{sc:WL-8.4} to third-order statistics. The
galaxy-galaxy-shear correlation, and the galaxy-shear-shear
correlations have been considered by Schneider \& Watts
(\cite{ScWa04}). These correlation functions have been related to the
underlying bispectrum of the dark matter and the third-order bias and
correlation functions, and appropriate aperture statistics have been
defined, that are related in a simple way to the bispectra and the
correlation functions.

In fact, integrals of these higher-order correlations have probably
been measured already. As shown in Fig.\ts\ref{fig:McKay1}, galaxies
in regions of high galaxy number densities show a stronger, and more
extended galaxy-galaxy lensing signal than more isolated
galaxies. Hence there is a correlation between the mean mass profile
around galaxies and the local number density of galaxies, which is
just an integrated galaxy-galaxy-shear correlation. In fact, such a
correlation is only first order in the shear and should therefore be
much easier to measure than the shear 3PCF. Furthermore, the
galaxy-shear-shear correlation seems to be present in the cosmic shear
analysis of the COMBO-17 fields by Brown et al.\ (2003), where they
find a stronger-than-average cosmic shear signal in the A901 field,
and a weaker cosmic shear signal in the CDFS, which is a field
selected because it is rather poor in brighter galaxies.

\subsection{\llabel{WL-9.3}Influence of LSS lensing on lensing by
clusters and galaxies}
The lensing effect of the three-dimensional matter distribution will
contaminate the lensing measurements of localized objects, such as galaxies
and clusters. Some of the associated effects are mentioned in this section.

\subsubsection{Influence of cosmic shear on strong lensing by galaxies.}
The lensing effect of foreground and background matter in a strong
lensing system will affect the image positions and flux ratios. As
this 3-D lensing effects are not recognized as such in the lens modelling, a
`wrong' lens model will be fitted to the data, in the sense that the
mass model for the lensing galaxy will try to include these additional
lensing effects not associated with the galaxy itself. In particular,
the corresponding predictions for the time delays can be affected
through this effect.

Since the image separation of strong lens systems are less than a few
arcseconds, the lensing effect of the LSS can be well approximated by
a linear mapping across this angular scale. In this case, the effect
of the 3-D matter distribution on the lens model can be studied
analytically (e.g., Bar-Kana 1996). The lens equation resulting from
the main lens (the galaxy) plus the linearized inhomogeneities of the
LSS is strictly equivalent to the single-plane gravitational lens
equation without these cosmological perturbations, and the mass
distribution of the equivalent single-plane lens can be explicitly
derived (Schneider 1997). For example, if the main lens is described
by elliptical isopotential curves (i.e., elliptical contours of
the deflection potential $\psi$) plus external shear, the equivalent
single-plane lens will be of the same form. The orientation of the
ellipticity of the lens, as seen by the observer, will be rotated by
the forground LSS by the same angle as the potential of the equivalent
lens, so that no observable misalignment is induced. This equivalence
then implies that the determination of the Hubble constant from
time-delay measurements is affected by the same mass-sheet degeneracy
transformation as for a single plane lens.

\subsubsection{LSS effects on the mass determination of clusters.}
The determination of mass parameters of a cluster from weak lensing is
affected by the inhomogeneous foreground and background matter
distribution. The effect of local mass associated with a cluster (e.g.,
filaments extending from the cluster along the line-of-sight) will bias the
mass determination of clusters high, since clusters are likely to be located in
overdense regions of the LSS, though this effect is considerably smaller than
claimed by Metzler et al.\ (2001), as shown by Clowe et al.\ (\cite{CdeLK}). 

Hoekstra (2001, \cite{HoekNFW}) considered the effect of the LSS on the
determination of mass parameters of clusters, using either SIS or NFW
models. For the SIS model, the one parameter characterizing this mass profile
($\sigma_v$) can be obtained as a linear estimator of the shear. The
dispersion of this parameter is then the sum of the dispersion caused by
the intrinsic ellipticity of the source galaxies and the cosmic shear
dispersion. For the NFW model, the relation between its two parameters
($M_{200}$, the mass inside the virial radius $r_{200}$, and the concentration
$c$) and the shear is not linear, but the effect of the LSS can still be
estimated from Monte-Carlo simulations in which the cosmic shear is assumed to
follow Gaussian statistics with a power spectrum following the Peacock \&
Dodds (1996) prescription. 

For the SIS model, the effect of the LSS on the determination of $\sigma_v$ is
small, provided the cluster is at intermediate redshift (so that most source
galaxies are in the background). The noise caused by the finite ellipticity in
this case is almost always larger than the effect by the LSS. There is an
interesting effect, however, in that the relative contribution of the LSS and
shape noise changes as larger aperture fits to the SIS model are considered:
The larger the field over which the shear is fitted to an SIS model, the
larger becomes the impact of cosmic shear, and this increase compensates for
the reduced shape noise. In effect, cosmic shear and shape noise together put
an upper limit on the accuracy of the determination of $\sigma_v$ from shear
data. The same is true for the determination of the mass parameters of the NFW
model, as shown in Fig.\ts\ref{fig:noise-NFW}. The uncertainties of the mass
parameters of NFW profiles are about twice as large as if the effects from the
LSS are ignored, whereas the effect is considerably smaller for the
one-parameter model of the SIS.  One should also note that a decrease of the
shape noise, which can be obtained by using data with a fainter limiting
magnitude, yields an increase of the noise from the LSS, since the fainter
galaxies are expected to be at higher redshift and therefore carry a larger
cosmic shear signal.  For low-redshift clusters, these two effects nearly
compensate.

\begin{figure}
\bc
\includegraphics[bb = 82 314 570 658, clip, width=9cm]{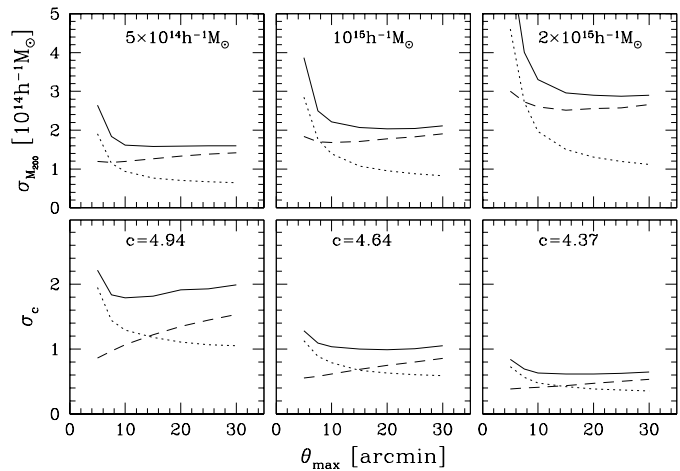}
\ec
\caption{The dispersion of the determination of the mass and concentration of
  three NFW halos at redshift $z_{\rm d}=0.3$. These parameters were derived
  by fitting an NFW shear profile to the shear simulated from an NFW halo with
  parameters indicated in the figure and adding shape noise and noise from
  cosmic shear. The outer angular scale over which the fit was performed is
  $\theta_{\rm max}$. Dotted curves show the effect from shape noise alone,
  dashed curves show the dispersion from cosmic shear, and the solid curves
  contain both effects. Surprisingly, the accuracy of the NFW parameters does
  not increase once $\theta_{\rm max}\sim 15'$ is reached, as for larger
  radii, the cosmic shear noise more than compensates for the reduced
  ellipticity 
  noise. Another way to express that is that the lensing signal at very large
  distance from the halo center is weaker than the rms cosmic shear and
  therefore does not increase the signal-to-noise any more (from Hoekstra\
  \cite{HoekNFW}) }
\flabel{noise-NFW}
\end{figure}

\subsubsection{The efficiency and completeness of weak lensing cluster
  searches.} 
We take up the  brief discussion at the end of Sect.\ts\ref{sc:WL-5.8} about
the potential of deriving a shear-selected sample of galaxy clusters. The first
studies of this question were based on analytical models (e.g., Kruse \&
Schneider 1999) or numerical models of isolated clusters (Reblinsky \&
Bartelmann 1999). Those studies can of course not account for the effects of
lensing by the LSS. Ray-tracing simulations through N-body generated LSS were
carried out by Reblinsky et al.\ (1999), White et al.\ (\cite{WvWMack}),
Hamana et al.\ (\cite{HaTaYo}), Vale \& White (2003), Hennawi \&
Spergel (2004)
and others. In these cosmological simulations, halos were identified based on
their 3-D mass distribution. They were then compared to the properties of the
lensing results obtained from ray tracing, either by considering the (smoothed)
surface mass density $\kappa$ (that could be obtained from a mass
reconstruction from the shear field) or by studying the aperture mass $M_{\rm
  ap}$ which can be obtained directly from the shear. In both cases, noise due
to the finite intrinsic source ellipticity can be added.

The two basic quantities that have been investigated in these studies are {\it
  completeness} and {\it efficiency}. Completeness is the fraction of dark
matter halos above some mass threshold $M_{\rm min}$
that are detected in the weak lensing data, whereas efficiency is
the fraction of significant lensing detections that correspond to a real halo.
Both of these quantities depend on a number of parameters, like the mass
threshold of a halo and the limiting significance $\nu$ of a lensing detection
[in the case of the aperture mass, this would correspond to
  (\ref{eq:MapSN})], as well as on the choice of the filter function
  $Q$. Hennawi \& Spergel (2003) have pointed out that even without
  noise (from observations or intrinsic galaxy ellipticities), the
  efficiency is limited to about 85\% -- even under these idealized
  condition, the selected sample will be contaminated by at least 15\%
  of spurious detections, generated by projection effects of the LSS.

To compare these predictions with observations, the six highest-redshift EMSS
clusters were all detected at high significance with a weak lensing analysis
(Clowe et al.\ 2000).  Clowe et al.\ (\cite{CloweEdics}) have studied 20
high-redshift clusters with weak lensing techniques. These clusters were
optically selected and are expected to be somewhat less massive (and
potentially more affected by foreground galaxies) than the EMSS clusters. Only
eight of these 20 clusters are detected with more than $3\sigma$ significance,
but for none of them does the SIS fit produce a negative $\sigma_v^2$. Only
for four of these clusters are the lensing results compatible with no shear
signal.

\section{\llabel{WL-10}Concluding remarks}
Weak lensing has become a standard tool in observational cosmology, as we have
learned how to measure the shape of faint galaxy images and to correct them
for distortions in the telescope and camera optics and for PSF effects. These
technical issues are at the very center of any observational weak lensing
research. It appears that at present, the accuracy with which shear can be
measured is sufficient for the data available today, in the sense that
statistical uncertainties are likely to be larger than potential inaccuracies
in the measurement of unbiased shear estimates from faint images. This,
however, will change quickly. The upcoming large cosmic shear surveys will
greatly reduce statistical uncertainties, and then the accuracy of shear
measurements from the data will be the essential limiting factor. Alternatives
to KSB have been developed, but they need to undergo thorough testing before
becoming a standard tool for observers. It should also be noted that the KSB
method is applied differently by different groups, in particular with regards
to the weighting of galaxies and other details. What is urgently needed is a
study in which different groups apply their version of KSB to the same data set
and compare the results. Furthermore, starting from raw data, the specific
data reduction methods will lead to slightly different coadded images, and
shear measurements on such differently reduced imaged should be
compared. These technical issues will be a central challenge for weak lensing
in the upcoming years. 

The ongoing and planned wide-field imaging surveys mentioned at the end of
Sect.\ts\ref{sc:WL-7.7} will allow us to investigate several central questions
of cosmology. The two aspects that I consider most relevant are the
investigation of the equation-of-state of the Dark Energy and the relation
between galaxies and the underlying dark matter distribution. The former
question about the nature of Dark Energy is arguably the central challenge of
modern cosmology, and cosmic shear is one of the very few methods how it can
be studied empirically. The relation between dark matter and galaxies is
central to our understanding of how galaxies form and evolve, and
galaxy-galaxy lensing is the only way how this relation can be investigated
without a priori assumptions. 

\begin{figure}
\bc
\includegraphics[bb=12 28 500 500, clip, width=11cm]{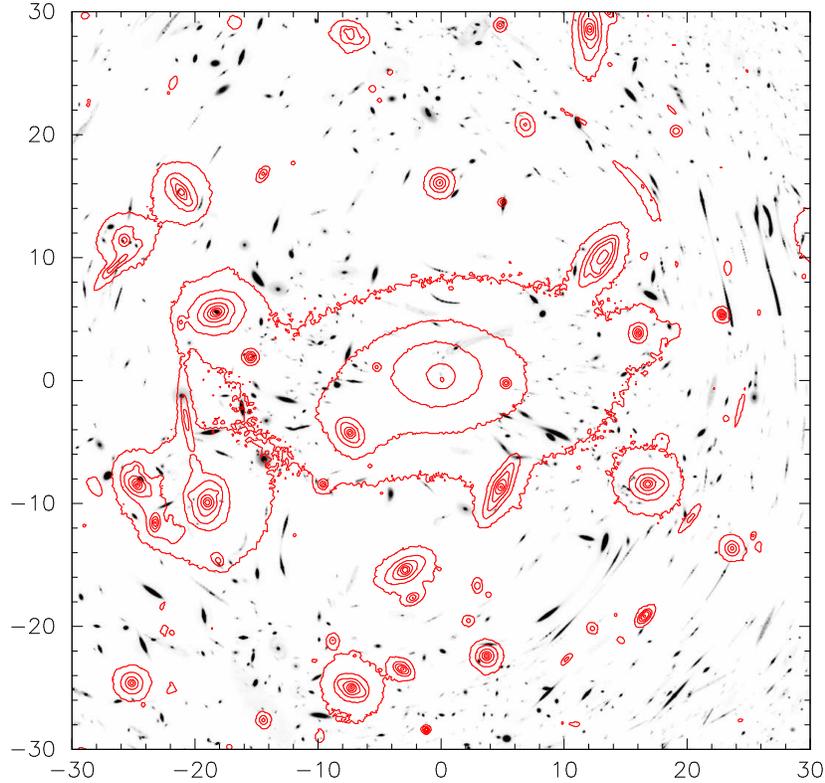}
\ec
\caption{Simulated image of lensed features in the very central part
of the massive cluster A2218, as observed with the future JWST. For
these simulations, the mass profile of the cluster as constrained from
HST observations and detailed modelling (Kneib et al. 1996) has been
used. The number density of (unlensed) sources was assumed to be
$4\times 10^6 {\rm deg}^{-2}$ down to K=29. The redshift distribution
assumed is broad and extends to redshift $z\sim10$ with a median value
$z_{\rm med}\sim 3$.  The brighter objects (cluster galaxies and
brightest arcs) seen by HST are displayed as contours, to make the
faint galaxy images visible on this limited dynamic range
reproduction. An enormous number of large arcs and arclets are seen;
in particular, numerous radial arcs can be easily detected, which will
allow us to determine the `core size' of the cluster mass
distribution. Due to the broad redshift distribution of the faint
galaxies, arcs occur at quite a range of angular separations from the
cluster center; this effect will become even stronger for
higher-redshift clusters. It should be noted that this 1 arcminute
field does not cover the second mass clump seen with HST; an JWST
image will cover a much larger area, and more strong lensing features
will be found which can then be combined with the weak lensing
analysis of such a cluster.  For this simulation, a pixel size of
$0\arcsecf06$ was used; the JWST sampling will be better by a factor of
2 (from Schneider \& Kneib 1998)} \flabel{A2218-NGST}
\end{figure}

Essentially all weak lensing studies today have used faint galaxies as
sources, since they form the densest source population currently
observable.  The uniqueness of faint optical galaxies will not stay
forever, with the currently planned future instruments. For example,
there is a rich literature of weak lensing of the cosmic microwave
background which provides a source of very accurately known
redshift. Weak lensing by the large-scale structure enhances the power
spectrum of the CMB at small angular scales, and the Planck satellite
will be able to measure this effect. In particular, polarization
information will be very useful, since lensing can introduce B-modes
in the CMB polarization. The James Webb Space Telescope, with its
large aperture of 6.5\ meters and its low temperature and background
will increase the number density of observable 
faint sources in the near-IR up to
$5\mu{\rm m}$ to several hundred per square arcminute, 
many of them at redshifts beyond 3,
and will therefore permit much more detailed weak lensing studies, in
particular of clusters (see Fig.\ts\ref{fig:A2218-NGST}; an observation of
this huge number of arcs and multiple images will answer questions about the
mass distribution of clusters that we have yet not even dared to ask). The
envisioned next generation radio telescope Square Kilometer Array will
populate the radio sky with very comparable source density as
currently the deepest optical images. Since the beam (that is, the
point-spread function) of this radio interferometer will be known very
accurately, PSF corrections for this instrument will be more reliable
than for optical telescopes.
Furthermore, higher-order correlation of the shear field with sources in the
field will tell us about non-Gaussian properties of galaxy-matter
correlations and biasing, and therefore provide important input into models of
galaxy formation and evolution.

\subsubsection{Acknowledgements.}
I enjoyed the week of lecturing at Les Diablerets a lot. First, I thank Georges
Meylan and his colleagues Philippe Jetzer and Pierre North for organizing this
school so efficiently and smoothly; to them also my sincere apologies for not
finishing these proceedings much earlier -- but sure enough, lecturing is more
fun than preparing lectures, and certainly very much more fun than writing
them up into something to be 
published.  Second, I thank my fellow lecturers Chris Kochanek and Joe
Wambsganss for discussions, great company and good spirits. Third, and
foremost, my compliments to the students who patiently sat through these
lectures which, I'm sure, were not always easy to follow; nevertheless I hope
that they grasped the essential points, and that these lecture notes help to
fill in the details. These notes would not have been possible without the many
colleagues and students with whom I had the privilege to collaborate over the
years on various aspects of weak lensing, among them Matthias Bartelmann,
Marusa Bradac, Doug Clowe, J\"org Dietrich, Thomas Erben, Bernhard Geiger,
Hannelore H\"ammerle, Bhuvnesh Jain, Martin Kilbinger, Lindsay King, Martina
Kleinheinrich, Guido Kruse, Marco Lombardi, Yannick Mellier, Hans-Walter Rix,
Mischa Schirmer, Tim Schrabback, Carolin Seitz, Stella Seitz, Patrick Simon,
Ludovic van Waerbeke and Anja von der Linden.
This work was supported by the German Ministry for
Science and Education (BMBF) through the DLR under the project 50 OR
0106, by the German Ministry for Science and Education (BMBF) through
DESY under the project 05AE2PDA/8, and by the Deutsche
Forschungsgemeinschaft under the project SCHN 342/3--1.

\end{document}